\documentclass[12pt,twoside]{mitthesis}
\usepackage{lgrind}
\usepackage{cmap}
\usepackage[T1]{fontenc}
\usepackage{graphicx}  
\usepackage{dcolumn}   
\usepackage{bm}        
\usepackage{verbatim}   
\usepackage{bbold}
\usepackage{color}
\usepackage{amsmath}
\usepackage{xcolor}
\usepackage{amsthm}
\usepackage{ulem}
\usepackage{amssymb}
\usepackage{amsmath,amssymb}
\usepackage{graphicx}
\usepackage{braket}
\usepackage{subcaption}
\usepackage{float}
\usepackage{relsize}
\usepackage[colorlinks=false,urlcolor=blue,urlbordercolor={0 1 1}]{hyperref}

\def\all{all}
\ifx\files\all  \else 

\def\a{\alpha}

\def\d{\delta}
\def\D{\Delta}

\def\s{\sigma}
\def\L{\Lambda}
\def\e{\epsilon}
\def\be{\begin{equation}}
\def\ee{\end{equation}}
\def\bea{\begin{eqnarray}}
\def\eea{\end{eqnarray}}

\def\>{\rangle}
\def\<{\langle}
\newcommand{\dbar}{\mathchar '26 \mkern-11mu d}

\def\BSCCO{\text{Bi}_2\text{Sr}_2\text{Ca}\text{Cu}_2\text{O}_8}

\theoremstyle{definition}

\theoremstyle{remark}

\begin{document}

\title{Exploring Strongly Interacting Gapless States: Cuprates, Pair Density Waves, and Fluctuating Superconductivity}

\author{Zhehao Dai}
\prevdegrees{B.S., Tsinghua University (2014)}
\department{Department of Physics}

\degree{Doctor of Philosophy}

\degreemonth{May}
\degreeyear{2020}
\thesisdate{May 15, 2020}


\supervisor{Patrick A. Lee}{William and Emma Rogers Professor of Physics}

\chairman{Nergis Mavalvala}{Associate Department Head}

\maketitle



\cleardoublepage
\setcounter{savepage}{\thepage}
\begin{abstractpage}
%
%
%
We study the physical property of pair density wave (PDW) and fluctuating PDW, and use it to build an effective theory of the strongly interacting pseudogap phase in cuprate high temperature superconductors. In Chapter~\ref{chap:opticalconductivity}, we study how Fulde-Ferrell state, the simplest form of PDW, responds to incident light. The collective motion of the condensate plays a key role; gauge invariance guides us to the correct result. From Chapter~\ref{chap:PDWpseudogap} to Chapter~\ref{chapter: fluctuating PDW in cuprates}, we construct a pseudogap metallic state by considering quantum fluctuating PDW. We analyze a recent scanning tunneling microscope (STM) discovery of period-8 density waves in the vortex halo of the d-wave superconductor. We put it in the context of the broader pseudogap phenomenology, and compare the experimental results with various PDW-driven models and a charge density wave (CDW) driven model. We propose experiments to distinguish these different models.  We present the Bogoliubov bands of PDW. We discuss fluctuating PDW from the general perspective of fluctuating superconductivity. We discuss how Bogoliubov bands evolve when the superconducting order parameter is fluctuating. We compare theoretical predictions with existing experiments on angle-resolved photoemission spectroscopy (ARPES), infrared conductivity, diamagnetism, and lattice symmetry breaking.

The material presented here is based on Ref.~\cite{dai2017opticalconductivity,dai2018STM,dai2019pseudogap}. Ref.~\cite{dai2019loop} is not discussed in this thesis but was completed during my time at MIT.

\end{abstractpage}


\cleardoublepage

\section*{Acknowledgments}
My time at MIT has been intellectually stimulating and rewarding. It is a pleasure working with so many talented and passionate people. 

I am incredibly grateful to my advisor Professor Patrick A. Lee. His intuition and intimacy with solid state physics deeply influenced me. I was attracted by his wisdom in his class; I continued the journey with him and learned a great deal under his mentorship during the past six years. I will always remember my eye-opening experiences brought by him, especially the occasions when he shared his insights on recent experiments and when a quick derivation of him swept all of my doubts away. I am thankful for his incredible guidance and unconditional trust. He gives me the freedom to explore whatever I found the most exciting at the moment. I thank him for teaching me the power of intuitive thinking on the hardest problems in our field, for freeing me from the constraint of mathematical rigorous, for showing me the way he thinks about experiments, and for being an invaluable mentor.

I am privileged to benefit from the mentorship of many others at MIT. I am grateful to Adam Nahum for a remarkable collaboration experience, for the thought-provoking conversations we had, and for his friendship. He seeks for the extreme clarity and seemingly impossible connections in challenging theoretical problems. I learned quite a bit of statistical physics and quantum criticality from him. I thank him for showing me the broadness of theoretical physics and the power of deep thinking. I have the privilege to collaborate with Professor T. Senthil, who has native wit on effective field theory. A large portion of this thesis would have been impossible without his insights. I am often astonished by his precise intuition on emergent phenomena, and the effectiveness of his simple approach. I learned a lot from Prof. Xiaogang Wen through his classes and various conversations. I thank him for sharing his vision of physics and his adventurous thoughts, and for the willingness to discuss a variety of topics at length. Thanks also to Professor Liang Fu for many inspiring discussions.

Many students and postdocs at MIT have significantly shaped my intellectual trajectory and made my life at MIT very enjoyable. I thank my collaborator Yahui Zhang for his wisdom and determination. I thank Liujun Zou for sharing his creative understanding and for being a good friend. I thank Xueda Wen and Jieqiang Wu for our theoretical physics swimming group and for the stimulating conversations on the way to Z center. Thanks also to Huitao Shen, Wenjie Ji, Brian Skinner, Zhen Bi, Debanjan Chowdhury, Hamed Pakatchi, Michael Pretko, Sagar Vijay, Dan Mao, Dominic Else, Jonathan Ruhman, Inti Sodemann, Samuel Lederer, Juven Wang, Michael DeMarco, Noah Yuan, Yang Zhang, Zheng Zhu, Sherry Chu, Sungjoon Hong, Jacob Colbert, Olumakinde Ogunnaike, and Zhiyu Dong for many illuminating discussions and for their encouragement.

Many thanks to my undergraduate advisor, Professor Zhengyu Weng, who introduced me to the subject of cuprate high temperature superconductors. His dedication to the hardest problems greatly influenced me.

I am deeply grateful to my wife Zhujun Shi, for her constant love and the enormous emotional and intellectual support she provides throughout my time in Cambridge. This thesis is dedicated to her.

Lastly, I thank my parents and my grandparents for making me all that I am, and for their enduring love.


\pagestyle{plain}
\tableofcontents
\newpage
\listoffigures

\chapter{Introduction}
\label{chap:intro}

Electrons in a solid are inevitably interacting. The Coulomb interaction between electrons are by definition the same order of magnitude as the interaction between electrons and ions, which is usually also comparable with the kinetic energy of electrons. In many solids, the interaction is surprisingly innocuous. Although the wavefunction and the energy of electrons are modified by the interaction, Landau Fermi liquid theory tells us that below certain energy scale, all excitations of the interacting system are in one to one correspondence to that of a free electron system. This is the story of most metals. Even when the interaction cause a phase transition, Landau's symmetry breaking paradigm comes into rescue: it is often sufficient to take a mean field treatment that reduces the interactions between each individual pairs of electrons into an effective background potential. However, when the interaction dominates, interesting new states emerge.

A particularly elegant set of examples of strongly interacting states are fractional quantum Hall states~\cite{TsuiStormer,laughlin1983anomalous}. For electrons confined in a thin layer and subject to strong magnetic field, their motions are restricted to little cyclotron orbitals with identical kinetic energies, leaving only the Coulomb interaction at play. In response, the electrons develop highly entangled patterns to avoid each other, as a side effect, showing precisely quantized Hall response.

Materials with partially occupied inner shell orbitals provide another playground of strong interaction. These electrons are more localized to the ions compared to electrons in the outer shell. When we have exactly one electron per unit cell, these electrons get the chance to each occupy their own orbital, leaving only their spin to fluctuate locally, forming the so-called Mott insulator~\cite{mott1937discussion}. These insulators are impossible in weakly interacting systems, where Fermi liquid theory predicts a metal with a half-filled Brillouin zone (B.Z.).

The fractional quantum Hall states belong to the category of gapped states. In order to excite an electron, the energy we pay need to be above a threshold. At low temperatures, there is nothing but the vast emptiness besides the elegant order of the ground states. On the contrary, a gapless state has active modes at arbitrarily low temperatures. For  example, the spin waves in a Mott insulator above the antiferromagnetic background. Yet, the true excitement comes when we remove some of the electrons so that the rest can hop again, producing a complex pattern of entangled positions and spins.

The goal of this thesis is to explore a strongly interacting gapless state in cuprates. In cuprates, electrons in d atomic orbitals form a Mott insulator. However when we remove some of the electrons (hole doping), an interesting gapless state interpolating between the Mott insulator and the usual Fermi liquid emerges. This state seems to have gapless electronic excitations located only on segments in the B.Z. instead of a closed Fermi surface, as if a Fermi surface is half-destroyed by a gap, hence the name `pseudogap'~\cite{timusk1999pseudogap,tallon2001doping,damascelli2003angle}.

Traditional perturbation theory which people rely on to describe both the standard model of our universe and conventional metals fails to address emergent phenomena like this. Even for the well-understood effective interaction of electron spins in the Mott insulator, there is no diagrammatic perturbative description of the usual kind~\cite{anderson2000brainwashed}. To penetrate the barrier between the microscopic Hamiltonian and the emergent phenomena, we take a phenomenological approach. We make the assumption that the mysterious pseudogap is a consequence of a fluctuating pair density wave (PDW), an emergent tendency to form electron pairs at nonzero momenta, and attempt to explain as much of the pseudogap phenomenology as possible based on this assumption. We now present basic properties of cuprates and key ingredients of our proposal.

\section{Cuprate high-temperature superconductors}

Since the discovery of cuprate high temperature superconductors, their unconventional properties have been recognized as the central problem of strongly interacting electronic systems. The first cuprate superconductor is discovered in lanthanum barium copper oxide (LBCO)~\cite{bednorz1986possible}. After that, many other cuprates with similar properties are identified, including but not limited to $\text{La}_{2-x}\text{Sr}_{x}\text{Cu}\text{O}_{2}$ (LSCO), $\text{Bi}_{2}\text{Sr}_{2}\text{Ca}_{n-1}\text{Cu}_{n}\text{O}_{2n+4+x}$ (BSCCO, $n=1,2,3$), $\text{Hg}\text{Ba}_{2}\text{Ca}_{n-1}\text{Cu}_{n}\text{O}_{2n+2+x}$ (HgBCCO), and $\text{YBa}_{2}\text{Cu}_{3}\text{O}_{7-x}$  (YBCO). The common ingredient responsible for most of the unconventional properties is the copper oxide layer, where the copper atoms form a (approximate) square lattice, and an oxygen atom sits in the middle of any two neighboring copper atoms. 

People were first attracted by the high superconducting transition temperature ($\text{T}_\text{c}$), which can be above 100K, but soon realized that the `normal' phases out of which the superconductor emerges are far more mysterious than the superconducting phase itself~\cite{keimer2015quantum}. For the superconducting phase, the d-wave pairing order itself becomes clear over time, although the microscopic pairing mechanism is still under debate, but for the normal phases, the so-called `pseudogap' and `strange metal', despite firmly established experimental results, even a phenomenological description is hard to get, let alone microscopic mechanisms.

\begin{figure}[htb]
\begin{center}
\includegraphics[width=2.9in]{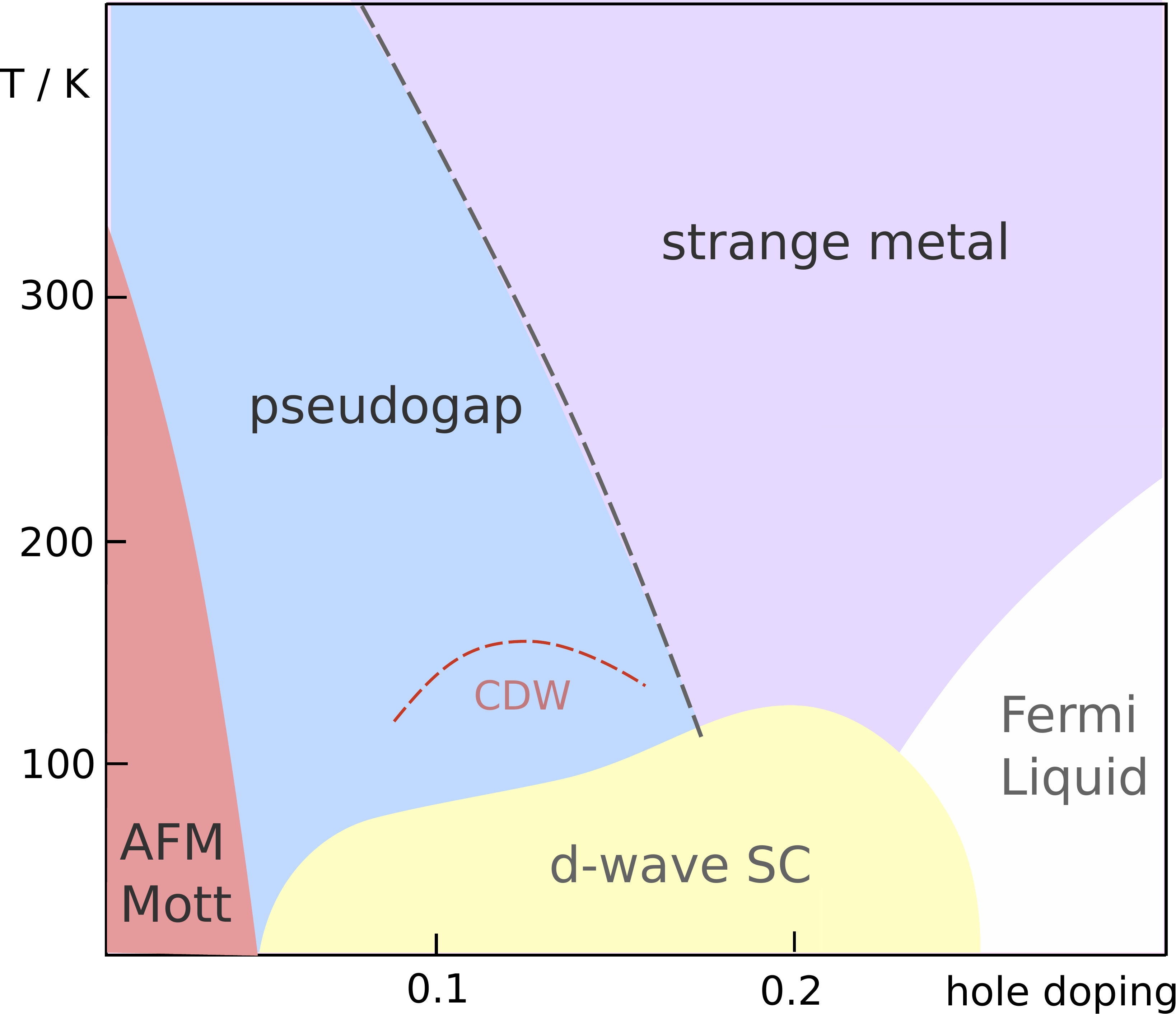}
\caption[Sketch of the phase diagram of YBCO]{Sketch of the phase diagram of YBCO. Phase diagram of other cuprates are qualitatively similar. The dashed red line represents the onset of the short-range CDW.}
\label{fig:cupratesketch}
\end{center}
\end{figure}

In Fig.~\ref{fig:cupratesketch} we sketch the phase diagram of YBCO. Without doping, the mother compound has one active electron per unit cell; they are localized to form an antiferromagnetic Mott insulator due to strong repulsion. For hole doping above 5\%, cuprates begin superconducting below a doping-dependent $\text{T}_\text{c}$; above $\text{T}_{c}$, several different normal phases appear in turn as the hole doping increases. The simplest normal phase is a Fermi liquid with a large Fermi surface, which appears at large doping. Intuitively, we have removed enough electrons to overcome the jamming near the integer filling. 

Between the Mott insulator and the Fermi liquid, the normal regions become weird. The fan centered around 19\% doping (Fig.~\ref{fig:cupratesketch}) is given the name `strange metal'. In this region, a large Fermi surface is detected, but the electrical conductivity grows linearly in temperature, suggesting a breakdown of the celebrated Fermi liquid theory, which predicts a quadratic temperature dependence. On the lower left of the strange metal region (below a temperature scale $\text{T}^*$) is the `pseudogap' region. It gets its name because part of the large Fermi surface seem to acquire a gap while the remaining part exists as disconnected gapless segments, the so-called `Fermi arcs'~\cite{damascelli2003angle}. In any fermionic theory we know, a Fermi surface is a closed curve separating occupied states from unoccupied states. The apparent `Fermi arcs' are totally unexpected.

T. Senthil once commented that cuprates are both blessed and cursed. Blessed because the high $\text{T}_\text{c}$ attracts attention, cursed because the high $\text{T}_\text{c}$ obscures the more interesting physics of the normal states. Indeed, one way to view the phase diagram is that the interesting pseudogap and strange metal, which could potentially become quantum ground state surrender to the mundane instability towards superconductivity.

With the hope to reveal the non-superconducting normal ground state behind cuprates, workers apply magnetic field bigger than the upper critical field ($\text{H}_\text{c2}$) to kill the superconductivity at low temperature~\cite{proust2019remarkable}. Given the unconventional pseudogap properties above $\text{T}_\text{c}$, it is surprise to discover that the transport properties seems to obey standard Fermi liquid behavior for $\text{H}>\text{H}_\text{c2}$ between 5\% and 19\% doping: the resistivity increases as $\text{T}^2$, the ratio between the thermal conductivity and the electrical conductivity obeys Wiedemann Franz law (indicating the transport is dominated by quasi-electrons), even quantum oscillations show up at higher field. The only unconventional feature is that the quantum oscillation, specific heat and Hall number indicates a small Fermi surface instead of a large Fermi surface (Sec.~\ref{sec:pseudogapintro}).

In cuprates we have this interesting hierarchy of energy scale. At the scale of the effective hopping $t$ and the anti-ferromagnetic coupling $J$ ($\sim$ 2000K), the $t-J$ model is generally believed to be a good description~\cite{PhysRev.115.2,anderson1987resonating,PhysRevLett.58.2794,varma1987charge,PhysRevB.37.3759}. Experimentally observed electron spectrum matches with the prediction of the $t-J$ model. Small-size numerical calculation and cold atom experiments can also help understand the theory. Down to the pseudogap scale ($\sim$ 300K), the electronic spectrum become unconventional; due to the strong interaction, it is so far practically impossible to extract predictions of the $t-J$ model in a controlled manner down to this energy scale. At low temperatures (roughly below $\text{T}_\text{c}\sim 100K$), we surprisingly recover conventional behaviors, like the superconductivity and the Fermi-liquid behavior at high fields. Apart from these phenomena, the linear T resistivity in the strange metal fan extends all the way from the lowest temperature to the scale of $J$ with approximately the same linear coefficient, which is perhaps the biggest mystery in cuprates.

Adding to the complexity, multiple experiments found evidences that the pseudogap crossover temperature $\text{T}^*$ is associated with sharp symmetry breaking transitions
~\cite{sato2017thermodynamic,HsieNaturePhysicshzhao2017global,shekhter2013bounding}.
Short range and long range charge density waves (CDW) are also seen universally inside the pseudogap region~\cite{blackburn2013x,ghiringhelli2012long,BlancoPhysRevB.90.054513,PhysRevB.96.134510,JulienNature477191wu2011magnetic,wu2013emergence,ZX1science350949gerber2015three,changNatureComm72016magnetic,ZX2PNAS11314647jang2016ideal}. However, none of these orders seem to solve the puzzle of the pseudogap.

Given the complexity, we take the approach of phenomenological theory to attack the pseudogap problem. That is to postulate the existence of certain state or certain dominant order, and attempt to explain as much of the pseudogap phenomenology as possible based on the postulate. For reasons that will be explained below, our postulate is that the origin of the pseudogap is a quantum disordered (fluctuating in space and time) pair density wave (PDW). With this assumption, we aim to provide a unified picture that connects the unconventional spectral properties above $\text{T}_\text{c}$, the gap and the `arc', with the Fermi-liquid like transport at low temperature and high field.

\section{Pair density wave}

Pair density wave (PDW) is defined as a superconducting order in electronic systems where electron pairs condense at non-zero momenta. Using $c_p$ as the electron annihilation operator at momentum $p$, a PDW order is defined by $\<c_{p}c_{-p+Q}\>\neq 0$, for some $Q\neq 0$. By definition it breaks both spatial translation symmetry and charge conservation.

\begin{figure}[htb]
\begin{center}
\includegraphics[width=4in]{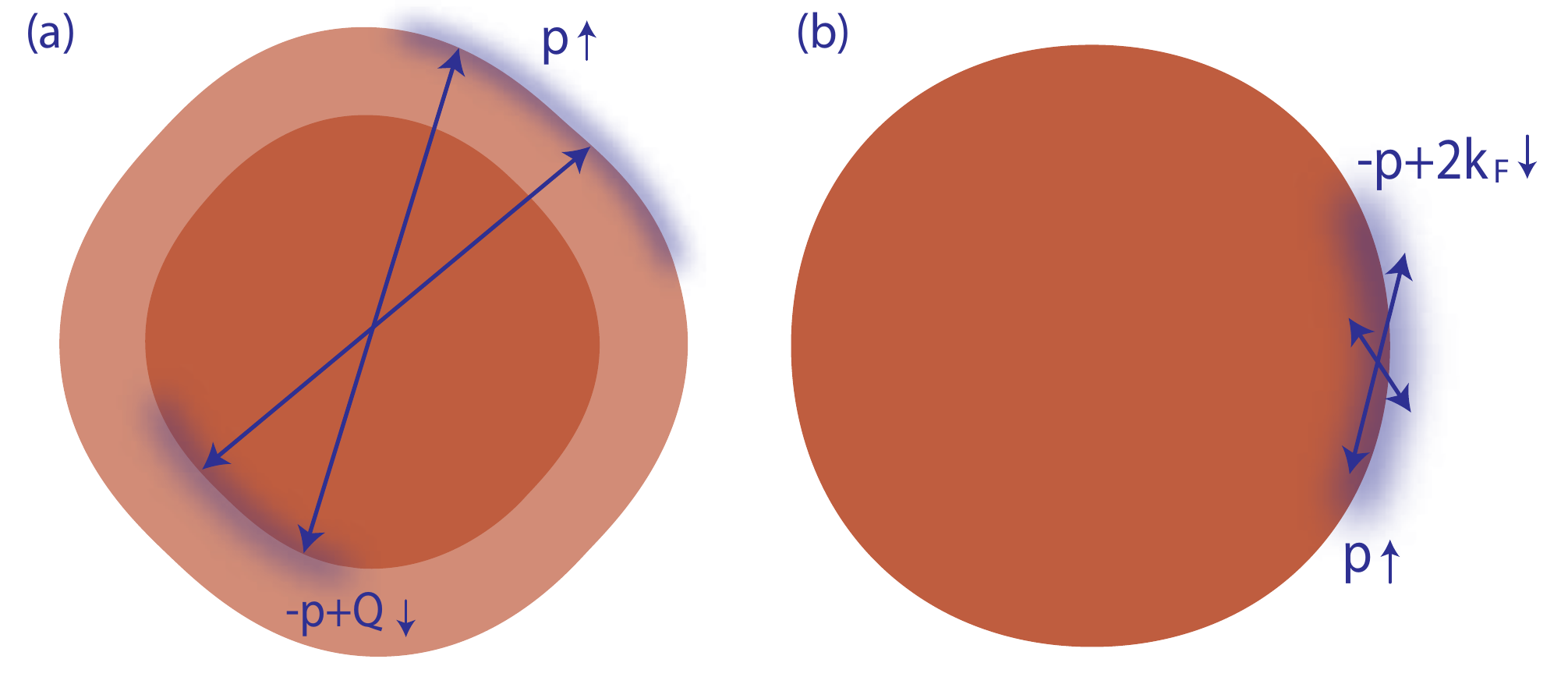}
\caption[Two examples of finite-momentum pairing]{Two examples of finite-momentum pairing. (a) FFLO pairing. The dark orange region is occupied by both spins, while the light orange region is occupied by up spin only. The blue shaded regions on the Fermi surface are gapped out by pairing. (b) Amperean pairing. A different pairing mechanism without spin-splitting, where the vicinity of a hot spot on the Fermi surface is gapped out, and the pairing momentum is close to $2k_{F}$.}
\label{fig:pairing}
\end{center}
\end{figure}

PDW was first proposed theoretically by Fulde and Ferrell~\cite{fulde1964superconductivity} and by Larkin and Ovchinnikov~\cite{larkin1965inhomogeneous} as a way to overcome the Pauli limiting effect of a magnetic field. They consider a setting where down spin and up spin electrons are slightly split in energy (Zeeman splitting); therefore they have slightly different Fermi surfaces (Fig.~\ref{fig:pairing}(a)). This can be realized by magnetic impurity or by perpendicular magnetic field in two-dimensional materials. For some combination of Zeeman splitting and temperature, they found a PDW order is preferred over both the normal state and the (translation-invariant) BCS superconductor. Intuitively, this is because finite-momentum pairing has the advantage of making pairs from only electrons close to the Fermi surfaces in the presence of Zeeman splitting (Fig.~\ref{fig:pairing}(a)). Fulde and Ferrell proposed a PDW with only one wave vector $Q$, as in Fig.~\ref{fig:pairing}(a). Larkin and Ovchinnikov proposed that electron pair condense at both $Q$ and $-Q$, hence generating a charge density wave (CDW) at momentum $2Q$.

Different from a conventional superconductor, a PDW state often has gapless Fermi surfaces of Bogoliubov quasiparticles. We can see this feature in FFLO states. In Fig.~\ref{fig:pairing}(a), only the blue shaded region acquires a superconducting gap because of the momentum-matching. The rest of the Fermi surface connects into the reflection of the Fermi surface by PDW momenta to form Bogoliubov Fermi surfaces of the PDW state. Because of these residue Bogoliubov Fermi surfaces, a PDW state has specific heat and spin susceptibility similar to a normal metal despite its superconducting nature.

In their original proposal, FFLO states occupy only a small region in the splitting-temperature phase diagram. Since they proposed this exotic order, it has proven to be very challenging to realize such an order experimentally~\cite{RevModPhys.76.263}. 40 years after the theoretical proposal, evidences of FFLO states have been found in $\text{CeCoIn}_{5}$ in 2003~\cite{bianchi2003possible}, and in $\kappa\text{-(BEDT-TTF)}_{2}\text{Cu(NCS)}_{2}$ in 2014~\cite{mayaffre2014evidence}. There are also theoretical and experimental efforts to realize FFLO states in cold atom systems~\cite{PhysRevLett.101.215301,chen2012classification,liao2010spin}. We shall discuss the mean-field description of the FF state in electronic systems and its optical response in Chapter~\ref{chap:opticalconductivity}.

In FFLO states, the pairing momenta is determined by the Zeeman splitting, which is usually very small in electronic systems. On the contrary, in cuprates, especially in the pseudogap region, PDWs without Zeeman splitting, whose periods are between 6 to 8 lattice spacing are currently under investigation~\cite{agterberg2020physics}. The discussion of PDW in the context of cuprates has a long history, which we briefly review in Chapter~\ref{chap:PDWpseudogap}. From Chapter~\ref{chap:PDWpseudogap} to Chapter~\ref{chapter: fluctuating PDW in cuprates}, we shall focus on the proposal of Lee~\cite{PhysRevX.4.031017}. Lee suggests the pairing of two electrons moving in the same direction (Fig.~\ref{fig:pairing}(b)) near the anti-nodal region of cuprates. Such pairs can condense at $Q\simeq 2k_F$. Although Lee and coworkers suggests the mechanism of Amperean pairing as a possible microscopic origin \cite{PhysRevLett.98.067006,PhysRevX.4.031017}, the proposal is mainly phenomenological, motivated by ARPES results and the observation of CDW in cuprates. For this purpose, the coexistence of pairing gap and gapless Fermi surface in PDW plays a crucial role.  

\section{Fluctuating superconductivity}

While PDW provides a new perspective in understanding the mysterious pseudogap behavior, what we really want to understand is a fluctuating PDW instead of a long-range PDW. For reasons that will be explained later, we expect the PDW to not have long range order or ordered at most in a small region in the temperature-doping-field phase diagram, but a fluctuating PDW amplitude, disordered due to limited phase stiffness, can exist in the large pseudogap region and account for the unconventional experimental observations.

Fluctuating PDW is a special case of fluctuating superconductivity; and we discuss it from this more general perspective. In principal, the term `fluctuating superconductivity' can refer to any quantum state or thermal ensemble where a superconducting order parameter has long correlation length but is not ordered. In practice, this terminology is only useful if the fluctuating superconducting order parameter serves as an organising principle to understand the physical system and help predict physical observable.

The fluctuation of superconductivity further divides into two classes: thermal fluctuation and quantum fluctuation. In the first case, the superconducting order parameter has long range order at zero temperature, but the long range order is destroyed by thermal excitations. In the second case, the superconducting order parameter does not have long range order even at zero temperature, because the ground state contains quantum superposition of different order parameter configurations. In this thesis, we focus on the second case: we consider a non-superconducting ground state close to a PDW ordered state. The strong amplitude of the PDW order parameter greatly changes the local electron spectrum, despite the fluctuation of its phase over long time and space separation.

A number of questions immediately come to mind: what kind of states can we get out of quantum fluctuation of a superconducting order? Does the pairing gap survive under the fluctuation? If the correlation length of the superconducting order parameter is sufficiently long, is the electron spectrum of the state similar to that in the long-range ordered superconductor? If so, how can a non-superconducting state have a superconducting spectrum? Does the point of view of fluctuating superconductivity help predicts low-energy, long-range properties beyond the correlation length of the superconducting order? Many of the questions are not fully understood even in the simple case of fluctuating conventional s-wave superconductor. For fluctuating PDW, the coexistence of pairing gap and gapless Bogoliubov Fermi surface presents new challenges to the problem.

\section{Plan of this thesis}

We organize this thesis as follows. In Chapter~\ref{chap:opticalconductivity}, we familiarize ourselves with PDW by studying the simplest example, the Fulde-Ferrell state. we explore the role of the quasiparticle excitation and the collective motion in response to incident light. In Chapter~\ref{chap:PDWpseudogap}, we introduce the phenomenology of the pseudogap, briefly review the history of PDW in the context of cuprates, and give an overview of the fluctuating PDW state we construct to explain the pseudogap. In Chapter~\ref{chapter: PDW band}, we present the band structure of bi-directional PDWs with commensurate and incommensurate periods, laying the foundation for later chapters. In Chapter~\ref{chap:STMPDW}, we analyze an encouraging experimental finding of period-8 density wave, directly related to our postulate of PDW. In Chapter~\ref{chap:fluctuatingSC}, we solve a warm-up problem of fluctuating s wave superconductor. During the process, we develop a useful way to think about quantum fluctuating superconductivity in general. In Chapter~\ref{chapter: fluctuating PDW in cuprates} we use ideas developed in previous chapters to build a theory of the pseudogap and compare it with experiments.

\chapter{Gap equation and Optical Response of Fulde-Ferrell state}\label{chap:opticalconductivity}

\section{Introduction} \label{introduction}
Pair density waves (PDW) occur when Cooper pairs condense at nonzero momenta. The first example of PDW is the Fulde-Ferrell-Larkin-Ovchinnikov state (FFLO), where finite-momentum pairing is preferred in a certain range of the Zeeman splitting \cite{fulde1964superconductivity,larkin1965inhomogeneous}. 
More recently, experimental evidence of FFLO states has been found in 
$\text{CeCoIn}_{5}$ 
\cite{bianchi2003possible} and $\kappa\text{-(BEDT-TTF)}_{2}\text{Cu(NCS)}_{2}$ \cite{mayaffre2014evidence}, 
and possible mechanisms stabilizing PDW have been proposed in high-T$_\text{c}$ cuprates \cite{berg2009striped,PhysRevX.4.031017}. 
Unlike conventional BCS superconductors, these phases with PDW usually have partially-gapped Fermi surfaces, almost normal specific heat and anisotropic electromagnetic response. Although many of the physical properties of PDW are well-established, to the best of our knowledge, the optical conductivity from PDW have not yet been addressed. The purpose of the present chapter is to report the unconventional features in the optical conductivity and to discuss its potential applications in various experimental systems. Most of the results presented here apply to a general class of PDW, but we mainly focus on Fulde-Ferrell (FF) states, where quantitative comparison might be made with experiments in the near future.

It is well-known that a single-band BCS superconductor, in the clean limit, has no optical absorption across the superconducting gap \cite{mahan2013many}. This absence of absorption is not protected by the symmetry of the Hamiltonian but by a special feature of the BCS ground state: single-particle states in the original band carrying opposite currents are always simultaneously occupied (or unoccupied), hence the ground state is an exact eigenstate of the current operator and the matrix element for AC absorption $\langle \text{excited state} |\ \vec{\mathbf{j}}\ |\text{G.S.}\rangle$ (often called the `coherence factor') vanishes.
However, this is not the case for finite-momentum pairing. Although the ground state has zero average current, it is no longer an eigenstate of the current operator. Finite-momentum Cooper pairs are in general optically active and they give rise to the dominant contribution to the AC conductivity in the energy range comparable to the pairing gap.

It is worth mentioning that the ground state generally involve PDW with multiple pairing momenta if finite-momentum pairing is favorable. For example, if we have Cooper pairs condensing at momentum $Q$, it's natural to have another pairing momentum $-Q$. The two pairing terms together cause the folding of the Brillouin zone (B.Z.), hence charge density waves (CDW) at momenta $2Q$, $4Q$ etc  \cite{larkin1965inhomogeneous}. It is also possible to have pairing momenta in different directions generating complex incommensurate patterns above the original lattice. However, for simplicity, we focus on the case with only one pairing momentum (FF pairing), a `pure PDW' with no charge modulation. The optical absorption from PDW with multiple pairing momenta should be qualitatively similar for frequencies around the pairing gap. This `pure PDW' with only a phase modulation in the pairing order parameter breaks the lattice translation symmetry, but it is actually invariant under the combination of a gauge transformation and the lattice translation. Note that the absolute phase is not a physical observable, only the phase difference is. More physically, even though we can take momentum from the condensate, we can only do so by taking charge 2e pairs at momentum $Q$. Thus the phase modulation of PDW can only be detected when tunneling electron pairs into or out of the system. Despite the phase modulation, every charge-neutral operator in this state is invariant under the lattice translation, for example, the charge density, current density, and the spin density. This is very different from the optical absorption of CDW only, where translation symmetry is broken for every charge-neutral operator.

One important thing in calculating optical conductivity is maintaining gauge invariance in the self-consistent main-field approximation. This issue was first discussed in BCS superconductors by Nambu \cite{nambu1960quasi}, and recently studied in strongly interacting superconductors \cite{he2015establishing,boyack2016gauge}. The key step is to carry out the vertex correction that is consistent with the gap equation \cite{nambu1960quasi,schrieffer1964theory,he2015establishing,boyack2016gauge}. We followed Nambu's approach and gave an explicit formula for optical conductivity in systems with simple electron-electron interactions. One subtlety in this calculation is that, in order to have non-zero AC conductivity, we must break Galilean symmetry explicitly in the electronic Hamiltonian. This is because the current operator coincides with the momentum operator in a Galilean symmetric system making the linear response to an uniform electromagnetic field trivial. This issue is discussed in more detail after a brief review on finite-momentum pairing.

\section{Finite-momentum pairing and gap equation}

We start by briefly reviewing the mean-field treatment of finite-momentum pairing, especially the diagrammatic interpretation of the mean field gap equation, which turned out to be useful in calculating linear response functions.

We consider a (2+1)-dimensional system with Hamiltonian 
\bea 
H=\sum\epsilon_{p,\sigma}\psi^{\dagger}_{p,\sigma}\psi_{p,\sigma} + \sum\lambda_{k}\psi_{p+k,\sigma}^{\dagger}\psi_{p'-k,\sigma'}^{\dagger}\psi_{p',\sigma'}\psi_{p,\sigma},
\eea 
where the four-Fermion interaction might be mediated by phonon or other more exotic mechanisms. To describe a state with finite-momentum pairing, it is convenient to introduce the 2-component Nambu spinor:
\bea 
\Psi_{p}=(\psi_{p+Q/2,\uparrow},\ \psi_{-p+Q/2,\downarrow}^{\dagger})^{T},
\eea 
where $Q$ is the paring momentum which should be determined self-consistently to minimize the energy of the mean-field ground state, as shown in references \cite{fulde1964superconductivity,larkin1965inhomogeneous}. The four-Fermion interaction can then be written as
\bea
\sum_{p,p',k}\lambda_{k}[\Psi^{\dagger}_{p+k}\tau_{3}\Psi_{p}][\Psi^{\dagger}_{p'-k}\tau_{3}\Psi_{p'}].
\eea 
The mean field Hamiltonian for finite-momentum pairing is:
\begin{equation}
\label{eq:mean field Hamiltonian}
H=\sum_{p}\Psi_{p}^{\dagger}\left(\begin{array}{cc}
\epsilon_{p+Q/2,\uparrow} & \Delta_{p}\\
\Delta_{p} & -\epsilon_{-p+Q/2,\downarrow}
\end{array}\right)\Psi_{p}
\end{equation}
We would like to point out an important difference with the BCS pairing. In the BCS case, the diagonal terms are always equal with opposite signs, so are the two eigenvalues. However, this superficial `particle-hole' symmetry is broken in the FF state. We may even have an `unpaired region' in the B.Z. where the two eigenvalues are of the same sign. For convenience, define $\bar{\epsilon}_{p}\equiv(\epsilon_{p+Q/2,\uparrow} + \epsilon_{-p+Q/2,\downarrow})/2$, $\epsilon'_{p}\equiv(\epsilon_{p+Q/2,\uparrow} - \epsilon_{-p+Q/2,\downarrow})/2$, and $\delta_{p}\equiv\sqrt{\bar{\epsilon}_{p}^{2}+\Delta_{p}^{2}}$. The two eigenvalues are given by
\begin{equation}
E_{p}^{\pm}=\epsilon'_{p}\pm\delta_{p}
\end{equation}
The unpaired region is where $\delta_{p}<|\epsilon'_{p}|$. The boundary of this region where $\delta_{p}=|\epsilon'_{p}|$ is the `Fermi Surface' left after FF pairing and the shift in momentum. Optical absorption occurs in the `paired region' when the frequency of light matches the splitting between the two bands $2\delta_{p}$.

The Nambu spinor introduced above allows us to treat the pairing gap on an equal-footing with the self-energy correction, and the conventional mean field gap equation can be understood as a Hatree-Fock approximation \cite{nambu1960quasi,schrieffer1964theory}. We approximate the four-Fermion interaction by a quadratic term and demand that, to the first order, the remaining interaction does not modify the propagator:
\begin{eqnarray}
\left\{\begin{array}{ll}
G(p)=1/(p_{0}-H_{0}(p)-\Sigma(p) + isgn(p_{0})0^{+})\\
0=-\Sigma(p) + i\int\frac{d^{3}k}{(2\pi)^3}\lambda_{k}\tau_{3}G(p-k)\tau_{3}
\end{array}\right.
\label{eq:mean field Green}
\end{eqnarray}
where $G(p)$ is the mean-field Green's function of the Nambu spinor, $p_{0}$ is the temporal component of the momentum, $H_{0}(p)\equiv \epsilon'_{p} + \bar{\epsilon}_{p}\tau_{3}$ is the Hamiltonian for the original band, and $\Sigma(p)\equiv\Delta_{p}\tau_{1}$ is the pairing term. We have ignored the diagonal self-energy correction in $\Sigma(p)$ since it is not important for our purpose.

\begin{figure}
\begin{center}
\includegraphics[width=3.5in]{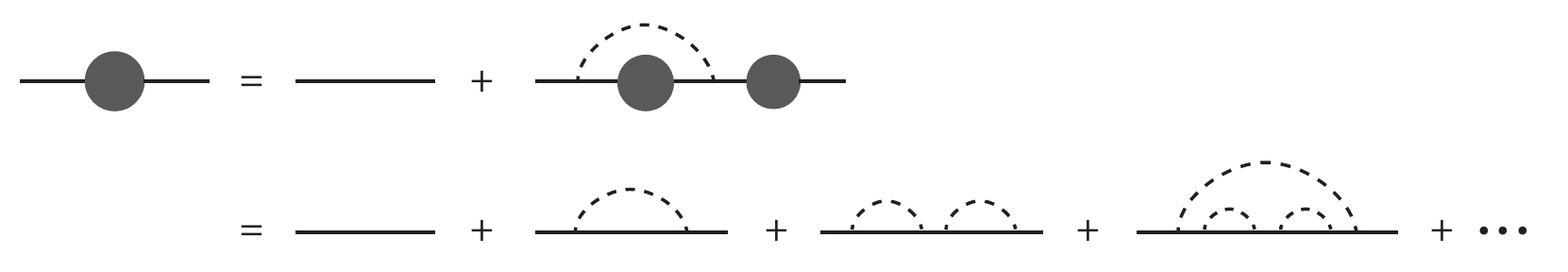}
\caption[Diagrammatic representation of the self-consistent mean field equation of the FF state]{The self-consistent equation of the mean field Green's function, and the diagrams included in this approximation. The solid line represents the 2-component Nambu spinor, and the dashed line represents the electron-electron interaction mediated by a boson, e.g. phonon. We have ignored the correction of the interaction, since it is not important for our purpose. All diagrams without the crossing of the interaction line is included.}%
\label{fig:selfenergy}
\end{center}
\end{figure}

This approximation is equivalent to summing over all Feynman diagrams without crossing in calculating the Green's function, as shown in Fig. \ref{fig:selfenergy}.

When the four-Fermion interaction has no momentum dependence near the Fermi surface, both $\lambda_{k}$ and $\Delta_{p}$ can be approximated by constants, and we arrive at the familiar gap equation after integrating out $k_{0}$:
\begin{equation}
\Delta=-\lambda\int_{\text{paired}}\frac{d^{2}\vec{\mathbf{p}}}{(2\pi)^{2}}\frac{\Delta}{2\sqrt{\bar{\epsilon}_{p}^{2} + \Delta^{2}}}
\label{eq:familiar gap equation}
\end{equation}
This gap equation is almost the same as the BCS gap equation, except the integral is restricted in the `paired region'.

\section{Vertex correction and gauge invariant electromagnetic response}

We are now ready to study the electromagnetic response of PDW. Following the Peierls substitution, we change $\epsilon_{p,\sigma}$ in the total Hamiltonian into $\epsilon_{p+eA,\sigma}$, where $\vec{\mathbf{A}}$ is the magnetic vector potential. We restrict ourself to the single band near the Fermi level, and focus on the limit of a weak and uniform external field as in the case of infrared absorption. Under these restrictions, the current operator $\vec{\mathbf{j}}\equiv-\partial H/\partial\vec{\mathbf{A}}$ can be written as
\begin{eqnarray}
\vec{\mathbf{j}}&=&\sum_{p,\sigma}\psi_{p,\sigma}^{\dagger}[-e\vec{\mathbf{v}}_{p,\sigma}-e^{2}\mathbf{m}_{p}^{-1}\vec{\mathbf{A}}]\psi_{p,\sigma}\label{eq:current}\\
&\equiv&\sum_{p}\Psi_{p}^{\dagger}[-e\vec{\mathbf{v}}_{1}(\vec{\mathbf{p}})\mathbb{1} - e\vec{\mathbf{v}}_{2}(\vec{\mathbf{p}})\tau_{3} -e^{2}\mathbb{m}_{p}^{-1}\vec{\mathbf{A}}]\Psi_{p}\label{eq:current,Nambu spinor}
\end{eqnarray}
where $\vec{\mathbf{v}}_{p,\sigma}\equiv\nabla_{p}\epsilon_{p,\sigma}$ is the band velocity and $\mathbf{m}_{p}\equiv(\nabla_{p}\nabla_{p}\epsilon_{p,\sigma})^{-1}$ is the effective mass tensor. $\vec{\mathbf{v}}_{1}(\vec{\mathbf{p}})$, $\vec{\mathbf{v}}_{2}(\vec{\mathbf{p}})$ and $\mathbb{m}_{p}$ are defined by the equation above and they depend on the pairing momentum. The current operator at zero field is usually called the paramagnetic current, and we would like to write the spatial components together with the temporal component $j_{0}=\sum_{p,\sigma}-e\psi_{p,\sigma}^{\dagger}\psi_{p,\sigma}$ as:
\begin{eqnarray}
j^{P}_{\mu}&=&\sum_{p}\Psi_{p}^{\dagger}\gamma_{\mu}(\vec{\mathbf{p}})\Psi_{p},\label{eq:paramagnetic current}\\
\gamma_{\mu}(\vec{\mathbf{p}})&\equiv&-e(\tau_{3},\ \vec{\mathbf{v}}_{1}(\vec{\mathbf{p}})\mathbb{1} + \vec{\mathbf{v}}_{2}(\vec{\mathbf{p}})\tau_{3})\label{eq:bare vertex}
\end{eqnarray}
The part of current proportional to $\vec{\mathbf{A}}$ in Eq. (\ref{eq:current}) and (\ref{eq:current,Nambu spinor}) is called the diamagnetic current, which does not contribute to the real part of the conductivity at any finite frequency.

Naively, one would like to plug the paramagnetic current and the mean-field excited states into the Kubo formula:
\begin{equation}
\label{eq: bare result}
\text{Re}\,\sigma_{ii}=\frac{\pi}{\omega}\sum_{n}|\langle 0|j^{P}_{i}|n\rangle|^2\delta(\omega-E_{n}+E_{0})
\end{equation}
where $i$ denotes the spatial components, and $0$ ($n$) denotes the ground state (excited states). This approach corresponds to plugging the mean-field Green's function into the bubble diagram without doing other corrections.

As explained in the introduction, the matrix element $\langle 0|j^{P}_{i}|n\rangle$ vanishes identically for BCS pairing, but not for finite-momentum pairing. Thus we expect a nonzero AC conductivity for a state with PDW. However the bare result given by the `mean-field-version' of Eq. (\ref{eq: bare result}) can not be trusted for at least two reasons: (1) This approach violates gauge invariance, specifically the Ward-Takahashi identity between the vertex and the Green's function \cite{nambu1960quasi,schrieffer1964theory}. (2) The result given by Eq. (\ref{eq: bare result}) is always nonzero for any finite-momentum pairing, but the AC conductivity should be exactly zero if the electronic Hamiltonian is Galilean invariant.

The latter statement may not be immediately obvious, especially in the case with spontaneous symmetry-breaking. So we give a careful explanation in this paragraph. When the energy band is parabolic, the current operator is proportional to the kinetic momentum operator: 
$\langle\vec{\mathbf{j}}(t)\rangle=-e\langle\vec{\mathbf{P}}(t)\rangle/m - ne^{2}\vec{\mathbf{A}}(t)/m$, where $\vec{\mathbf{P}}$ is the canonical momentum per unit volume. Since $\vec{\mathbf{P}}$ commutes with the Hamiltonian under uniform perturbation, its average value remains zero all the time. Thus the linear response is trivial and we got $\sigma(\omega)=ie^{2}n/m(\omega + i0^{+})$. We can see that there is only a delta function in the real part of the conductivity, and this derivation holds regardless of whether the ground state is a symmetry-breaking state or not.

\begin{figure}
\begin{center}
\includegraphics[width=4in]{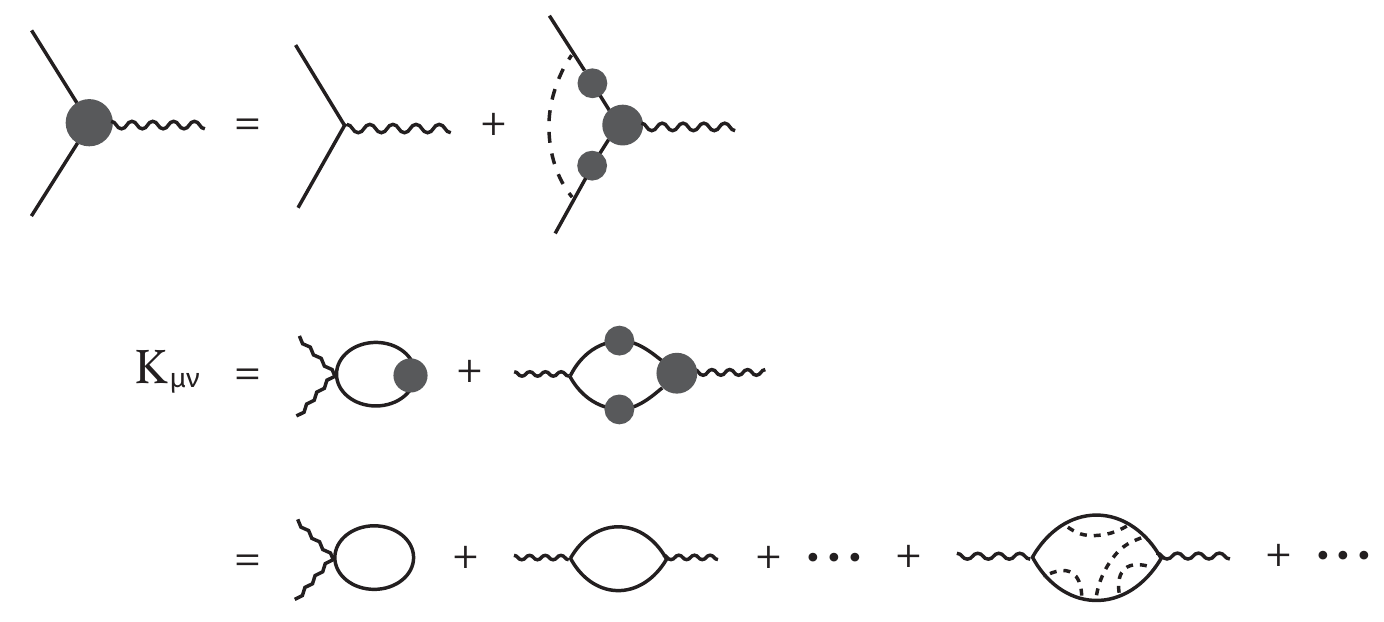}
\caption[Self-consistent vertex correction of the FF state]{The self-consistent vertex correction and the diagrams included in the corrected electromagnetic response function $K_{\mu\nu}$ (defined as $j_{\mu}=K_{\mu\nu}A_{\nu}$). The solid line represents the Nambu spinor, the dashed line represents the electron-electron interaction and the curly line represents the electromagnetic field. The second diagram on the first line of $K_{\mu\nu}$ is the paramagnetic response $P_{\mu\nu}$.}%
\label{fig:vertex}
\end{center}
\end{figure}

The inconsistencies (1) and (2) can be solved by a well-known technique in QED, first introduced to superconductors by Nambu to restore the gauge invariance in the BCS formalism \cite{peskin1995introduction,nambu1960quasi,schrieffer1964theory}.
The key observation is that, whenever an electron-photon vertex appears in a chain of electron lines, we can always form a `gauge-invariant subgroup' of diagrams by considering all different places to insert the corresponding photon line along this chain. And the Ward-Takahashi identity is automatically preserved if we sum over all diagrams in this subgroup. 
As discussed in the previous section, the mean field Green's function contains all diagrams without crossing. Following the diagrammatic technique, if we plug the mean field Green's function into the bubble diagram, we are forced to include all corrections to the bubble diagram without crossing. This can be done by introducing a corrected electron-photon vertex, as shown in Fig. \ref{fig:vertex}. Those diagrams containing a 2-electron-2-photon vertex correspond to the average value of the diamagnetic current, which does not contribute to the imaginary part of the response function (real part of the conductivity) at any finite frequency, so we focus on the paramagnetic part of the response function (defined as $j^{P}_{\mu}=P_{\mu\nu}A_{\nu}$):
\begin{eqnarray}
P_{\mu\nu}=-i\int\frac{d^{3}p}{(2\pi)^{3}}\text{Tr}[\gamma_{\mu}(p,p')G_{p'}\Gamma_{\nu}(p',p)G_{p}]
\end{eqnarray}
where $\gamma_{\mu}(p,p')$ ($\Gamma_{\mu}(p,p')$) is the bare (corrected) vertex of the 2-electron-1-photon interaction. $\Gamma_{\mu}(p,p')$ is given by a self-consistent equation as depicted in Fig. \ref{fig:vertex}: 
\begin{eqnarray}
\Gamma_{\mu}(p',p)&=&\gamma_{\mu}(p',p)+\nonumber\\
i\int\frac{d^{3}k}{(2\pi)^{3}}&\lambda_{k}&\tau_{3}G(p'\!-\!k)\Gamma_{\mu}(p'\!-\!k,p\!-\!k)G(p\!-\!k)\tau_{3}
\label{eq: vertex correction}
\end{eqnarray}
We are interested in the case $\vec{\mathbf{p}}=\vec{\mathbf{p}}'$, and we have $\gamma_{\mu}([p_{0}+\omega,\vec{\mathbf{p}}],[p_{0}, \vec{\mathbf{p}}])=\gamma_{\mu}(\vec{\mathbf{p}})$ as shown in Eq. (\ref{eq:paramagnetic current}) and (\ref{eq:bare vertex}).

Eq. (\ref{eq: vertex correction}) can be solved analytically when the four-Fermion interaction has no momentum dependence near the Fermi surface. If we further assume the pairing gap $\Delta$ is much smaller than the band width, the self-consistent vertex acquires a simple form:
\begin{eqnarray}
\label{eq:corrected vertex}
&\vec{\mathbf{\Gamma}}&=-e(\vec{\mathbf{v}}_{1}(\vec{\mathbf{p}})\mathbb{1}\! +\! \vec{\mathbf{v}}_{2}(\vec{\mathbf{p}})\tau_{3} + 2i\Delta I(\vec{\mathbf{v}}_{2})\tau_{2}/\omega I(1)),\\
&I&(f)\equiv \int_{\text{paired}}\frac{d^{2}\vec{\mathbf{p}}}{(2\pi)^{2}}\frac{f(p)}{\delta_{p}(\omega-2\delta_{p})(\omega+2\delta_{p})}
\label{eq:integral}
\end{eqnarray}
where $I(f)$ is a linear functional defined by the integral which appears repeatedly in the remaining part of the chapter. Finally the corrected optical conductivity is given by
\begin{eqnarray}
\text{Re}\,&\sigma_{ij}&(\omega>0)= -\text{Im}\,P_{ij}(\omega>0)/\omega\\
&=&-\frac{4e^{2}\Delta^{2}}{\hbar\omega}\text{Im}\left[I(v_{2i}v_{2j})-I(v_{2i})I(v_{2j})/I(1)\right]
\label{eq: final results}
\end{eqnarray}
Note that we have omitted the infinitesimal imaginary part of $\omega$ in the integral \ref{eq:integral} since the pole structure in retarded response functions is different from that in path integrals, and $\omega$ should always be replaced by $\omega + i0^{+}$ for retarded response. When $\omega>0$, the imaginary part of the integral is given by
\begin{equation}
\text{Im}\,I(f)=-\pi\int_{\text{paired}}\frac{d^{2}\vec{\mathbf{p}}}{(2\pi)^{2}}\frac{f(p)}{4\delta_{p}^{2}}\delta(\omega-2\delta_{p})
\end{equation}
which is proportional to the joint density of states (JDOS) in the paired region. We found that the first term in Eq. (\ref{eq: final results}) is nothing but the bare result given by the `mean-field-version' of Eq. (\ref{eq: bare result}), while the second term is given by the vertex correction. As discussed before, only those points in the `paired region' of the B.Z., where the frequency matches the band splitting, contribute to the real part of the optical conductivity. For a given $\omega$, these points lie on arcs in the B.Z.

Another important ingredient in Eq. (\ref{eq: final results}) is $\vec{\mathbf{v}}_{2}$. Recall that $\vec{\mathbf{v}}_{2}$ is defined by Eq. (\ref{eq:current}) and (\ref{eq:current,Nambu spinor}). In the case of FF pairing, when the pairing momentum is much smaller than the Fermi momentum, we have
\begin{equation}
v_{2i}(\vec{\mathbf{p}})=(\mathbf{m}_{p}^{-1})_{ij}Q_{j}/2 + O(Q^{2})
\label{eq:Q dependence}
\end{equation}
As discussed above, gauge invariance is guaranteed in this formalism. Furthermore, we found that the problem regarding Galilean symmetry is automatically solved: if the band is parabolic, $\vec{\mathbf{v}}_{2}=\vec{\mathbf{Q}}/2m=\text{const.}$, hence $v_{2i}$ and $v_{2j}$ can be dragged out of the integral in Eq. (\ref{eq: final results}), and the vertex correction cancels the bare result. However there is no exact Galilean symmetry in real solids, and Eq. (\ref{eq: final results}) and (\ref{eq:Q dependence}) shows that the optical conductivity from PDW is proportional to $Q^{2}$. We refer the readers to the appendix for more details on the Ward-Takahashi identity, the vertex correction and the final result for optical conductivity.

\section{Results for tight-binding bands}
\begin{figure}
\begin{center}
\includegraphics[width=3.5in]{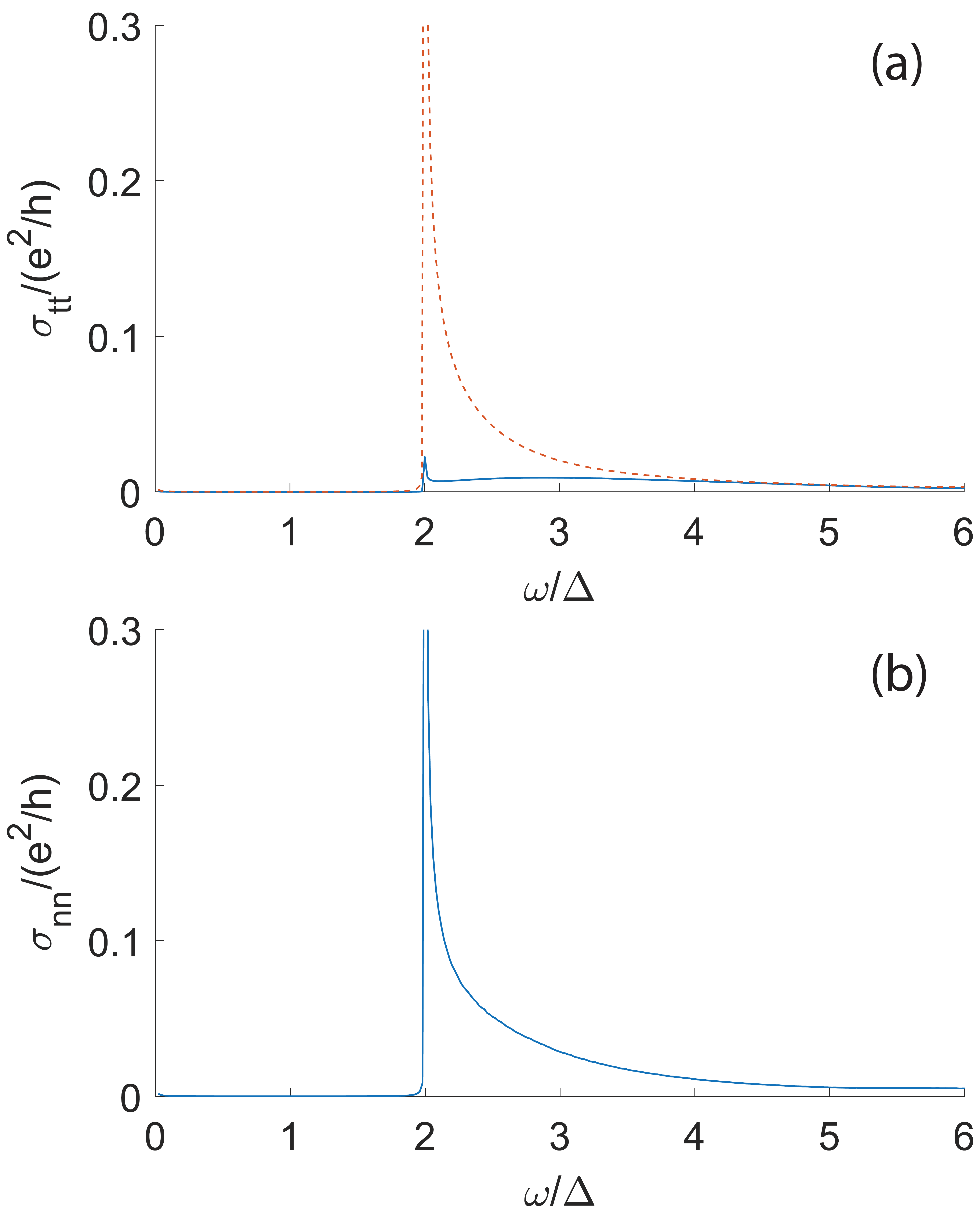}
\caption[Optical conductivity of the FF state calculated for tight-binding bands]{Optical conductivity of the FF state calculated for tight-binding bands on a 2-dimensional square lattice, $t_{2}/t_{1}=0.35$. The spin splitting is set to be $0.4t_{1}$, which is about four percent of the band width, and the pairing momentum is $(0.1/a, 0.1/a)$. (a) Conductivity in the direction of the pairing momentum. The dashed orange line is the bare result and the blue line is the corrected result. (b) Conductivity in the perpendicular direction. The vertex correction is identically zero in this direction by symmetry.}%
\label{fig:result}
\end{center}
\end{figure}
We have calculated the optical conductivity of FF states explicitly for tight-binding bands with NN hoping $t_{1}$ and NNN hoping $t_{2}$ on a square lattice. The result shown in Fig. \ref{fig:result} is for $t_{2}/t_{1}=0.35$, spin splitting $0.4t_{1}$, at half-filling. The pairing momentum is $(0.1/a,0.1/a)$, where $a$ is the lattice constant. AC conductivity shows up in both Fig. \ref{fig:result}a and Fig. \ref{fig:result}b above $2\Delta$ and there are divergent peaks right at $2\Delta$ (although the divergence of the blue curve in Fig. \ref{fig:result}a appears to be small, it is guranteed to be a true divergence by analytical analysis of Eq. (\ref{eq: final results})) due to the corresponding divergence in the JDOS. As mentioned in the previous section, for a given $\omega$, only the arcs in the B.Z. satisfying the frequency-matching condition contribute to AC absorption. When $\delta\omega\equiv\omega-2\Delta\simeq 0$, the frequency-matching condition $\omega=2\delta_{p}$ gives $\bar{\epsilon}_{p}=\sqrt{\omega^{2}/4-\Delta^{2}}\propto\sqrt{\delta\omega}$, then the JDOS is $N(0)d\bar{\epsilon}_{p}/d\omega\propto 1/\sqrt{\delta\omega}$, where $N(0)$ is the density of states (DOS) of the normal metal. Hence the $1/\sqrt{\delta\omega}$ divergence in the optical conductivity at $2\Delta$. This divergence has the same form of the divergence in the DOS and JDOS of s wave BCS superconductors, but the real part of the AC conductivity is identically zero in BCS superconductors for any band structure, as explained in the introduction.

The effects of the vertex correction on divergent peaks depend on the type of divergence as well as the details of the band structure, and can be dramatically different in different situations. If there is a single singularity of the JDOS on the frequency-matching arc giving the dominant contribution, we can replace $\vec{\mathbf{v}}_{2}$ by its value at the singularity, and it is clear from Eq. (\ref{eq: final results}) that the vertex correction completely cancels the divergence in the bare result. However, the divergence at $2\Delta$ is due to the whole arc in the paired region satisfying $\bar{\epsilon}_{p}\simeq 0$, and it remains divergent after the vertex correction. The ratio between the corrected result (shown as blue line in Fig. \ref{fig:result}a) and the bare result (dashed orange line in Fig. \ref{fig:result}a) depends on the variance of $\vec{\mathbf{v}}_{2}$ on the frequency-matching arc.  We found that in the current example, the divergence in the conductivity along the pairing momentum $\sigma_{tt}$ is strongly suppressed by the vertex correction, whereas there is no vertex correction at all in the perpendicular direction since the perpendicular component of $\vec{\mathbf{v}}_{2}$ is odd under the reflection over $(\pi,\pi)$.

\section{Discussion}

We have shown that there is nonzero AC absorption from PDW if we break Galilean symmetry explicitly in the electronic Hamiltonian (which is usually the case in solids). When the pairing momentum $Q$ is much smaller than the Fermi momentum $p_{F}$ and the pairing gap $\Delta$ is much smaller than the band width $W$, the AC conductivity is proportional to $(Q/p_{F})^{2}W/\Delta$. We estimated the typical optical conductivity in $\kappa\text{-(BEDT-TTF)}_{2}\text{Cu(NCS)}_{2}$ around the frequency of the pairing gap and away from the divergent peak to be at the order of $0.01\text{e}^{2}/\text{h}$, based on the recent experiment \cite{mayaffre2014evidence}. However since a direct measure of the pairing momentum and the pairing gap is still missing, it is hard to give a more accurate estimation. Vertex correction plays an important role in this AC absorption, and dramatically changes the behavior of the optical conductivity in the direction of the pairing momentum. 

This nonzero absorption could be used as an experimental evidence for PDW. Furthermore, the various features discussed in the previous section can help determine the pairing gap and the direction of the pairing momentum in experiments. We have focused on the case with only one pairing momentum in the present chapter, and we have ignored the momentum dependence of the pairing gap near Fermi surface in the explicit calculation. The results for more general PDW should be similar, but we would like to discuss some possible differences in this paragraph. (1) A weak momentum dependence of the pairing gap introduces a cutoff to the $1/\sqrt{\delta\omega}$ divergence at $\omega=2\min[\Delta_{p}]$, whereas a strong momentum dependence completely destroys the $1/\sqrt{\delta\omega}$ behavior and leaves only a finite jump. (2) When the PDW state has more than one pairing momenta, one or more CDW will be generated by the interference, and there will be nonzero absorption below the `pairing gap' $2\min[\Delta_{p}]$. The magnitude of this `in gap' absorption increases with the magnitude of the CDW. (3) We have not discussed the effect of impurities so far. Since there is a finite density of states left at Fermi level, there will be a Drude peak coexisting with the absorption we discussed when the inverse of the mean free time of electrons is smaller than the pairing gap. Whereas in the opposite limit, even BCS superconductors have nonzero optical absorption above the gap \cite{mahan2013many} and there is no sharp feature for PDW.

\chapter{Pseudogap and Pair Density Wave}\label{chap:PDWpseudogap}

\section{Pseudogap}\label{sec:pseudogapintro}

The pseudogap phase has long been considered a central puzzle in the study of the cuprate high temperature superconductors\cite{keimer2015quantum}. After decades of studies, experimental results are rich and well established, but the underlying physics become even more mysterious over time.

Along the doping axis, the pseudogap occupies an intermediate region, taking YBCO as an example, it starts from 5\% hole doping to 19\% hole doping. At 0\% doping, we have one active electron per copper atom, the electrons stuck by repulsion to form a Mott insulator with antiferromagnetic spin order. The system remains antiferromagnetic with zero or negligible conductivity until around 5\% doping. On the other hand, above 19\% doping, the electrons become fully mobile again, namely, we have a large Fermi surface whose area matches the total number of electrons, and Fermi liquid behavior is observed at the superconducting transition temperature. However, in the pseudogap region, between 5\% and 19\% doping, only part of the large Fermi surface remains gapless. The material is metallic above the superconducting transition temperature, but part of the Fermi surface in the direction of $(0,\pi)$ and $(\pi,0)$ (the anti-nodal region) are gapped (seen from ARPES), and the rest of the Fermi surface exists as disconnected segments, the so-called Fermi arcs~\cite{damascelli2003angle} (for recent data in Bi2212, see Ref.~\cite{hashimoto2014energy,chen2019incoherent}). This weird spectral feature is challenging our fundamental understanding of metals.

The word `pseudogap' often directly refers to the anti-nodal electron gap. It starts out being very large (several hundreds meV) near the Mott insulator, but it persists in the anti-nodal region for intermediate doping, where it co-exists with superconductivity. In this thesis the term pseudogap phase refers to this intermediate doping regime, roughly in the range between $p$ = 0.08 to 0.19 in YBCO, where the pseudogap itself is about 80 meV or less. This regime has been under intense study and is commonly considered to be a central puzzle in the cuprate high-Tc problem~\cite{keimer2015quantum}. At around a temperature scale $\text{T}^*>\text{T}_\text{c}$, which depends on the doping, the spectral gap is observed to be gradually filled in, crossing over to the large Fermi surface.

The pseudogap is different from the d-wave superconducting gap in several aspects. First it leaves four gapless arcs instead of four nodes. Moreover, a detailed ARPES investigation on Bi2201~\cite{ZX1science350949gerber2015three} (where $\text{T}^*$ and $\text{T}_\text{c}$ has a large separation) reveals that the minimum of the gap along the cut $k_y=\pi$ is not at the Fermi surface but shifted outside. On the other hand, it has similarities with a superconducting gap. For example, STM experiments report that the local density of states (momentum integrated) is roughly particle-hole symmetric near the Fermi surface.

Even though the spectral feature of the pseudogap region is quite prominent, it is hard to interpret theoretically. This is partly because the basic concepts we used to described the phenomena, like Fermi surface, gap, and even metal, are only vaguely defined at finite temperature. At finite temperature any gap is partially filled in, and any material has nonzero conductivity. Thus the pseudogap phenomena is a quantitative feature instead of a sharply-defined qualitative distinction. In order to sharply define the pseudogap phenomena, we can either look for materials where $\text{T}^*\gg \text{T}_\text{c}$, or kill the superconductor by magnetic field and study the pseudogap at zero temperature. The latter approach has been very fruitful in the last twenty years (for a recent review, see Ref~\cite{proust2019remarkable}). We briefly review the experimental results.

The high-field low temperature phase diagram is universal for many different cuprate families, but the critical doping differ by a few percentage among different families. For simplicity we shall refer to the doping in YBCO in this paragraph. For magnetic field $\text{H}>\text{H}_\text{c2}$, the ground state is always metallic above 5\% doping. The pseudogap metallic ground state occupies the doping range from 5\% to 19\%. Above 19\% we have the usual Fermi liquid with a large Fermi surface. The critical doping 19\% is found consistently from specific heat, Hall number, conductivity and thermal conductivity~\cite{proust2019remarkable}. Across the entire doping range in the pseudogap phase, Wiedemann-Franz law is satisfied~\cite{PhysRevX.8.041010,PhysRevB.93.064513}. This surprising finding suggests that despite the mysterious anti-nodal gap and the Fermi arcs, the low-energy transport may be dominated by conventional quasi-electrons. In the middle of this range, roughly between 8\% and 16\%, the pseudogap coexists with long range and static short range CDW~\cite{blackburn2013x,ghiringhelli2012long,BlancoPhysRevB.90.054513,PhysRevB.96.134510,JulienNature477191wu2011magnetic,wu2013emergence,ZX1science350949gerber2015three,changNatureComm72016magnetic,ZX2PNAS11314647jang2016ideal}. In this range, the resistivity is found to increase as $T^2$ at low-temperature~\cite{proust2016fermi}, consistent with Fermi liquid behavior. The Hall number is small and positive~\cite{leboeuf2007electron}, indicating a small electron-like pocket occupying a few percent of the B.Z.. Furthermore, quantum oscillations is observed at high field and identified with small electron-like pockets~\cite{doiron2007quantum,PhysRevB.98.140505,sebastian2011quantum,ramshaw2011angle}. Another important phenomenology is that for doping near 1/8, the superconductivity is suppressed by an unexpectedly small magnetic field of about 20T ~\cite{chang2012decrease,grissonnanche2014direct,Julien2arXivzhou2017spin}, suggesting that the high-field ground state and the d-wave superconductor have very similar ground state energy near this doping. In the pseudogap phase but without CDW, Hall number changes dramatically to small negative numbers, suggesting a hole pocket whose area equals the hole doping.

In this thesis, we shall focus on the middle range where pseudogap coexists with CDW. The zero-field electron spectrum suggests a highly unconventional state while the low temperature transport results consistently indicating a Fermi liquid with a small electron pocket. Is the mysterious pseudogap phase just a Fermi liquid after all? Where do the electrons go besides those contribute to low-energy specific heat and transport? These are the main themes of the rest of the thesis. 

Before introducing various theoretical proposals and our proposal based on the assumption of fluctuating PDW, we briefly review recent experiments that demonstrate the pseudogap is not only a crossover but a genuine phase transition in the temperature-doping phase diagram. Some form of broken crystalline symmetry has been shown to occur from ultrasound attenuation \cite{shekhter2013bounding}, second harmonic generation\cite{HsieNaturePhysicshzhao2017global}, and the anisotropy of the spin susceptibility\cite{sato2017thermodynamic,Matsuda2unpublished}. Just below this temperature, neutron scattering has detected the onset of intra-cell magnetic moments\cite{bourges2011novel} which have been interpreted in terms of orbital loop currents\cite{varma2006theory}, even though this experimental finding has recently been challenged, at least in the case of YBCO\cite{bourges2017comment}. At lower temperatures, short range order charge density wave (CDW) order emerges, often, but not always, suppressed by the onset of superconductivity\cite{blackburn2013x,ghiringhelli2012long,BlancoPhysRevB.90.054513,PhysRevB.96.134510}.  In high magnetic field the CDW order in YBCO dramatically increases its range, as seen in NMR\cite{JulienNature477191wu2011magnetic,Julien2arXivzhou2017spin,wu2013emergence}. X ray scattering reveals that it is unidirectional and becomes stacked in phase between layers\cite{changNatureComm72016magnetic,ZX1science350949gerber2015three,ZX2PNAS11314647jang2016ideal}.  There seems to be two distinct forms of CDW co-existing, one long ranged ordered and uni-directional, while the other is short ranged and exists in both directions. It is quite mysterious why they have the same incommensurate period. Adding to this complexity, a recent STM experiment detected CDW with period 8a co-existing with the previously observed period 4a CDW in the “halo” surrounding the vortex core\cite{edkins2018magnetic}.

It has proven to be extremely challenging to develop a theoretical picture to describe this rich and unexpected set of phenomena. Theoretical efforts can be roughly divided into two classes. The first involves microscopic theories that start with a model Hamiltonian such as the Hubbard model and attempt to solve for the low energy properties. Due to the complexity of the strong correlation problem, progress along this line has been made mainly with numerical methods. Approximate methods such as cluster DMFT (dynamical mean field theory) have shown that the Hubbard model indeed exhibit a  phase  where anti-nodal gap and near nodal gapless  carries co-exists and that this state undergoes d wave paring at low temperature~\cite{PhysRevLett.110.216405}. Other methods such as DMRG (density matrix renormalization group)~\cite{PhysRevB.98.140505}, Monte Carlo studies of projected wavefunctions~\cite{himeda2002},  exact diagonalization~\cite{PhysRevLett.113.046402} and other cluster embedding methods~\cite{zheng2017stripe}, provide information mainly on the ground state and its competitors. There appear to be a general consensus that while the d wave superconductor is a favored ground state, there exists a large variety of states that are very close in energy~\cite{zheng2017stripe}. These include various density waves with energy that is surprisingly insensitive to the period. 

A second line of attack is to do phenomenological theory. Here one postulate the existence of certain state or certain dominant order, and attempt to explain as much of the pseudogap phenomenology as possible based on the postulate. In view of the large variety of observations, even this is a highly nontrivial task.

Even though CDW orders generally appear in the pseudogap phase, there is now general agreement that the anti-nodal gap is not caused by the charge order. One reason is that if the anti-nodal gap were to come from CDW, the electron spectrum would be very different from what is observed in ARPES. In the CDW model~\cite{PhysRevX.4.031017}, as we move from the anti-nodal region to the nodal region, the gap should close by unoccupied states coming down towards the Fermi energy. On the contrary, ARPES~\cite{he2011single} reports that the gap is closed by occupied states coming up towards the Fermi energy.

There is a large literature on the origin of the anti-nodal gap and the Fermi arc, ranging from fluctuating anti-ferromagnet~\cite{PhysRevB.94.205117}, spiral spin density wave~\cite{PhysRevLett.117.187001},  
Umklapp scattering of a pair of electrons~\cite{robinson2019anomalies,PhysRevB.73.174501,PhysRevB.95.201112},
spinon gap in a gauge theory formulation~\cite{RevModPhys.78.17}, to fluctuating superconductivity of some kind~\cite{emery1995importance,PhysRevX.4.031017}. 
Fluctuating d-wave superconductivity was a popular starting point. 
It assumes that in the underdoped region, due to the small superfluid stiffness, phase fluctuations greatly suppress the superconducting $\text{T}_\text{c}$ and the pseudogap is due to a large pairing amplitude that survives up to high temperature~\cite{emery1995importance}. 
However a d wave pairing gives rise to nodal points and it is not easy to obtain Fermi arcs in this scenario. We also introduced above that ARPES reported differences between the pseudogap near the anti-node and an usual d-wave gap.

For reasons that will be explained below, our postulate is that the origin of the anti-node gap is from a quantum disordered (fluctuating in space and time) pair density wave (PDW).

\section{Pair density waves in cuprates}

The notion of PDW in the context of cuprates has a long history. Himeda, Kato and Ogata \cite{himeda2002}  found in 2002 by projected Monte Carlo studies that the PDW is the preferred ground state in the presence of stripe order. Starting from the standard stripe picture \cite{tranquada1995jm}of a period 8 spin density wave (SDW) and a period 4 CDW, they found that the d wave superconductor is more stable if the sign of the order parameter is reversed at the hole poor region of the CDW, leading to a period 8 PDW. We shall refer to this state as the stripe-PDW. They proposed that if the stripe-PDW is stacked perpendicular to each other from one layer to the next, the resulting state has drastically reduced Josephson coupling and may explained the disappearance of the Josephson plasma edge observed in Nd doped LaSr2CuO4 (LSCO)\cite{tajimaPRL862001c}. Strong anisotropy in the transport properties was discovered in the LBCO $\text{La}_{2-x}\text{Ba}_x\text{CuO}_4$ system\cite{PhysRevLett.99.067001} and since that time the theory of layer de-coupled PDW and related phases has been greatly advanced.\cite{PhysRevLett.99.127003,PhysRevB.79.064515} For a review, see  Ref.~\cite{berg22009NTPhysstriped}.

The next development is the introduction of a Landau theory description. \cite{PhysRevLett.99.127003,PhysRevB.79.064515,agterberg2008dislocations,berg1NatPhys2009charge} Agterberg and Tsunetsugu\cite{agterberg2008dislocations} described the coupling of PDW with various subsidiary orders such as CDW and magnetization waves. By examining the interplay between the PDW vortex and the dislocation in the CDW, they showed that it is possible to suppress the PDW order by phase fluctuations, while the subsidiary CDW order remains long ranged. Berg, Fradkin and Kivelson\cite{berg1NatPhys2009charge}  constructed a phase diagram using renormalization group arguments which include regions in parameter space where the primary PDW order is destroyed while CDW and a novel charge 4e superconductor survive. Berg et al \cite{berg22009NTPhysstriped}suggested that the stripe PDW may have a more general applicability than the low temperature behaviors in the LBCO family, ie, it may be behind the pseudo-gap phase. Part of their argument is based on the spectral property of such a uni-directional PDW. We comment that while this state produces what looks like a Fermi arc, the gap is actually small near the antinode in the direction perpendicular to the stripe orientation\cite{PhysRevB.77.174502,berg22009NTPhysstriped}.  This kind of two gap structure has difficulties  with STM and ARPES data.

Stimulated by a detailed angle resolved photo-emission (ARPES) study of the single layer cuprate Bi2201\cite{heSci3312011single}, Lee \cite{PhysRevX.4.031017} proposed that the unusual features of the spectra can be explained by postulating a bi-directional PDW state as the underlying state of the pseudogap. The pairing is produced by singlet pairing of electrons with momenta $K_i+p$ and $K_i-p$ where the $K_i$’s are located at or near the Fermi surface at the anti-nodal points.  This gives rise to a bi-directional PDW. The pair carries momenta $P\hat{x}$ and $-P\hat{x}$ which equal twice the momentum K near the $(\pi, 0)$ antinode and are along the x-axis. There is a similar pair $P\hat{y}$ and $-P\hat{y}$ which are along the y-axis. There are 4 order parameters: $\D_{P\hat{x}}$, $\D_{-P\hat{x}}$, $\D_{P\hat{y}}$ and $\D_{-P\hat{y}}$. While Lee proposed using the idea of Amperean pairing\cite{PhysRevLett.98.067006} as the microscopic origin of the PDW, most of the paper was phenomenological, and explored the consequences of an assumed PDW. As such many of the conclusions are quite general. 

Nevertheless we would like to emphasize that the motivation for introducing the bi-directional PDW is different from that for the uni-directional PDW\cite{berg22009NTPhysstriped,fradkin2015colloquium}, which is rooted in the phenomena observed in the LSCO/LBCO family at relatively low temperatures. Our view is that the recently discovered CDW which survives up to 150K are distinct from the stripe physics associated with LSCO/LBCO. The wave-vector decreases with increasing doping, whereas the stripe wave vector increases linearly up to about 0.125 doping and saturate, following the Yamada plot\cite{yamada1998PRB57doping}. For YBCO the period is incommensurate and close to 3, very different from the period 4 CDW associated with 1/8 doping in LSCO. Finally there is no sign of the SDW that is “intertwined” with the stripes. 

As phenomenology the bi-directional PDW produces the pseudogap at the antinodes and the Fermi arcs near the nodes. (strictly speaking these are the electron-like segments of closed orbits made up of Bogoliubov quasi-particles.) It also produces two important features of the antinodal gap. It explains why the gap closes at the end of the Fermi arcs with states moving up from lower energy (see Fig.~\ref{Fig: PDW 3-band illustration}), while a CDW-generated gap will necessarily close by a state coming down in energy. As opposed to conventional pairing, the spectrum is not particle-hole symmetric at each k point, which explains why the momentum of the minimum gap is shifted away from the original Fermi surface. In addition, CDW at wave-vectors $\vec{Q}=2P\hat{x}$ and $2P\hat{y}$ naturally emerges as subsidiary orders, making it unnecessary to postulate the CDW order as a separate instability.

The states at the Fermi arcs play two important roles. First they greatly suppress the superfluid density and therefore the phase stiffness, so that the PDW is subject to strong phase fluctuations over most of the phase diagram in the H-T plane. Secondly the normal state gives rise to a linear term in the entropy, which lowers the free energy and stabilizes it at finite temperatures, even if it is not the true ground state at zero magnetic field. In addition, in the superconducting state, a CDW with period $P\hat{x}$ and $P\hat{y} (=\vec{Q}/2)$ naturally appears if the PDW phase is pinned to that of the d wave pairing and reference was made to an STM experiment on YBCO where CDW at $Q$ and $Q/2$ have been reported\cite{Yeh1,Yeh2}, where $Q=0.28 (2\pi/a)$ matches what is now determined by X-ray scattering. We shall come back to the STM experiment in Chapter~\ref{chap:STMPDW}.

We should point out that other workers have also associated PDW with the pseudogap phenomenon. Zelli , Kallin and Berlinsky\cite{ZelliQtmOscPhysRevB.86.104507} used the quasi-particle orbits produced by an uni-directional PDW order to produce quantum oscillations. A related proposal was recently made by M. Norman and J.C. Davis.\cite{Normanunpublished} We will comment on this below. Yu et al\cite{yuPNAS126672016magnetic} have interpreted their high magnetic field phase diagram in terms of a possible PDW. Two distinct pair fluctuation lifetimes have been reported in tunneling experiments, possibly indicative of the presence of two kinds of superconductors\cite{koren2016observation}. Other papers consider a PDW with the same wave-vector and on equal footing as the CDW and are less relevant to the present discussion\cite{PepinPhysRevB.90.195207,WangPhysRevLett.114.197001}.

\section{Subsidiary orders of the pair density wave}

Since the composite CDW and an additional orbital current order of the PDW will play an essential role in our discussion, we give a more detailed explanation here.
 
Our construction assumes bi-directional PDWs with wave-vectors $P\hat{x}$ and $P\hat{y}$ which are characterized by four PDW order parameters, $\D_{P\hat{x}}$, $\D_{-P\hat{x}}$, $\D_{P\hat{y}}$, and $\D_{-P\hat{y}}$, with  equal amplitudes. One notices immediately that a term in the Landau free energy that couples linearly to density wave order is allowed by symmetry: $\rho_{2P\hat{x}}\D_{P\hat{x}}\D^*_{-P\hat{x}}$ This means that an ordered PDW with wave-vector $P\hat{x}$ necessarily induces a secondary order of CDW at wave vector $2P\hat{x}$. Perhaps less obvious is the notion that even if the primary order is fluctuating in space and time, a static and long range CDW order can also be induced, under the right circumstances. Consider the case when the phases of $\D_{P\hat{x}}$ and $\D_{-P\hat{x}}$ are wildly fluctuating but the relative phase between them is not. The linear coupling term will induce long range CDW order. Whether this happens or not depends on detailed choices of model parameters and this kind of phase diagram has been explicitly demonstrated in special cases~\cite{agterberg2019review,agterberg2008dislocations,berg2009charge}. This kind of possibility has been given the name vestigial order in a related disorder-driven case~\cite{nie2014quenched}, but we will continue to use the term composite order in this thesis.

For the bi-directional PDW, a second possibility exists, ie CDW at wave-vector $P\hat{x}+P\hat{y}$ may be induced by the term:
$\rho_{P\hat{x}+P\hat{y}}(\D_{P\hat{x}}\D^*_{-P\hat{y}}+ \D^*_{-P\hat{x}}\D_{P\hat{y}})$. However, such a CDW has not been seen experimentally. Fortunately, Ref.~\cite{agterberg2008dislocations} has provided an explanation. They pointed out that there is another term that couples to an orbital magnetization density wave (MDW) which takes the form: $M_{P\hat{x}+P\hat{y}}\cdot i(\D_{P\hat{x}}\D^*_{-P\hat{y}} -  \D^*_{-P\hat{x}}\D_{P\hat{y}})$. The MDW involves orbital current at a finite wave-vector which produces an orbital magnetization. Note that the magnetization comes from orbital current and not spin, because the PDW order is a total spin singlet and will not couple to the spin degree of freedom in the absence of spin-orbit coupling. It turns out the  two terms inside the parenthesis in these Landau free energy terms either add or cancel each other, depending on their relative phase. Since CDW at $P\hat{x}\pm P\hat{y}$ is not observed, we assume the PDW order parameters have the phases that $\D_{P\hat{x}}\D^*_{-P\hat{y}}= - \D^*_{-P\hat{x}}\D_{P\hat{y}}$, such that the contribution to CDW cancels out, but MDW is stabilized at this wave-vector. The MDW may be detectable by neutron scattering, as will be discussed in Sec.~\ref{subsection:Symmetry breaking in the pseudopgap phase.}. For more discussion on the coupling between PDW and CDW/MDW, see Appendix~\ref{Appendix: Symmetry of the fluctuating PDW state}.

Besides the MDW at momentum $P\hat{x} \pm P\hat{y}$ and the CDW at momentum $2P\hat{x}, 2P\hat{y}$, there are other composite orders of PDW that only breaks discrete symmetries, such as time reversal or mirror reflection~\cite{agterberg2008dislocations,berg2009charge}. We discuss these discrete symmetries briefly in Sec.~\ref{subsection:Symmetry breaking in the pseudopgap phase.}.

\section{Fluctuating pair density wave state as the pseudogap ground state}

So far, we have not introduced what we mean by `fluctuating PDW'. It may refer to any state with a long but finite correlation length of the PDW order parameters. In order to reveal as much physical properties of the pseudogap as possible from the assumption of the fluctuating PDW, it is crucial to give a concrete description of how we destroy the long range order of the PDW while preserving the desired features of the PDW. Only after this process, we can use the proposal of fluctuating PDW to describe the metallic pseudogap phase. This is the topic of Chapter~\ref{chap:fluctuatingSC} and~\ref{chapter: fluctuating PDW in cuprates}

Now we describe our basic postulate for the pseudogap phase. We assume the existence of robust amplitudes of the four PDW order parameters, $\D_{P\hat{x}}$, $\D_{-P\hat{x}}$, $\D_{P\hat{y}}$, and $\D_{-P\hat{y}}$ over a large part of the doping and temperature range that is associate with the pseudogap. We assume the overall superconducting phase of the four parameters are subject to strong quantum phase fluctuations even at zero temperature. We further assume that the relative phases of these four order parameters are perfectly locked at low temperature, so that the CDW at twice the PDW wave-vector and MDW at 
$P\hat{x}\pm P\hat{y}$ are generated as long range ordered composite orders. In anticipation of what follows, we emphasize that the MDW (as a composite order) plays no role in producing the anti-nodal gap, but it will play an important role in determining the size of the reduced Brillouin zone (BZ) due to the increase periodicity. We claim that after quantum disordering the PDW, we get a metallic state with only a small electron pocket made of the nodal gapless arcs, and the rest of the electron exist in the form an insulator of electron pairs. We compare theoretical predictions with experiments in Sec.~\ref{Sec: broader aspects experiments}.

We will mostly focus on a description of the zero temperature quantum state that emerges once the d wave pairing state is destroyed by a magnetic field. The focus on a quantum state and its low lying excitations allow us to make sharp statements. On the other hand, we will also make some qualitative predictions at zero field and finite temperature, taking advantage of the fact that the pseudogap scale $T_{PG}\gg T_{c}$.

\chapter{Static Pair Density Wave Bands in Cuprates}
\label{chapter: PDW band}

A PDW condensate is a bath of charge 2e bosons carrying specific nonzero momenta. It mixes an electron with a hole, like regular superconductivity, but only at shifted momenta. To illustrate the PDW we consider in cuprates, we first sketch the band structure along the cut $k_y = \pi$, considering the effects of x-directional PDW and y-directional PDW separately.

\begin{figure}[htb]
\begin{center}
\includegraphics[width=0.8\linewidth]{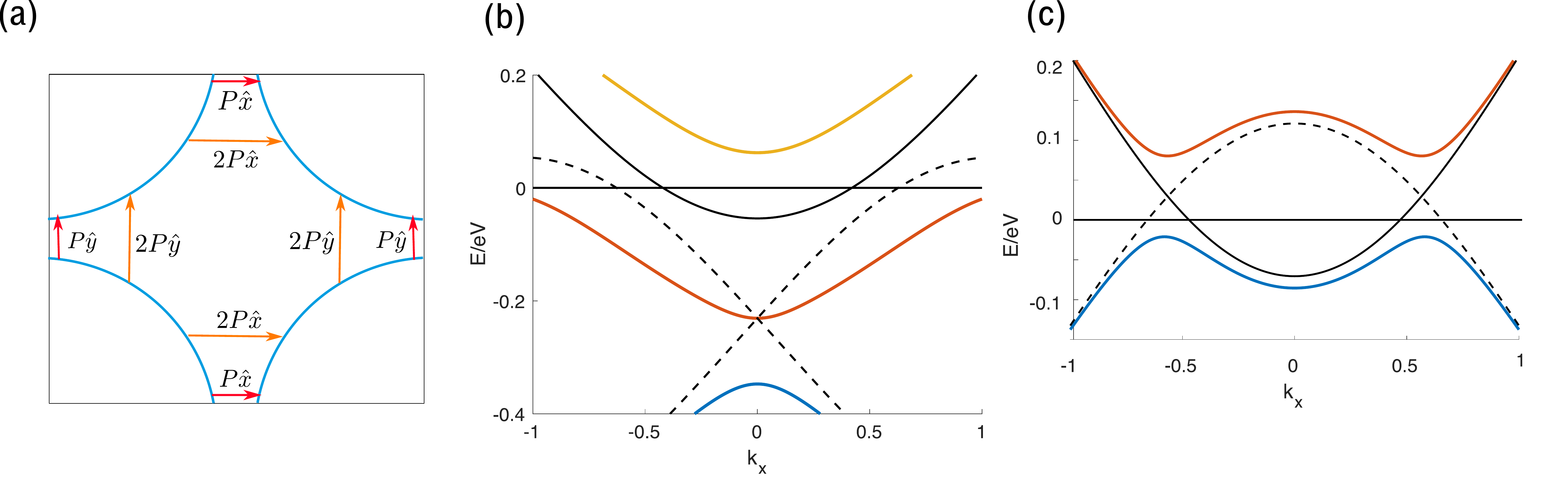}
\caption[Illustration of PDW pairing in cuprates]{(a) Illustration of the PDW momenta ($\pm P\hat{x}$ and $\pm P\hat{y}$) and the secondary CDW momenta ($\pm 2 P\hat{x}$, and $\pm 2P\hat{y}$) on the large Fermi surface. (b) Effects of the x-directional PDW along the line $k_y = \pi$ (c) Effects of the y-directional PDW along the line $k_y = \pi$. The original electron band ($\e_k$) is shown as the solid black lines. PDW reflected bands ($-\e_{\pm P\hat{x}-k}$ and $-\e_{\pm P\hat{y}-k}$) are shown as the dotted black lines. The hybridized Bogoliubov bands are shown in colors.}
\label{Fig: PDW 3-band illustration}
\end{center}
\end{figure}

Fig.~\ref{Fig: PDW 3-band illustration}(b) illustrates effects of x-directional PDW. We plot the energy of $c_{\vec{k}}$ (the original electron) as the solid black line, and energy of $c^{\dagger}_{\pm P\hat{x} - \vec{k}}$ as the dashed black lines. PDW hybridizes these three bands into the red and blue bands below the Fermi energy, and the yellow band above the Fermi energy. Fig.~\ref{Fig: PDW 3-band illustration}(c) illustrates the mixing between $c_k$ and $c^{\dagger}_{\pm P\hat{y} -k}$ under y-directional PDW. In this case $c^{\dagger}_{P\hat{y} -k}$ and $c^{\dagger}_{-P\hat{y} -k}$ happen to be degenerate, and the electron band effectively couples to only their equal-weight superposition. Hybridization of the electron band and this superposition gives the red band and the blue band.
\footnote{The asymmetric superposition of $c^{\dagger}_{\pm P\hat{y}-k}$ does not couple to the electron; therefore appears to stay gapless. But this is an artifact of the 3-band approximation. For example, the coupling between this band and $c_{\pm 2P\hat{x}+k}$ can gap it} 
For bidirectional PDW, PDW in x-direction and PDW in y-direction together open a gap at antinodes (if the PDW amplitude is big enough). Which one dominates depends on details of the band structure, and the pairing momentum.

Different from what is reported in Ref.~\cite{PhysRevX.4.031017} (where the effect of the y-direction PDW was not considered), we find that y-directional PDW generically contributes more to the spectral gap at or near $k_y=\pi$. This feature can also be seen in the recent work of Tu and Lee~\cite{tu2019}. In this scenario, as we gradually increase the PDW amplitude, the Fermi surface is gradually pushed towards larger absolute value of $k_x$ before the gap opens (Fig.~\ref{Fig: PDW 3-band illustration}(c)), while if the x-directional PDW dominates, we would see the Fermi surface pushed towards smaller $k_x$ and disappear at zero momentum (Fig.~\ref{Fig: PDW 3-band illustration}(b)).
In either case, as we move from $ky = \pi$ to $k_y = \pi/2$, at some point, PDW stops to provide a full gap. Because of the momentum-mismatch, PDW barely do anything to nodal electrons. For more details, see Ref.~\cite{PhysRevX.4.031017,PhysRevB.77.174502}. We remark that the addition of the y-direction PDW contribution shown in Fig.~\ref{Fig: PDW 3-band illustration}(c) has the desirable feature that the gap opens up for smaller pairing amplitude compared with the contribution from x-direction PDW alone.

In the analysis presented above, we have ignored higher order effects of PDW. For example, $c^{\dagger}_{P\hat{x} - k}$ also mixes with $c_{k - 2P\hat{x}}$. In general, we should consider the mixing between all of $c_{k + mP\hat{x} + nP\hat{y}}$ ($m+n$ even) and $c^{\dagger} _{-k + m'P\hat{x} + n'P\hat{y}}$ ($m'+n'$ odd).

As we consider this higher order mixing, the commensurability between the PDW momentum and the original lattice comes into play. In the rest of this section, we first discuss the band structure of a commensurate period-6 PDW and then briefly discuss the band structure of an incommensurate PDW close to period-8.

\section{Commensurate period-6 pair density wave}
\label{subsection: Mean-field PDW bands in cuprates}

\begin{figure}[htb]
\begin{center}
\includegraphics[width=0.7\linewidth]{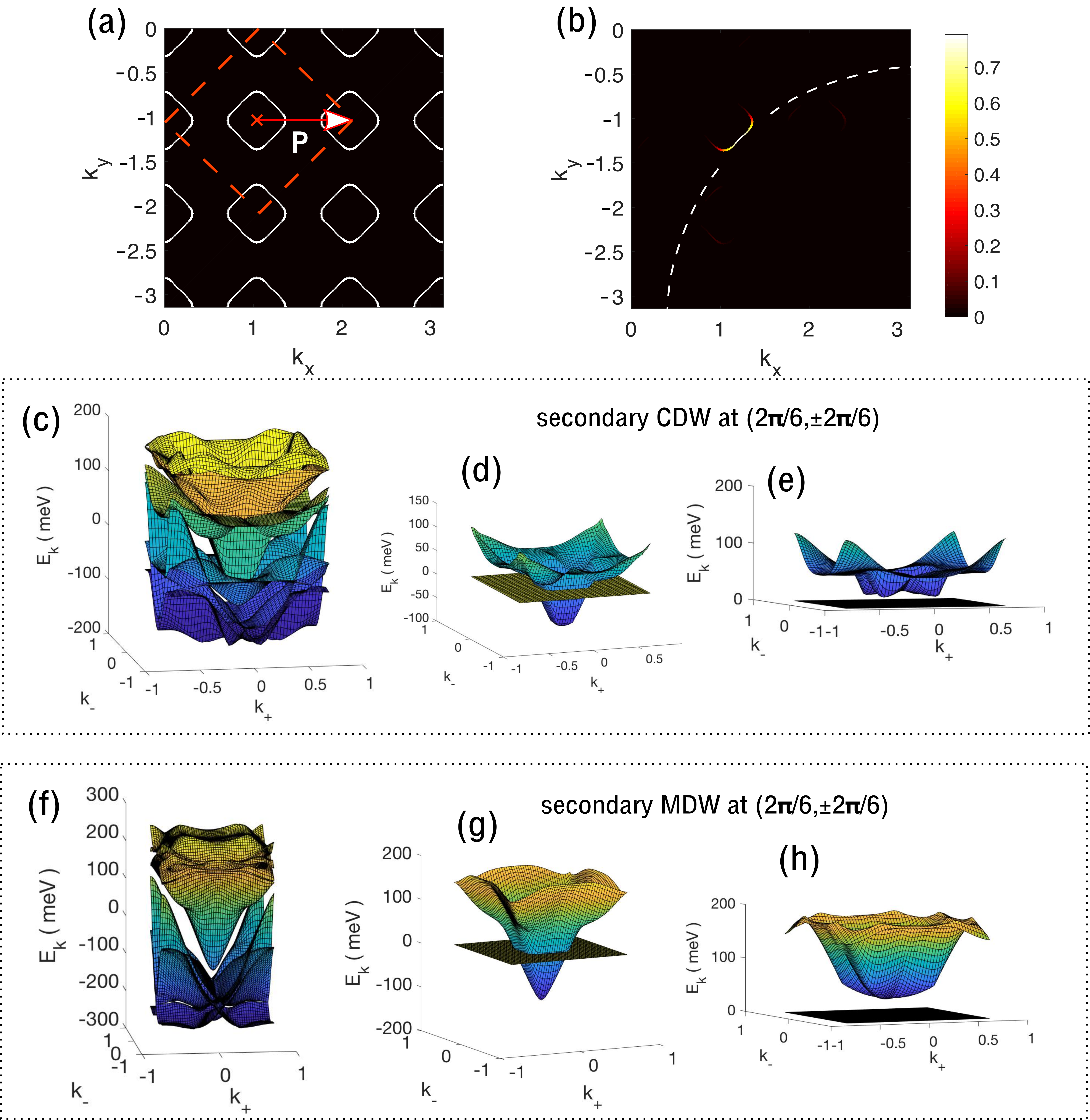}
\caption[Bogoliubov bands of commensurate PDW]{Bogoliubov bands of commensurate PDW. We use the mean-field Hamiltonian in Eq.~\ref{Eq: PDW mean field}, with hopping parameters $t = 154\text{meV}, t_p = -24\text{meV}, t_{pp} = 25\text{meV}, t_{ppp} = -5\text{meV}$ (see Eq.~\ref{Eq: tightbindingenergy}), chemical potential $\mu = -126\text{meV}$, PDW momentum $2\pi/6$, and PDW order parameter $|\D_P| = 40\text{meV}$. The original B.Z. is reduced to the small B.Z. spanned by $(\pi/3,\pm\pi/3)$. There are 36 bands coming from 18 electron bands and 18 hole bands in the reduced B.Z. Fig. (a): The gapless Fermi pocket. We plot the right-lower quadrants of the original B.Z., The red dashed line represents the reduced B.Z. The arrow represents PDW momentum ($P\hat{x}$ and $P\hat{y}$ are identical in the reduced B.Z.). Fig. (b): The spectral weight of zero-energy fermions in the original B.Z.. The white dashed line illustrates the large Fermi surface. Note that the new Fermi surface are mainly composed by the nodal portion of the original Fermi surface; its shape is barely changed by the PDW. Fig. (c): Bogoliubov bands close to Fermi energy. The PDW amplitudes are $\D_{P\hat{x}} = \D_{-P\hat{x}} = \D_{P\hat{y}} = \D_{-P\hat{y}}= 40\text{meV}$. This choice of phase produces CDW order at $(P,\pm P)$. $k_+$ and $k_-$ run between $\pm\pi/3\sqrt{2}$ along the diagonals. Bogoliubov bands appear in pairs: Each pair of bands have identical shape, they are related by a flip in energy (similar to the BCS bands) and a further shift by the PDW momentum. Fig. (d): the gapless band in Fig. (c). The horizontal plane represents the Fermi energy. Fig. (e): the first gapped band in Fig. (c). Fig. (f/g/h), the same as Fig. (c/d/e), except for $\D_{P\hat{x}} = \D_{-P\hat{x}} = \D_{P\hat{y}} = 40\text{meV}, \D_{-P\hat{y}}= -40\text{meV}$. This produces a magnetization density wave (MDW) state which orders at $(P,\pm P)$ and breaks time-reversal symmetry.}
\label{Fig: Bogoliubov bands}
\end{center}
\end{figure}

In this section, we focus on the commensurate case with $P = 2\pi/6$, which is relevant to YBCO near 8\% doping (period-3 CDW)~\cite{PhysRevB.96.134510}. The reduced B.Z. of non-superconducting density waves is spanned by $P\hat{x} \pm P\hat{y}$, with an area equal to 1/18 of the original B.Z. (red dashed square in Fig.~\ref{Fig: Bogoliubov bands}(a)). The 4 PDW momenta are all $(\pi,\pi)$ in the reduced B.Z.. The Hamiltonian we consider is 

\bea
H &=& \sum_{\vec{k},\s} \e_{\vec{k}} c^{\dagger}_{\vec{k},\s}c_{\vec{k},\s}\nonumber\\ 
&+& \sum_{\vec{k}} \D_{P\hat{x}}(\vec{k}) c_{\vec{k},\uparrow}c_{-\vec{k} + P\hat{x},\downarrow} + \D_{-P\hat{x}}(\vec{k}) c_{\vec{k},\uparrow}c_{-\vec{k} - P\hat{x},\downarrow} \nonumber\\
&+& \sum_{\vec{k}} \D_{P\hat{y}}(\vec{k}) c_{\vec{k},\uparrow}c_{-\vec{k} + P\hat{y},\downarrow} + \D_{-P\hat{y}}(\vec{k}) c_{\vec{k},\uparrow}c_{-\vec{k} -P\hat{y},\downarrow}\nonumber\\
&+& h.c.,
\label{Eq: PDW mean field}
\eea
where $\vec{k}$ runs in the original B.Z., and $\e_{\vec{k}}$ is the tight-binding dispersion: 

\bea
\label{Eq: tightbindingenergy}
\epsilon_k = &-&2t (\cos(k_x)+\cos(k_y)) - 4t_p\cos(k_x)\cos(k_y) \nonumber\\ &-&2t_{pp}(\cos(2k_x) + \cos(2k_y)) - \mu \nonumber\\ &-&4t_{ppp}(\cos(2k_x)\cos(k_y)+\cos(2k_y)\cos(k_x)).
\eea
For the choices of $t$, $t_p$, $t_{pp}$, $t_{ppp}$ and $\mu$, see the description of Fig.~\ref{Fig: Bogoliubov bands}. We choose a locally d-wave form factor for the PDW:

\bea
\label{Eq: PDW form factor}\D_{\vec{P}}(\vec{k}) = 2\D_{\vec{P}}[\cos(k_x-P_x/2) - \cos(k_y-P_y/2)]
\eea
Note that we do not include secondary orders like the CDW and the MDW explicitly in the Hamiltonian. However, the ground state of the Hamiltonian shows these secondary orders, for example $\<c_{\vec{k}}c^{\dagger}_{\vec{k}+\vec{P}\hat{x}+ \vec{P}\hat{y}}\>\neq 0$ and $\<c_{\vec{k}}c^{\dagger}_{\vec{k}+2\vec{P}\hat{x}}\>\neq 0$. These orders appear automatically as higher order effects of the PDW amplitudes.

As a general feature of the Nambu spinor representation, Bogoliubov bands of PDW shows up in pairs; each band has a partner that is flipped in energy and shifted by the PDW momentum.\footnote{For incommensurate PDW, we usually make an cutoff of higher order mixing which breaks this formal particle-hole symmetry (as shown in Fig.~\ref{Fig: PDW 3-band illustration}).} Of the 18 pairs of bands (coming from 18 electron bands and 18 hole bands), only 1 pair is gapless, giving 2 identical gapless Bogoliubov pockets in the reduced B.Z., shown in Fig.~\ref{Fig: Bogoliubov bands}(a).
\footnote{ All other bands are gapped out by the PDW as long as the PDW has a large amplitude and is bi-directional. See the description under Fig.~\ref{Fig: Bogoliubov bands} for details. Alternatively, we can reduce the PDW gap but explicitly add CDWs at momentum $2P$ to achieve similar results. On the other hand, bi-directional PDW is crucial in order to have only one pair of gapless bands. For a previous study of the band structure of unidirectional PDW with composite orders, see Ref.~\cite{PhysRevB.77.174502}. } 
However, the 2 pockets represent the same excitations. Counting the degrees of freedom, there is only one gapless pocket per spin. The reason is that the Nambu spinor representation shifts the down spin electrons by the PDW momentum, causing a superficial doubling. Physically, there are 2 pockets related by $(\pi,\pi)$ in the reduced B.Z. because momenta is conserved only up to $(\pi,\pi)$ when PDW is ordered. We shall see in Sec.~\ref{subsection: gapless sector} that after disordering the PDW, only the pocket at the center of the B.Z. left. The other pocket becomes a broad 2-particle continuum with a small gap.

Fig.~\ref{Fig: Bogoliubov bands}(b) shows the spectral weight of zero-energy electrons in the original B.Z.. We can see that gapless excitations come solely from nodal electrons along the original Fermi surface; anti-nodal electrons are all gapped. The CDW generated by the PDW connects the gapless arcs to form a closed pocket. Note that the effect of zone-folding in electron spectral function is visible only at the tips of the nodal arc, due to the fact that the CDW amplitude is much smaller than the hopping. On the contrary, if we were to gap out anti-nodal electrons by only CDW, we would need a CDW amplitude comparable to the hopping, resulting in an unrealistically large mixing between $c_k$ and $c_{k+2P}$.

By the approximate $C_4$ symmetry of the $\text{CuO}_2$ plane, we assume the 4 PDW order parameters in Eq.~\ref{Eq: PDW mean field} have about the same amplitude. However, different choices of the 4 phases give different ground-state energies and symmetries~\cite{agterberg2008dislocations}. Of the 4 phases, we can use the $U(1)$-charge symmetry to fix one. In the limit that the PDW wavelength is much bigger than the lattice spacing, we can use continuous translation in x and y direction to fix two more phases. In this case, the only nontrivial phase is $e^{i\theta}\equiv \D_{P\hat{x}}\cdot\D_{-P\hat{x}}/(\D_{P\hat{y}}\cdot\D_{-P\hat{y}})$. Time reversal symmetry requires it to be 1. Any other choice breaks time reversal (spontaneously). Fig.~\ref{Fig: Bogoliubov bands}(c) and Fig.~\ref{Fig: Bogoliubov bands}(f) shows the 8 bands close to the Fermi energy for $\theta = 0$ and $\theta = \pi$ correspondingly. The time-reversal invariant case ($\theta = 0$) has a CDW at momentum $(2\pi/6, \pm 2\pi/6)$ (App.~\ref{Appendix: Symmetry of the fluctuating PDW state}), which is apparently excluded by current experiments. The time-reversal breaking case ($\theta = \pi$) has a more stable band structure with a larger gap for the gapped bands (Fig.~\ref{Fig: Bogoliubov bands}(h)). In this case, the secondary order generated by PDW at momentum $(2\pi/6, \pm 2\pi/6)$ is purely current modulation without charge modulation. This orbital magnetization density wave (MDW) may also break the mirror symmetry along the diagonal. In each case, the specific band gap depends on the band structure and PDW order parameters, but the nodal pocket and the shape of bands are more robust. See Ref.~\cite{agterberg2008dislocations} and App.~\ref{Appendix: Symmetry of the fluctuating PDW state} for details on the symmetry of the commensurate and incommensurate PDW.
\clearpage

\section{Incommensurate pair density wave}\label{Sec: PDW with long range order}

When the PDW is incommensurate to the original B.Z., we need to set a cutoff in momentum-space calculation. It was previously reported in Ref.~\cite{PhysRevX.4.031017} by one of the author that a 5-band model describing the mixing of $c_{\vec{k}}$, $c^{\dagger}_{-\vec{k}+P\hat{x}}$, $c^{\dagger}_{-\vec{k}-P\hat{x}}$, $c_{\vec{k}+2P\hat{x}}$ and $c_{\vec{k}-2P\hat{x}}$ (similarly in y direction) produce Bogoliubov pockets with predominant electron weight on one side and predominent hole weight on the other side. In order to capture the effect of B.Z. folding caused by the subsidiary CDW, we increase the cutoff, and include the mixing among $c_{k+2mP\hat{x}+2nP\hat{y}}$ for $m,\ n$ up to $\pm 2$ (for details, see Appendix A). We used the Hamiltonian in Eq.~\ref{Eq: PDW mean field}, the PDW form factor in Eq.~\ref{Eq: PDW form factor} with $\D = 45$meV, the band structure in Appendix A and CDW momentum $2P\simeq0.28(2\pi/a)$ measured in Ref.~\cite{PhysRevB.96.134510}. This choice of the PDW momentum is relevant to Hg1201 and BSCCO.

We found that, the electron-like part of the 4 Bogoliubov pockets recombine into a predominantly electron-like pocket just like the pocket we get for the commensurate period-6 PDW. The only difference is that the incommensurate density waves produce higher-order repetitions of the pocket all over the original B.Z. (light blue segments in Fig.~\ref{Fig: PDW band stucture, pocket}). The spectral weights of these repetitions decrease fast as we go to higher-order mixing. We believe that this pocket formed by mainly electron like segments will give rise to quantum oscillations.

We would like to mention that as we increase doping, the 4 copies of the electron pockets in Fig.~\ref{Fig: PDW band stucture, pocket}(b) touch each other. In some parameter range, Fermi surface topology changes, and a hole pocket forms in the middle. This Lifshitz transition is predicted for Hg1201 at $10\%$ doping in Ref.~\cite{PhysRevB.96.134510}, and for YBCO at a larger doping. However, distinguishing subtle changes of Fermi surface topology is beyond the scope of our discussion.

\begin{figure}[htb]
\begin{center}
\includegraphics[width=6in]{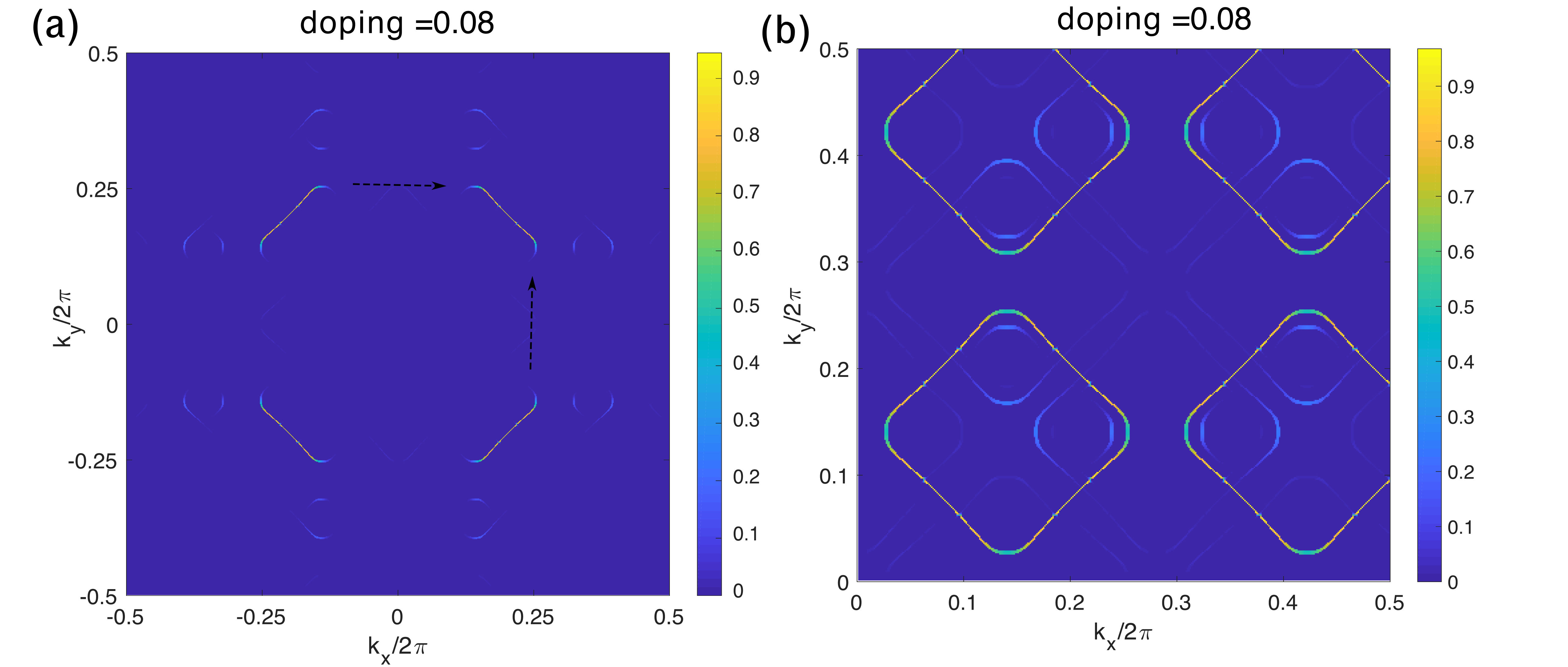}
\caption[Fermi pocket from incommensurate PDW]{Band structure of the Bogoliubov quasi-particle and possible Fermi pockets in a PDW state. (a)Electron weight on the Fermi-pocket of Bogoliubov quasi-particle. We used the band structure in Appendix A, CDW momentum $Q_x=Q_y=0.28(2\pi/a)$ measured in Ref.~\cite{PhysRevB.96.134510} PDW order parameter $\D_{Q/2}=45$meV, no explicit CDW order parameter in mean field Hamiltonian, and plotted the electron weight at Fermi energy and each momentum $k$ in the B.Z. (For details, see Appendix A). Electron weight is large on 4 ``arcs'' in the nodal direction. The anti-nodal direction is gapped out by PDW. (b) Details of the reconstructed electron-like pocket after B.Z. folding caused by CDW. We plotted the total electron weight at momenta up to $Q_x$ and $Q_y$. This pocket is formed by 4 segments with electron weight $>80\%$. It has the same shape as the Harrison-Sebastian pocket. Physically there is only one pocket, others are its copy shifted by $Q_x$ and $Q_y$. we only show the upper right quadrant of the B.Z.}
\label{Fig: PDW band stucture, pocket}
\end{center}
\end{figure}

\chapter{Short-range Pair Density Wave in the Vortex Halo}\label{chap:STMPDW}

In 2018, a scanning tunneling microscopy (STM) experiment reports the observation of charge density wave (CDW) with period of approximately 8a in the halo region surrounding the vortex core~\cite{edkins2019magnetic}, in striking contrast to the approximately period 4a CDW that are commonly observed in the cuprates. 

In the d-wave superconducting state, charge conservation symmetry is already broken; therefore a CDW and a PDW no longer have any symmetry distinction. However, we can still ask which of them is the primary order in the vortex halo. From the Landau theory point of view, it is most likely that the primary order parameter has an energy functional that favors nonzero amplitude and the subsidiary order parameter is only nonzero due to linear coupling with the primary order. Thus we have the CDW-driven scenario and the PDW-driven scenario. The PDW-driven scenario further divides into different possibilities regarding why the PDW appears in the vortex halo.

The main theme of this thesis, which we explain in Chapter~\ref{chap:PDWpseudogap}, is exploring the fluctuating PDW state as a mother state behind the pseudogap phenomena. In this chapter, we take a step back from our main theme, and address the adequacy of each of the following scenarios as the explanation of the double period CDW, put in the broader context of the pseudogap phenomenology. (We shall refer to the previously observed CDW momentum as $Q$, and the new CDW momentum as $Q/2$.)
\begin{enumerate}
    \item The Q/2 CDW is the primary order, while the Q CDW is subsidiary.
    \item The Q/2 PDW is a competing order, or an example of “intertwined order” where several order parameters such as PDW, CDW, SDW and d-wave pairing are intimately related to each other. In this picture, the PDW exists only in the vortex halo and vanishes outside.
    \item The PDW is the primary order, the “mother state” that exists at a high energy scale and lurks behind a large segment of the phase diagram in the temperature/magnetic field plane. In order to explain the pseudogap at the anti-nodes the PDW is assumed to be bi-directional.  While its order is destroyed by phase fluctuations, there are several subsidiary orders that emerge at lower temperatures which account for the observed complexity of the phase diagram. We shall also include a discussion of the canted PDW (see Fig.~\ref{Fig: PDW weird}).
\end{enumerate}

Throughout this chapter we assume the PDW to be bi-directional. A recent paper by Wang et al.\cite{wang2018} addresses issues related to the PDW in the STM experiment and there are similarities and differences with the present work. They consider the d-wave superconductivity and the PDW as competing states inside the vortex halo and construct a sigma model description combining the two orders. They focus their calculations to an uni-directional PDW. They argue against the persistence of the PDW outside the vortex halo. As such their picture is closer in spirit to scenario (2) as outlined above.

We study the charge density wave structures near the vortex core in these models. We emphasize the importance of the phase winding of the d-wave order parameter.  The PDW can be pinned by the vortex core due to this winding and become static. Furthermore, the period 8 CDW inherits the properties of this winding, which gives rise to a special feature of the Fourier transform peak. We propose it as the key experimental signature that can distinguish between the PDW-driven scenario from the more mundane option that the period 8 CDW is primary. We discuss the pro’s and con’s of the options considered above. 

Finally we attempt to place the STM experiment in the broader context of pseudogap physics of underdoped cuprates. We relate the STM observation to the unusual properties of X-ray scattering data on CDWs carried out to very high magnetic field. We discuss properties of the possible high-field pseudogap ground state for $H>H_\text{c2}$.

\section{Recent STM results on period-8 density wave}


First we give a short summary of the recent low temperature STM experiment in $\BSCCO$\cite{edkins2019magnetic}. The doping is about 0.17.  At zero field patches of 4a CDWs are observed. These appear locally uni-directional and have $d$ form factors. The correlation length is very short, about twice of the lattice spacing. At a finite field of 8.25T, by subtracting off the zero field data, period 4a and period 8a CDWs are revealed in the ``halo'' region around the vortex core. These appear to be bi-directional and have s-wave form factors. The signals are symmetric when the voltages are reversed. We distinguish bi-directional from "checkerboard" order, which consists of local patches of uni-directional stripes. From the widths of the Fourier transform peaks, the correlation length of the 8a and 4a CDW is about 8a and 4a respectively, comparable to their wavelengths.
By examining the signals that are odd upon reversing the voltage, another 4a CDW is found which has $d$ form factors. Its correlation length is about 5a and it is uni-directional, running in the same direction from vortex to vortex.

Purely on symmetry grounds, the observation of period 8a bidirectional charge order in the presence of a background superconductor implies that there are also period-8 modulations in the pair order parameter. Specifically if the Fourier component $\rho_{Q/2}$ of the density at a wave vector $Q/2$ is non-zero, then it implies a non-zero Fourier component $\Delta_{Q/2} \sim \Delta_d \rho_{Q/2}$ in the pairing order parameter (where $\Delta_d$ is the order parameter for the standard $d$-wave superconductor). An important question then is whether the observed period-8 modulations are driven primarily by the pinning of soft fluctuations of $\rho_{Q/2}$ (and $\Delta_{Q/2}$ is a subsidiary) or whether the driver is pinning of soft fluctuations of $\Delta_{Q/2}$ (and the observed $\rho_{Q/2}$ is a subsidiary).  We will call the former CDW-driven and the latter PDW-driven. Clearly this is not a symmetry-based distinction and it is natural to wonder if the question is meaningful at all.  However we will argue in this chapter that there are, in fact, two distinct possibilities for the observed period-8 charge order which have distinct experimental signatures. It is natural to associate these two distinct possibilities with the (looser) distinction between the CDW-driven and the PDW-driven mechanisms.

\section{Basic features of bi-directional pair density wave}

In this section, we explore the implications of the PDW-Driven scenario, and contrast it with the CDW-driven scenario. We will particularly emphasize the two distinct structures of the period-8 charge order and their experimental distinctions. 

\begin{figure}[htb]
\begin{center}
\includegraphics[width=3in]{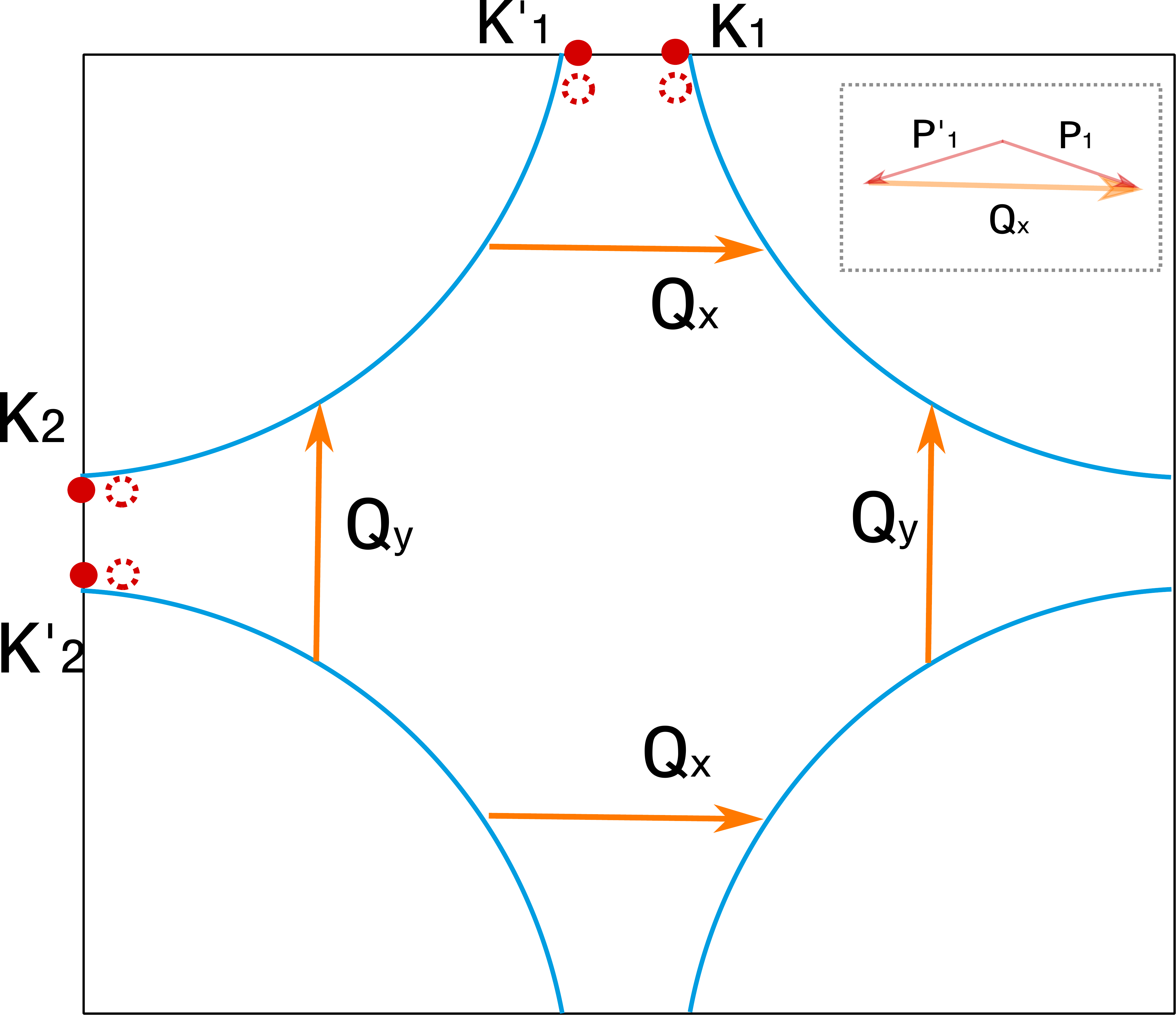}
\caption[Illustration of the bare Fermi surface, CDW momenta and PDW momenta]{Illustration of the bare Fermi surface, CDW momenta and PDW momenta. CDW momenta $Q_x$ and $Q_y$ are shown as yellow arrows. PDW momenta are $P_1 = 2K_1$, $P'_1 = 2K'_1$ in x direction, and $P_2 = 2K_2$, $P'_2 = 2K'_2$ in y direction. The CDW is a subsidiary order of the PDW, its momenta $Q_x = P_1 - P'_1$, $Q_y = P_2 - P'_2$. We consider two scenarios: (1) $K_i$ and $K'_i$ are located right at B.Z. boundary (solid red dots). $P_1 = -P'_1=Q_x/2$, $P_2 = -P'_2 = Q_y/2$. (2) $K_i$ and $K'_i$ are slightly shifted (dotted red circles); $P_1$ and $P'_1$ have a small y component, as shown in the inset figure (The small y component is exaggerated). } 
\label{Fig: PDW weird}
\end{center}
\end{figure}

The new CDW recently found in $\BSCCO$ has a momentum close to $2\pi/8$, half of the momentum of the well-known short-range CDW at zero field. In the PDW-driven scenario, we consider a bi-directional PDW order with the same momentum, that is roughly the momentum between tips of the bare Fermi surface in the anti-nodal direction\cite{PhysRevX.4.031017}. Bi-directional PDW state with such a momentum is previously proposed by one of the authors \cite{PhysRevX.4.031017}. Following this proposal, we write down a mean field Hamiltonian

\bea
H &=& \sum_{k,\s} \e_k c^{\dagger}_{k,\s}c_{k,\s}\nonumber\\ 
&+& \sum_{k} \D^*_{P_1}(k) c_{k,\uparrow}c_{-k + P_1,\downarrow} + \D^*_{P'_1}(k) c_{k,\uparrow}c_{-k + P'_1,\downarrow} \nonumber\\
&+& \sum_{k} \D^*_{P_2}(k) c_{k,\uparrow}c_{-k + P_2,\downarrow} + \D^*_{P'_2}(k) c_{k,\uparrow}c_{-k + P'_2,\downarrow}\nonumber\\
&+& h.c.
\label{Eq: long range PDW mean field}
\eea
We used the notation: $P_1 = 2K_1,\ P'_1 = 2K'_1$ --- as shown in Fig.~\ref{Fig: PDW weird} $K_1$ and $K'_1$ are located at or near the Fermi surface at anti-nodal points, generically incommensurate with the B.Z.; Similarly, $P_2 = 2K_2,\ P'_2 = 2K'_2$. The four PDW order parameters generate CDW orders $\rho_{Q_x}$ and $\rho_{Q_y}$ in second order perturbations even though we do not include them explicitly in the Hamiltonian. 
\bea
\rho_{Q_x}\sim \D_{P_1}\D^*_{P'_1},\ \rho_{Q_y}\sim \D_{P_2}\D^*_{P'_2}.
\label{Eq: subsidiary doubled CDW}
\eea 
We associate this subsidiary CDW as the well-known short-range CDW at zero field; it has momenta $Q_x = P_1 - P'_1$, $Q_y = P_2 - P'_2$, with magnitude $Q\simeq 2\pi/4$ in the recent STM experiment. In principle, we can also add a CDW in (1,1) direction, e.g. $\rho \sim \D_{P_1}\D^*_{P'_2} + \dots$. However, this CDW is absent in the recent STM experiment; we explain the reason in detail in the next subsection.

Naively one may expect that if the PDW has local $d$ form factor, the CDW generated by Eq.\ref{Eq: subsidiary doubled CDW} has $s$ form factor. This argument is not generally correct,  because $s$ and $d$ form factor for a finite-momentum order parameter has no sharp symmetry distinction \footnote{In momentum space, there are two amplitudes $A^x_a$ and $A^y_a$ at momentum $\mathbf{Q_a/2}$ which correspond to density waves in $x$ bond and $y$ bond.  Here $a$ denotes $x$ or $y$: $\mathbf{Q_x/2}=(\frac{2\pi}{8},0)$ and $\mathbf{Q_y/2}=(\frac{2\pi}{8},0)$. The definition currently used by the community is to define $A^x_a\pm A^y_a$ as the s/d-wave component. However, under $C_4$ rotation, $A^x_x$ transforms to $A^y_y$. Therefore the current definiton of s/d-wave form factor is not related to symmetry and generallly they should be mixed. An alternative definition of s vs d-wave component is $A^x_x \pm A^y_y$, which is related to the C4 rotation around a particular reference point. However, this definition may not be very useful because if we shift the reference point by half of the period in one direction, what we would define as d-wave would becaome s-wave.}

It is a local property, which is not captured by the long wavelength description of a Landau order parameter. In fact, when we solve our mean field Hamiltonian with only the $d$ wave PDW as input, the CDW that emerges at $Q$ is predominantly $d$ wave. In view of the experimental observation of the $s$ symmetry CDW near the vortex core, this may simply indicate that the mean field theory is not adequate to give a microscopic description. Nevertheless, we want to convey the message that this result shows that it is entirely possible that a $d$ wave CDW can emerge as a subsidiary order.

We define the common phases $\theta_{\text{P},x},\ \theta_{\text{P},y}$ and relative phases $\phi_x,\ \phi_y$ of the PDW order parameters, and the phases of Q CDW order parameters as

\bea
\D_{P_1} = |\D_{P_1}|e^{i(\theta_{\text{P},x} + \phi_x)}&,&\ \D_{P'_1} = |\D_{P'_1}|e^{i(\theta_{\text{P},x} - \phi_x)}\nonumber\\
\D_{P_2} = |\D_{P_2}|e^{i(\theta_{\text{P},y} + \phi_y)}&,&\ \D_{P'_2} = |\D_{P'_2}|e^{i(\theta_{\text{P},y} - \phi_y)}\nonumber\\
\rho_{Q_x} = |\rho_{Q_x}|e^{i\gamma_x}&,& \ \rho_{Q_y} = |\rho_{Q_y}|e^{i\gamma_y}, 
\eea
As shown in Eq.\ref{Eq: subsidiary doubled CDW}, $\gamma_x = 2\phi_x$ and $\gamma_y = 2\phi_y$ are the phase difference between PDW order parameters, hence the phases of the subsidiary CDW order parameter
\footnote{There is a redundancy in this definition: we can shift $\theta_x$ and $\phi_x$ ($\theta_y$ and $\phi_y$) both by $\pi$ without changing any physical order parameter. Thus, $\phi_x$ ($\phi_y$) is determined only up to $\pi$ without reference to the choice of $\theta_x$ ($\theta_y$).}; 
they are proportional to the shift of density wave pattern in real space. On the other hand, $\theta_{\text{P},x}$ and $\theta_{\text{P},y}$ carry charge 2 under external electromagnetic field; when coexist with uniform d-wave superconductivity $|\D_d|e^{i\theta_d}$, the relative phases $\theta_{\text{P},x}- \theta_d$ and $\theta_{\text{P},y} - \theta_d$, together with $\phi_x$ and $\phi_y$ determines the spatial pattern of new CDW orders with momenta $P_1, P'_1, P_2$, and $P'_2$, which are close to or equal to $Q/2$.

We consider two scenarios: (1) $K_i$ and $K'_i$, $i = 1,2$ are located at the boundary of B.Z., shown as solid red dots in Fig.~\ref{Fig: PDW band stucture, pocket}(a): $2K_1 = -2K'_1 = P_1 = -P'_1 = Q_x/2,\ 2K_2 = -2K'_2 = P_2 = -P'_2 = Q_y/2$ (2) $K_i$ and $K'_i$ are slightly shifted, shown as dashed red dots. The shifts in momenta can be either positive or negative, giving a $Z_2$ order parameter in each direction. We refer to this scenario as canted PDW. This possibility was discussed in Ref.~\cite{Agt2PhysRevB.91.054502} in relation with loop current. It has a potential ability to account for T-reversal breaking and nematicity. Regarding the recent STM experiment, these two scenarios give similar predictions. We focus on the first scenario and comment on the second when necessary.

Unlike the pairing in a conventional superconductor, where electrons forming a Cooper pair have opposite momenta and opposite velocity, this finite-momentum pairing groups electrons with momenta $K_i + \d k$ and $K_i - \d k$, (similarly, $K'_i + \d k$ and $K'_i - \d k$) and it has a strong effect only when these two momenta are both close to the Fermi surface. As a result, it opens a gap only in the anti-nodal direction (shown in Fig.\ref{Fig: PDW band stucture, pocket}), and leaves a gapless surface of Bogoliubov quasi-particle in the nodal direction. 

\subsection{Static short range pair density wave}
\label{subsec:staticshortrangePDW}

In this subsection, we discuss the situation where a short-range PDW coexists with d-wave superconductivity. We focus on the setup of the recent STM experiment where a period-8 density wave was found in the vortex halo of d-wave superconductor. To simplify the discussion, we consider the simplest scenario: $P_1 = -P'_1 = Q_x/2$, $P_2 = -P'_2 = Q_y/2$. We have 4 PDW order parameters: $\D_{\pm Q_x/2}$ and $\D_{\pm Q_y/2}$.

We consider the following couplings between PDW, d-wave, and CDW order parameters in a Landau theory in translation-invariant systems. We can write them in momentum space as

\bea
\D F = &-&a\rho_{Q_x}\D^*_{Q_x/2}\D_{-Q_x/2} - b \rho_{Q_x}[\D^2_d\D^{*2}_{Q_x/2}+\D^{*2}_d\D^{2}_{-Q_x/2}]\nonumber\\
&-& c\rho_{Q_x/2} [\D^*_d\D_{-Q_x/2} + \D^*_{Q_x/2}\D_d] -\dots ,
\label{Eq: Landau theory, momentum space}
\eea
where $a$, $b$,  $c$ are real coupling constants. For simplicity, we write down only couplings in x direction. Couplings in y direction are similar. These momentum-space couplings are conceptually helpful, but the strong breaking of translation symmetry introduced by the vortex core brings in new physics that are better captured by a real-space analysis.

Before we start, it is important to note that what the experimentalists found is not a long-range PDW or CDW. Instead, STM experiment identified a static short-range charge order that lives only inside the vortex halo, with the apparent correlation length comparable to its-wavelength. Theoretically, a ``short-range order'' naturally fluctuate with time; the existence of static short-range order raises many questions. What pins the phases of the order parameters? Why does it appear only in vortex halo? One may tend to think of a phase competition between the uniform d-wave superconductivity and the PDW, so that the latter may be greatly enhanced near the vortex core. However, a phase competition alone does not explain why the short-range order is static.

The answer of these questions may lie in the following observation: just like the way spatial inhomogeneity pins short-range CDWs, a spatial pattern of superconductivity close to the vortex core pins a short-range PDW. This static PDW then extends to a larger region with radius defined by its correlation length $\xi_P$. Outside $\xi_P$, there is still a PDW amplitude fluctuating with time, but the time average decays exponentially.

For concreteness, we choose the origin to be the center of the vortex, $(r,\theta)$ to be the polar coordinate, $(x,y)$ to be the Cartesian coordinate, and write down the following ansatz for the amplitude of the d-wave SC and the PDW:

\bea
\label{Eq: d-wave amplitude in real space}
\D_d(\mathbf{r}) &=& |\D_d(r)|e^{i\theta_d}e^{i\theta}\\
\D_\text{PDW}(\mathbf{r}) &=& 2|\D_{Q_x/2}| e^{-r/\xi_P} e^{i\theta_{P,x}}\cos(Qx/2 + \phi_{x})\nonumber\\
+ &2&|\D_{Q_y/2}| e^{-r/\xi_P} e^{i\theta_{P,y}}\cos(Qy/2 + \phi_{y}),
\label{Eq: PDW amplitude in real space}
\eea
where $|\D_d(r)|= r/\sqrt{r^2 + r^2_\text{core}}$. $e^{i\theta}$ encodes the $2\pi$ phase winding of the d-wave amplitude. We have three length scales. The radius of the vortex core: $r_\text{core}\simeq 3a$, the period of the PDW: $4\pi/Q\simeq 8a$, and the radius of vortex halo, where field-enhanced CDWs are found: we identify the halo size as $r_\text{halo}\sim\xi_P\sim 4\pi/Q$. A usual Landau theory with slowly-varying order parameters implicitly assumes that $r_\text{core}\gg 4\pi/Q$, $\xi_P\gg 4\pi/Q$. However, we are in the opposite limit: $4\pi/Q \sim \xi_P > r_\text{core}$. 

Since $\xi_P$ and $4\pi/Q$ are close to each other, and they are one order of magnitude larger than the lattice constant, we do not separate the exponential decay of order parameters $\D_{\pm Q_x/2}$ ($\D_{\pm Q_y/2}$) from the oscillatory part $\cos(Qx/2 + \phi_{x})$ ($\cos(Qy/2 + \phi_{y})$), as in a usual Landau theory. Instead, we take the ansatz in Eq.~\ref{Eq: d-wave amplitude in real space} and Eq.~\ref{Eq: PDW amplitude in real space}, and  write down their couplings in real space together with charge density profile $\rho(r)$. 

\bea
\label{Eq: Landau theory, real space}
\D F = &-&\int \{a\rho(\mathbf{r})\D_\text{PDW}(\mathbf{r}) \D^*_\text{PDW}(\mathbf{r})\nonumber\\
&+& b\rho(\mathbf{r})[\D^2_d(\mathbf{r}) \D^{*2}_\text{PDW}(\mathbf{r}) + \D^{*2}_d(\mathbf{r}) \D^2_\text{PDW}(\mathbf{r})]\nonumber\\
&+& c\rho(\mathbf{r})[\D^*_d(\mathbf{r})\D_\text{PDW}(\mathbf{r}) + \D_d(\mathbf{r})\D^*_\text{PDW}(\mathbf{r})]\nonumber\\
&+&s[\D^*_d(\mathbf{r})\D_\text{PDW}(\mathbf{r}) + \D_d(\mathbf{r})\D^*_\text{PDW}(\mathbf{r})]\}d^2 \mathbf{r}
\eea
We would like to remind the readers again that this free energy is not a Landau free energy in the usual sense, since we include the oscillatory part of the PDW explicitly in $\D_\text{PDW}(\mathbf{r})$.

The last term in Eq.~\ref{Eq: Landau theory, real space}: $-s\int \D^*_d(\mathbf{r})\D_\text{PDW}(\mathbf{r}) d^2 \mathbf{r} + c.c.$ is the lowest-order symmetry-allowed term that describes the phase locking between the PDW and the d-wave SC order parameter near a vortex core. In the case of spatially slowly-varying order parameters, this term usually vanishes because of momentum mismatch, eg. if the d-wave superconductivity has uniform amplitude. However, close to the vortex core, the rapid changing of d-wave amplitude strongly breaks translation symmetry. Furthermore the phase winds by $2\pi$ around the core, and near the core the winding is sufficiently rapid that it can phase match the finite wave-vector of the PDW. As a result the PDW is pinned to match the spatial pattern of vortex core so that free energy is minimized.

\begin{figure}[htb]
\begin{center}
\includegraphics[width=3in]{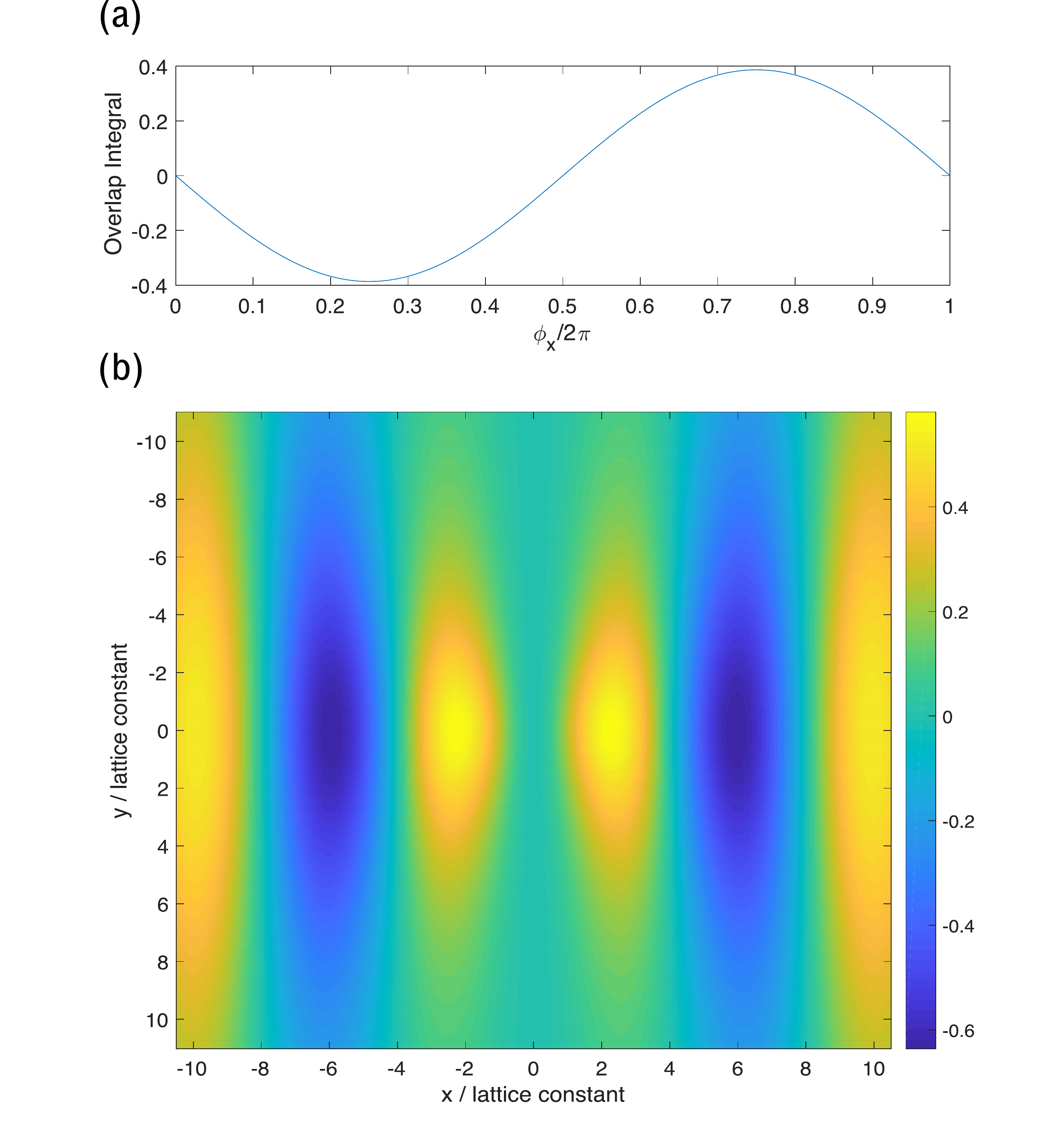}
\caption[Overlap integral between the PDW and the d-wave SC near a vortex core]{(a) Overlap integral $\int \D^*_d(\mathbf{r})\D_\text{PDW}(\mathbf{r}) d^2 \mathbf{r}$ as a function of $\phi_{x}$, for $\theta_x = 0$. We have set their maximum amplitude to 1 for both $\D_d(\mathbf{r})$ and $\D_\text{PDW}(\mathbf{r})$, and we normalize the integral by the overlap of the PDW with itself inside the vortex core of radius $3a$. $\phi_x$ is pinned to $3\pi/2$. The large overlap implies the real-space pattern of the PDW matches the pattern of d-wave vortex core almost perfectly at $\phi_x = 3\pi/2$ --- its amplitude is reduced only because d-wave amplitude is reduced in the vortex core. (b) The integrand $\D^*_d(\mathbf{r})\D_\text{PDW}(\mathbf{r})$ as a function of $\mathbf{r}$ near vortex core, for $\phi_{x} = 3\pi/2$, $\theta_x = 0$. Outside the vortex core, the integrand alternating between positive and negative because of momentum mismatch. However, within the first period of the PDW in the center, the integrand is always positive, giving a large overlap. This is because d-wave and PDW both change sign across the origin. The d-wave SC amplitude changes sign due to the $2\pi$ phase winding, and the PDW changes sign because of the $\sin(Qx/2)$ factor.} 
\label{Fig: pinning by overlap}
\end{center}
\end{figure}

Because of the phase winding, d-wave amplitude changes sign across the origin and the overlap integral is optimized when the PDW has the form $\sin(Qx/2)$ which also changes sign at the origin. Thus $\phi_{x}$ and $\phi_{y}$ are pinned to be $-\pi/2$.
Then the overall phase, $\theta_{P,x} = \theta_d,\ \theta_{P,y} = \pi/2 + \theta_d$, are pinned so that the overlap is a positive real number. This pinning mechanism completely fixes the phases of the PDW; a simple calculation of the overlap integral indicates the pinning is very effective in the vortex core. For details, see Fig.~\ref{Fig: pinning by overlap}. 

Of course, at the length scale of 10 lattice constants, everything except a microscopic model is merely an oversimplified illustration. Nonetheless, we believe this simple illustration captures the underlying physics of the phase-locking between d-wave and various PDW order parameters. This pinning mechanism is effective exactly because $4\pi/Q > r_\text{core}$ in the cuprates. In the opposite limit, d-wave order parameter changes slowly. According to a usual Landau theory, this coupling cancels out. In the remaining part of this section, we discuss the consequences of this phase-locking on subsidiary charge order. We confirmed these consequences by an exact diagonalization study in the next section.

Note that the PDW does not have a vortex. Since the PDW lives only in small patches, vortices are not required\cite{agterberg2015checkerboard}, and it is energetically favorable to not have vortices in the PDW-driven scenario.

This PDW order generates various CDWs in the vortex halo:

(1) bi-directional Q/2 CDW. According to Eq.~(\ref{Eq: Landau theory, momentum space} -~\ref{Eq: Landau theory, real space}), it has the following amplitude in real space
\bea
\label{Eq: CDW in real space}
\rho_{\a}(\mathbf{r}) =  F(r)\cos(\theta + \theta_d - \theta_{P,\a})\cos(Q_{\a}\cdot\mathbf{r} + \phi_{\a})
\eea
where $\a = x, y$, $Q_\a = Q_x, Q_y$ and $F(r) \sim 2c|\D_d(r)\D_{Q_{\a}/2}| e^{-r/\xi_P}$. The most interesting feature is that, apart from normal plane-wave factor, there is an additional factor $\cos(\theta + \theta_d - \theta_{P,\a})$ depending on the polar angle. A choice of the relative angle $\theta_d - \theta_{P,\a}$ selects a special angle along which $\rho_{\a}(\mathbf{r})$ vanishes. We point out that the pinning mechanism we discussed predicts that the amplitude $\rho_x$ vanishes in the vertical direction, when $\theta\sim\pm \pi/2$, while the amplitude $\rho_y$ vanishes in the horizontal direction, when $\theta\sim 0,\pi$. This choice restores C4 symmetry. Physically, this new feature originates from the $2\pi$ winding of d-wave order parameter. We can identify two contributions to $\rho_{Q/2}$: $\D^*_d\D_{Q/2}$ which carries -1 dislocation, and $\D_d\D^*_{-Q/2}$ which carries +1 dislocation. The interference of these two terms give rise to a nodal direction in real space. This is an important prediction in the PDW-driven scenario. On the contrary, in the CDW-driven scenario, it is energetically favorable to put the dislocation in the PDW amplitude, and the CDW amplitude is rather featureless. In the next section, we discuss the same feature in Fourier space, and propose follow-up experiments to distinguish the PDW-driven and the CDW-driven scenario.

(2) Q CDW. According to Eq.~\ref{Eq: Landau theory, momentum space} there are two contributions: 
\bea
\rho^A_Q \sim a\D^*_{-Q/2}\D_{Q/2},
\label{Eq: CDWA}
\eea
which we call $\text{CDW}_A$, and 
\bea
\rho^B_Q \sim b(\D^{*2}_d\D^2_{Q/2} + \D^{2}_d\D^{*2}_{-Q/2}),
\label{Eq: CDWB}
\eea 
which we call $\text{CDW}_B$, which we can think of as a harmonic of $\rho_{Q/2}$. $\text{CDW}_A$ does not rely on the phase-locking between the d-wave SC and the PDW; it is already pinned to be static short-range CDWs by impurities at zero magnetic field, and it persists above $\text{T}_\text{c}$. On the other hand, a static $\text{CDW}_B$ rely on the phase-locking. Similar to the Q/2 CDW, it is a superposition of +2 dislocation and -2 dislocation, and it exists only in vortex halos. In the case of spatially uniform PDW and CDW orders, there is no distinction between the two. However, in a spatially inhomogeneous situation such as what we encounter near the vortex core, there is a physical distinction. For example, $\text{CDW}_A$ may be extended in space while $\text{CDW}_B$ may be localized near the vortex core. In this case the two CDW may have different local form factors, such as d or s-wave. These form factors may in turn determine which one prefers to be bi-directional or uni-directional, because the coefficient of the quartic term that couples the amplitudes of the x and y oriented CDW may be different. In the STM data there already appears to be two kinds of CDWs , one pinned  to the vortex core and one which already exists at zero filed. We will make further use of this distinction in later discussions.

Naively, one would expect a CDW with momentum $(Q/2,Q/2)$ appears in the second order --- in real space this term may show up in the contribution $\rho(\mathbf{r})\sim a\D_\text{PDW}^*(\mathbf{r})\D_\text{PDW}(\mathbf{r})$. However, the pinning in the vortex core requires 

\bea
\D_\text{PDW}(\mathbf{r}) \sim e^{i\theta_d}(\sin(Qx/2) + i\sin(Qy/2)),\\
\D_\text{PDW}^*(\mathbf{r})\D_\text{PDW}(\mathbf{r}) \sim \sin^2(Qx/2) + \sin^2(Qy/2),
\eea

and the cross term $\sin(Qx/2)\sin(Qy/2)$ with momenta $(\pm Q/2, \pm Q/2)$ cancels out due to the $\pi/2$ relative phase. As a consequence, there is no $(2\pi/8, 2\pi/8)$ CDW in the leading order. In the fourth order, such a CDW is generated by the term $\D_{d}^{*2}(r)\D^2_\text{PDW}(r)$, but the amplitude is weak and subject to broadening effect given by dislocations. The absence of the $(2\pi/8, 2\pi/8)$ CDW is previously discussed in Ref.~\cite{agterberg2008dislocations}. It was pointed out that in the uniform case when the PDW does not have a vortex, the relative phase between the PDW in x and y direction determines whether the $(2\pi/8, 2\pi/8)$ CDW is present or not. If the phase is zero it is present, while if it is  $\pi/2$  bond currents are generated, producing a magnetization density wave at the same wave-vector instead. This magnetization density wave will be discussed in great detail in a later section. In the uniform case it is not known which phase is preferred. In our case we find that in the presence of a vortex, the phase choice $\pi/2$ is energetically favorable, therefore the $(2\pi/8, 2\pi/8)$ CDW is absent in the leading order. On the contrary, in the CDW-driven scenario, naively the $(2\pi/8, 2\pi/8)$ CDW is comparable to the $(2\pi/4, 0)$ CDW. The absence of a $(2\pi/8, 2\pi/8)$ Fourier peak in STM data is an evidence favoring the PDW-driven scenario.

Next, we would like to comment on the correlation length of the PDW in the recent STM experiment. In the PDW-driven scenario, as discussed above, the Q/2 CDW has $2\pi$ phase winding around the vortex core. A simple calculation shows that this phase winding broadened the Fourier peak by roughly a factor of 2. Thus the intrinsic correlation length of the Q/2 CDW and the PDW should be close to 16 lattice constants, a little smaller than the half of the distance between neighboring vortex cores.

We end this section with some comments on the implications if a canted PDW is present. While the CDW generated by Eq(2) retains the wave-vector Q along the x and y axes, the double period CDW generated by the analog of the third term in Eq.~\ref{Eq: Landau theory, momentum space} now has-wave-vector $P$ and $P'$. Similarly, its harmonic generated by the analog of the second term in Eq.~\ref{Eq: Landau theory, momentum space} have wave-vectors $2P$ and $2P'$. It is worth noting that we now have two distinct CDWs and the difference between the A and B type CDW is now a sharp one that can be made even in a uniform system. A second point is that there is now an additional pinning mechanism. The term $(\Delta_d e^{i\theta(\mathbf{r})})^2 (\Delta_P \Delta_{P'})^*$ is allowed if the local phase gradient matches the canting momentum $p = (P+P')/2$. This leads to a locking term at some distance from the vortex core where the phases are matched. The possible detection of the canting angle will be discussed in the next section.

With the above understanding of the PDW-driven scenario, we propose the following phenomenological picture explaining the recent STM experiment in $\BSCCO, \ 17\%$ doping, up to 8.5T:

\begin{itemize}
\item A short-range PDW is pinned by the vortex core and extends to its correlation length.
\item We estimate the intrinsic correlation length of the PDW to be 16 lattice constants. The period-8 CDW appears to have a shorter correlation length $\sim$ 8 lattice constants as determined from the width of the Fourier transform peak by fitting it to a Gaussian. Part of this width is not intrinsic and is due to the $2\pi$ phase winding.
\item The period 8 CDW produces as a harmonic a period 4 CDW, which we have labeled as $\text{CDW}_B$. Its width is subject to the same blurring as the period 8 CDW. On the other hand, the static PDW near the vortex core nucleates the period-4 $\text{CDW}_A$ by $\D^*_{-Q/2}\D_{Q/2}$, which is not affected by the phase winding around the vortex. These two CDWs may have different form factors and different asymmetry factors between x direction and y direction. However it is hard to extract their correlation length separately based on the current data, since their Fourier peaks mix together. The width of $2\pi/4$ Fourier peak translates to a correlation length around 4a. This serves as a lower bound of the intrinsic correlation lengths of $\text{CDW}_A$ and $\text{CDW}_B$.
\item At zero field, $\D_{-Q/2}$ and $\D_{Q/2}$ fluctuate with time, we rely on their relative phase
being pinned by spatial inhomogeneity to give a static $\text{CDW}_A$. This effect gives much weaker period-4 CDW puddles with a very short correlation length of order 2a. This CDW is unidirectional in each small puddle. We tentatively identify the unidirectional part of CDWs both in zero field and in the vortex core as $\text{CDW}_A$.
\item The static-PDW-enhanced correlation length of $\text{CDW}_A$ is enough to give some overlap between neighboring vortices. It is energetically favorable for the unidirectional part to align its direction and stretch its phase between vortices smoothly to gain overlap energy.
\item The PDW-driven model predicts the absence of a $(2\pi/8,2\pi/8)$ peak.
\item Given the strong pinning effect and relatively small correlation length, these CDWs may not be able to overcome the local pinning effect and become phase coherent between halos.
\end{itemize}

\section{Experimental proposal}
The disappearance of the $(\frac{2\pi}{8},\frac{2\pi}{8})$ CDW order is surprising for a CDW-Driven model while it can be naturally explained in the PDW-Driven model, as shown in last section. Despite this already existing evidence favoring the PDW-Driven model, more experimental predictions need to be tested to fully settle down this issue.  In this section we propose experiments to distinguish the PDW-Driven and the CDW-Driven scenario unambiguously. Besides, in the PDW-Driven scenario  our proposed experiment can extract the relative phase between PDW order parameters and the $d$ wave order parameter, which is physical.

The main prediction of the PDW-Driven scenario is that the CDW order parameter at $Q_x/2=(\frac{2\pi}{8},0)$ and $Q_y/2=(0,\frac{2\pi}{8})$ have the following profile as shown in Eq.~\ref{Eq: CDW in real space}
\begin{equation}
	\mathlarger \rho_{\mathbf {Q_{\a}/2}}(r,\theta)=e^{i\phi_a}F_P(r) \cos(\theta-\theta_a) 
	\label{eq:pdw_profile}
\end{equation}
where$(r,\theta)$ is the polar coordinate of real space around the vortex center and $a$ denotes $x$ or $y$ direction. 
 
$F_P(r)$ vanishes at $r=0$ and decays as $e^{-\frac{r}{\xi}}$ at large $r$.  It has its maximum at a nonzero distance to the center. $\theta_x=\theta_{P_x}-\theta_d$ and $\theta_y=\theta_{P_y}-\theta_d$ are the relative phases of PDW order parameters $\Delta_{\pm P_a}=|\Delta_{P_a}|e^{i\theta_{P_a}\pm i \phi_a}$ compared to the d-wave order parameter  $\Delta_D(r,\theta)=|\Delta_D|e^{i\theta_d}e^{i\theta}$

In contrast, the CDW-Driven scenario shows quite distinct profile of the period $8$ CDW order parameter:
 \begin{equation}
 	\mathlarger \rho_{\mathbf{Q_a/2}}(r,\theta)=e^{i\phi_a}F_c(r)
 	\label{eq:cdw_profile}
 \end{equation}
$F_c(r)$ has its maximum at $r=0$ and decays far away with $e^{-\frac{r}{\xi}}$. The CDW order parameter doesn't have angle dependence in this scenario.

\begin{figure}[H]
\centering
  \begin{subfigure}[b]{0.45\textwidth}
    \includegraphics[width=\textwidth]{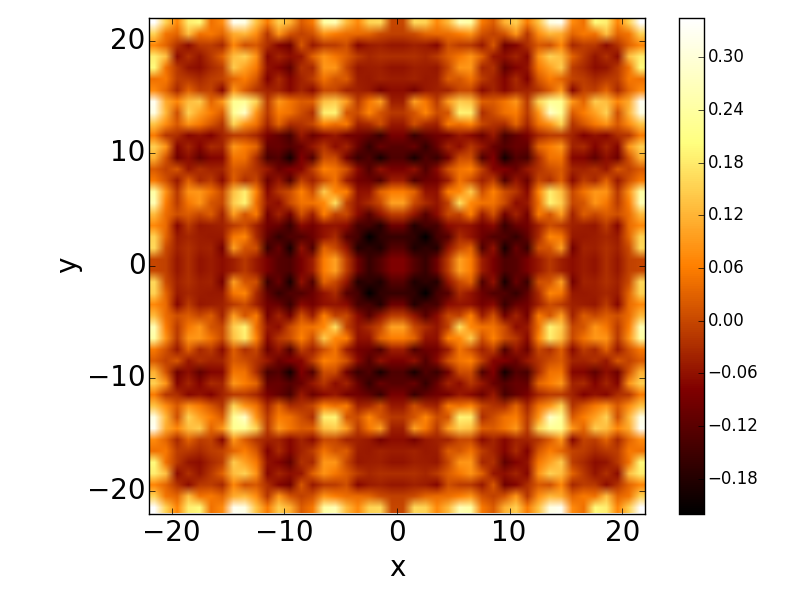}
    \caption{PDW-Driven}
    \label{fig:real_pdw}
  \end{subfigure}
  \begin{subfigure}[b]{0.45\textwidth}
    \includegraphics[width=\textwidth]{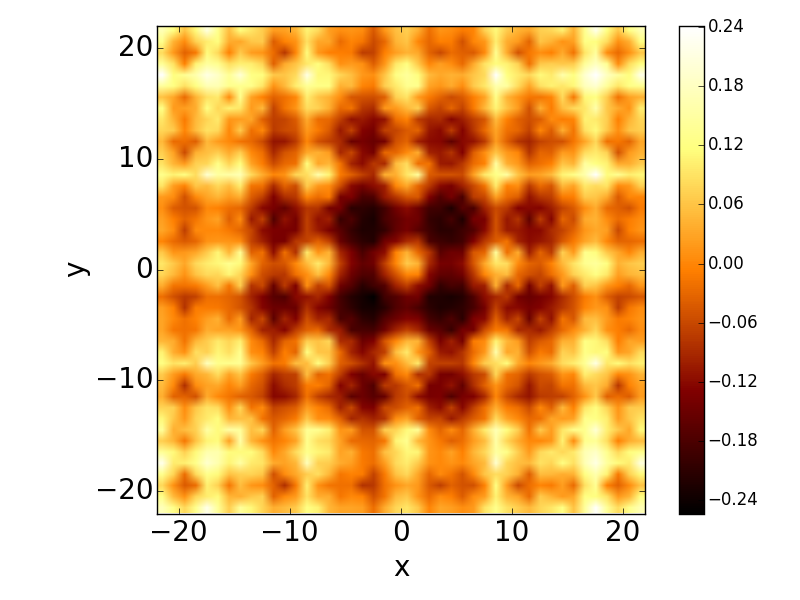}
    \caption{CDW-Driven}
    \label{fig:real_cdw}
  \end{subfigure}

  \caption[Real Space Plot of on-site LDoS]{Real Space Plot of on-site LDoS $\nu^E(\mathbf{r})$ at $E=30$meV for the PDW-Dirven and the CDW-Driven model.}
  \label{fig:real_space}
\end{figure}

 Clearly the CDW order parameter profile from the PDW-Driven and the CDW-Driven models have both different radius dependence and angle dependence. A real space plot of LDoS can be found in Fig.~\ref{fig:real_space}. The $\cos(\theta-\theta_a)$ factor in the PDW-Driven model means a superposition of strength $\pm 1$ dislocation of the CDW order parameter and in principle STM experiments can extract $\theta_a$.

 Here we will propose the following experimental predictions to distinguish the above two different CDW profiles. In the STM experiment, what is measured is the local density of states (LDoS) at a fixed energy $\nu(\mathbf{r},E)$.  For a fixed energy, $\nu^E(\mathbf{r})=\nu(\mathbf{r},E)$ has the same symmetry as density and we expect it to follow Eq.~\ref{eq:pdw_profile} and Eq.~\ref{eq:cdw_profile}.

Before going to specific predictions, it may be worthwhile to give one general suggestion to the data analysis procedure of experimental data. For both the PDW-Driven scenario and the CDW-Driven scenario, the phase of the CDW order with momentum $Q_a$ is expected to be locked to position of the vortex center. As a result, signals from different vortex halos are not coherent. Therefore, it's better to shift the position of each vortex center to the origin when doing Fourier Transformation for each vortex halo.  In this way we can make different vortex halos coherent and greatly enhance signals.

The following are predictions for the PDW-Driven scenario and how to detect it in experiment. As a benchmark, we show our numerical simulation data. We did a quantum-mechanical calculation with spatially varying PDW and d-wave SC amplitudes. The method of our simulation is summarized in Appendix C. Profile of the d-wave order parameter is $\Delta_D(r,\theta)\sim \frac{r}{\sqrt{r^2+r_0^2}}$ with vortex core size $r_0=3.5$ lattice constants. We used a profile of the PDW with $r$ dependence as $\Delta_P(r,\theta)\sim e^{1-\sqrt{r^2+\xi^2}/\xi}$ with correlation length $\xi=15$. In the following, local density of states $\nu^E(r)$ is obtained at fixed energy $E=30$ meV. Note we only show $d$ wave form of Bond LDoS because the CDW generated by our model is dominated by $d$ wave. However, we expect our predictions in the following sections do not rely on the form factor.

\subsection{Split peaks for period 8 CDW}

The first prediction for the PDW-driven scenario is that the peak at $\mathbf{Q_a/2}$  is split to two peaks in the direction decided by $\theta_a$.  

Recall that the density modulation $\mathlarger \rho(\mathbf r)=\int^0_{-\infty} dE \nu^E(\mathbf r)$ is given by the integral of LDoS $\nu^E(\mathbf r)$ over the occupied states. We define the slowly varying complex amplitude $\nu^E_{\mathbf{Q_a/2}}(\mathbf r)$ by writing the real space local DoS as $\mathlarger \nu^E(\mathbf r)=\sum_{a}\mathlarger \nu^E_{\mathbf{Q_a/2}}(\mathbf r)e^{\frac{1}{2}i\mathbf{{Q}_a}\cdot \mathbf{r}}+h.c.$. This is the analog of $\mathlarger\rho_{\mathbf{Q_a/2}}$ discussed in the last section.  We assume that $\nu^E_{\mathbf{Q_a/2}}(\mathbf r)$ has a similar real space profile as $\mathlarger\rho_{\mathbf{Q_a/2}}$ as given in Eq.~\ref{eq:pdw_profile}, i.e. it is confined to the vicinity of the vortex core and importantly, is proportional to $\cos(\theta-\theta_a)$. Recall that this factor encodes the phase winding of the d-wave superconductor and is therefore an important signature for the PDW-driven scenario. This assumption is supported by our numerical simulations, and will be  discussed and shown in greater detail later in Fig.~\ref{fig:pdw_angle} and Fig.~\ref{fig:local}.

We define $\tilde \nu^E(\mathbf q)$ to be the Fourier Transform of $\nu^E(\mathbf r)$. For $\mathbf q$ in the vicinity of $\mathbf{Q_a/2}$ we define
\begin{equation}
 	\tilde A_a(\mathbf q)=\tilde \nu^E(\mathbf q-\mathbf{Q_a/2})=\sum_{\mathbf{r}}\mathlarger \nu^E_{\mathbf{Q_a/2}}(\mathbf r)e^{-i \mathbf{q}\cdot \mathbf{r}}
 	\label{eq:conv}
 \end{equation}
Consider $a$ in the $x$ direction.  When $\theta_a=0$, it's easy to see that the absolute value of $\tilde A_a(\mathbf q)$ has two peaks in $x$ direction because of the $\cos \theta$ factor. 
This is because $\cos \theta=\frac{x}{\sqrt{x^2+y^2}}$ produces a line of zero in  $\nu^E_{\mathbf{Q_a/2}}(\mathbf r)$ along the $y$ direction through the vortex core. 
$\nu^E_{\mathbf{Q_a/2}}(\mathbf r)$ is odd under $x \rightarrow -x$ and as a result 
$\tilde A_a(\mathbf{q_x}=0)=0$ 
and $\tilde A_a(\mathbf q)$ has a splitting along the $\mathbf{q_x}$ direction.   
 The splitting is roughly $\delta q \sim \frac{1}{\xi}$. For general $a$ and general $\theta_a$, the line of zero in $\tilde A_a(\mathbf q)$ is rotated by an angle $\theta_a$.
  Therefore, the absolute value of $\mathlarger{\tilde \nu^E}(\mathbf q)$ should have two  peaks at $\mathbf{q}\approx\mathbf{Q_a/2}$ with the splitting in the direction of $\theta_a$.

This prediction is confirmed by our numerical simulation. Here we show two different phase choices for the PDW-Driven model.  The splitting of period $8$ peaks along the direction $\theta_a$ is very clear for PDW-Driven models while the CDW-Driven model shows regular peaks.

Therefore, we suggest to fit experimental data with a split-peak model.  In our simulation, if we choose the vortex center as the origin, we found that $\tilde \nu^E(\mathbf q)$ is dominated by real part. Thus it is better to plot only real part of $\tilde \nu^E(\mathbf q)$. Besides, there should be a sign change at $\mathbf q=(\frac{1}{8}\frac{2\pi}{a},0)$ if we plot Re$\nu^E(q_x)$ along the $q_y=0$ cut, as shown in Fig.~\ref{fig:fftx}. Again, this comes from the Fourier transformation of $\cos(\theta)$.

\begin{figure}[htb]
\centering
  \begin{subfigure}[b]{0.45\textwidth}
    \includegraphics[width=\textwidth]{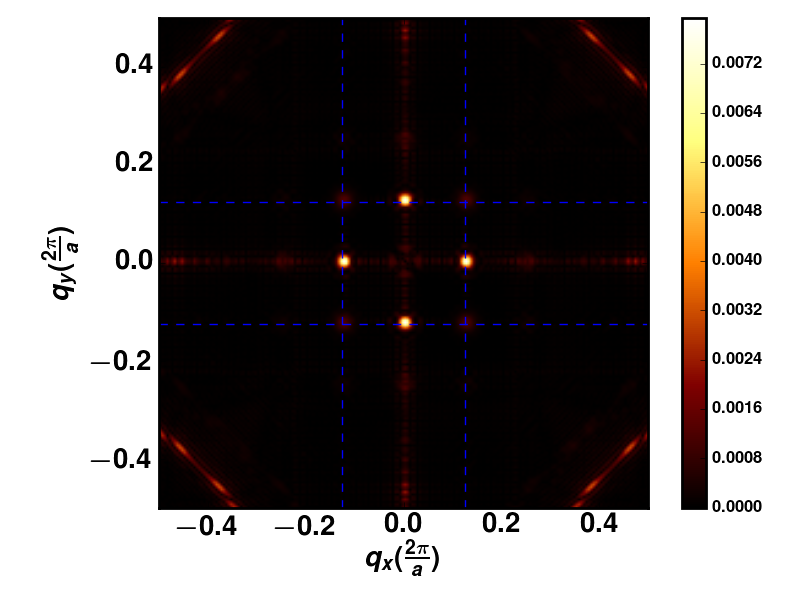}
    \caption{CDW-Driven}
    \label{fig:cdw_fft}
  \end{subfigure}

  \begin{subfigure}[b]{0.45\textwidth}
    \includegraphics[width=\textwidth]{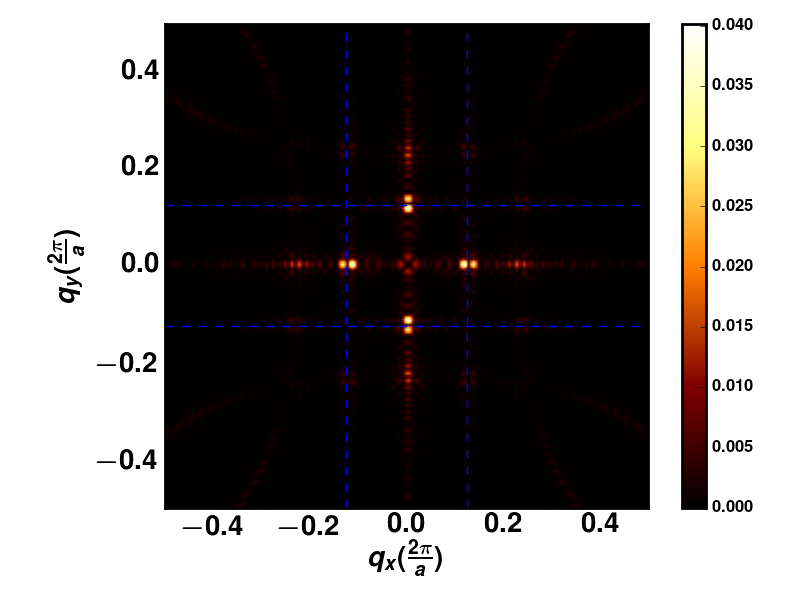}
    \caption{PDW-Driven: $\theta_x=0$ and $\theta_y=\frac{\pi}{2}$}
    \label{fig:phase0}
  \end{subfigure}
  \begin{subfigure}[b]{0.45\textwidth}
    \includegraphics[width=\textwidth]{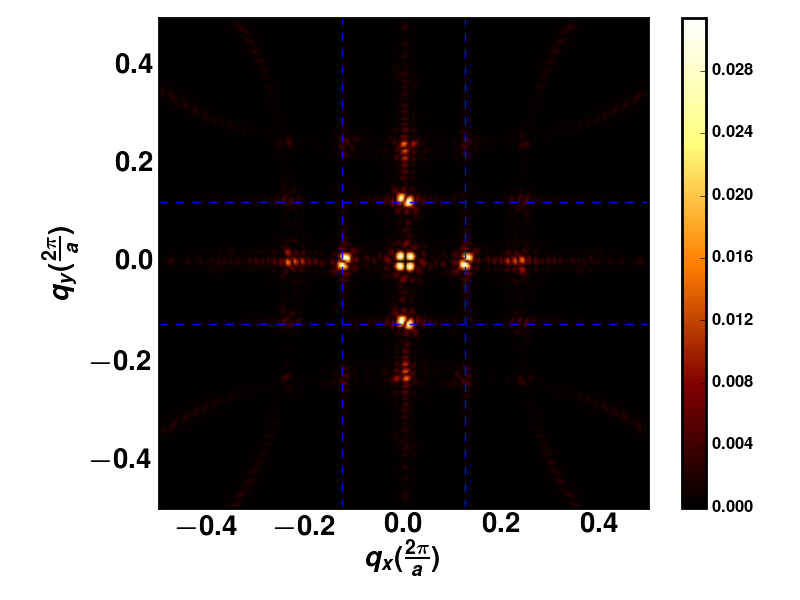}
    \caption{PDW-Driven: $\theta_x=\frac{\pi}{4}$ and $\theta_y=\frac{3\pi}{4}$}
    \label{fig:phase1}
  \end{subfigure}
  \caption[Regular and split Fourier peaks in LDoS]{$|\tilde \nu^E(q)|$ with $E=30$ meV for PDW-Driven and CDW-Driven Models.}
  \label{fig:fft}
\end{figure}

\begin{figure}[htb]
\centering
\includegraphics[width=0.45 \textwidth]{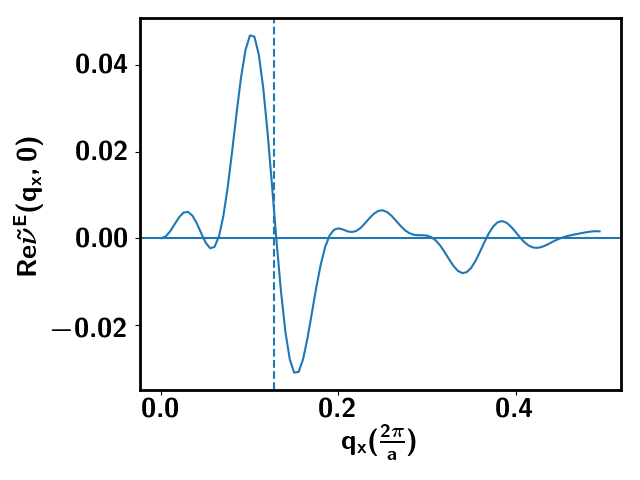}
\caption[Fourier transform of LDoS as a function of $q_x$]{$Re \tilde \nu^E(q_x,0)$ for the PDW-Driven model with $\theta_x=0$ and $\theta_y=\frac{\pi}{2}$; There is a clear sign change at $q_x=\frac{1}{8}\frac{2\pi}{a}$.}
\label{fig:fftx}
\end{figure}

 \subsection{Direct visualization of ``dislocation''}

 To have a direct visualization of profile shown in Eq.~\ref{eq:pdw_profile} for the PDW-Driven model, we need to extract the local CDW order parameter $\mathlarger \nu^E_{\mathbf{Q_a/2}}(x,y)$ from STM data $\nu^E(x,y)$.   For each position $(x_0,y_0)$, we construct a new image by multiplying a gaussian mask:
 \begin{equation}
 	\bar \nu^E(\mathbf{r};\mathbf{r_0})=e^{-\frac{|\mathbf{r}-\mathbf{r_0}|^2}{2W^2}} \nu^E(\mathbf{r})
 	\label{eq:local_order}
 \end{equation}
We found that $W=8$ is a good choice in our simulation. Then we can extract the local CDW order parameter $\mathlarger \nu^E_{\mathbf{Q_a/2}}(\mathbf{r_0})$ by a Fourier Transformation of $\tilde \nu^E(\mathbf{r};\mathbf{r_0})$:
 \begin{equation}
 	\mathlarger \nu^E_{\mathbf{Q_a/2}}(\mathbf{r_0})=\sum_{\mathbf{r}} \mathlarger{\bar \nu^E}(\mathbf{r};\mathbf{r_0})e^{-\frac{1}{2}i \mathbf{Q_a}\cdot \mathbf{r}}
 \end{equation}

 After extracting $\mathlarger \nu^E_{\mathbf{Q_a/2}}(\mathbf{r_0})$ for each position, we can easily visualize it and decide whether there is a  superposition of strength $\pm 1$ dislocations.

 The above algorithm can also be implemented by filter algorithm in momentum space directly as in Ref.~\cite{hamidian2016atomic}:
 \begin{equation}
 	\mathlarger \nu^E_{\mathbf{Q_a/2}}(\mathbf{r_0})=\sum_{\mathbf q} \tilde \nu^E(\mathbf q)G(\mathbf{Q_a/2}-\mathbf q)e^{-i\mathbf{(Q_a/2-q)}\cdot \mathbf r_0}
 \end{equation}
where the filter is $G(\mathbf q)=\sum_re^{-\frac{|\mathbf{r}|^2}{2W^2}}e^{-i \mathbf q \cdot \mathbf r}=e^{-\frac{W^2}{2}|\mathbf q|^2}$.

\begin{figure}[htb]
\centering
  \begin{subfigure}[b]{0.2\textwidth}
    \includegraphics[width=\textwidth]{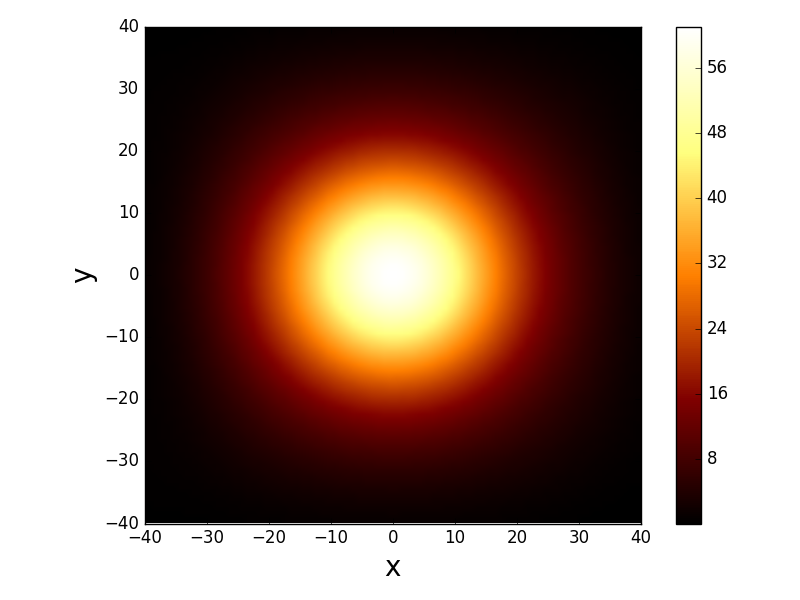}
    \caption{Local $|\mathlarger \nu^E_{\mathbf{Q_x/2}}|^2$; CDW-Driven Model}
    \label{fig:cdw_x}
  \end{subfigure}
  \begin{subfigure}[b]{0.2\textwidth}
    \includegraphics[width=\textwidth]{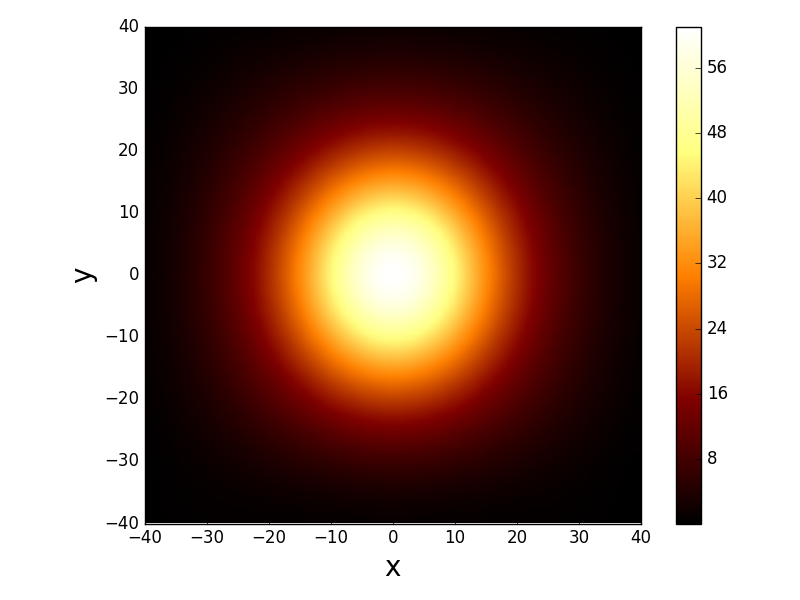}
    \caption{Local $|\mathlarger \nu^E_{\mathbf{Q_y/2}}|^2$; CDW-Driven Model}
    \label{fig:cdw_y}
  \end{subfigure}

  \begin{subfigure}[b]{0.2\textwidth}
    \includegraphics[width=\textwidth]{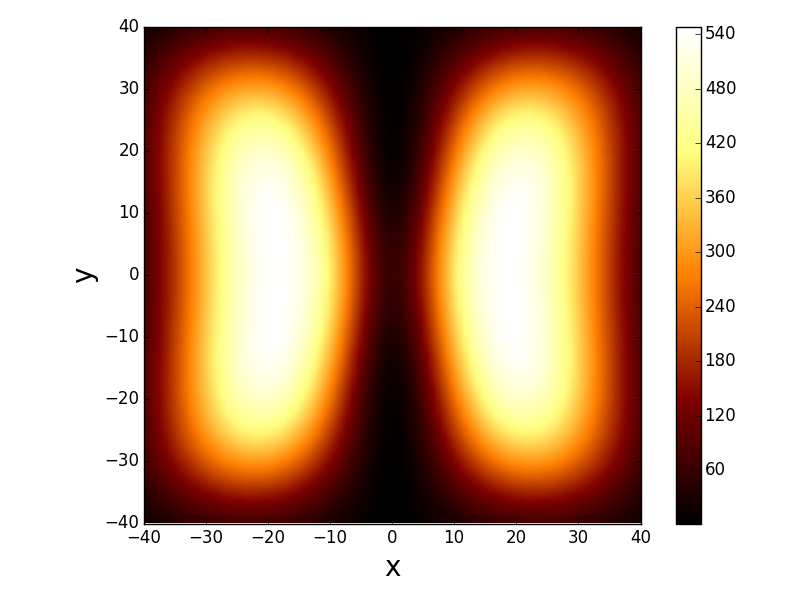}
    \caption{Local $|\mathlarger \nu^E_{\mathbf{Q_x/2}}|^2$; $\theta_x=0$ and $\theta_y=\frac{\pi}{2}$}
    \label{fig:phase0_x}
  \end{subfigure}
  \begin{subfigure}[b]{0.2\textwidth}
    \includegraphics[width=\textwidth]{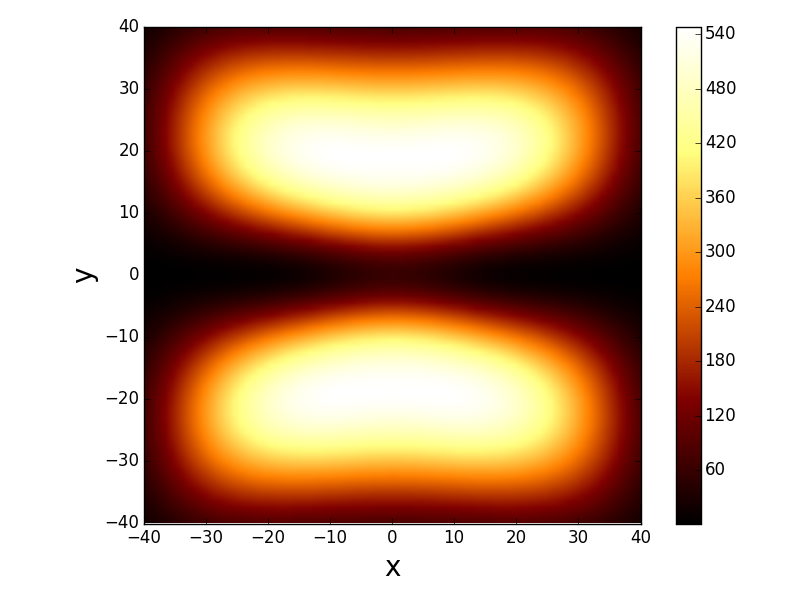}
    \caption{Local $|\mathlarger \nu^E_{\mathbf{Q_y/2}}|^2$; $\theta_x=0$ and $\theta_y=\frac{\pi}{4}$}
    \label{fig:phase0_y}
  \end{subfigure}

  \begin{subfigure}[b]{0.2\textwidth}
    \includegraphics[width=\textwidth]{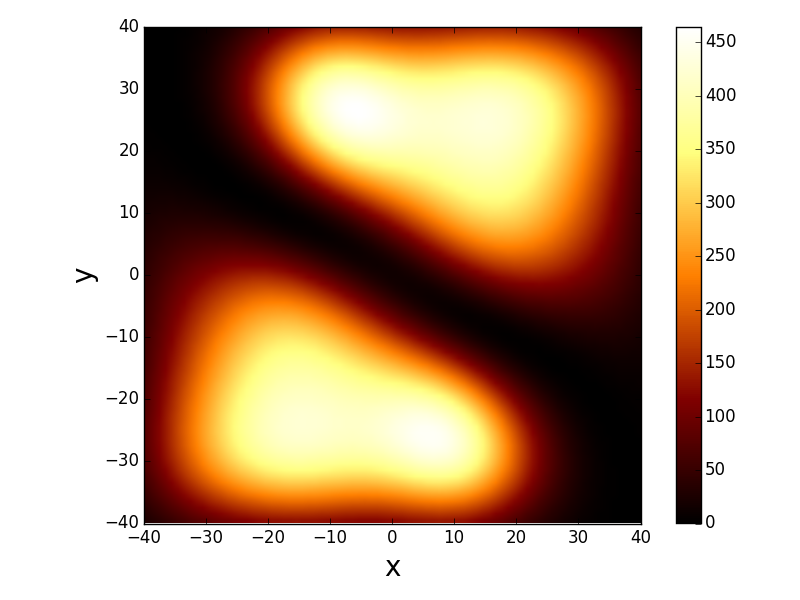}
    \caption{Local $|\mathlarger \nu^E_{\mathbf{Q_x/2}}|^2$; $\theta_x=\frac{\pi}{4}$ and $\theta_y=\frac{3\pi}{4}$}
  \end{subfigure}
  \begin{subfigure}[b]{0.2\textwidth}
    \includegraphics[width=\textwidth]{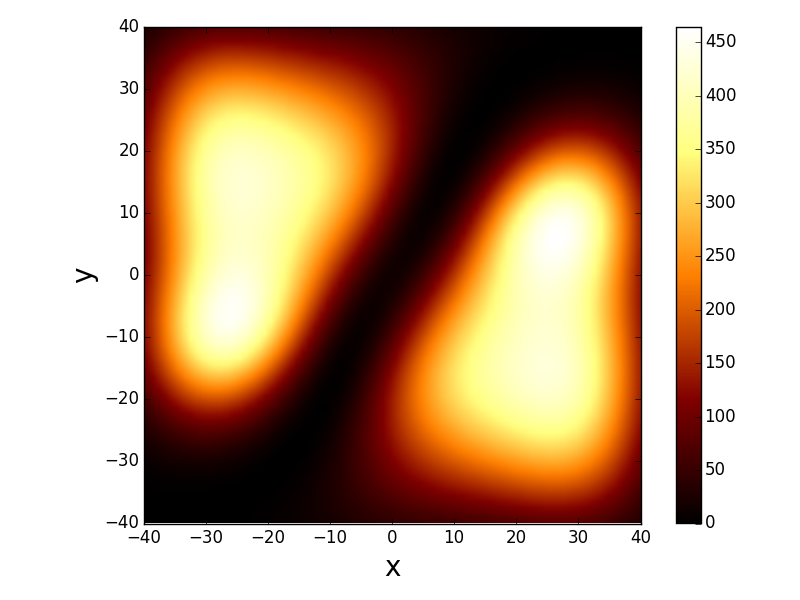}
    \caption{Local $|\mathlarger \nu^E_{\mathbf{Q_y/2}}|^2$; $\theta_x=\frac{\pi}{4}$ and $\theta_y=\frac{3\pi}{4}$}
  \end{subfigure}

  \caption[Local order parameter in real space]{$|\nu^E_{\mathbf{Q_a/2}}|^2$ from the CDW-Driven and PDW-Driven models. $(a)$ and $(b)$ are from the CDW-Driven model; Others are from PDW-Driven models. $E=30$ meV.}
  \label{fig:local}
\end{figure}

\begin{figure}[htb]
\centering{}
	\begin{subfigure}[b]{0.4\textwidth}
    \includegraphics[width=\textwidth]{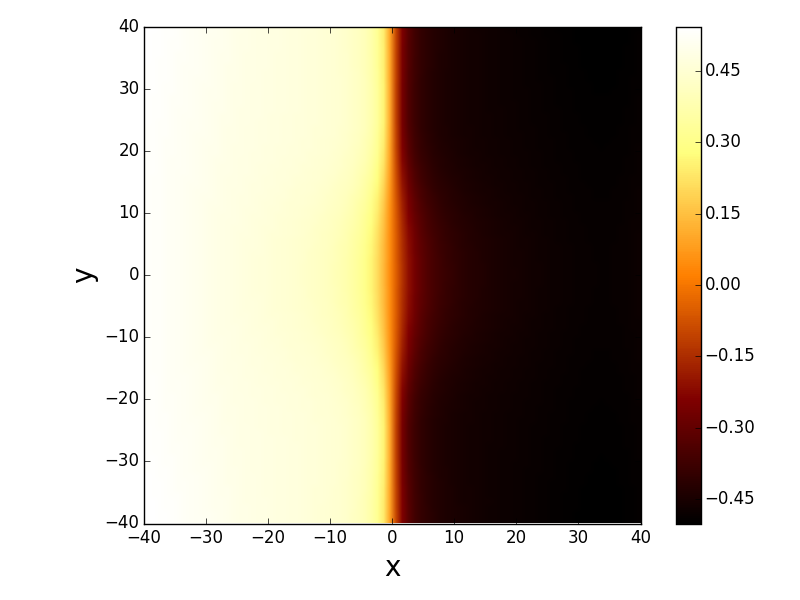}
    \caption{$\arg \nu^E_{\mathbf{Q_x/2}}(x)$. Phase of $\nu^E_{\mathbf{Q_x/2}}(x)$ jumps from $-\pi/2$ to $\pi/2$ across the line $x=0$. }
    \label{fig:phase_plot}
  \end{subfigure}

	\begin{subfigure}[b]{0.4\textwidth}
    \includegraphics[width=\textwidth]{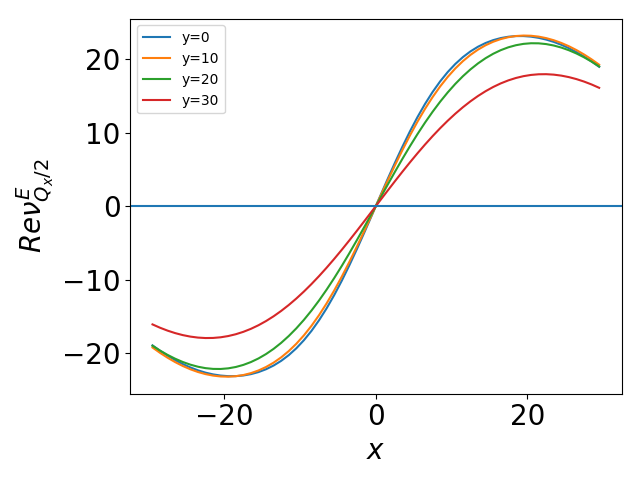}
    \caption{Re $\nu^E_{\mathbf{Q_x/2}}(x)e^{-i\phi_x}$ at fixed y.}
    \label{fig:fixy}
  \end{subfigure}
  \begin{subfigure}[b]{0.4\textwidth}
    \includegraphics[width=\textwidth]{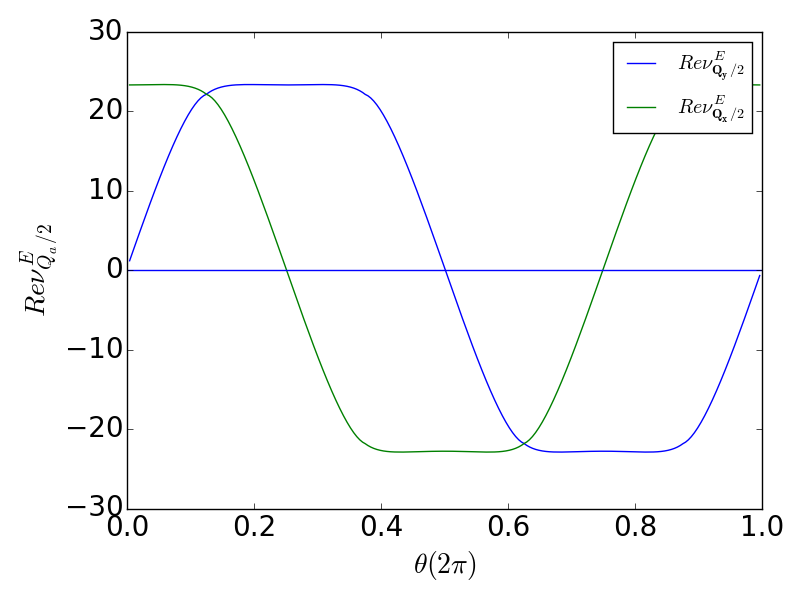}
    \caption{Re $\nu^E_{\mathbf{Q_a/2}}(\theta)e^{-i\phi_a}$ at $r=15$.There is a clear cosine-like dependence.}
    \label{fig:angle}
  \end{subfigure}

  \caption[Phase and amplitude of the local order parameter in real space]{$\nu^E_{\mathbf{Q_a/2}}$ for the PDW-Driven model with $\theta_x=0$ and $\theta_y=\frac{\pi}{2}$. }
  \label{fig:pdw_angle}
\end{figure}

In Fig.~\ref{fig:local}, we show visualization for simulated data of $|\mathlarger \nu^E_{\mathbf{Q_a/2}}|^2$ from both the CDW-Driven model and the PDW-Driven model. The distinction is very obvious. For the CDW-Driven model, $\nu^E_{\mathbf{Q_a/2}}$ has the maximal intensity at the vortex center. For the PDW-Driven model, $|\nu^E_{\mathbf{Q_a/2}}|$ vanishes along a line across the vortex center in the direction of $\theta_a\pm\frac{\pi}{2}$, in agreement with a $\cos(\theta-\theta_a)$ angle dependence. Across the dark line, phase of the local amplitude $\nu^E_{\mathbf{Q_a/2}}$ has a $\pi$ shift, as shown in Fig.~\ref{fig:phase_plot}.  We can see the phase of $\nu^E_{\mathbf{Q_a/2}}$ is $\phi_a$ or $\phi_a+\pi$.  Therefore we can remove the overall phase by  $\nu^E_{\mathbf{Q_a/2}}\rightarrow \nu^E_{\mathbf{Q_a/2}}e^{-i\phi_a}$ and make it real. Then angle dependence  $\nu^E_{\mathbf{Q_a/2}}\sim \cos(\theta-\theta_a)$ can be visualized directly in Fig.~\ref{fig:angle}. For an uni-directional PDW, Wang et al.\cite{wang2018} also noted the phase jump by $\pi$ by tracking the position of the DOS peaks in real space\cite{wang2018}. In Fig.~\ref{fig:fixy}, we plot $Re \nu^E_{\mathbf{Q_x/2}}(x)$ at fixed y. For $y=0$, $|\nu^E_{\mathbf{Q_x/2}}(x)|$ gives the radius dependence $F(r)$. We can see that the maximum is at finite $r$. However, our simulation may overestimate the maximum because of boundary effects due to the finite size.

Finally, we comment on challenges to apply this algorithm to real experimental data and possible ways to increase the signal to noise ratio. (1) The existence of multiple vortices and impurities modifies the $\cos(\theta-\theta_a)$ angle dependence. In general, there is no time reversal symmetry or any lattice symmetry left, and $\mathlarger \nu^E_{\mathbf{Q_a/2}}(\mathbf{r_0})$ is complex. Thus the line of zero we predicted in the simple model may not be exact. We still expect the real and imaginary parts of $\mathlarger \nu^E_{\mathbf{Q_a/2}}(\mathbf{r_0})$ to each have a line of zero but the lines will no longer coincide. As a result the line of zero's shown in  Fig.~\ref{fig:local}(c-f) will partially fill in. However, in the current experiment\cite{edkins2019magnetic}, the distance between neighboring vortices is roughly 3 times of the size of the halo; the distortion of the CDW profile by neighboring vortices is not significant. Furthermore, the phase locking mechanism we discussed in the previous section predicts that the Fourier peak of the CDW around each vortex split in the same direction. Thus the split peak signal should be observable in the existence of multiple vortices. (2) There is a smooth background, which will add an offset to the $\cos(\theta-\theta_a)$ factor. If we assume the background is smooth, it can be subtracted with sophisticated data analysis technique. (3) Although it is not necessary to analyze each vortex separately, doing it may increase the signal to noise ratio. If we choose the origin of Fourier transformation to be the center of each vortex, our PDW-driven model predicts the Fourier amplitude of the period 8 CDW to be real. We expect the noise to have random phase, and plotting the real part of the amplitude instead of the absolute value can enhance the signal to noise ratio. 

\subsection{Magnetization density wave}
\label{subsec:mdwmoment}

In the PDW-Driven scenario, we will also get magnetization density wave. Orbital magnetic moment of each plaquette $M(\mathbf{r})$ can be estimated through the following equation:
\begin{align}
	M(\mathbf{r}+\frac{\hat{x}}{2}+\frac{\hat{y}}{2})&=\frac{a^2}{4}\big(I(\mathbf{r},\mathbf{r+\hat{x}})+I(\mathbf{r+\hat{x}},\mathbf{r+\hat{x}+\hat{y}})\notag\\
	&+I(\mathbf{r+\hat{x}+\hat{y}},\mathbf{r+\hat{y}})+I(\mathbf{r+\hat{y}},\mathbf{r})\big)
\end{align}
where $a=3.5\textup{\AA}$ is lattice constant. $I(\mathbf{r},\mathbf{r+\hat{r}_a})$ is the current going through the bond from $\mathbf{r}$ to $\mathbf{r+\hat{r}_a}$ where $a$ denotes $x$ or $y$.

\begin{figure}
\centering
  \begin{subfigure}[b]{0.4\textwidth}
    \includegraphics[width=\textwidth]{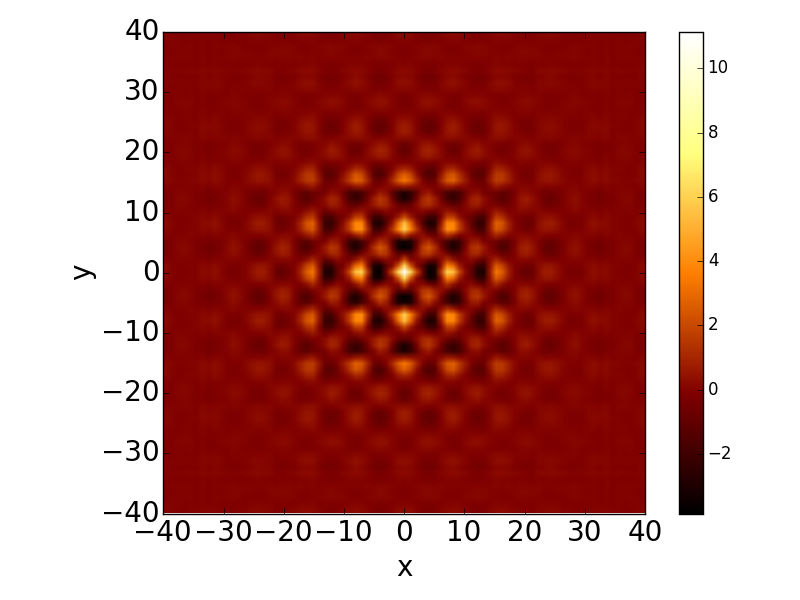}
    \caption{Real space pattern of magnetic moment $M(\mathbf{r})$ in unit of $10^{-3}\mu_B$.}
    \label{fig:fdw_real}
  \end{subfigure}
  \begin{subfigure}[b]{0.4\textwidth}
    \includegraphics[width=\textwidth]{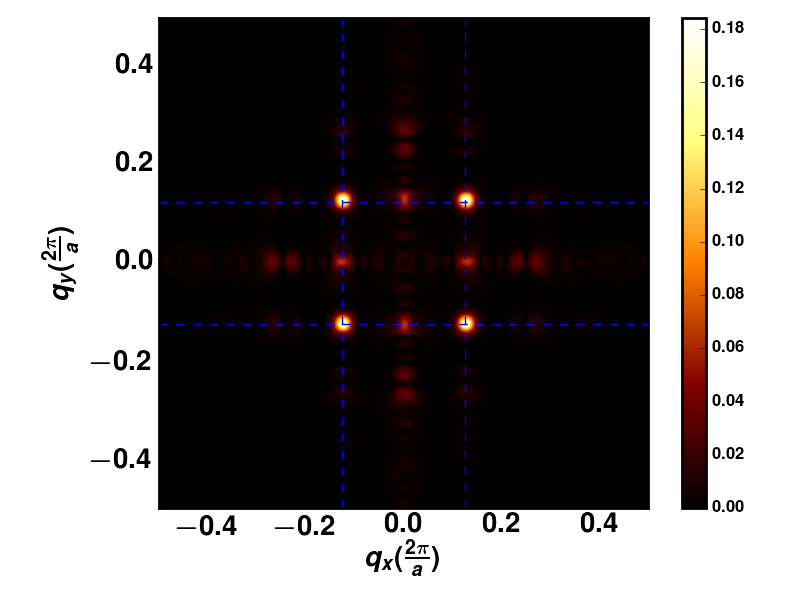}
   \caption{Magnetic moment $M(q)$ in momentum space.}
    \label{fdw_fft}
  \end{subfigure}

  \caption[Magnetization Density Wave pattern from the PDW-Driven model in vortex halo]{Magnetization Density Wave pattern from the PDW-Driven model in vortex halo.}
  \label{fig:fdw}
\end{figure}

$M(\mathbf{r})$ has density wave with momentum $\mathbf{Q_x/2}=(\frac{2\pi}{8},0)$, $\mathbf{Q_y/2}=(0,\frac{2\pi}{8})$. There are also density waves along the diagonals $Q_{\pm,\pm}=(\pm \frac{2\pi}{8},\pm \frac{2\pi}{8})$. Real space and momentum space pattern of magnetic moment are shown in Fig.~\ref{fig:fdw}.  Amplitude of density wave at momentum $(\frac{2\pi}{8},\frac{2\pi}{8})$ is around $0.005\mu_B$ and may be possible to be detected by neutron scattering experiments. The observation of the magnetization density wave at this-wave-vector offers the opportunity to definitively settle the question of uni-directional vs bi-directional PDW.

\subsection{Other types of pair density wave}
Our discussion mainly focuses on bidirectional PDW models. However, other types of PDW states have been proposed before. In this section, we show signatures for a Unidirectional PDW model and a canted PDW model. Therefore STM experiments can rule out or support these kinds of PDW models.

\begin{figure}[htb]
\centering
    \includegraphics[width=0.4\textwidth]{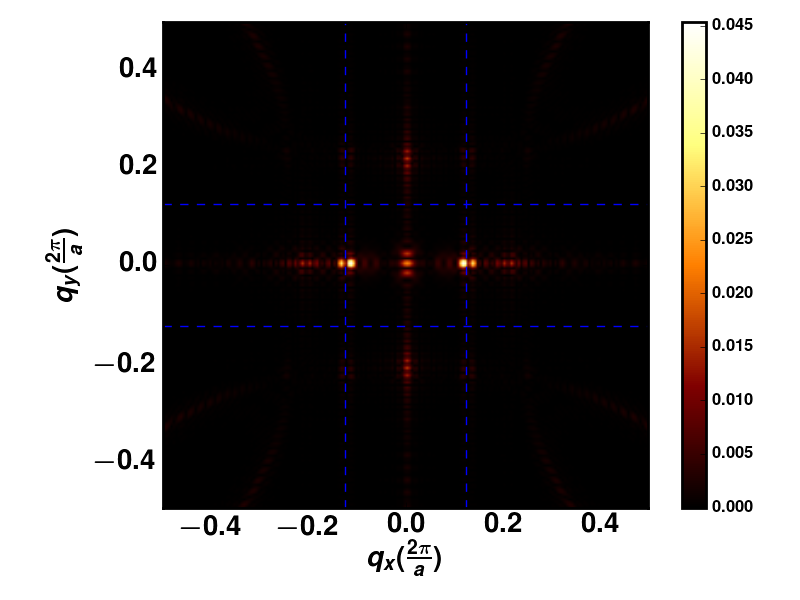}
    \caption[Fourier peak of LDoS for unidirectional PDW]{$|\tilde \nu^E(q)|$ for the unidirectional PDW with phase $\theta_x=0$.}
    \label{fig:uni_fft}
 \end{figure}

For the unidirectional PDW shown in Fig.~\ref{fig:uni_fft} with only $x$ component, Fourier Transform data only show peak at $\mathbf{Q_x/2}$, not at $\mathbf {Q_y/2}$. There is still split of peak consistent with our previous discussions for bidirectional PDW.

\begin{figure}[htb]
 \centering
    \includegraphics[width=0.4\textwidth]{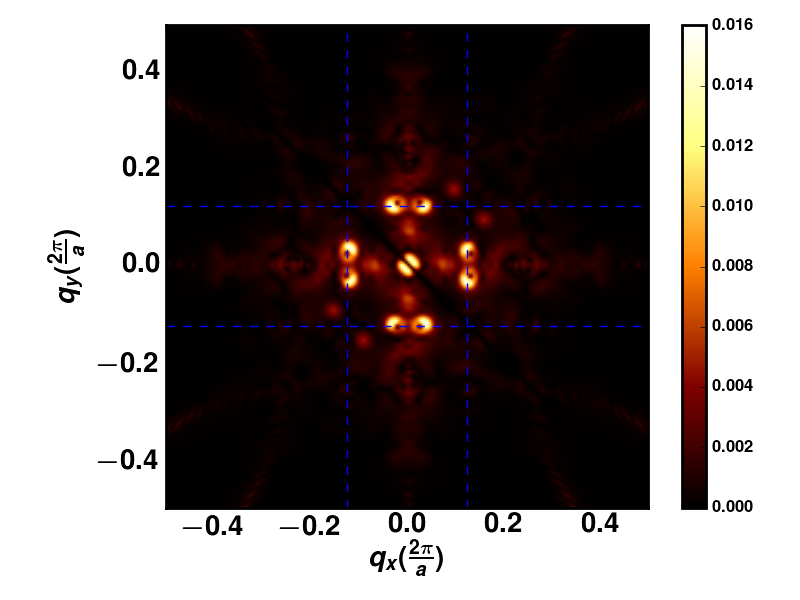}
    \caption[Fourier peak of LDoS for canted PDW]{$|\tilde \nu^E(\mathbf{q})|$ for the canted PDW with shifted momentum $p=0.03*2\pi/a$. Phase of the PDW is $\theta_x=0$ and $\theta_y=\frac{\pi}{2}$.}
    \label{fig:canted_fft}
\end{figure}

For the canted PDW, we expect the peak in $\nu^E(q)$ deviates from $(1,0)$ and $(0,1)$ direction. For the canted PDW model with shifted momentum $p=0.03*2\pi/a$: $\mathbf{P_1}=(\frac{2\pi}{8},p)$, $\mathbf{P'_1}=(-\frac{2\pi}{8},p)$ and $\mathbf{P_2}=(p,\frac{2\pi}{8})$,$\mathbf{P'_2}=(p,-\frac{2\pi}{8})$  this shift shows up in Fig.~\ref{fig:canted_fft}.  Because of condition $\tilde \nu^E(q)=\tilde \nu^E(-q)^*$, we see double peak with shift $p$. In experiment it may be better to detect this feature with complex amplitude $\tilde \nu^E(q)$ instead of intensity $|\tilde \nu^E(q)|$.


\begin{figure}[htb]
\centering
  \begin{subfigure}[b]{0.22\textwidth}
    \includegraphics[width=\textwidth]{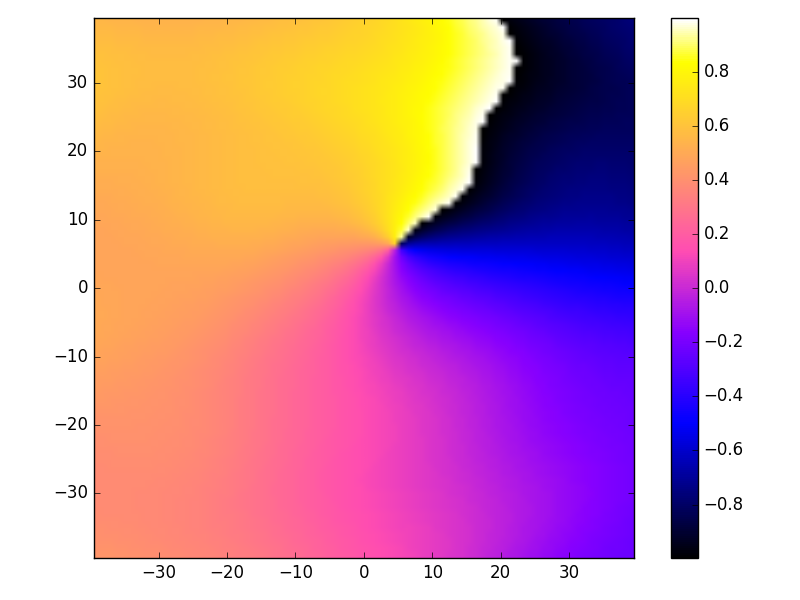}
    \caption{$P=(\frac{2\pi}{8},p)$}
  \end{subfigure}
  \begin{subfigure}[b]{0.22\textwidth}
    \includegraphics[width=\textwidth]{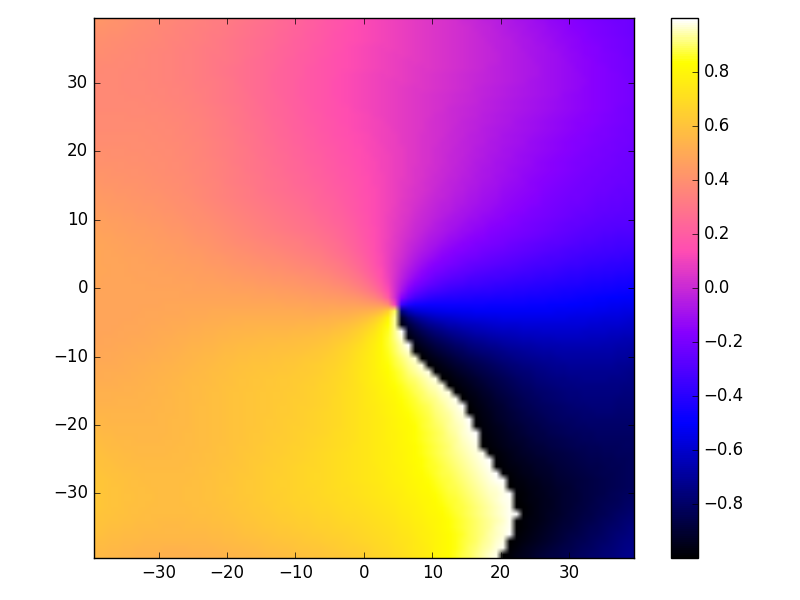}
    \caption{$P=(\frac{2\pi}{8},-p)$}
  \end{subfigure}

  \caption[Phase of the local order parameter in real space for canted PDW]{$\arg \nu^E_P(\mathbf{r})$ for the canted PDW in unit of $\pi$.}
  \label{fig:vortex_canted}
\end{figure}

If we can decide the value of shift momentum $|p|$ from Fourier Transformation data, then we can extract local order parameter $\nu^E_{P}(\mathbf{r})$ with $P_\pm=(\frac{2\pi}{8},\pm p)$ following Eq.~\ref{eq:local_order}. It turns out that $P=(\frac{2\pi}{8},p)$ has an anti-vortex while $P=(\frac{2\pi}{8},-p)$ has a vortex, as shown in Fig.~\ref{fig:vortex_canted}.

\begin{figure}[htb]
\centering
  \begin{subfigure}[b]{0.22\textwidth}
    \includegraphics[width=\textwidth]{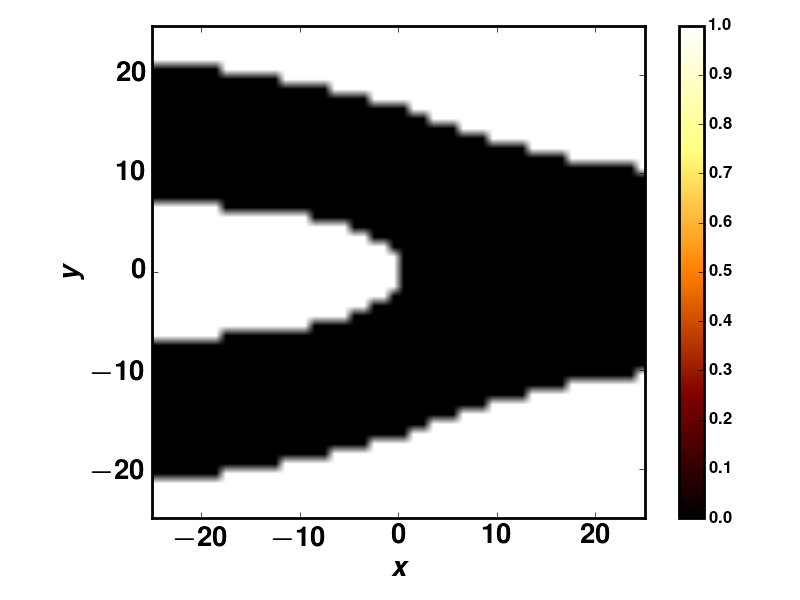}
    \caption{$\arg \cos(\theta-p y)$ in Eq.~\ref{eq:simulated_canted}. }
    \label{fig:simulate_canted_phase}
  \end{subfigure}
  \begin{subfigure}[b]{0.22\textwidth}
    \includegraphics[width=\textwidth]{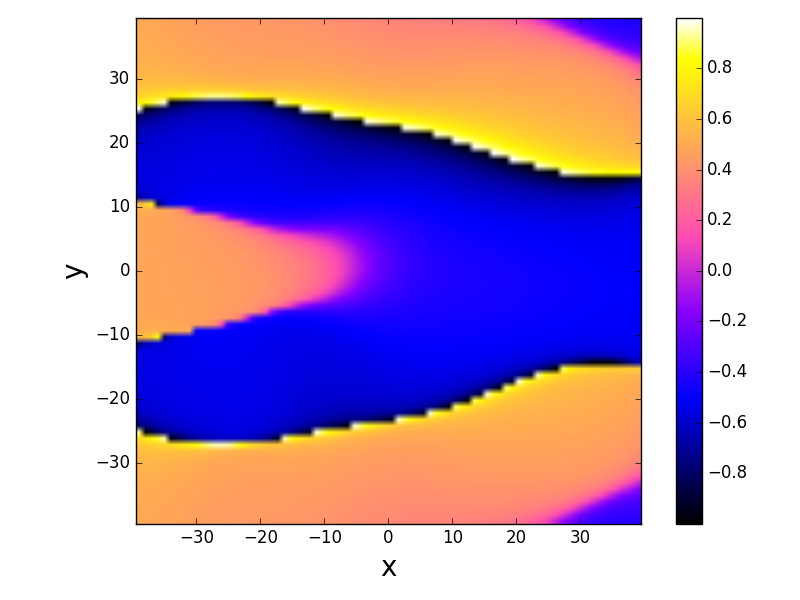}
    \caption{$\arg \nu^E_{P_0}(\mathbf r)$ from Canted PDW Driven model.}
    \label{a_canted_unshift_plot}
  \end{subfigure}

  \caption[Visualization of the local order parameter for canted PDW]{Visualization of $\arg \nu^E_{P_0}(\mathbf r)$ for canted PDW Driven model. $P_0=(\frac{2\pi}{8},0)$ and shifted-momentum is $p=0.03*2\pi$.}
  \label{fig:canted-unshift_plot}
\end{figure}

If momentum resolution is not good enough to decide the value of $p$, we propose to visualize $\nu^E_{P_0}(\mathbf(r))$ with  $P_0=(\frac{2\pi}{8},0)$. If it is ordinary PDW-Driven, we get similar plot as in Fig.~\ref{fig:phase_plot}. If it is canted PDW-Driven, we will get strange position dependence of $\arg \nu^E_{P_0}(\mathbf r)$  like in Fig.~\ref{a_canted_unshift_plot}. This is a signature of the canted PDW and it's consistent with the following equation:

\begin{equation}
	\nu^E_{P_0}(\mathbf r)\sim \cos(\theta-py)
	\label{eq:simulated_canted}
\end{equation}

\section{Summary}
We now summarize some of the conclusions from the discussion in previous sections. The observation of period 8a bidirectional charge order in the vortex halo directly  means that there are induced order parameters
$\rho_{Q_x/2}, \rho_{Q_y/2}$.  In the presence also of a non-zero superconducting order parameter $\Delta_d$ of the usual $d$-wave superconductor, the period-8 charge order necessarily implies that there are also period-8 modulations in the superconducting order parameter $\Delta_{Q_x/2}, \Delta_{Q_y/2}$, {\em i.e}, Pair Density Wave order at the same period. Given this obvious equivalence in the superconductor between charge and pairing modulations, it may seem to be a moot question whether what is observed is primarily charge order or pair order at period-8. Nevertheless we have shown that there are two distinct possibilities for the observed period-8 order which naturally correspond to two distinct driving mechanisms. 

In the CDW-driven scenario, we simply postulate that there are slow fluctuations of a previously unidentified period-8 CDW in the uniform superconductor. In the vicinity of the vortex the breaking of translational symmetry and the weakening of the superconducting order may then pin the fluctuations of the period-8 CDW and lead to static ordering. Period-4 charge order then appears as a subsidiary order. In this scenario it is natural to expect that the phase of the induced CDW order does not wind on going around the vortex core. 

In the PDW-driven scenario on the other hand, we postulate that there are slow fluctuations of period-8 PDW that are pinned in the vortex halo. The induced period-8 CDW then will have a strength $\pm 1$ dislocation centered at the vortex core. More precisely the induced period-8 CDW will be a superposition of a configuration with a strength $+1$ dislocation and one with a strength $-1$ dislocation. This leads to  a rather different spatial profile for the induced period-8 CDW. A further difference is that there are now two distinct kinds of induced period-4 CDW orders which we have referred to as CDW$_A$ and CDW$_B$. The  CDW$_A$ pattern has no winding around the vortex core while the CDW$_B$ pattern is a superposition of strengths-$\pm 2$ dislocations.

We discussed the extent to which existing data supports either scenario. In particular in the PDW-driven scenario there is a natural explanation for the absence of peaks at $2\pi \left(\frac{1}{8}, \frac{1}{8}\right)$ as reported in the experiments. It is however important to analyze the data more carefully to clearly establish which of these scenarios is realized, and we described a number of distinguishing features. Most importantly the spatial profile  of the induced charge orders due to the dislocation structure in the PDW-driven scenario should be discernible using the methods we describe.

 Note that within either of these scenarios there is no general reason for a predominantly $d$-form factor period-$8$ charge order to induce only an $s$-form factor period-4 charge order \cite{Note1}. From our numerical simulation of $d$ wave PDW coexistence with uniform $d$ wave superconductor, the period $8$ CDW we get is actually dominated by $d$ wave, instead of $s$ wave from naive expectation. Thus we do not have a natural explanation of the observations on form factors in the experiments.

A further question that one can ask is whether the fluctuation order that is pinned on the halo is unidirectional or bidirectional. The observed period-$8$ modulations are apparently  bidirectional. The simplest explanation therefore is that the ``parent' order is also bidirectional. However one may postulate that there are domains of different unidirectional patches within the vortex halo. This may be easy to check in the STM data. 
 
Finally an important question is whether the period-8 PDW (if it is really the driver) is merely a competing/intertwined order with the standard $d$-wave superconductor or whether it is a ``mother" state with a very large amplitude that controls the physics up to a much larger energy scale than the standard $d$-wave order itself. Just based on the STM experiments alone there does not seem to be any clear way to answer this question. 
 
However in the following section, by combining with information from other existing experiments,  we will provide suggestive  arguments in favor of a mother PDW state.  

\section{A broader perspective on pair density wave and its relation to the pseudogap state of cuprates}

In this section we take a broader perspective and ask whether the message learned from the STM data on Bi-2212 can inform us on anomalies observed in other cuprates and more generally on the pseudogap itself. We shall assume that  the data are described by the fluctuating PDW ("mother state") scenario and we shall assume that scenario continues to hold in other under-doped cuprates. We focus our attention on YBCO where extensive data on the CDW up to high magnetic field is available\cite{changNatureComm72016magnetic,ZX1science350949gerber2015three,ZX2PNAS11314647jang2016ideal}. The picture that emerges from these studies is that SRO CDW appears below about 150K over a doping range between x=0.08 and 0.16\cite{BlancoPhysRevB.90.054513}. This SRO CDW has very weak interlayer ordering centered around L=1/2 where L is the c axis-wave-vector in reciprocal lattice unit. These peaks grow with decreasing temperature but their strength  weaken and their in plane linewidth broaden below $\text{T}_c$. These peaks occur along both a and b axes. Above a field of 15 to 20T, a uni-directional CDW emerges and rapidly becomes long range along the b axis. The onset of long range ordered CDW is consistent with earlier NMR data.\cite{wu2013emergence,wu2015incipient} At the same time, the SRO CDW remains along both a and b axes. Thus the high magnetic field data shows that there are two kinds of CDW with the same incommensurate period which does not change with magnetic field. As the experimentalists remarked\cite{ZX1science350949gerber2015three,ZX2PNAS11314647jang2016ideal}, this is very puzzling because having the same incommensurate wave-vector suggests the two kinds of CDW share a common origin. 

\begin{figure}[htb]
\centering
\includegraphics[width=0.6\linewidth]{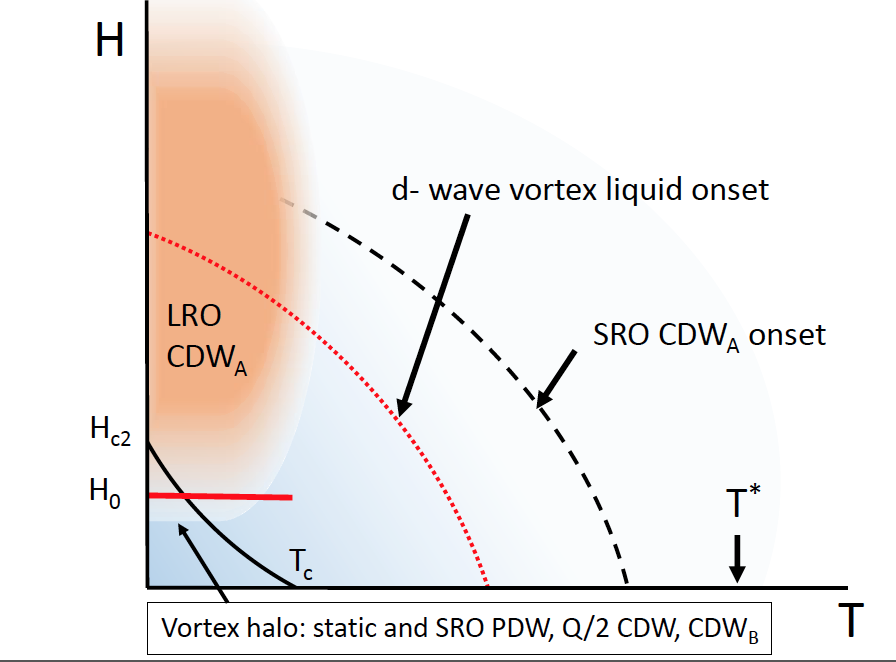}
\caption[H-T Phase Diagram for an underdoped cuprates]{H-T Phase Diagram for an underdoped cuprates. The light blue shading indicates that a fluctuating PDW is pervasive over a large segment of the $H-T$ plane for underdoped Cuprates. Dashed line indicates the onset of short range ordered CDW at wave-vector Q. It is a subsidiary order of the PDW which we refer to as $\text{CDW}_A$. The solid red line marks the magnetic field $H_0$ as defined in Eq.(14) in terms of the coherence length $\xi_P$ of the PDW which marks the size of the vortex halo. It is closely related to the field $H_{c2}$ which marks the onset of a vortex solid phase and LRO superconductivity. Within this phase and inside the vortex halo we expect the pinned static  PDW,  Q/2 CDW as well as its harmonic, a  wave-vector Q CDW which we refer to as $\text{CDW}_B$. The $\text{CDW}_B$ short range order state may extend to higher magnetic field much beyond $H_{c2}$. The dotted red line indicates the onset of a vortex liquid phase. The brown area indicates the appearance of long range ordered type A CDW with wave-vector Q. }
\label{fig:phase_diagram}
\end{figure}

If we interpret the observed CDW as subsidiary to a fluctuating PDW, the latter must exist above the CDW onset at 150K and most likely above $\text{T}^*$ which is taken as the thermodynamic signature of the pseudogap. Similarly we take the viewpoint that quantum oscillations require the existence of bi-directional CDW\cite{sebastian2012towards},which implies that fluctuating PDW extends to magnetic fields of 100T and beyond. By continuity we expect fluctuating PDW to cover a large segment of the H-T plane, as shown in Fig.~\ref{fig:phase_diagram}. The PDW must be strongly fluctuating in time, because there is no sign of superconductivity from transport measurements outside of a limited region near $\text{T}_\text{c}$ and $\text{H}_{c2}$. However, diamagnetic signals are observed over a much larger regime\cite{yuPNAS126672016magnetic}, a point which we shall return to later. Nevertheless, our picture is that the subsidiary orders such as CDW can be more robust and make their presence felt. This is particularly true of $\text{CDW}_A$ (see Eq.~\ref{Eq: CDWA}) which does not require d-wave pairing for its presence. So we assign $\text{CDW}_A$ to be the SRO CDW which onsets below 150K, as shown by the dashed line in Fig.~\ref{fig:phase_diagram}.

Below $\text{T}_\text{c}$ the phase stiffness of the LRO d-wave robs oscillator strength from the PDW, diminishing its already weak phase stiffness even further. This explains the reduction of the CDW strength below $\text{T}_\text{c}$. On the other hand, we saw in Sec~\ref{subsec:staticshortrangePDW} that in a magnetic field a vortex can pin the PDW to form a static but short range halo around the core. This in turn induces a CDW at wave-vector Q/2 and its harmonic $\text{CDW}_B$. All these states are located roughly inside the superconducting region as indicated in Fig.~\ref{fig:phase_diagram}. Of course being tied to the vortices mean that the strengths of these states are proportional to the magnetic field. Note that we expect the  d-wave phase stiffness to be  reduced inside the halo while that of the PDW to be strengthened. 

We define the field $H_0$ as 
\bea
H_0=\phi_0/(2\pi\xi^2_P)
\eea
where $\phi_0 = hc/2e$ is the flux quanta in a superconductor, $\xi_P$ is the correlation length of the pinned PDW. The $2\pi$ in the denominator has been inserted to make this equation resemble the definition of $H_{c2}$ and the exact numerical factor should not be taken seriously. The point is to provide a scale for the  field where the pinned PDW starts to strongly overlap. For $H>H_0$, the d-wave superconductor is being squeezed out and the PDW phase regains its stiffness. It eventually becomes depinned as the d-wave pairing diminishes and resumes its dynamical fluctuation. In this region the $\text{CDW}_A$ grows  in strength and coherence, recovering the growth with decreasing temperature that was interrupted by the onset of $\text{T}_\text{c}$ for  $H<H_0$.
The fact that the LRO CDW is uni-directional even though the PDW is bi-directional can be rationalized by the following argument. There is a term in the Landau free energy $\gamma_{a,b} |\rho_{Q_x}|^2 |\rho_{Q_y}|^2$, where $a,b=x,y$ labels the Cu-O bond in x direction and y direction. As we discussed previously, the local s-wave and d-wave form factors related to these two bonds do not have symmetry distinction, but there are still 2 degrees of freedom in each unit cell, and they may behave differently. In the channel where $\gamma$ is large and positive, the free energy strongly prefers uni-directional order; in the channel where $\gamma$ is small, we can have bidirectional CDWs. In YBCO the presence of the chain already broke tetragonal symmetry to begin with, making it even more plausible that the order grow strongly in one direction. On the other hand, the term $\Delta_P \Delta_{-P}^* \rho_Q$ is linear in $\rho_Q$, meaning that some SRO is likely generated in the orthogonal direction. We shall return to this point later.

Returning to the region below $\text{H}_{c2}$ we expect to find the pinned PDW and the CDW with period Q/2 as static but short range ordered. This is because the static order of the Q/2 CDW requires the static order of d-wave pairing as well as PDW. The Q/2 CDW should persist to lower field with decreasing amplitude. It may be expected to have correlation length similar to that found in the STM experiment, which we estimate to be about 16 lattice spacings. It will of course be of great interest to search for this by X-ray scattering. On the other hand, the period  Q $\text{CDW}_B$ can be thought of as a harmonic of the period Q/2 CDW, but it can exist even in its absence. Thus we expect it to exist up to higher field. We do not know exactly how high a field it can persist to, but it cannot go above the d-wave vortex liquid regime. It is worth noting that in practice there can be remnants of static pinned vortices even above $\text{H}_\text{c2}$. Yu et al.\cite{yuPNAS126672016magnetic} reported hysteretic behavior which extends to very high field at low temperatures, leading them to identify a second vortex solid regime. The existence of some form of bi-directional CDW that persists up to high field at low temperature is important in order to explain the quantum oscillations. We believe the LRO unidirectional CDW cannot by itself give rise to quantum oscillations, but the combination with some SRO CDW in the direction perpendicular to it may be sufficient. This can come from the bi-directional $\text{CDW}_B$ discussed above if it persists to high field,  or it is possible that a short range order $\text{CDW}_A$ is generated along direction $a$ at higher field as explained earlier. 


In support of the picture outlined above, we note that there is extensive NMR data showing that $\text{H}_0$ is typically 5 to 10T below the $\text{H}_{c2}$ as measured by transport\cite{Julien2arXivzhou2017spin,wu2013emergence}. Thus there is a close relationship between $\text{H}_{c2}$ and the vortex halo size as defined by the size of the pinned PDW. We also recall that the CDW that we identify as type A in Bi-2212 is uni-directional, which agrees with this assignment for YBCO. We note that the Bi-2212 sample used has a doping of 0.17 which lies on the upper end of the observability of CDW in YBCO samples. The $\text{H}_{c2}$ and corresponding $\text{H}_0$ are expected to be very high. So the 8.25T used in the STM experiment is expected to be far below the regime where $\text{CDW}_A$ can achieve long range order.

In Fig.~\ref{fig:phase_diagram} we add the line $\text{H}_0$ to a phase diagram in the H-T plane for under-doped Cuprates, following the proposal of Yu et al\cite{yuPNAS126672016magnetic}. The resistive $\text{H}_{c2}$ is the boundary of the vortex solid and marks the resistive transition. (To avoid cluttering, we did not show the emergence of a second vortex solid regime mentioned earlier that extends to high field at low temperature\cite{yuPNAS126672016magnetic}.)  The key point made by Yu et al. is that there is a large region of vortex liquid in the phase diagram where there is strong superconducting amplitude. The evidence for this is a strong diamagnetic signal. Given the small size of the true vortex core where the d-wave coherence peak is destroyed, it is reasonable to  interpret the vortex liquid as a region of strong d-wave superconducting amplitude with dynamical vortices that persists to very high field.  It is less certain how high in temperature the d-wave vortex liquid extend. It is possible that the diamagnetic signal may come from PDW fluctuations at high fields\cite{yuPNAS126672016magnetic,PhysRevX.4.031017}. Thus the location of the dotted line in Fig.~\ref{fig:phase_diagram} that indicate the extent of d-wave vortex liquid is quite uncertain, especially in the temperature direction.

\begin{figure}[htb]
\centering
\includegraphics[width=0.4\linewidth]{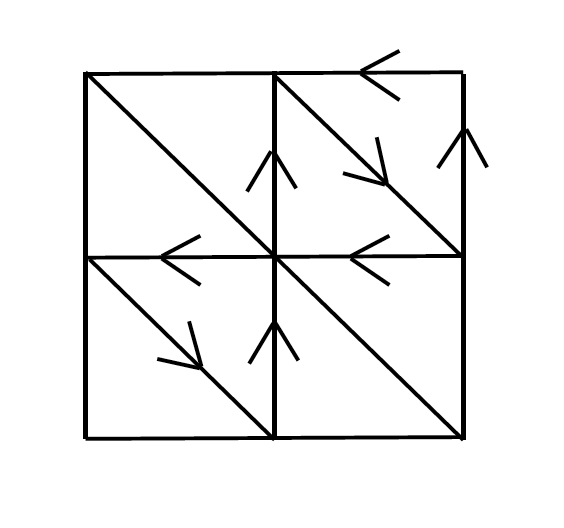}
\caption[Illustration of the Loop Current produced by the canted PDW]{Illustration of the Loop Current produced by the canted PDW.}
\label{fig:loop_current}
\end{figure}



We should mention that similar CDW has been seen in the Hg-based compound. Here the doping range extends further down to $x$ of order 0.06 and up to about 0.12. Another difference is that there is no clear suppression of the CDW at $\text{T}_c$. Instead its strength seems to saturate. It should be noted that unlike YBCO, this is a tetragonal system. From existing X ray data, it is not known whether the CDW is bi-directional or uni-directional. Apart from these differences, the observations seem to fit into the same phase diagram shown in Fig.~\ref{fig:phase_diagram}.

Finally we comment on the symmetry breaking observed at the $\text{T}^*$ lines which lies at a temperature above the onset of SRO CDW. This seems to be associated with breaking a lattice symmetry, perhaps a kind of nematic order. Importantly, a recent experiment on the anisotropy of the spin susceptibility\cite{Matsuda2unpublished} found  the nematic axis to be along the diagonal in a single layered Hg-based compound, while it is along the bond direction in YBCO\cite{sato2017thermodynamic}. This would rule out nematicity based on CDW which should be along the bond direction in a single layer tetragonal system. The observation in YBCO can be understood from the stacking of two orthogonal directions of diagonal nematicity in each layer. Such nematicity agrees with the symmetry of the orbital current model\cite{aji2009quantum}.  As mentioned earlier, in the PDW model it was pointed out by Agterberg et al.\cite{Agt2PhysRevB.91.054502} that adding canting to the PDW model as described earlier has the same symmetry as the orbital current model. The four different combinations of (p1,p2) give rise to a 4 state clock model. Fluctuations between (1,1) and (-1,-1) restores time reversal symmetry but gives rise to a diagonal breaking of nematic symmetry, just like the orbital current model. Indeed a canted PDW model will carry intra-cell  currents as shown in Fig.~\ref{fig:loop_current}, which is the closest we can get to Varma’s model in a single band model. As seen in this figure, the current can be understood as supercurrent running along x and y, with a return current along one of the diagonal bond. In fact we find that such a current pattern emerges from the PDW model. Without self-consistent determination of the mean field ground state, there is a net current along x and y, which presumably will be fixed by {}a proper return current in a self-consistent mean field theory. However, the current we find is very small, on the order of $10^{-3} t$ on each bond. This gives rise to a moment of about $10^{-3} \mu_B$ which is too small compared with the 0.1 $\mu_B$ reported by neutron scattering.  We note on general ground that the orbital current in the PDW model must be small. Let us define the canted component of the wave-vector as $p = (P + P')/2$.  The supercurrent can be estimated from the product of the phase gradient which is $p$ and the spectral weight, which is $x/m$ where $1/m$ is proportional to $ta^2$. Thus we expect the maximal supercurrent to be $x|p|t$ where $p$ is in reciprocal lattice units. Since $|p|$ should be less than $|P|$, we expect $x|P|$ to be less than $10^{-2}$ and similarly for the moment in units of $\mu_B$. Thus it is unlikely that the canted PDW model can account for the orbital current observed by neutron. However, it potentially can explain the onset of diagonal nematicity at $\text{T}^*$.

\section{A fluctuating pair density wave ground state?}
\label{sec:outlook}

Finally we call attention to the most interesting part of the phase diagram, the region at zero temperature and above $H_{c2}$. What is the nature of this state? How can our analysis of the vortex halo help us understand this high-field ground state?


In this chapter, we have been focused on the point of view that the PDW is the primary order, the 'mother state' behind the pseudogap phenomena. The phase of the PDW order parameter is fluctuating elsewhere, but pined by the vortex core of the d-wave superconductor to be static. The paper by Wang et al.~\cite{wang2018} presents a different point of view. In their sigma model, the PDW is competing with the d-wave superconductor; therefore present only in the vortex halo where the d-wave amplitude is suppressed. As far as the STM data are concerned, both points of views seem to work, but we believe the sigma model picture will run into some difficulty if we ask the question of what happens when the vortex halos overlap. Clearly d-wave superconductivity will be destroyed. The relatively large size of the vortex halo will explain why the destruction of d-wave superconductivity occurs at an unexpectedly low magnetic field. However, in the sigma model picture, it is difficult to see how one can avoid the conclusion that the resulting state is a long range ordered PDW and therefore a genuine superconductor. In contrast, in the fluctuating PDW scenario, the static PDW will be liberated and becomes freely fluctuating again once the pinning due to the d-wave phase winding disappears for H greater than $\text{H}_{c2}$ and d-wave pairing is killed.  

By following this logic, the fluctuating PDW point of view leads us naturally to the following question. What is the nature of this high field state? Is it a metallic state? If so, is it a Fermi liquid? Is the dissipation due to the metallic state responsible for quantum disordering the PDW? Is the electron spectrum similar to the static PDW spectrum? If so, how can a metallic state having a Bogoliubov-like band? 

We would like to point out that the name 'fluctuating PDW state' itself does not tell us the physical properties it has except that the PDW order parameter has a long correlation length. In order to clarify the nature of this state and expose as much of its physical properties as possible, we need a concrete construction of the quantum ground state. This the goal of the next two chapters. For simplicity, we shall focus on the commensurate case, starting with the static PDW bands introduced in Sec.~\ref{Eq: PDW mean field} and drive it to a non-superconducting state by phase fluctuation.

A central question regarding the fluctuating PDW state is how the electron spectrum evolves as we disorder the PDW. The similarity between the PDW bands and ARPES data in cuprates is impressive~\cite{PhysRevX.4.031017}. PDW gaps out antinodal fermions and leaves a gapless `arc' in the nodal direction. However, the conceptual question about how the spectral feature of PDW, which is a superconducting order, can be used to explain the spectrum of the metallic pseudogap state has yet to be addressed. In the next chapter, we first answer this question in a simpler setting of a fluctuating s wave superconductor. The insights gained from this simple situation guide us in the construction of the fluctuating PDW state.

\chapter{Quantum Fluctuating s-wave Superconductor}\label{chap:fluctuatingSC}

At the end of the last section, we discuss the necessity of a theory of quantum fluctuating pair density wave. Needless to say, trying to fluctuate the PDW state with 18 pairs of bands, introduced in Sec.~\ref{Eq: PDW mean field}, is a formidable task. However, the central problems we want to solve, namely, whether the antinodal PDW gap persists when the long-range order is destroyed and how the electron spectrum evolves, are universal to any fluctuating superconducting order. In this chapter, we take a detour from pair density waves to discuss a much simpler fluctuating s-wave superconductor, where the central problems are most prominent and easier to solve.

For an s-wave superconductor, there are two possibilities when superconductivity is quantum disordered at zero temperature. If the pairing is weak (BCS limit), the electron gap is a collective phenomena, and it may vanish immediately when superconductivity is disordered. Alternatively, if the pairing amplitude is large, and the superconductivity is destroyed by the phase fluctuation of its order parameter, it is known that a pairing gap of electron can survive even if the phase coherence of superconductivity is destroyed~\cite{PhysRevB.39.2756,PhysRevLett.47.1556,bouadim2011single}.

The persistence of the electron gap is most obvious in the BEC limit. In this limit, electrons form tightly bound pairs, and the single-electron gap is just the binding energy of the pair, which is well-defined even for a single pair (like a molecule), therefore independent of whether the pairs condense or not. When the condensate is destroyed by phase fluctuations, the bosonic pairs may form a Mott insulator (if its density happens to be commensurate with the underlying lattice), a Wigner crystal that further breaks translation symmetries, or simply pinned by disorders. No matter which quantum state the bosonic pairs go to, the single-electron gap always persists when the superconductivity is disordered.

In this chapter, we consider the simple case that we have 2 electrons per unit cell. In the BEC limit the superconducting phase is essentially the superfluid phase with 1 boson per unit cell (in average). Increasing the repulsion of the pairs, we can disorder the superconductor to get a bosonic Mott insulator, which is adiabatically connected to the atomic insulator with one pair per unit cell. This quantum phase transition is well studied \cite{sachdev_2011}; the effective theory is the 2+1D XY model. Only the bosonic gap closes at the transition.

However, electrons in a solid seldom form tight pairs. We are interested in the more realistic intermediate pairing regime, where the pairing amplitude is comparable or smaller than the Fermi energy but not too small. When we gradually reduce the pairing amplitude from the BEC limit to the intermediate pairing regime, by continuity, we expect the same transition from the superconductor to the paired insulator still exists, and the electron gap is nonzero across the transition. We argue below that this intermediate regime, where the electron gap remains nonzero on the disordered side is relevant to cuprates.

In the intermediate pairing regime, difficulties arise when we try to understand the electron spectrum when the superconductivity is disordered. Fermionic excitations in a superconducting state are Bogoliubov quasi-particles which are superpositions of electrons and holes. When the superconductivity is quantum disordered but close to the superconductor-insulator phase boundary, we expect by continuity that the insulator should have a band structure close to the Bogoliubov band. Can an insulator have a Bogoliubov-like band? Does it violate charge conservation? 

In the rest of this subsection, we solve this puzzle of Bogoliubov bands and build intuition on the pairing induced insulator in the intermediate pairing regime. We first analyze it theoretically and then verify the results by a simple numerical calculation.

To the best of our knowledge, the momentum-dependent electron spectrum has not been discussed even in this simple case of fluctuating s wave superconductor; therefore our analysis should be of broader interests.

\section{Fluctuating s-wave superconductor}\label{subsection: gapped sector}

For concreteness, we imagine a metal with 2 bands per spin, each half-filled, to give 2 electrons per unit cell. Under s-wave pairing, the Fermi surface is fully gapped. We then disorder the bosonic pair at low energy while maintaining the pairing to get the bosonic Mott insulator. On the insulating side, close to the transition (where the boson gap closes), we are in the limit that the gap for charge 2e bosonic excitations (which we call $\D_b$) is much smaller than the gap for charge e fermionic excitations (which we call $\D_f$), and they are both smaller than the Fermi energy:
\bea 
\D_b \ll \D_f< E_F.
\eea

For energy scales much smaller than $\D_f$, we cannot excite any fermion; the system is effectively a bosonic system, and all charges are carried by bosons in the low-energy effective description. We then tune the boson interaction at this length scale to drive it to a Mott insulator with a small gap $\D_b$. Note that this procedure can be done most effectively when the range of interaction is comparable to the size of the boson. More physically, each bosonic pair we consider in cuprates spans around 4 lattice spacing, comparable to the MDW enlarged unit cell, but still has considerable overlap with neighboring pairs. We are naturally in the limit where a Mott gap starts to be possible, and it has to be small if there is any. 

Note that we cannot get the desired insulator by treating pairing perturbatively. If we start from a Fermi liquid, and calculate the self energy correction by coupling to a small-gap charge-2e boson, we can at most get a Fermi surface with reduced spectral weight~\cite{PhysRevB.79.245116}. The reason is simply that to connect the unoccupied electrons well-above the Fermi level, and the occupied electrons well-below the Fermi level, the real part of the corrected self energy must change sign by going through zero, hence giving a Fermi surface.
\footnote{In principal, the self energy may also diverge, as the BCS self energy, but it is not possible when the boson is gapped. In fact, such a divergence signals the breakdown of the perturbation.}

In fact, the key feature that makes this insulator easy to understand is precisely that the charge 2e boson gap $\D_b$ is much smaller than the fermion gap $\D_f$. We may compare this feature with a superconductor, where $\D_b=0$ (ignore Coulomb interaction), or with a free-electron insulator, where the lowest bosonic excitation is just the 2-electron excitation at the band minimum, hence $\D_b = 2\D_f$. Interestingly, this pairing-induced insulator is adiabatically connected to a trivial band insulator, but energetically closer to a superconductor.

When the pair excitation gap is much smaller than the single fermion gap, band theory cannot give a satisfactory description. As an effective field theory, we use a complex boson field $\phi$ to describe low energy pair excitations, and a fermion operator to create a gapped unpaired electron. At low energy, the bosonic action should be quadratic in time since it has integer filling per unit cell~\cite{sachdev_2011}.

\bea
\label{Eq:bosonLagrangian}
\mathcal{L}_b &=& \frac12|\partial_t\phi|^2 - \frac12 v_b^2|\nabla\phi|^2 - \frac12 \D_b^2|\phi|^2\\
\mathcal{H}_b &=& \sum_k E^b_k (a^\dagger_ka_k + b^{\dagger}_kb_k),
\eea
where we use canonical quantization to write $\phi(p) = \frac{1}{\sqrt{E^b_p}}(a_p + b_{-p}^{\dagger})$, and $E^b_k = \sqrt{\D_b^2 + v_b^2k^2}$ for small $k$. $\phi_p$ carries charge 2e; $b_p$ and $a_p$ are the annihilation operators of the bosonic pair and the vacancy of pair.

\begin{figure}[htb]
\begin{center}
\includegraphics[width=0.3\linewidth]{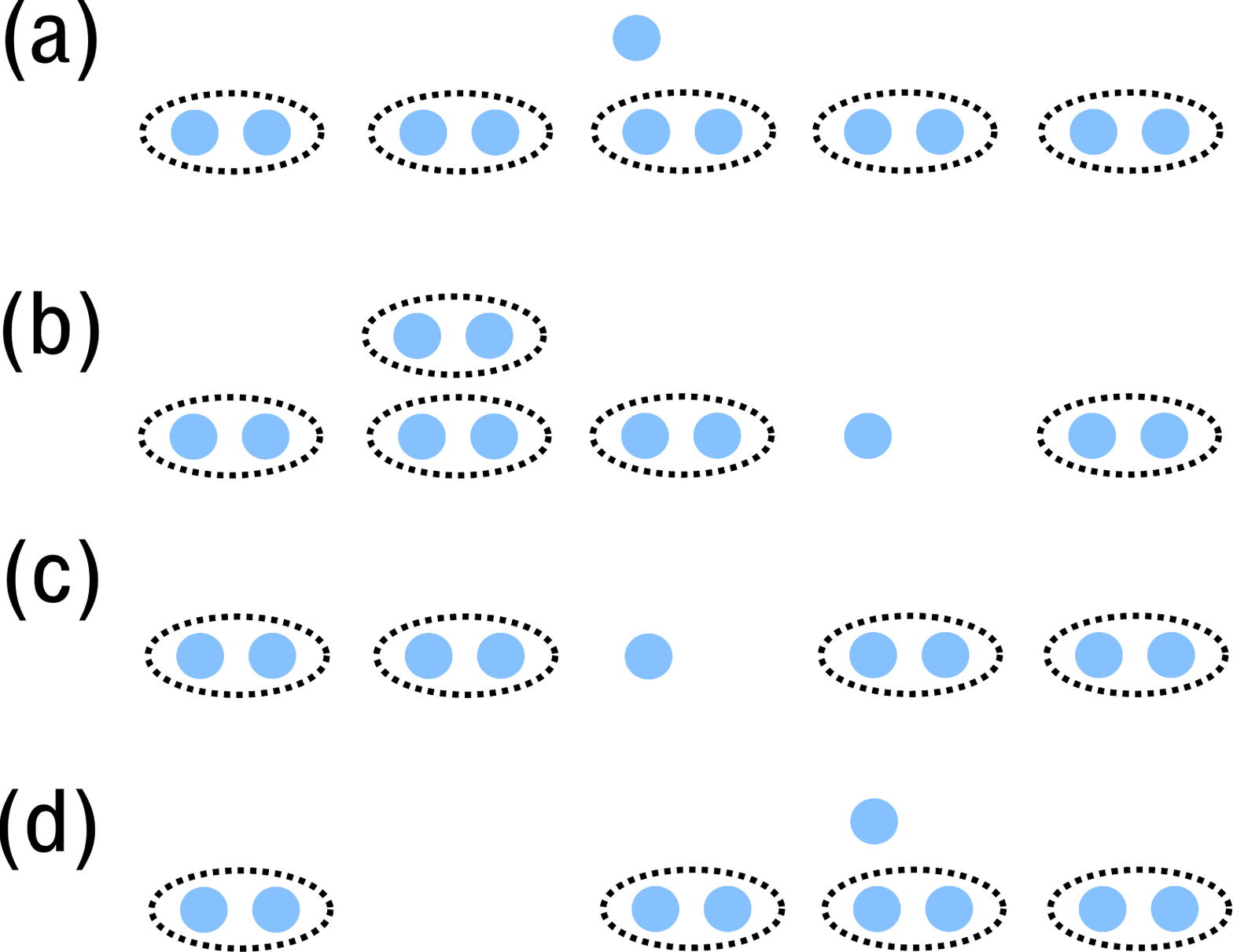}
\caption[Sketch of fermionic excitations in a quantum-fluctuating superconductor]{Fig. (a) and (b): sketch of excitations created by adding an electron. Fig. (c) and (d), sketch of excitations created by removing an electron. In ARPES experiments, the incident photon may break a pair and create a hole (Fig. (c)); it may then decay into the continuum of an electron and a boson vacancy as illustrated in Fig. (d).}
\label{Fig: electron hole continuum}
\end{center}
\end{figure}

As illustrated in Fig.~\ref{Fig: electron hole continuum}, the basic excitations in this system are electrons, holes, pairs and vacancies of pairs. Contrary to our usual intuition, pairs and vacancies of pairs are well-defined quasi-particles in this insulator for they are the lowest charged excitations. For energy scales below $\Delta_f$, the bosonic theory in Eq.~\ref{Eq:bosonLagrangian} is the complete description of low energy excitations.

Since a fermion cannot decay into a boson, electron excitations and hole excitations can still be quasiparticles even though $\D_f$ is much larger than $\D_b$. However, the electron and hole spectra are strongly affected by the low-energy boson; therefore they are very different from the spectrum of a band insulator. As illustrated in Fig.~\ref{Fig: electron hole continuum}(a) and Fig.~\ref{Fig: electron hole continuum}(b), when we add an electron to the system, it may either be a single electron (Fig.~\ref{Fig: electron hole continuum}(a)), or split into a hole and a pair (Fig.~\ref{Fig: electron hole continuum}(b)). Since these two configurations have the same electric charge, an eigen-state of the charge e excitation is always a mixture of the two. In fact, the single electron in Fig.~\ref{Fig: electron hole continuum}(a) is just the special case of Fig.~\ref{Fig: electron hole continuum}(b), where the hole and the pair overlap. Thus, whether the addition of an electron creates a quasiparticle excitation depends on whether the hole and the pair in Fig.~\ref{Fig: electron hole continuum}(b) form a bound state. The physics for removing an electron is similar, as illustrated in Fig.~\ref{Fig: electron hole continuum}(c-d). This line of thinking is particularly useful in the current case, where the boson gap is small. Since the energy of the bosonic pair is small around zero momentum, if the electronic excitation has lower energy than the hole excitation at momentum $k$, the electronic excitation likely form a quasiparticle, but the hole excitation is no longer a quasiparticle: it decays into the two-particle continuum with an electron near momentum $k$ and a boson near momentum $0$.

\begin{figure*}[htb]
\begin{center}
\includegraphics[width=0.8\linewidth]{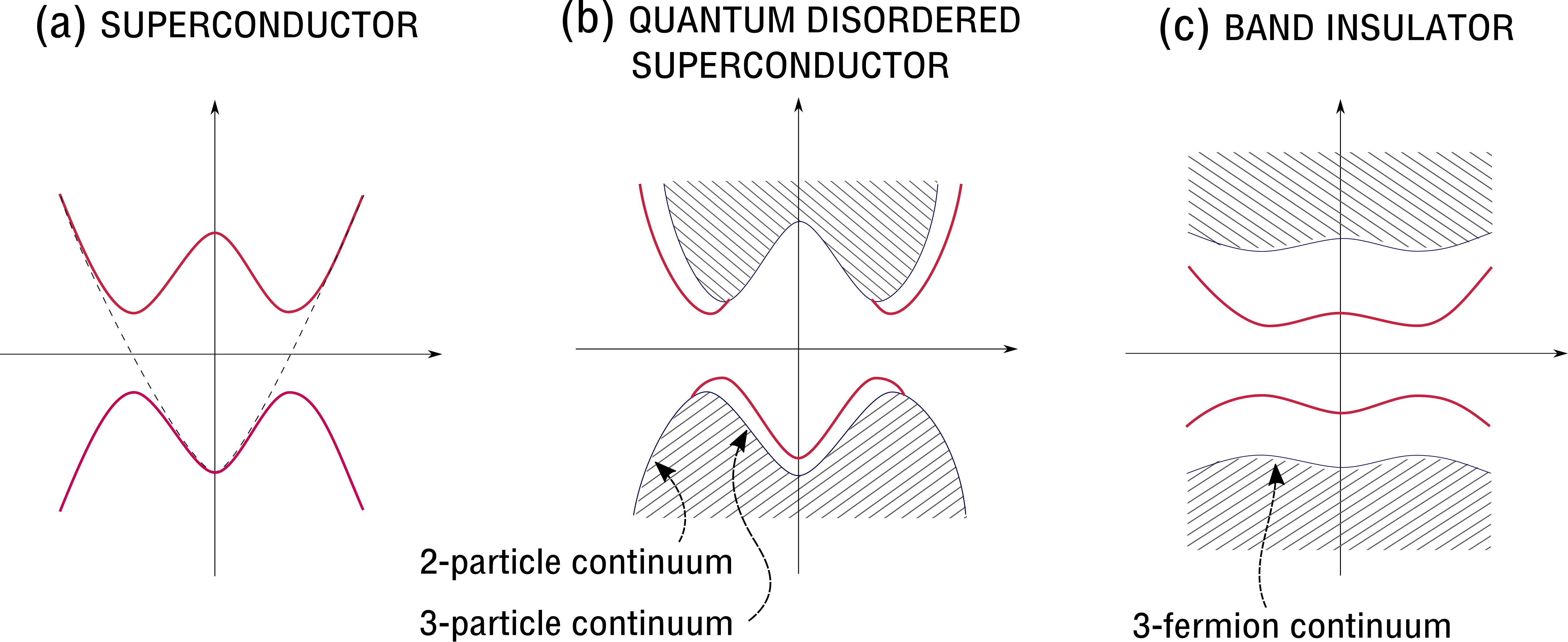}
\caption[Evolution of fermionic excitation from an s wave superconductor to an insulator]{Evolution of fermionic excitation from an s wave superconductor to an insulator. Fig. (a): The BCS band of an s wave superconductor (solid red line). The original electron band before pairing is shown as the dashed line. Fig. (b): Electron band (solid red line) and the boson-fermion continuum (shaded area) when the superconductor is quantum disordered but close to the transition point, $\D_b \ll \D_f < E_F$. The multi-particle continuum here plays a more important role than in usual insulator, because the bosonic pair has a small energy gap when it is close to condensing. The quasi-electron band and the quasi-hole band (solid red line), together with the k-dependent threshold of the 2-particle continuum (shaded) together resembles the BCS band. Fig. (c): electron and hole band in a usual band insulator (solid red line) and the 3-fermion continuum (shaded area). We can smoothly interpolate between Fig. (b) and Fig. (c): as we increase the boson gap, the boson-fermion continuum gradually separates from single-fermion excitations. Eventually, the electron band has little resemblance of the BCS band, the boson fades into the 2-fermion continuum, and the boson-fermion continuum becomes the 3-fermion continuum.}
\label{Fig: fluctuating SC band}
\end{center}
\end{figure*}

In order to understand the fermionic spectrum of the insulator in the limit $\D_b\ll \D_f < E_F$ , we first look at the BCS bands of the superconductor.

\bea
H_\text{f,BCS} = \sum_k (c^\dagger_{k\uparrow}, c_{-k,\downarrow})
\left( \begin{array}{cc}
\epsilon_k & \D_f \\
\D_f & -\epsilon_{-k}
\end{array} \right)
\left(\begin{array}{c}
c_{k,\uparrow}\\ c^\dagger_{-k,\downarrow}
\end{array} \right)
\eea

The fermionic excitations are Bogoliubov quasiparticles with energy
\bea
E^f_k = \sqrt{\e_k^2 + \D_f^2}
\eea

When the boson is barely disordered, we expect the fermionic spectrum to roughly follow the Bogoliubov bands but with two important changes: (1) excitations should now carry definite charges, (2) there may not be quasiparticle excitations at all momenta in this strongly interacting limit. No matter whether there is a quasiparticle or not at a specific momentum $k$, there is always an energy threshold for manybody states with charge $\pm e$ and momentum $k$. When there is a quasi-electron, there is a single state at the threshold instead of a continuum of states; in this case, we define the excitation energy of the quasi-electron to be $E^{e}_k$. Similarly, we define the excitation energy of the quasi-hole to be $E^{h}_k$, if it exists at momentum $k$. By definition, $E^{e}_k, E^{h}_k > 0$. To be consistent with conventions in free electron band theory, we plot $E^{e}_k$ and $-E^{h}_k$, to put charge e excitations in the upper-half plane, and charge -e excitations in the lower-half plane (Fig.~\ref{Fig: fluctuating SC band}). 

When the pairing is smaller than the band width, by continuity, we postulate Fig.~\ref{Fig: fluctuating SC band}(b) as the band structure of the insulator. For momenta  away from the band minimum and larger than the original Fermi momentum, we have the usual electron as a quasi-particle, with energy $E^{e}_k$ slightly distorted from the dispersion of the metal by pairing (Fig.~\ref{Fig: fluctuating SC band}(b), solid red curve in the upper plane). 
\footnote{It may decay into 3 fermions when $E^{e}_k>3\D_f$, but we ignore this usual decaying process for now.} 
There is no way to excite a hole at these \textit{unoccupied} momenta, but we can create an electron and remove a zero-momentum pair, hence a 2-particle continuum for hole excitations starting roughly from the energy $E^{e}_k + \D_b$.
\footnote{Here we assume the boson velocity is not too small, so the energy for bosonic excitation is small only near zero momentum.}
Similarly, for momenta smaller than the original Fermi momentum and away from the band minimum, we have quasi-holes with the energy $E^{h}_k$ (Fig.~\ref{Fig: fluctuating SC band}(b), solid red curve in the lower plane) and a 2-particle continuum for electron excitations starting roughly from $E^{h}_k + \D_b$. Near the band minimum (at the original Fermi surface), we should have at least one of the quasi-electron and quasi-hole, because the lowest fermionic excitation cannot decay into other particles. Since, the electron and hole dispersion are approximately symmetric near the band minimum, we should have a range where quasi-electron and quasi-hole coexist. 

As we follow the electron band from outside the Fermi surface to inside the Fermi surface (in Fig.~\ref{Fig: fluctuating SC band}(b)), the quasi-electron excitation starts to transition from a single electron depicted in Fig.~\ref{Fig: electron hole continuum}(a) to a bound state of hole and continuum depicted in Fig.~\ref{Fig: electron hole continuum}(b). After passing the band minimum, the excitation energy goes up, and the bound state become weaker, and finally the hole and pair no longer bind together, and the quasi-electron fades into the 2-particle continuum. The unbinding transition happens when $E_k^e = \text{min}_p\{E_p^h + E^b_{k-p}\}$. Deep in the Fermi sea, electron excitations do not make sense, and there is not even a resonance above the 2-particle continuum.

The quasi-particle band, together with the threshold of the 2-particle continuum resembles a BCS band. In addition, at energies $2\D_b$ above each quasi-particle excitation, we have a 3-particle continuum of one fermion and a particle-hole pair of bosons. Multi-particle continuum plays an important role in the insulator we discussed because of the small gap of the bosonic pair.

As we drive the insulator farther away from the critical point, the boson gap increases, and the fermion band gradually separates from the boson-fermion continuum. Eventually, the boson gap is so large that it fades into the 2-fermion continuum, and we arrive at a usual band insulator (Fig.~\ref{Fig: fluctuating SC band}(c)).

\begin{figure}[htb]
\begin{center}
\includegraphics[width=0.5\linewidth]{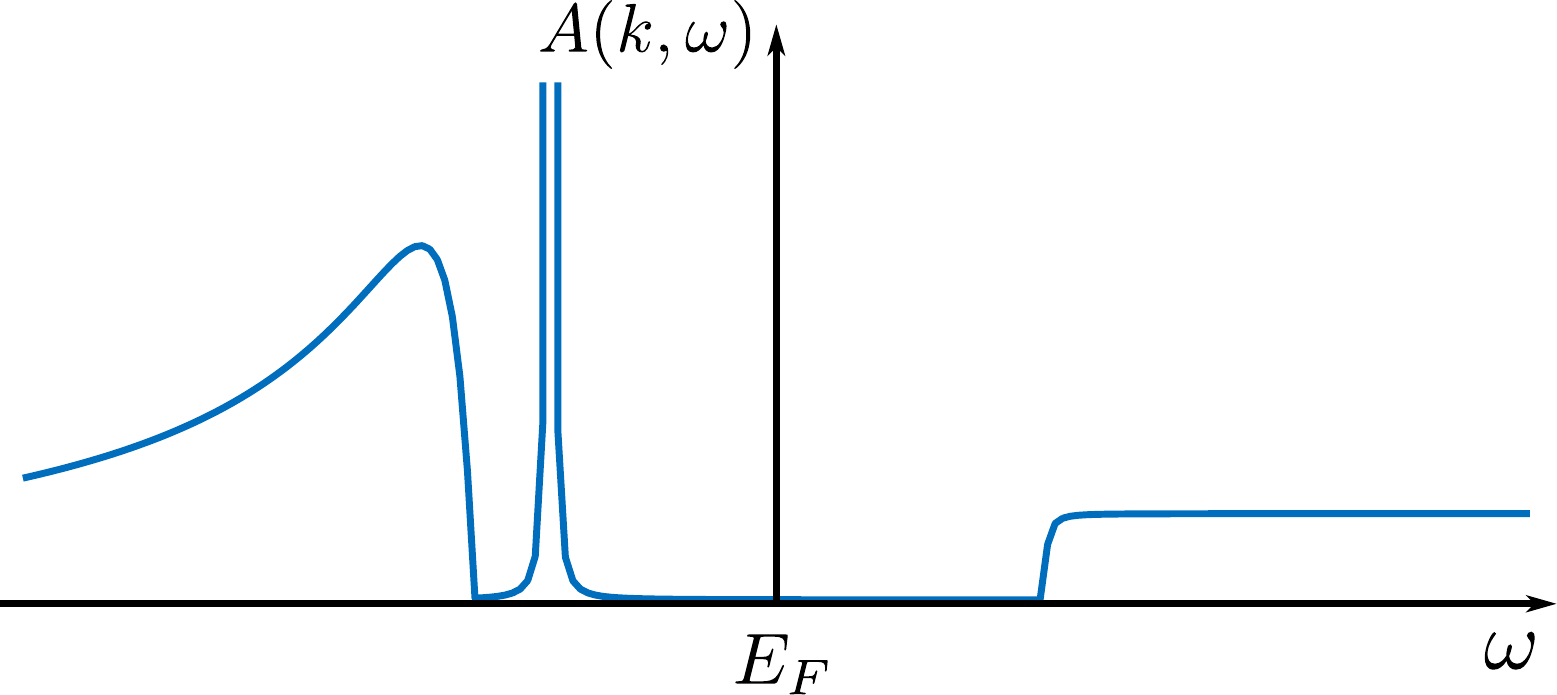}
\caption[Sketch of the electron spectral function at some $k<k_F$]{Sketch of the electron spectral function at some $k<k_F$. Below $E_F$ there is a quasi-hole peak (delta function  in the ideal case, broadened here for the purpose of illustration.) and a 3-particle continuum. Above $E_F$ there is a 2-particle continuum. More detailed calculations in Sec.~\ref{subsection: ARPES} show that the 2-particle continuum onsets as a step function, while the 3-particle continuum decreases as $1/\omega$ for large frequencies.}
\label{Fig:sketchspectrum}
\end{center}
\end{figure}

To further illustrate the unconventional spectral features of this pairing-induced insulator, we sketch the spectral function for a fixed momentum $k<k_F$, where only quasi-hole exists. See Fig.~\ref{Fig:sketchspectrum}. We shall discuss the spectral features of the multi-particle continuum in more details in comparison with ARPES in Sec.~\ref{subsection: ARPES}.

We would like to comment that we present a non-perturbative understanding of fluctuating orders, a way to open a gap on Fermi surface without breaking any symmetry. Our discussion is general; whether the resulting state is energetically favorable or not depends on details. With special care of the charge and momentum carried by the fluctuating boson, similar arguments apply to other fluctuating orders, e.g. PDW, CDW and SDW, if the boson gap is much smaller than the fermion gap. The common feature is that quasiparticle peaks exist only in part of the B.Z., and it must be replaced by boson-fermion continuum in the rest of B.Z.. For fluctuating PDW, the boson has a small energy near a finite momentum $P$; electron at momentum $k$ and hole at momentum $k-P$ compete: if one of them has smaller energy, the other likely falls into the boson-fermion continuum.

\section{Pairing-induced insulator in 1D}\label{subsection: 1D numerics}

\begin{figure}[htb]
\begin{center}
\includegraphics[width=0.7\linewidth]{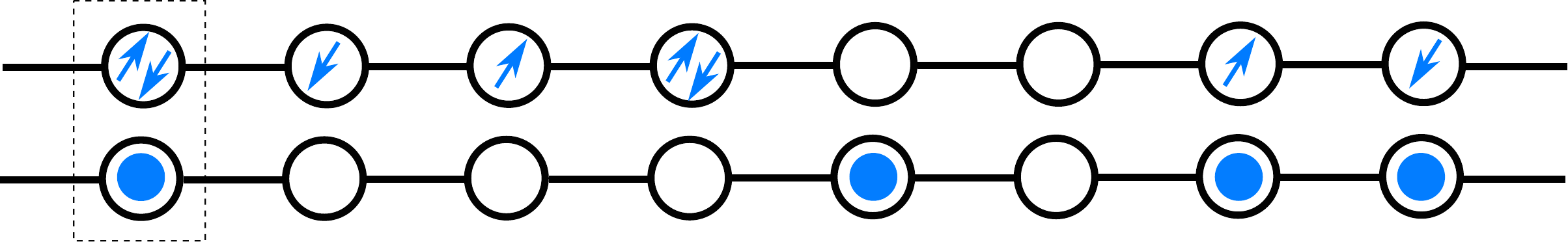}
\caption[1D boson-fermion model]{1D boson-fermion model. Blue dots represent charge-2 hardcore bosons, blue arrows represent spin-up and spin-down fermions}
\label{Fig: 1D lattice}
\end{center}
\end{figure}

To test the idea of pairing induced insulator and its electron spectral function discussed in Sec.~\ref{subsection: gapped sector}, we design a simple 1D model, with charge-1, spin-1/2 fermion $c_{i\s}$ and charge-2, hardcore boson $b_{i}$. As illustrated in Fig.~\ref{Fig: 1D lattice}, each unit cell can have a spin-up fermion, a spin-down fermion, and a hardcore boson, independently. The Hilbert space for each unit cell is 8-dimensional. We choose the Hamiltonian to be:
\bea
H = &-&t_c\sum_{\<ij\>, \s}c^{\dagger}_{i\s}c_{j\s} -t_b\sum_{\<ij\>}b^{\dagger}_{i}b_{j}\nonumber\\
&+& \Delta\sum_{i} b_{i}^{\dagger}c_{i\uparrow}c_{i\downarrow} + h.c. + U\sum_{i} P^{0,4}_{i}
\label{Eq: 1D Hamiltonian}
\eea
where $P^{0,4}_i$ is the projector that is 1 if the $i$th unit cell contains total charge 0 or 4. This Hamiltonian conserves The total charge
\bea
Q = \sum_{i} 2 b^{\dagger}_{i}b_{i} + \sum_{i,\s}c^{\dagger}_{i\s}c_{i\s}.
\eea
There is an overall particle-hole symmetry that pins the total filling to charge-2 per unit cell. (Both the fermion and the hardcore boson are, on average, half-filled.) If $\<b_i\>\neq 0$, the $c_{i\s}$ fermion forms a proximity-induced 1D superconductor~\footnote{In a pure 1D system, we never have $\<b_i\>\neq 0$, but at best a power-law order.}.
What interests us is that even with this purely 1D model, with $b_i$ disordered, the pairing term still opens a fermion gap, but drives the system into an insulating state (for a range of $U$). To make connection with real materials, we can think of the boson as describing well-developed fermion pair of another band. We use this fermion-boson model instead of an all-fermion model, both for numerical convenience, and to illustrate how boson and fermion exchange density dynamically.

The physics of the pairing can be understood as follows. In the free theory, $\D = U = 0$, the left-moving and right-moving electron operator $c_{L,\sigma}$ and $c_{R,\sigma}$ have scaling dimension $1/2$. Without further interaction, the hardcore boson corresponds to a free fermion under Jordan-Wigner transformation, and $b^{\dagger}\sim e^{i\phi}$ has scaling dimension 1/4.
\footnote{We can determine the scaling dimension of the boson operator by bosonization. Write $b^{\dagger}\sim  e^{i\phi}$, and the corresponding left-moving and right-moving fermion after Jordan-Wigner transformation as $f_{R/L}^{\dagger} = e^{i(\phi \pm \theta)}$. As free fermion operators, $f_L $ and $f_R$ have scaling dimension 1/2, and $f_Lf_R\sim e^{2i\phi}$ has scaling dimension 1. Thus, $b^{\dagger}\sim e^{i\phi}$ has scaling dimension 1/4.}
Thus the pairing interaction $b^{\dagger}c_{\uparrow}c_{\downarrow}$ has scaling dimension $5/4$ and is relevant. The gapless fermion is unstable to pairing. The pairing renormalizes the bare boson operator $b$ into $\tilde{b} \sim u b+ v c_{\uparrow}c_{\downarrow}$. A single electron with no partner to form a pair fails to make the superposition with the boson, resulting in a pairing gap. Below this pairing gap, the model is effectively a model of the renormalized boson. The renormalized boson takes the density of both the bare boson and the fermion pairs below Fermi surface, becoming filling 1 per unit cell at low energies. Adding infinitesimal $\Delta$ immediately draw the system from the independent boson-fermion Luttinger liquids, to a one-component bosonic Luttinger liquid at low energy. Whether the bosonic Luttinger liquid is stable depends on the renormalized bosonic repulsion.

\begin{figure*}[htb]
\begin{center}
\includegraphics[width=0.9\linewidth]{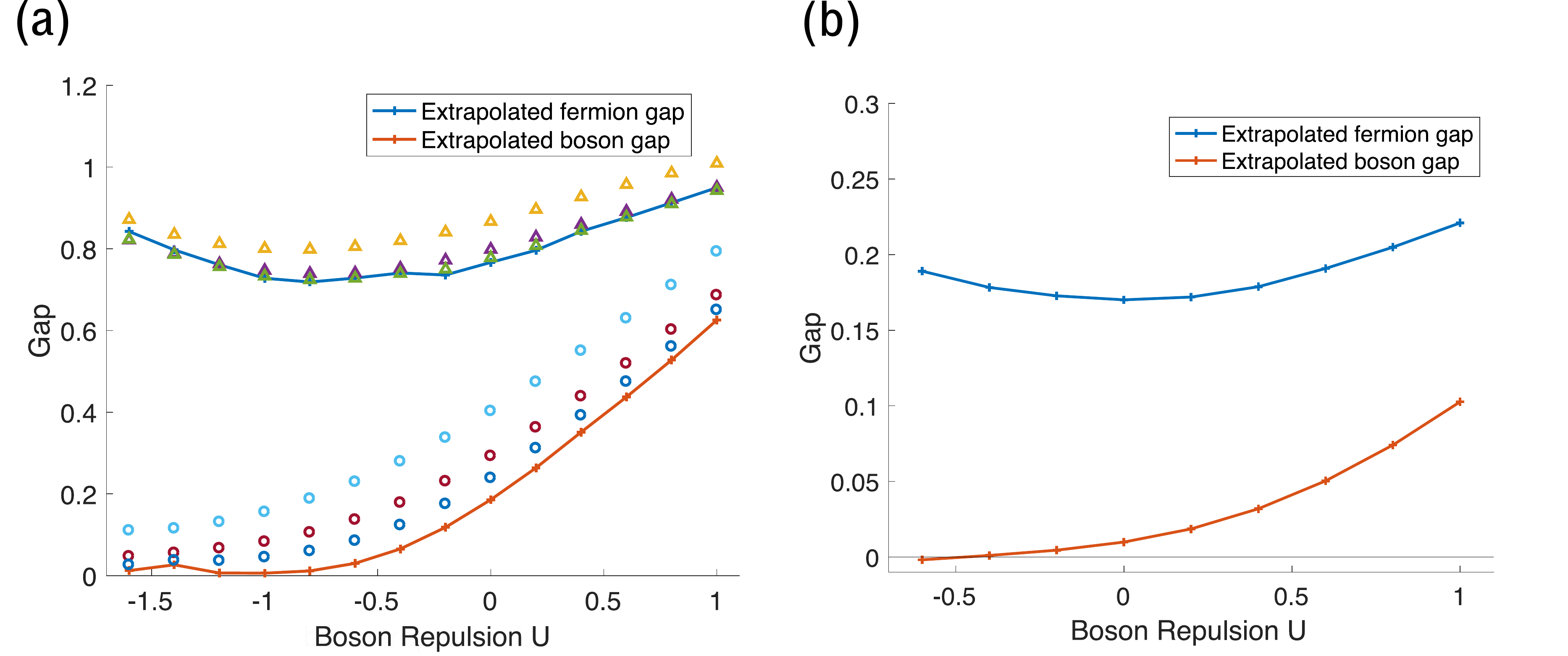}
\caption[Fermion and boson gap of the 1D model]{(a) fermion gap (blue `+') and boson gap (red `+') of the 1D model, extrapolated from finite size DMRG calculation with system size $L=10,20,40$, $t_b = t_c =1,\D=1.3$, total filling: charge-2 per unit cell. Fermion gaps for $L =10$ (yellow triangle), $L=20$ (purple triangle), $L=40$ (green triangle), and boson gaps for $L = 10$ (light blue circle), $L = 20$ (dark red circle), $L=40$ (dark blue circle) are shown for reference. (b) The same as (a) except for small pairing $\D=0.5$. The finite-size extrapolation is shown in Appendix~\ref{Appendix: DMRG}. In both cases, the ground state go through a transition from a bosonic Luttinger liquid to a bosonic Mott insulator.  Fermion gap stays open across the transition.}
\label{Fig: 1D fermion boson gap}
\end{center}
\end{figure*}

\begin{figure}[htb]
\begin{center}
\includegraphics[width=0.7\linewidth]{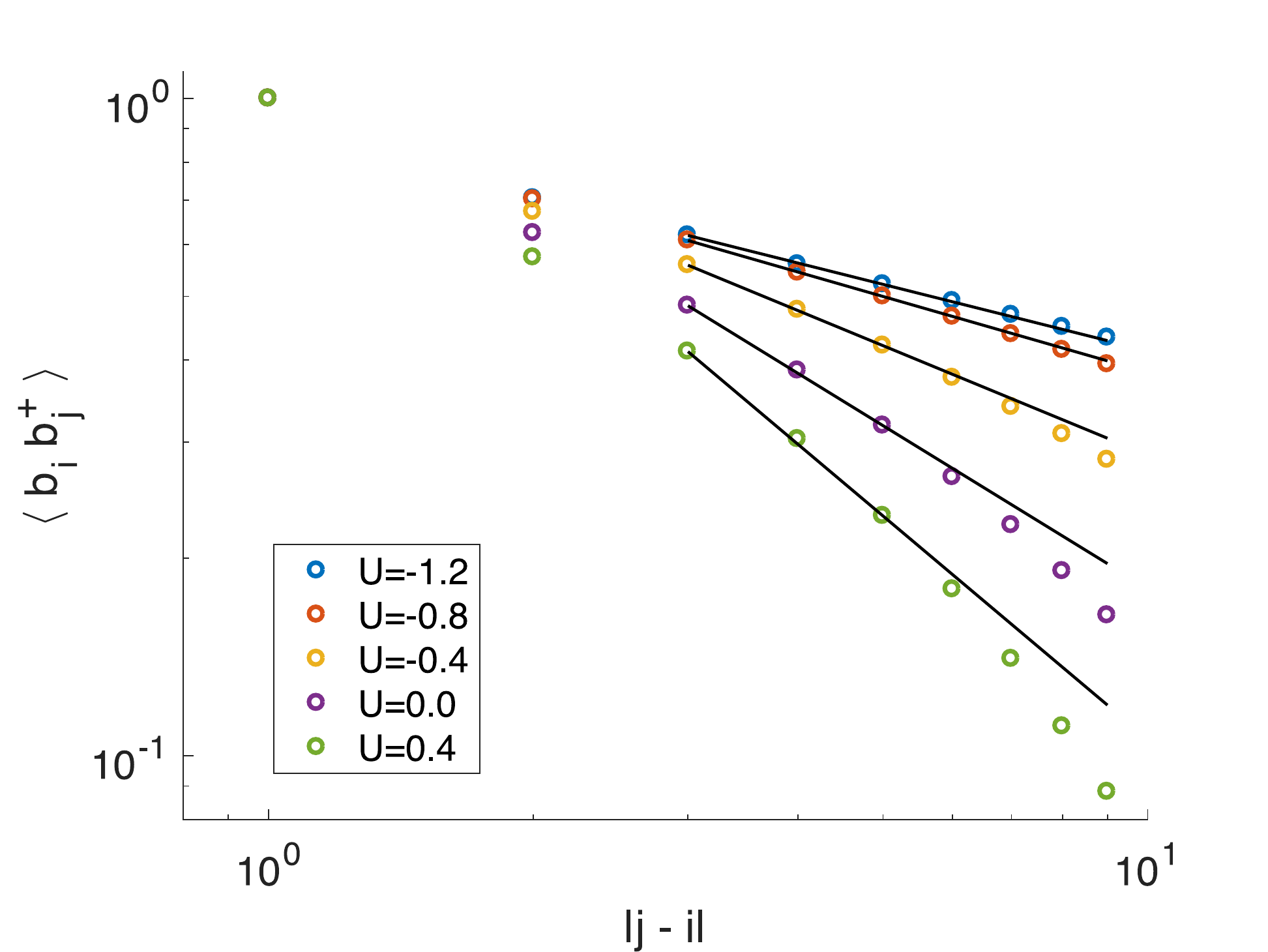}
\caption[Boson correlator of the 1D model]{Boson correlator $\< b_ib^{\dagger}_j\>$, in log-log scale. We use the ground state calculated by DMRG for $L=40$, fix $i=16$, and scan $j = 17,18,\dots, 25$. Black lines are guides to the eye. The correlator decays as power-law for $U=-1.2,-0.8$, but faster than power-law for $U=-0.4,0.0,0.4$, consistent with the gap calculated by DMRG.}
\label{Fig: 1D boson correlator}
\end{center}
\end{figure}

By tuning the bosonic Hubbard $U$, we can realize 3 different phases. For large repulsive $U$, we should have a bosonic Mott insulator in 1D, with charge 2 per unit cell. The state on each site is a superposition between the fermion pair and the bare boson.
(Since translation and particle-hole symmetry is maintained, the average occupation of the bare boson is 1/2 per site.)
For a range of attractive $U$, the renormalized boson forms a charge-2 Luttinger liquid. Single fermion is gapped, but the pair is gapless, realizing a Luther-Emery liquid. For large attractive $U$, we either have a CDW or phase separation. The charge on each site wants to deviate from 2, either smaller or larger. Note that no matter what $U$ is, single fermion is always gapped by the pairing. By design, the original boson itself has average filling $1/2$ and it is impossible to form a Mott insulator on its own. Seeing an insulator that preserves the translation symmetry implies that the boson has absorbed all the fermions to increase its effective filling to 1. We are interested in the transition between the Luther-Emery liquid and the Mott insulator, i.e., the emergence of the insulating phase with a small Mott gap.

We calculate the approximate ground state by DMRG for systems with length $L = 10,20,40$. We consider two cases, with large pairing ($t_b = t_c =1, \Delta = 1.3$) and relatively small pairing ($t_b = t_c =1, \Delta = 0.5$). In each case, we scan $U$ to drive the system from the bosonic Luttinger liquid to the pairing-induced insulator. For all parameters shown in Fig.~\ref{Fig: 1D fermion boson gap} and Fig.~\ref{Fig: 1D boson correlator}, we find that translation symmetry is preserved in the bulk. In the large pairing case (Fig.~\ref{Fig: 1D fermion boson gap}(a)), the extrapolated boson gap (red `+') is zero within the error bar for approximately $U\le -0.8$, and nonzero above that, indicating a continuous phase transition into an insulating ground state (see also the boson correlator in Fig.~\ref{Fig: 1D boson correlator}). On the other hand, the fermion pairing gap (blue `+') barely changes during the process, even deep in the insulating side. The pairing-induced insulating phase with $\Delta_b < \Delta_f$, which we are mostly interested in, is clearly present. The small pairing case ($\Delta=0.5$, Fig.~\ref{Fig: 1D fermion boson gap}(b)) shows the same physics. Note that the boson gap is still well-below the fermion gap even when the bare repulsion $U$ is much larger than the fermion gap, because the weakly bound renormalized boson feels a much smaller effective repulsion.  Theoretically, we know the renormalized boson goes through a KT transition at zero temperature in 1+1 dimension. We found the critical U to be around $-0.7$ for $\Delta = 1.3$, and $-0.2$ for $\Delta = 0.5$.

\begin{figure}[htb]
\begin{center}
\includegraphics[width=3in]{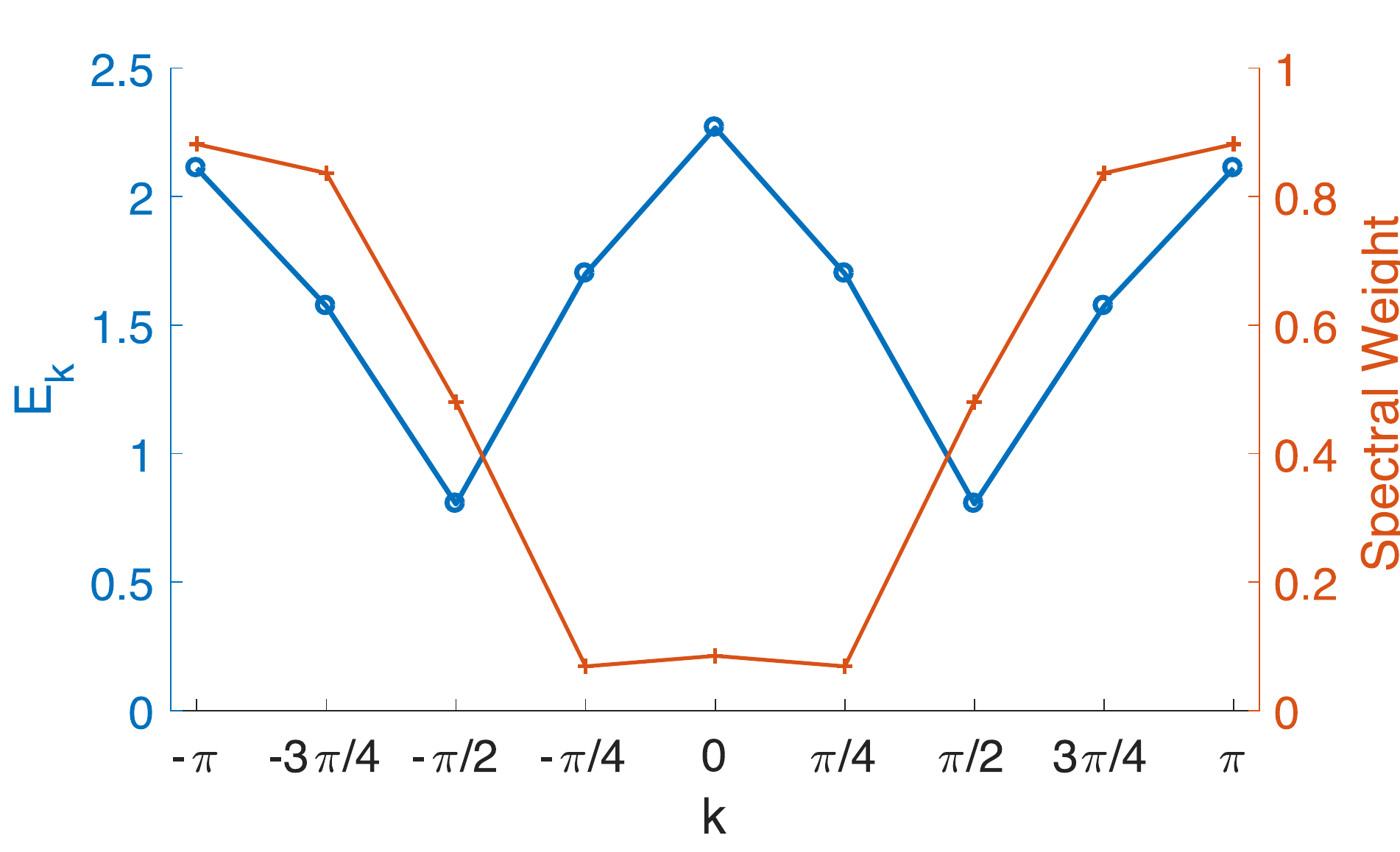}
\caption[Spectrum and spectral weight of the lowest fermionic excitations]{Spectrum (blue circles) and spectral weight (red `+') of the lowest charge +1 fermionic excitations, for $-\pi<k<\pi$, $t_b = t_c =1, \D = 1.3, U=-1$. As sketched in Fig.~\ref{Fig: fluctuating SC band}(b), the threshold of fermionic excitations roughly follows the Bogoliubov band. Fermion excitations outside the Fermi sea are quasiparticles. Inside Fermi sea, the thresholds represent 2-particle continuum with zero quasiparticle weight.}
\label{Fig: 1D fermion spectrum weight}
\end{center}
\end{figure}

Finally, we compute the energy threshold for charge-1 excitations at each momentum for $L=8$ (Fig.~\ref{Fig: 1D fermion spectrum weight}) by the Lanczos algorithm. The blue line shows its dispersion, which roughly follows the BCS curve. The red line shows the spectral weight of the excitation: $Z\equiv |\<n|c^{\dagger}_{k}|0\>|^2$. This confirms our physical picture as we illustrated in Fig.~\ref{Fig: fluctuating SC band}. We find that the state for the addition of a single fermion  has considerable overlap with the original fermion for $k>\pi/2$, where the free-fermion band is unoccupied; and vanishing overlap with the original fermion for $k<\pi/2$, where the excitation is essentially hole plus pair.

\chapter{Quantum Fluctuating Pair Density Wave in Cuprates}
\label{chapter: fluctuating PDW in cuprates}

In the previous section, we discuss the fluctuating s wave superconductor. We figure out how pairing gap survives when the superconducting order is quantum fluctuating. We analyze the evolution of the fermion spectrum from the Bogoliubov band of the superconductor to the charge-conserved band of the insulator. However, quantum disordering the PDW state still seems a daunting task.

Following the discussion in Sec.~\ref{sec:outlook}, we imagine increasing the magnetic field in the d-wave superconductor. Vortices carrying static short-range PDWs emerge one by one as we increase the field. The d-wave superconductor is gradually replaced by local patches of the vortex-core state, and is completely destroyed when the vortices overlap at $H=H_\text{c2}$. The high-field ground state should have a considerable PDW amplitude like in the vortex core. But without the vortex to pin it, the PDW naturally fluctuates. The goal of this chapter is to clarify the nature of this state and expose as much of its physical properties as possible. A number of questions immediately come to mind:

1.  Is this a metallic state?  If so, is it smoothly connected to a conventional metal? Do we need to appeal to exotic concepts such as electron fractionalization and topological order to describe this state? If there is a Fermi surface, does it obey Luttinger’s theorem?

2. Does this state have small electron pockets that is consistent with those ob-served in quantum oscillation experiments?

3.  Why is there no sign of superconducting fluctuations in transport data in this high field low temperature regime. Are there other signs of the superconducting fluctuations? Are there other data that can be explained by this point of view that are difficult to explain otherwise?

We give our theoretical construction in the next section, providing a possible set of answers of the questions above. For simplicity, we shall focus on the commensurate case, with period-6 PDW and period-3 CDW. (See Sec.~\ref{Eq: PDW mean field} for the band structure of the static PDW.) Our construction also applies to the incommensurate case with little change. We shall briefly comment on it in Sec.~\ref{subsec: Luttinger}. In the last section, we compare our theory with experiments.

\section{Constructing the fluctuating pair density wave state}\label{sec: constructingthefluctuatingPDWstate}

In this section, we describe a way to quantum disorder the PDW to arrive at the desired pseudogap ground state.

At the first glance, the task seems  overwhelming. The interplay between the phase fluctuation of superconductivity and gapless electron modes is usually difficult to deal with. Previous theoretical discussions on quantum disordering a zero-momentum d wave superconductor, which has gapless electron nodes, lead to models with fractional degrees of freedoms, and the discussions are yet to be settled (Ref.~\cite{PhysRevB.60.1654,PhysRevB.66.054535}). In our case the PDW has gaps only near the anti-nodes and a gapless region exists in the form of Fermi arcs. The gapless excitations seem to make the task even harder. However, it turns out that the composite order in the form of CDW comes to our rescue. The CDW connects the Fermi arcs and produces electron pockets, in the way proposed by Harrison and Sebastian~\cite{PhysRevLett.106.226402}.

An important bonus of this picture is that in the new reduced BZ, the only gapless excitations are those of the electron pockets and these are naturally decoupled from the Bogoliubov-like quasi-particles that are associated with the PDW pairing. This is the key insight that allows us to make progress on this long standing problem.


For anti-nodal electrons, the central puzzle is whether the anti-nodal gap persists as an electron gap when PDW is disordered, and if so, how to understand the gapped electron spectrum when PDW is disordered. In the previous section, we have solved this puzzle in fluctuating s-wave superconductors. Despite differences in pairing momentum and form factor, the physics of the electron gap and the interplay between electrons and pairs are essentially the same in the case of fluctuating pair density waves.

It is experimentally observed that cuprate high-temperature superconductors have a very short coherence length, about 4 lattice spacing. It suggests the size of a pair is roughly comparable with the distance between neighboring pairs and the size of the MDW enlarged unit cell we consider; therefore the Coulomb repulsion between neighbouring pairs may drive the pairs into a Mott insulating phase. We propose the scenario that the anti-nodal electron gap is preserved when PDW is disordered, and the electron pairs form a Mott insulator in the MDW enlarged unit cell without further symmetry breaking. In Sec.~\ref{subsec: Constructing the fluctuating PDW ground state} and Sec.~\ref{subsection: ARPES}, we apply the theory of a fluctuating fully gapped superconductor to describe the anti-nodal electron spectrum. 

Theoretically, the idea of a tight pair goes back to Anderson: roughly speaking, a hole in the $t-J$ model breaks a spin singlet nearby, two holes can avoid breaking two singlets by forming a pair, resulting in a pairing energy at a fraction of $J$. There has also been earlier discussions treating the anti-nodal pairs as bosonic preformed pairs that are coupled to the nodal electrons.~\cite{PhysRevB.55.3173}.  

Unlike antinodal electrons, which are strongly paired under PDW, nodal electrons barely couple to the PDW because of momentum mismatch. (The PDW momentum $P$ is about twice the anti-nodal Fermi momentum; as seen from Fig ~\ref{Fig: Bogoliubov bands}a, it is considerably larger than the momentum that can be formed with a pair of electrons in the small Fermi pocket.) 
The nodal `arcs' are cut out and reconnected by the secondary CDW and remains largely unchanged by the PDW. Therefore while they are in principle  Bogoliubov bands, the gapless nodal bands can be viewed as electron bands weakly coupled to the PDW condensate. When the PDW disorders, the nodal bands go back to a pure electron band.

For the gapless bands coming from nodal electrons, the Bogoliubov-band paradox shows up in a different way. In the mean-field calculation (Fig.~\ref{Fig: Bogoliubov bands}), there are 2 gapless bands, hence 2 pockets, with identical shape, shifted by the PDW momentum, but the 4 `arcs' on the original Fermi surface can only form one closed pocket. From the perspective of total gapless degrees of freedom, the 2 pockets in the ordered PDW state is actually one pocket per spin, the same as we expect for the Harrison-Sebastian pocket. This is because the Nambu spinor representation $(c_{k\uparrow}, c_{P-k\downarrow}^{\dagger})^\text{T}$ already includes both spins, and puts down spin at shifted momenta. However, in the PDW-ordered state, due to the small but nonzero mixing of $c_{k\uparrow}$ and $c_{P-k\downarrow}^{\dagger}$, the gapless fermions acquire a nonzero spectral weight at PDW-shifted momenta, which should be absent in the PDW-disordered ground state. As we disorder the PDW, we need to explain how this extra spectral weight disappears. The answer is also rooted in the interplay between the bosonic pair and the electron, which we discuss in Sec.~\ref{subsection: gapless sector}.

In summary, by disordering the PDW, we arrive at a metallic state with a small electron pocket in the B.Z. folded by CDW and MDW. The extra charge density is carried by paired electrons which form a Mott insulator in the enlarged unit cell. The antinodal pairing gap is maintained. The state we are describing is adiabatically connected to a conventional small-pocket Fermi liquid with a large insulating gap of antinodal electrons.

We arrange this section as follows. In Sec.~\ref{subsection: gapless sector} we discuss the physics of the gapless sector. In Sec.~\ref{subsec: Luttinger}, we discuss how electrons partition their density into the gapped antinodal insulator and the gapless pocket, specifically, how Luttinger's theorem can be satisfied. In Sec.~\ref{subsec: Constructing the fluctuating PDW ground state} we synthesize understandings of simple models into a construction of the fluctuating PDW ground state in cuprates.

\subsection{Gapless sector: electron pocket}\label{subsection: gapless sector}

\begin{figure}[htb]
\begin{center}
\includegraphics[width=0.6\linewidth]{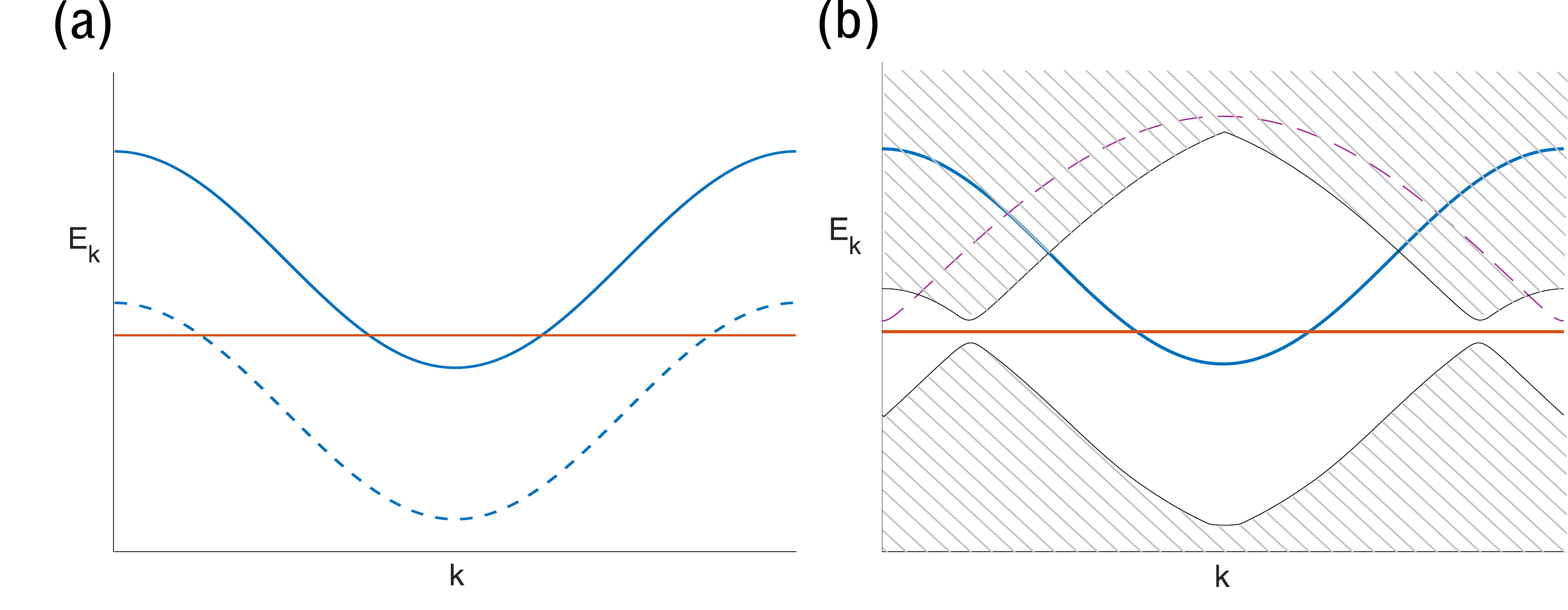}
\caption[Gapless band in PDW-ordered and PDW-disordered states]{Gapless band in (a) PDW-ordered, (b) PDW-disordered state. Solid blue lines in (a) and (b) represent the bare electron dispersion. Solid orange lines in (a) and (b) represent Fermi energy. The dashed blue line in (a) represents the PDW-reflected band. The dashed purple line in (b) represents boson dispersion. The upper/lower shaded area in (b) represents 2-particle continuum of charge $\pm1$, which is calculated from the assumed fermion dispersion (solid blue curve) and boson dispersion (dashed purple curve).}
\label{Fig: gapless band continuum}
\end{center}
\end{figure}

In the previous two subsections, we use simple models to illustrate the physics relevant to the gapped sector of the fluctuating PDW. We introduce the low-energy effective theory, the boson theory, of the quantum-disordered superconductor, and analyze the influence of the small-gap boson on gapped electrons. 

In this subsection, we use the following model to illustrate the physics of the gapless sector in the fluctuating PDW state.

\bea
H &=& \sum_{k}\e_k c_k^{\dagger}c_{k} + \sum_{k} (E^b_k)^2|\phi_{k}|^2  \nonumber\\
&+& \lambda\sum_{k,q}\phi_{\pi+q}c_{k\uparrow}c_{\pi -k-q\downarrow} + h.c.\\
&=& \sum_{k}\e_k c_k^{\dagger}c_{k} + \sum_{k} E^b_k (b_k^{\dagger}b_k + a_k^{\dagger}a_k)  \nonumber\\
&+& \lambda\sum_{k,q}\frac{1}{\sqrt{E^b_p}}(a_{\pi+q} + b_{\pi-q}^{\dagger})c_{k\uparrow}c_{\pi -k-q\downarrow} + h.c.,
\eea
where $\phi(p) = \frac{1}{\sqrt{E^b_p}}(a_p + b_{-p}^{\dagger})$ is the relativistic boson field describing fluctuating PDW, as introduced in Sec.~\ref{subsection: gapped sector}.
We assume the bare electron has a small pocket at the center of the B.Z., with a dispersion of the solid blue curve in Fig.~\ref{Fig: gapless band continuum}. The bosonic pair ($b_k$) and vacancy of pair ($a_k$) are related by approximate particle-hole symmetry near its superconductor-insulator transition. We assume their band minimum is at momentum $\pi$. We also assume their dispersion is given by the dashed purple curve in Fig.~\ref{Fig: gapless band continuum}(b). In the third term, we are interested in small $q$, and those $k$ around 0 and $\pi$.

If the boson condense at $\pi$-momentum, $\<\phi_\pi\>\equiv\phi_{s}\neq 0$, we can rewrite the fermion in Numbu basis, $\Psi_k\equiv (c_k,c_{\pi - k}^{\dagger})^{\text{T}}$. At the mean-field level

\be
H_f = \sum_k \Psi_k^{\dagger}
\left( \begin{array}{cc}
\e_k & \lambda\phi_s^*\\
\lambda\phi_s & -\e_{\pi - k}
\end{array} \right) \Psi_k
\ee
Since $\e_k$ and $-\e_{\pi - k}$ always have a large difference (Fig.~\ref{Fig: gapless band continuum}(a), solid blue line and dashed blue line), the coupling barely does anything. The band structure is the original electron band plus the reflected band. Due to the small mixing between the two bands,  the new gapless pocket at $\pi$ gains a small electron weight.

If the boson disorders, to the first order, the coupling can be ignored and the fermion maintains its bare single-band dispersion, with only one gapless pocket (Fig.~\ref{Fig: gapless band continuum}(b)). However, the reflected band maintains its presence at finite energy. We can create a hole of the solid blue band and a pair in the dashed purple band to make a 2-particle continuum for electronic excitation. The energy of the two-particle excitation at momentum $k$ can be $|\e_q| + E^b_{k-q}$ for every momentum $q$ such that $\e_q <0$ (so that we can excite a hole at momentum $q$). We calculated possible values of the two-particle excitation energy from the assumed boson and fermion distribution, and illustrate them as the shaded region in the upper half plane. Similarly, there is a two-particle continuum of an electron and a vacancy of pair. The two-particle continuum is strictly gapped since the boson is gapped. When $\D_b$ is small, part of the threshold of the continuum roughly resembles the reflected band shown in Fig.~\ref{Fig: gapless band continuum}(a). The rest of the threshold follows the boson dispersion. 

\subsection{Partition of the electron density and the Luttinger's theorem}
\label{subsec: Luttinger}

The reader may reasonably worry about the abrupt nodal-anti-nodal partition, for there is no sharp distinction between nodal and anti-nodal electrons on the original Fermi surface. Furthermore, for the above construction to work we need to partition the charge density, so that the bosonic pair is at commensurate density to form a Mott insulator, and the gapless pocket satisfies Luttinger's theorem. But the nodal electron pocket we start with is given by a mean-field PDW, which is a pairing state and does not satisfy Luttinger's theorem automatically. 

Our justification of this partition is twofold. First, the CDW descending from PDW cut the original Fermi surface into separate bands, so there is a natural distinction between nodal and anti-nodal electrons; second, the partition of density between the gapless fermion and the boson is a property of the energetics of the manybody ground state, which the mean-field PDW fails to address. Here we can only argue that such a partition is locally stable. Let us imagine that at some density, the gapless Fermi pocket  satisfies Luttinger's theorem in the reduced B.Z., consequently, the boson has integer filling consistent with the requirement of a Mott insulator. At low energies, the boson sector and the fermion sector effectively decouple. As we dope the system away from that density, it is energetically favorable for the extra electrons/holes to enter the gapless sector to avoid paying the Mott gap. Thus, the boson-fermion phase we considered is stable in a range of doping. Whether underdoped cuprates choose to partition its density this way, however, is an energetic question that can be tested only experimentally.

Next we check whether the available expereimental data are consistent with Luttinger's theorem. 
Although STM reports commensurate CDW of period 4 in a range of underdoped $\text{Bi}_2\text{Sr}_2\text{CaCu}_2\text{O}_{8+x}$ (Bi2212), resonant x-ray scattering and non-resonant hard x-ray diffraction report an incommensurate CDW in YBCO, with period smoothly passing through 3, and in $\text{HgBa}_2\text{CuO}_{4+\delta}$ (Hg1201), with period smoothly passing through 4.\cite{PhysRevB.96.134510} Whether a specific cuprate has incommensurate or commensurate CDW may depend on details like the strength of lattice-pinning, but the existence of CDW seems to be universal. Since Luttinger's theorem is a well defined concept only for commensurate superlattices, we restrict ourselves to commensurate CDW and PDW here. The incommensurate case will be viewed as comprising of commensurate domains. 

To compare with experiments, we identify the CDW momentum measured experimentally as twice the PDW momentum, and we check whether the pocket size measured from quantum oscillation obeys Luttinger's theorem at the specific doping when the CDW is commensurate. This kind of data is available only for the YBCO and Hg1201 systems, and within error bar, both YBCO and Hg1201 pass the test. According to Ref.~\cite{PhysRevB.96.134510}, in YBCO, the CDW has momentum about $0.33*2\pi$ at 8\% doping, where the electron pocket is about  1.5\% of the original B.Z.,  accommodating 3\% of the electron density. The rest of the density, 0.92-0.03 = 0.89 per unit cell, is consistent with 16/18 = 0.89, ie 8 charge $2e$ bosons per MDW unit cell (which is 18 times the original unit cell).
\footnote{Equivalently, we can count the charges relative to half-filling, and say there is a pair of holes per MDW unit cell. These two countings are equivalent because the area of the MDW unit cell is an even multiple of the area of the original unit cell.}
In Hg1201, the experimental data is limited and we follow Ref.~\cite{PhysRevB.96.134510} to use their numbers based on the use of a parametrized band structure which they found to be in excellent agreement with the data.  The CDW has momentum about $0.25*2\pi$ at 12\% doping, where the electron pocket is about 4\% of the original unit cell, the rest of the density, 0.88 - 0.08 = 0.80 per unit cell, is consistent with 26/32 = 0.81, ie 13 bosons per MDW unit cell (which is 32 times of the original unit cell). 

The doping at which the CDW is commensurate can be determined experimentally with an error bar of roughly $1\%$, which is inherited from the error bar of the CDW momentum~\cite{PhysRevB.96.134510}. This uncertainty gives an uncertainty of the expected Fermi surface area, which is about $10\% \sim 15\%$ of the folded B.Z. We note that the test of Luttinger theorem is most sensitive to the doping density at a given commensurate doping, and the pocket size is only a small correction.  Thus Luttinger's theorem poses a highly nontrivial test to candidate theories as long as the doping density at a commensurate CDW momentum is  known with reasonable accuracy.

To further illustrate the nontriviality of the Luttinger theorem test, we note that the choice of the MDW unit cell is crucial. Since only  CDWs at 2P have been observed, one might be tempted to choose  $2P\hat{x}$ by $2P\hat{y}$ as the reduced BZ instead. In this case the real space unit cell is half the size of the MDW unit cell and we will have 6.5 bosons per unit cell for the Hg1201 case. This violates the integer density condition for the bosonic Mott insulator. In other words, if we form a bosonic Mott insulator in the CDW superlattice,  Luttinger's theorem will be strongly violated.

Finally, we comment on the incommensurate case~\cite{Dominic}. When the MDW momenta $P\hat{x}\pm P\hat{y}$ is incommensurate with the original primitive vectors $\frac{2\pi}{a}\hat{x},\ \frac{2\pi}{a}\hat{y}$. We do not have a single well-defined B.Z. However, we can peak any two of the above reciprocal vectors to form a B.Z.. For example,  we get a B.Z. with area $2P^2$ from $P\hat{x}+P\hat{y}$ and $P\hat{x}-P\hat{y}$, the original B.Z. with area $(2\pi/a)^2$, and a B.Z. with area $\frac{2\pi P}{a}$ from $P\hat{x}+P\hat{y}$ and $\frac{2pi}/{a}\hat{x}$. If the total density is an even integer times any of the three B.Z. areas we can in principle have an atomic insulator. For example, just start with an atomic insulator whose B.Z. is spanned by the two selected momenta, and slightly perturb the position of the electrons according two the other two momenta. Putting several atomic insulators together, we see an insulating state is possible whenever

\bea
\text{electron density} = 2m\times 2P^2 + 2n\times\frac{2\pi P}{a} +2k\times(\frac{2\pi}{a})^2,
\eea
where $m,n,k$ are integers. Comparing to the commensurate case, we have less restrictions. If $m,n,k$ can be arbitrarily large, we have little predicting power. In reality, we expect these numbers to be not too large. 

Theoretically, we can always let the bosons form a Wigner crystal that further breaks translation symmetry, thus adding new reciprocal momenta. However, we restrict ourselves to the scenario that disordering the PDW does not further break translation since there is no experimental evidence for the new momenta.

\subsection{Fluctuating pair density wave state in cuprates}\label{subsec: Constructing the fluctuating PDW ground state}

Now, we are ready to apply insights acquired from the simple models to construct the fluctuating PDW state for cuprates.

Under the assumption that the pseudogap is a fluctuating PDW gap, we estimate relevant energy scales as follows. The anti-nodal fermion gap in Bi2212 near 12\% doping, measured by ARPES and STM, is around $60\ $meV. We identify it with $\Delta_f$ in previous theoretical analysis. As we move to the nodal direction, the fermion gap decreases. From the mean field calculation, the lowest gapped band has a gap around $30$ meV. The boson gap has not been measured yet, and we roughly estimate it as follows. Without other obvious velocity scale, we assume the boson velocity to be similar to the \textit{anti-nodal} Fermi velocity. Therefore  $\Delta_b \sim \Delta_f\cdot(\ \text{coherence length}\ /\ \text{correlation length})$, which is between $10$ meV and $30$ meV.

Of the 36 bands (18 pairs of bands) in the mean-field PDW ansatz, 2 are gapless. In the MDW reduced B.Z., the PDW momentum is $(\pi,\pi)$. We apply the theory in Sec.~\ref{subsection: gapless sector} to the gapless bands. After disordering the PDW, the 2 Bogoliubov bands become 1 gapless electron band plus 1 gapped electron-boson continuum. As we discussed in Sec.~\ref{subsec: Luttinger}, the Fermi pocket automatically adjust its area to satisfy Luttinger's theorem, in order to avoid paying the Mott gap of the bosonic sector. On the other hand, the 34 gapped bands are more complicated than the simple model we have in Sec.~\ref{subsection: gapped sector}. The difference is the existence of many low-lying gapped bands. Thus even though the boson gap is smaller than the anti-nodal gap, it may be larger than the gap of low-lying electrons. However, the picture that  all these fermions are gapped
and that at low enough energy, the bosonic pairs carry all the charges of the gapped bands is unchanged. At the energy scale of $20$meV, we start to see both fermionic excitations that break pairs and bosonic excitations that move the pair as a whole. Similar to the fluctuating s wave superconductor discussed in Sec.~\ref{subsection: gapped sector}, as we disorder PDW, a Bogoliubov band of ordered PDW evolves into quasi-electron band in part of the B.Z. and  hole-pair continuum elsewhere. Roughly speaking, the Bogoliubov bands coming from the original electron bands become quasi-electron excitation with a 3-particle continuum at slightly higher energy; the Bogoliubov bands coming from PDW-reflected bands become a broad 2-particle continuum with no well-defined quasi-particle (Fig.~\ref{Fig: ARPES th}(a)). This dichotomy is too crude if a large number of bands have similar energy. Generically, the single-particle Green's function mixes multi-boson-fermion contributions from the boson band and all of the fermion bands. Due to the low-energy boson, low-energy two-particle continuum is abundant in the B.Z.

Due to the coexistence of the gapped and gapless sector, and the presence of many low-lying gapped fermion bands, the quasi-particles we discussed previously may be considerably broadened. First, we discuss the fate of the boson. The boson near the PDW momentum cannot decay into the nodal gapless band because of momentum mismatch, otherwise the gapless band would be gapped by PDW in the first place; nor can it decay into the anti-nodal fermions if its energy is smaller than the anti-nodal gap. However, the boson may decay into low-lying gapped fermions: their energy gaps could be comparable (depending on details of the band structure), and the momenta of low-lying fermions cover the majority of the reduced B.Z.. However, the decaying rate should be parametrically small because it relies on the small CDW amplitudes to match the momentum. Thus, even though the boson may not have infinite lifetime, they may still be sharp excitations near the PDW momentum. Second, for the fate of the anti-nodal fermions, since it has a large gap, apart from the boson-fermion continuum we discussed before, the quasi-particle peak itself is also severely broadened by decaying into 3 gapless/small-gap fermions. We shall analyze these spectral features with ARPES and infrared absorption data in the next section.

\section{Broader aspects and experimental implications}\label{Sec: broader aspects experiments}

So far, we have been focusing on the high-field ground state of underdoped cuprates. However, the phenomena we discussed, including the anti-nodal fermion gap, the decrease of fermionic carrier density, and the nodal gapless fermions are also present in the zero-field pseudogap. 
In the limit that the pseudogap transition temperature $\text{T}^*\gg T_{c}$ (the superconducting transition temperature), which is achieved in a range of doping, the superconducting phase occupies only a small region of the temperature-field phase diagram, on top of the pseudogap phenomena. In that limit, it is reasonable to expect the pseudogap physics at temperature $T_\text{c}<T\ll \text{T}^*$ connects smoothly to the zero-temperature, $H > H_c$ pseudogap ground state we present. Therefore we also compare our theoretical predictions with zero-field finite-temperature data. 

Many finite-frequency spectral properties of the pseudogap is maintained below $\text{T}_\text{c}$. For these properties, we may still use the predictions of our boson-fermion model. However, approaching $\text{T}^*$, the system crosses over to the strange-metal region, where our model does not apply.

On the other hand, it is interesting to discuss fluctuating zero-momentum superconductivity (SC) and fluctuating PDW in a unified picture, and compare their properties. As discussed before, we model the system as nodal electron pocket plus antinodal gapped excitations effectively described by bosonic pairs. The bosonic pair has a local band minimum at finite momentum, which we identified as fluctuating PDW. At low magnetic field and low temperature, cuprates become d-wave superconductors; therefore, the bosonic pair should have another local band minimum at zero-momentum, which closes at $\text{T}_\text{c}$ to give the superconductivity. In the normal state, the 2 band minima of the bosonic Mott insulator give fluctuating PDW and fluctuating SC correspondingly.

The fluctuating SC associated with zero-momentum boson differs from the fluctuating PDW in many aspects. Since it actually orders below $\text{T}_\text{c}$, its fluctuation depends sensitively on temperature. As the first approximation, we may ignore the quantum fluctuation of zero-momentum boson and describe the thermal fluctuation by classical statistical mechanics. On the contrary, since the PDW boson maintains a finite gap everywhere in the phase diagram, thermal fluctuations are largely suppressed. Moreover, the zero-momentum boson decays into the gapless nodal pocket in the normal state, resulting in a considerable dissipation, whereas the PDW boson is immune from that decaying channel and stays relatively sharp because of momentum mismatch. Our discussion on the quantum fluctuation of the PDW is very different from the conventional dissipative Ginzburg-Landau formulation. In that formulation, pairing correlator decays exponentially in real time due to dissipation, $\<\D^*(r,t)\D(r,0)\>\sim e^{-t/\tau}$. However, pairing correlator at the same location oscillates in time in our model, $\<\D^*(r,t)\D(r,0)\>\sim e^{i\D_b t}/t$, with negligible exponential decaying at low temperature, just as every gapped bosonic system. Due to this difference, fluctuating SC, which is close to the conventional thermal fluctuation, produces large Nernst signal and diamagnetism, while the fluctuating PDW boson gives sharper features in spectroscopic measurements. We would like to point out here that the correlator  $\<\D^*(r,t)\D(r,0)\>$ is in principle measurable by tunneling experiments, and a concrete scheme has recently been proposed~\cite{PhysRevB.99.035132}.

Both fluctuating SC and fluctuating PDW modify the spectral function of electrons. On the gapless PDW pocket, the superconducting gap is purely due to d-wave SC; near the antinode, their effects mix together. The combined effect depends on the relative strength of the two, which varies with chemical formula, temperature, and momentum. When $\text{T}^*\gg\text{T}_\text{c}$, we expect the anti-nodal gap to come mainly from fluctuating PDW. Below $\text{T}_\text{c}$, ordered superconductivity gaps out low-lying fermions, hence the reduction of decaying channel for anti-nodal fermions, and the emergence of a sharper anti-nodal peak. As discussed below, this picture is consistent with the data on the single layer Bi2201. On the other hand, for Bi2212 close to optimal doping (still underdoped), a sharp quasiparticle peak emerges from a relatively broad continuum just below $\text{T}_\text{c}$, and the spectral weight of the peak is apparently proportional to the superfluid density~\cite{feng2000signature,PhysRevLett.87.227001}. This behavior cannot be explained by the fluctuating PDW alone. We also notice that we do not have a clear separation of scale in this situation: $\text{T}^*$ is only two times $\text{T}_\text{c}$. We leave further discussion of Bi2212 to future works.

Underdoped Bi2201, consists of single $\text{CuO}_2$ layers separated far away from each other, has $\text{T}^*$ much bigger than $\text{T}_\text{c}$. It is ideal for analyzing pseudogap effect due to the lack of interlayer splitting and large separation between $\text{T}^*$ and $\text{T}_\text{c}$~\cite{hashimoto2014energy,he2011single}. It has the fermion spectrum closest to what we expect from fluctuating PDW alone. We discuss it in Sec.~\ref{subsection: ARPES}. For other spectroscopic probes, like infrared conductivity and density-density response, we expect to see contributions from fluctuating PDW at $\omega > 2\D_b \sim 40$meV, and contributions from SC at lower frequencies (Sec.~\ref{subsection: infrared}). 

Both fluctuating SC and fluctuating PDW contribute to diamagnetism and Nernst effect. It is well known that as temperature approaches $\text{T}_\text{c}$, the diamagnetism and Nernst signal from fluctuating SC diverges~\cite{PhysRevLett.95.247002,PhysRevLett.88.257003,PhysRevB.73.024510,PhysRevLett.89.287001,larkin2008fluctuation,PhysRevLett.96.147003,PhysRevLett.99.117004,PhysRevB.73.094503}. In contrast, the fluctuating PDW contributions are far less dramatic unless the corresponding boson gap decreases substantially in high fields.

In the following parts of this section, we use our boson-fermion model to work out signatures of the fluctuating PDW. We compare theoretical results with experiments on ARPES, infrared absorption, density-density response, diamagnetism and Nernst effect.

\subsection{ARPES}\label{subsection: ARPES}

As we discussed in Sec.~\ref{subsection: gapped sector} and Sec.~\ref{subsec: Constructing the fluctuating PDW ground state}, the fluctuating PDW state naturally has both charge $\pm2$e bosons and charge $\pm$e electrons/holes at low energy. Their interplay produce unconventional ARPES signal. Since the charge $\pm2$e boson is cheap, when we kick out an electron from the sample, the hole may decay into a charge -2e boson and a charge e electron. In analogy to Fig.~\ref{Fig: fluctuating SC band}(b), the threshold to create a hole excitation at momentum $k$ roughly follows the Bogoliubov bands of PDW, but only in a part of the B.Z. the threshold corresponds to quasi-hole excitations. The other part of the Bogoliubov bands, which comes mainly from PDW reflection, is replaced by a blurred 2-particle continuum of an electron and a small-gap charge -2e boson. Furthermore, wherever we have a sharp quasi-particle in the spectrum, we can add a charge +2e boson and a charge -2e boson to make a 3-particle continuum with the same charge, at the same momentum, and with energy only 2$\Delta_b$ higher. The spectral features of these multi-particle continuum with total charge $-e$, which can be probed by ARPES, are easily calculated by considering the decay rate (the imaginary part of the self-energy) using Fermi's Golden rule or simple dimensional analysis. Consider the simplest coupling $\delta H_1 =\lambda_1 \phi c c + h.c.$ and $\delta H_2 =\lambda_2 \phi^*\phi c^{\dagger}c$, where $\phi(p) = \frac{1}{\sqrt{E^b_p}}(a_p + b_{-p}^{\dagger})$ is the relativistic boson field (see Sec.~\ref{subsection: gapped sector}), with momenta close to the PDW momentum $P$, and $E^b_p = \sqrt{|v_b(p-P)|^2 + \Delta_b^2}$.

\bea
\label{Eq: 2 particle continuum}\text{Im}\Sigma_{2p}(q,\omega) &\propto& \int \dbar^2 p \ \frac{1}{E^b_p}\delta(\omega - E^b_p -E^e_{q-p})\\
&\propto& \theta (\omega - \Delta^{(2)}_q)
\eea
\bea
\text{Im}\Sigma_{3p}(q,\omega) &\propto& \int \ \frac{\dbar^2 p_1\ \dbar^2 p_2}{E^b_{p_1}E^b_{p_2}}\delta(\omega - E^b_{p_1} - E^b_{p_2} -E^h_{q-p_1-p_2})\nonumber\\
\label{Eq: 3 particle continuum}&\propto& (\omega - \Delta^{(3)}_q)\ \theta(\omega - \Delta^{(3)}_q),
\eea
when $\omega - \Delta^{(3)}_q \gg \Delta_b$.

We use the shorthand $\dbar^2 p\equiv \frac{dp_xdp_y}{(2\pi)^2}$. $E^e_{k}/E^h_k$ represents the dispersion of the quasi-electron/ quasi-hole. $\D^{(2)}_q$ ($\D^{(3)}_q$) is the energy threshold to create 2(3) particles at momentum q: $\D^{(2)}_q\equiv \text{min}_{p_1}[E^b_{p_1} + E^e_{q-p_1}]$, $\D^{(3)}_q\equiv \text{min}_{p_1,p_2}[E^b_{p_1} + E^b_{p_2} + E^h_{q-p_1-p_2}]$. When the boson gap is small, and the boson velocity is comparable to the Fermi velocity near the antinode, $\D^{(2)}_q$ and $\D^{(3)}_q$ roughly follows the Bogoliubov bands of PDW.

The main message is that whenever we have a PDW reflected band, we should see a step function in spectral function (Eq.~\ref{Eq: 2 particle continuum}); and whenever we have a (PDW-modified) quasi-hole, we should see a spectral function 

\bea
A(\omega) = \text{Im} \frac{1}{\omega - E^h_q - i(\omega - \Delta^{(3)}_q)\ \theta(\omega - \Delta^{(3)}_q) - i\Gamma},\ \ \ 
\eea
which has a quasi-hole peak together with a 3-particle continuum (Eq.~\ref{Eq: 3 particle continuum}). The spectral signature is a relatively sharp onset of peak at $E^h_q$, but a long $1/(\omega - \Delta_q^{(3)})$ tail above the 3-particle threshold. 

When $\D_b$ is small, $\D^{(3)}_q\simeq E^h_q$, the quasi-hole peak merges with the 3-particle continuum, and

\bea 
A(\omega) \sim \frac{\theta(\omega - E^h_q)}{\omega - E^h_q}
\eea

\begin{figure}[htb]
\begin{center}
\includegraphics[width=0.7\linewidth]{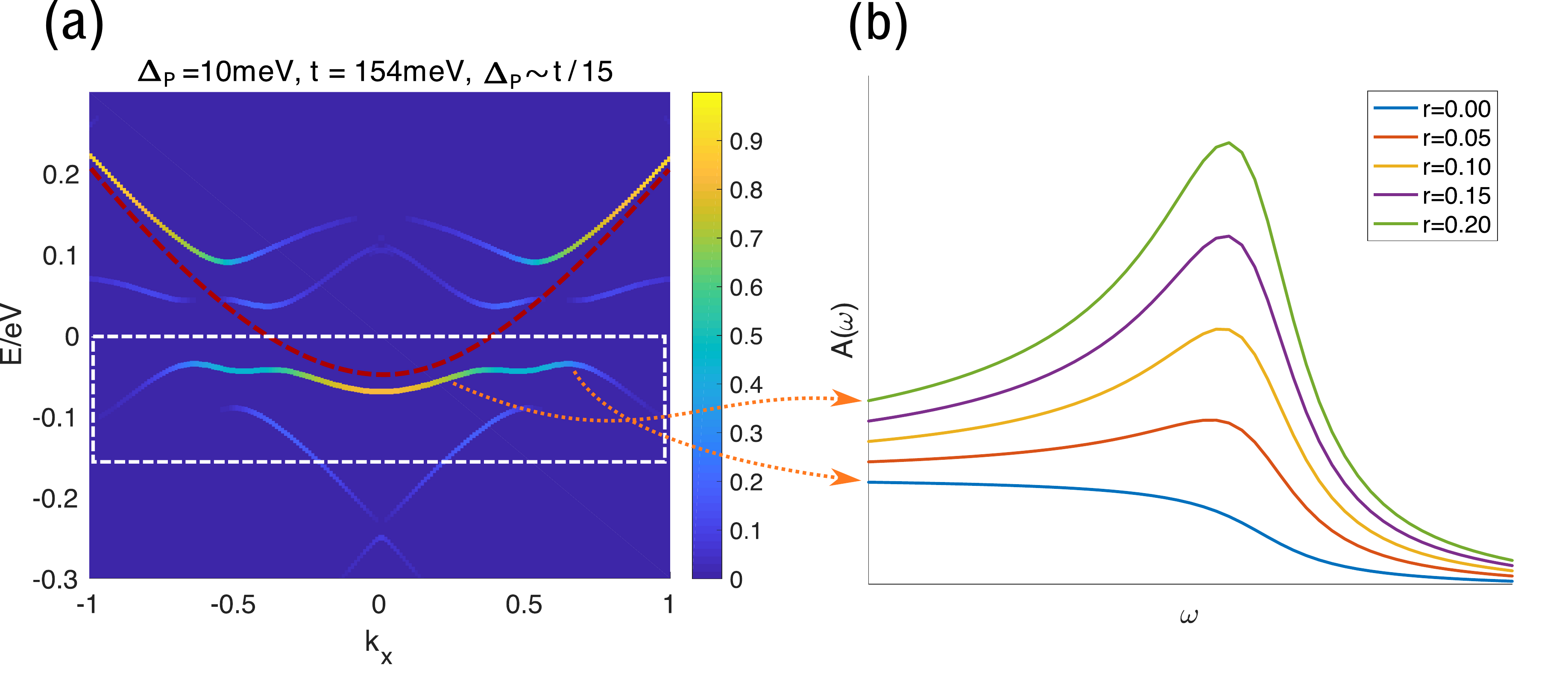}
\caption[Electron spectral function for quantum fluctuating PDW]{(a) Mean-field PDW spectrum along the line $ky=\pi$. PDW momentum $2\pi/6$, PDW pairing $\Delta_P = 10\text{meV}$. (see Eq.~\ref{Eq: PDW mean field} and Eq.~\ref{Eq: PDW form factor} for definition). We use tight-binding band with $t=154\text{meV}, t_p = -24\text{meV}, t_{pp} = 25\text{meV}, t_{ppp} = -5\text{meV}$, chemical potential $\mu = -126\text{meV}$. Color plot represents the spectral weight in mean-field calculation. The dashed red line illustrates the original electron band. The dashed white box shows the range of energy probed by ARPES in Ref.~\cite{he2011single}.  (b) Illustration of the evolution from a 3-particle continuum  to a broad 2-particle continuum of the fluctuating PDW.  We use the ratio $r$ to interpolate  between the two as defined in  Eq.~\ref{Eq: step plus peak}.}
\label{Fig: ARPES th}
\end{center}
\end{figure}

\begin{figure}[htb]
\begin{center}
\includegraphics[width=0.7\linewidth]{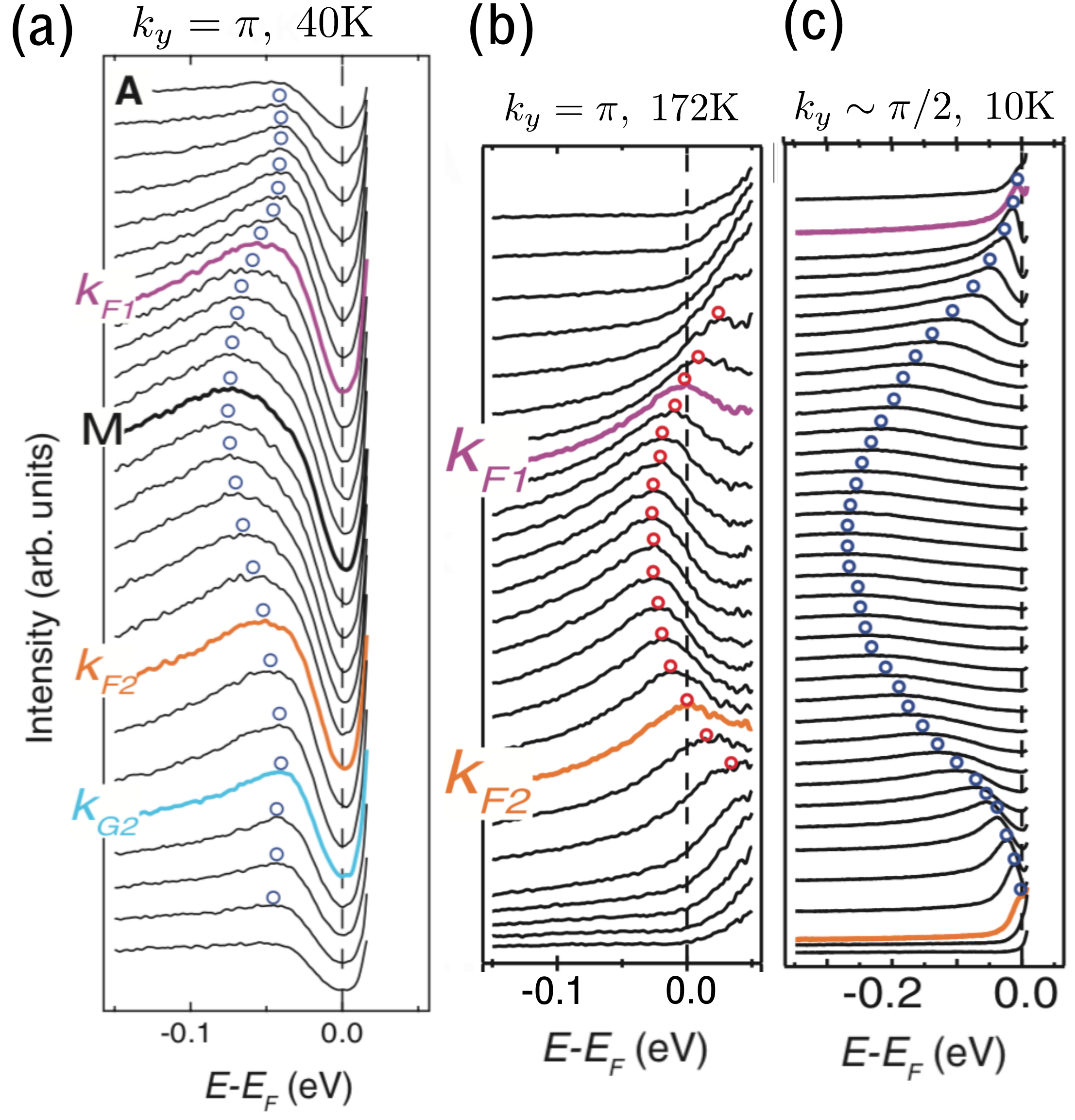}
\caption[ARPES results for Bi2201]{(a) Fig. 4A of Ref.~\cite{he2011single}. spectral function along the cut $k_y = \pi$, below $\text{T}^*$ and above $T_\text{c}$ (40K). The M point refers to $k_x=0$, $k_{F1}$ and $k_{F2}$ are roughly at $k_x=-0.2 \pi$ and $0.2 \pi$ respectively. (b) Fig. 2A of Ref.~\cite{he2011single}. The same as (b), except at temperature above $\text{T}^*$ (172K). (c) Fig. 2N of Ref.~\cite{he2011single}. spectral function along a cut $k_y \sim\pi/2$, at 10K.}
\label{Fig: ARPES exp}
\end{center}
\end{figure}

It's important to know whether ARPES can resolve the boson gap. In Sec.~\ref{subsec: Constructing the fluctuating PDW ground state}, we estimate the boson gap to be $10$meV to $30$meV from the correlation length of PDW. The state of art synchrotron ARPES has an energy resolution of a meV, which can in principle resolve the boson gap. However, the anti-nodal quasi-electron peak is at high energy, suffers from substantial broadening through the process of decaying into gapless/small-gap fermions. When the broadening of quasi-electron peak is comparable $\D_b$, the single-particle peak merges with the 3-particle continuum. We just see a broadened $\theta(\omega - E^h_q)/(\omega-E^h_q)$ peak, as if the boson is gapless.

Fig.~\ref{Fig: ARPES th}(a) shows the mean-field spectrum of bidirectional-PDW with relatively small PDW gap, along the cut $k_y = \pi$. To compare with ARPES results (Fig.~\ref{Fig: ARPES exp}(a), reproduced from Ref.~\cite{he2011single}), we focus on the energy-momentum range in the white box, where the mean-field spectral weight  concentrates on a single Bogoliubov band. Comparing with Fig.~\ref{Fig: PDW 3-band illustration}, we find that a simple 2-band calculation with only y-directional PDW captures main features in this energy-momentum range. This is in contrast with the discussion in ~\cite{PhysRevX.4.031017} which focused on the x-directional PDW. Here we find that the x-directional PDW helps increase the band gap, and produce a flat shoulder near the band minimum.

The sharp spectral function in the mean-field calculation is greatly transformed by the PDW fluctuation. For $k_x < k_F$, the Bogoliubov band follows the original electron band (dashed red line). We expect a broadened $\theta(\omega - E^h_q)/(\omega-E^h_q)$ peak just above the quasi-particle energy. (green line in Fig.~\ref{Fig: ARPES th}(b)). At large $k_x$, the Bogoliubov band is far from the original band of the metal; it largely comes from PDW-reflected bands, which we expect to be a 2-particle continuum when PDW is fluctuating, consequently a (broadened) step function in ARPES. (blue line in Fig.~\ref{Fig: ARPES th}(b)).  Going from small $k_x$ to large $k_x$, we expect the hole excitation created by ARPES to gradually mix with boson-electron bound state, until some $k>k_F$, where the boson and electron no longer bound together. The spectral feature is that a quasiparticle resonance disappears (from the green line to blue line in Fig.~\ref{Fig: ARPES th}(b)) right at the onset of the step-function.

Phenomenologically, we can write the electron annihilation operator as
\bea 
c_k = r_1\tilde{c}_k + r_2\sum_{q}\phi^*_{q}\tilde{c}^{\dag}_{k-q}
\eea 
The first term produces a broad quasi-hole resonance $\theta(\omega - E^h_q)/(\omega-E^h_q)$, and the second term produces a step-function background $\theta(\omega - E^h_q)$. Just to illustrate the qualitative trends, we plot (lorentzian broadened)
\bea
\label{Eq: step plus peak}A(\omega)\propto \theta(\omega - \e_q) + r\theta(\omega - \e_q)/(\omega-\e_q)
\eea
where $r\equiv r_1/r_2$, with gradually increasing $r$ in Fig.~\ref{Fig: ARPES th}(b). In general, $r_1$ and $r_2$ depends on energy and momentum. We know qualitatively how they changes, but near the antinode, we have no reliable way to calculate their energy-momentum dependence. However, when $k\gg k_F$, in the limit $E^0_k\gg \omega, E^h_k$, where $E^0_k$ is the dispersion of the original band without PDW (dashed red line in Fig.~\ref{Fig: ARPES th}(a)), we can treat PDW perturbatively, and the spectral function from the 2-particle continuum is given by

\bea
A(\omega) &\sim& \text{Im}\frac{1}{\omega - E^0_k - i|\Delta|\theta(\omega - E^0_k)}\nonumber\\
&\sim& \frac{|\Delta|}{(E^0_k)^2}\theta(\omega - \e_k)
\eea
Thus the height of the step function quickly decays as we move farther away from $k_F$.

Experimental results along the same cut in Bi2201, just above $\text{T}_\text{c}$, is shown in Fig.~\ref{Fig: ARPES exp}(a)~\cite{he2011single}. Following the peaks of the spectral functions (blue dots), we see the gap minimum is not at the original Fermi surface ($K_{F1}$ and $K_{F2}$), but shifted outward in momentum ($K_{G2}$), consistent with PDW~\cite{PhysRevX.4.031017}. Moreover, the entire frequency dependence of electron spectral function matches with our expectation of the fluctuating PDW (Fig.~\ref{Fig: ARPES th}(b)). As shown in Fig.~\ref{Fig: ARPES exp}(a), when scanning from large $k_x$ to small $k_x$, we first encounter a step function that onsets at about 20meV and when k is less than the Fermi momentum,  a broad resonance emerging just above the step function. This is as expected from the transition from a bound state of boson and electron into a quasi-hole. Identifying the ARPES results with spectral functions of fluctuating PDW, we get an upper bound of the boson gap, $\D_b \lesssim 20$ meV, consistent with our previous estimation.

There are concerns on whether the step-function background in Bi2212 is intrinsic or an artifact of ARPES due to disorder induced scattering that mixes different momenta~\cite{PhysRevB.69.212509}. However, at least in Bi2201, the step-functions we analyzed appear only in the anti-nodal region (for comparison with the nodal region, see Fig.~\ref{Fig: ARPES exp}(c)), and disappear above $\text{T}^*$ (Fig.~\ref{Fig: ARPES exp}(b)), providing strong evidence that they are intrinsic and related to the pseudogap. We also notice that these step functions start at around $20$ meV below Fermi energy, different from the step functions that start right at Fermi energy in Bi2212.

Bi2201 is ideal for analyzing the pseudogap for the large separation between $\text{T}_\text{c}$ and $\text{T}^*$ even close to optimal doping, and for the lack of bilayer splitting~\cite{hashimoto2014energy,he2011single}. We found the anti-nodal spectrum of Bi2201 fitted best with a relatively small PDW pairing, $\D\sim t/15$. We also notice that if pairing were to be increased to $\D\sim t/4$, the band structure is no longer captured by a simple 2-band hybridization: there are many bands sharing small spectral weights. Considering PDW fluctuation, the spectral function may just be a featureless continuum above PDW gap. This large-pairing scenario may be the case for other cuprates with larger $\text{T}_\text{c}$ and $\text{T}^*$.

\subsection{Infrared conductivity and density-density response}
\label{subsection: infrared}

Cuprates have a flat ab-plane infrared conductivity plateau, which differs from a Drude peak that decays as $1/\omega^2$ at high frequencies~\cite{puchkov1996pseudogap,RevModPhys.77.721}. As temperature lowers, the low-frequency peak become narrower, and the conductivity shows an upturn in the infrared region, starting roughly at $40$meV. This extra infrared conductivity have never been throughly understood. Ref.~\cite{puchkov1996pseudogap,PhysRevB.90.014503,PhysRevB.57.R11089,PhysRevLett.81.4716} attempt to explain it by electron scattering with charge-neutral boson. However, we find that it matches well with the conductivity of a charge 2e boson.

Consider a free boson with charge $e^*$, minimally coupled to electromagnetic field.

\bea
\mathcal{L} = \frac{1}{2}|(\partial_{t} + ie^{*}V)\phi|^2 - \frac{1}{2}\sum_{i=1,2}v_b^2|(\partial_{i} + ie^{*}A_i)\phi|^2\nonumber\\ - \frac{1}{2}|\D_b|^2|\phi|^2,\ \ 
\eea

where the momentum of the boson is measured from the PDW momentum. By canonical quantization, $E^b_p = \sqrt{\D_b^2 + v_b^2 p^2}$, $\phi_p = \frac{1}{\sqrt{E^b_p}}(a_{p} +b_{-p}^{\dagger})$, and

\bea
j_{i} = \frac{\delta\mathcal{L}}{\delta A^{i}} = \sum_{p} \frac{e^{*}v_b^2}{E_p} p_{i}(a^{\dagger}_{-p} + b_p)(a_{-p} + b_p^{\dagger})
\eea

By Kubo formula

\bea
\text{Re} \sigma_{xx}(\omega) &=& \frac{\pi}{\omega}\sum_{n}|\<n|j_x|0\>|^2\delta(\omega - (E_n - E_0))\\
&=& \frac{(e^*)^2v_b^4\pi}{\hbar\omega}\int \dbar^2 p\ \frac{p_{x}^2}{(E^b_p)^2}\ \delta(\omega - 2E^b_p)\\
&=& \frac{(e^*)^2}{16\hbar}(1-4\Delta_b^2/\omega^2)\theta(\omega - 2\Delta_b)
\eea

\begin{figure}[htb]
\begin{center}
\includegraphics[width=0.7\linewidth]{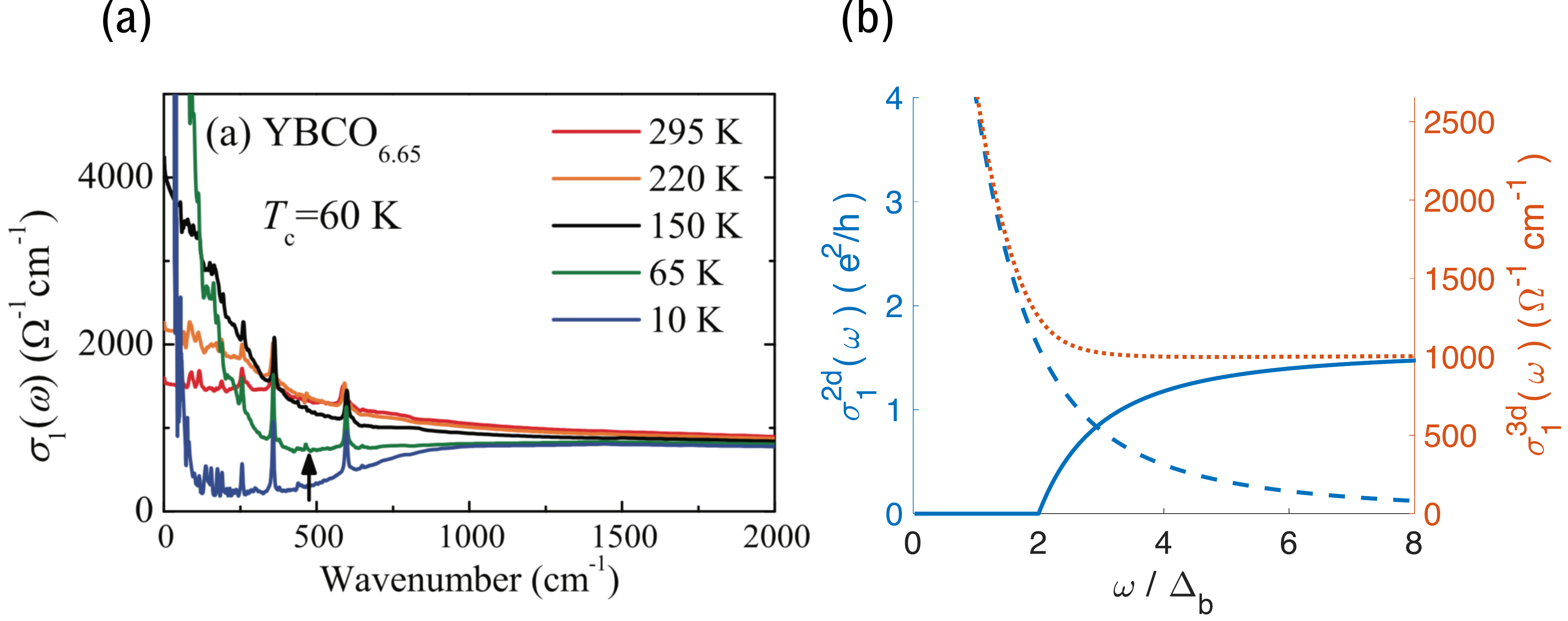}
\caption[Infrared conductivity: theory vs. experiment]{(a) Real part of infrared conductivity measured from reflectance (Fig. 3(a) of Ref.~\cite{PhysRevB.90.014503}). (b) Solid blue curve: AC conductivity of a free charge 2e boson with gap $\D_b$. We calculated the 2D conductivity of each layer, and converted it to a 3D conductivity using the lattice parameter of YBCO. The conductivity of the free relativistic boson saturates at $\frac{\pi}{2}e^2/h$ when $\omega\gg\D_b$, which corresponds to $1.0\times10^3\ \Omega^{-1}\text{cm}^{-1}$. Dashed blue curve: a Drude peak. Dashed orange curve: the sum of the boson conductivity (blurred by a Lorentzian) and the Drude peak.}
\label{Fig: infrared}
\end{center}
\end{figure}

We plot the result for $e^* = 2e$ as the solid blue curve in Fig.~\ref{Fig: infrared}(b), and convert the 2D conductivity to the 3D in-plane conductivity using the lattice constant of YBCO. The optical conductivity of the boson depends on its dispersion and interaction, hence non-universal. However, the linear onset of conductivity at $\omega\simeq 2\D_b$, namely $\sigma_{xx}\propto (\omega - 2\D_b)\theta (\omega - 2\D_b)$, is universal for a gapped boson. The onset is linear because of the combination of a constant density of states in 2D and an absorption matrix element $\sim\text{velocity}^2\sim\delta\omega$, for $\omega\simeq 2\D_b$. For a free relativistic boson with charge 2e, the conductivity at high frequency saturates at $\sigma_{xx} = \frac{\pi}{2}e^2/h$, independent of its gap or velocity. The linear onset of the conductivity together with the saturation value of an order 1 number times $e^2/h$ are signatures of a gapped relativistic particle. Interactions and changes in dispersion modify the order 1 number, but does not change the qualitative features of the conductivity. (For detailed explanation and calculation, see  Ref.~\cite{sachdev_2011}).

Surprisingly, the infrared conductivity plateau around 12\% doping is almost exactly $\frac{\pi}{2}e^2/h$ per $\text{CuO}_2$ layer, the same as the free boson, both in YBCO and in Bi2212. (It changes a little with doping. See Fig.~\ref{Fig: infrared} for comparison with YBCO. See Fig.6 of Ref.~\cite{puchkov1996pseudogap} for Bi2212.) Moreover, the frequency dependence of bosonic conductivity matches well with the conductivity upturn at low temperature. If we add a Drude peak to the bosonic conductivity, we reproduce the flat infrared conductivity observed at higher temperature. 

The extra infrared conductivity provide evidence for the charge 2e boson. However, the numerical agreement may not be taken too seriously, for the interaction between bosons and fermions may modify the result. Experimentally, the infrared plateau extends to frequency as high as $400$meV~\cite{PhysRevB.43.7942}, where our boson fermion model does not apply. We cannot explain the high-energy behavior of the plateau, but we suspect that the boson contribution connects to the incoherent part of the spectral weight (also seen in ARPES) to give the long plateau.

Note that even though the conductivity upturn is prominent only below $\text{T}_\text{c}$, it has little to do with the absorption across the SC gap. As discussed in Ref.~\cite{PhysRevB.90.014503},  features of SC is around 100$\text{cm}^{-1}\sim 12$meV, five times smaller than the frequency scale of the upturn. Although not fully understood, ordered SC seems to make the low-energy peak narrower without changing the conductivity upturn starting from $40$meV. If we associate the infrared conductivity upturn to the PDW boson, the boson gap should be $20$meV, consistent with our previous estimation.

Unlike s-wave SC, fermions gapped by PDW  absorb light across the pairing gap even in the clean limit~\ref{chap:opticalconductivity}. However, this is much smaller than the bosonic contribution,  according to the estimation in Chapter~\ref{chap:opticalconductivity}, which found $\sigma^{2D}_f \sim \frac{e^2}{h} (a/\lambda)^2 E_f/\D_f\sim \frac{1}{10}e^2/h$, where $a$ is size of the original unit cell, $\lambda\sim 8a$ is the wavelength of PDW. The absorption due to  the gapped fermion bands give various tiny peaks from $50$meV to $200$meV, which may be too small to identify. The delta function peaks observed experimentally are mostly due to optical phonons.

The same phenomena is also observed in density-density response. By current conservation, we expect
\bea
\text{Im}\,\Pi(q\sim 0, \omega) &=& \text{Im}\,\<\rho\rho\> = \text{Im}\,\<jj\>\cdot q^2/\omega^2\nonumber\\
&=& \text{Re}\,\sigma(\omega)\cdot q^2/\omega\\
\text{Im}\,\Pi(q\sim 0, \omega) &\sim& \frac{\pi}{2}\frac{e^2}{h}\frac{q^2}{\omega},\text{ in mid-infrared}
\eea
Abbamonte's group measured the density-density response in cuprates~\cite{mitrano2018anomalous,husain2019crossover}. Below 100meV, they claim the signal is dominated by phonon. Between 100meV and 1eV, at optimal doping, they report an unusual $\text{Im}\,\Pi$ independent of $\omega$. In overdoped samples, $\text{Im}\,\Pi$ decreases as $\omega$ decreases to 100meV. However, in underdoped samples, $\text{Im}\,\Pi$ increases as $\omega$ decreases to 100meV~\cite{husain2019crossover}. While this upturn is unusual in metallic states, here it is simply required by current conservation (see Eq. 27) to be consistent with the infrared conductivity.

Finally, we discuss c-axis conductivity. For bilayer cuprates like YBCO and Bi2212, $\text{CuO}_2$ layers are organized as closed bilayers with several atomic layers between neighbouring bilayers. Given the experimental fact $\sigma_{zz}\gg \omega\epsilon_0$ in the mid-infrared, the measured conductivity away from resonant peaks is mainly determined by inter-bilayer hopping instead of intra-bilayer hopping. Physically, the intra-bilayer hopping is so effective that most of the voltage drop are on the barrier between neighbouring bilayers. Across this barrier of 3 or 4 atomic layers, pair hoping is much smaller than single-fermion hopping. Therefore, we expect tunneling of the small-gap fermion to  dominate the measured c-axis conductivity.

\subsection{Remnants of superconductivity}

Long-range ordered PDW breaks charge conservation and is a superconducting order. Being close to the long-range PDW, the fluctuating PDW state has properties reminiscent of a superconductor. In this subsection, we briefly discuss the diamagnetic response, Nernst effect, and DC conductivity of the fluctuating PDW state. In short, fluctuating PDW gives a diamagnetic susceptibility inversely proportional to the boson gap without increasing the DC conductivity. This is because the bosons transit from a superconductor into an insulator instead of a metal. Nernst effect comes from thermally excited PDW bosons, which are suppressed when $T<\D_b$. Experimentally observed diamagnetism and Nernst signal near $\text{T}_\text{c}$ comes mainly from fluctuating zero-momentum SC. Due to the boson gap, the contribution from fluctuating PDW  is smaller and less sensitive to temperature.

We start from diamagnetism. We calculate the current response to the vector potential $j_i(\omega,q) = K_{ij}A_{j}$, at $\omega = 0, q = q_y\hat{y}$. In this setting, magnetic susceptibility of the boson $\chi_b = -K_{xx}/q_y^2$.

The current operator at finite $q$ is
\bea
j_i(q) = \sum_p e^*v^2_b(p_i + q_i/2)\phi^*(p)\phi(-p-q) \nonumber\\
+ \sum_p(e^*)^2v^2_b\phi^*(p)\phi(-p)A_i(q)
\label{Eq: current}
\eea

The response of the first term is given by Kubo formula

\bea
\text{Re} R_{xx} &=& \sum_n|\<n|j_x(q)|0\>|^2\frac{-2}{E_n-E_0}\\
&=& (e^*)^2v_b^4\int_0^{\Lambda}\frac{-2p_x^2\ \dbar^2p}{E^b_p E^b_{p+q}(E^b_p+E^b_{p+q})}
\eea
We expand the expression in $q_y$, the constant term is canceled b{}y the second term of Eq.~\ref{Eq: current}, and the quadratic term gives us magnetic susceptibility

\bea
\chi_b &=& -\text{Re}\frac{R_{xx}(q_y) - R_{xx}(0)}{q_y^2}\\
&=& -(e^*)^2v_b^4 \int\frac{p_x^2}{(E^b_p)^3}\left(-\frac{5}{2}\frac{v_b^4 p_y^2}{(E^b_p)^4} + \frac{3}{4}\frac{v_b^2}{(E^b_p)^2}\right)\dbar^2p\nonumber\\
\label{Eq: diamag}&=& -\frac{e^2v_b^2}{6\pi\D_b}\\
&=& \chi_f \frac{2mv_b^2}{\Delta_b},
\eea
for $e^*=2e$, where $\chi_f$ =  $e^2/12 \pi m$ stands for Landau diamagnetic susceptibility for 2D free fermion with mass $m$. $\chi_b^{3D} = \chi_b/d$, where $d$ is the average distance between $\text{CuO}_2$ layers. This result holds for temperature and Landau-level splitting smaller than the boson gap. We note that compared with $\chi_f$ 
, Eq.~\ref{Eq: diamag} is enhanced by the ratio $\frac{2mv_b^2}{\D_b}$. There has been report of a significant amount of diamagnetism in underdoped YBCO at low temperatures at 40T magnetic field which is much larger than the transport $\text{H}_\text{c2}$~\cite{yuPNAS126672016magnetic}. Our Eq.~\ref{Eq: diamag} involves the boson velocity $v_b$ which is not known, but the predicted diamagnetism should be temperature dependent on the scale of the boson gap.

When the temperature is comparable to or larger than the boson gap, with external magnetic field, bosons exhibit Nernst effect. Under temperature gradient and magnetic field, thermally excited charge 2e and charge -2e bosons drift in different directions, giving a net electric current. 

For temperature smaller than the boson gap and the lowest fermion gap, and away from the superconducting dome, DC conductivity, Hall conductivity, specific heat and quantum oscillation comes solely from the small electron pocket. This decrease of fermionic carrier density at low energy is the main consequence of the fluctuating PDW. However, it is hard to describe how conductivity changes as we enter the pseudogap region from high temperature, since we do not have a theory for the strange metal. 

\subsection{Symmetry breaking in the pseudopgap phase.}
\label{subsection:Symmetry breaking in the pseudopgap phase.}

\begin{figure}[htb]
\begin{center}
\includegraphics[width=0.3\linewidth]{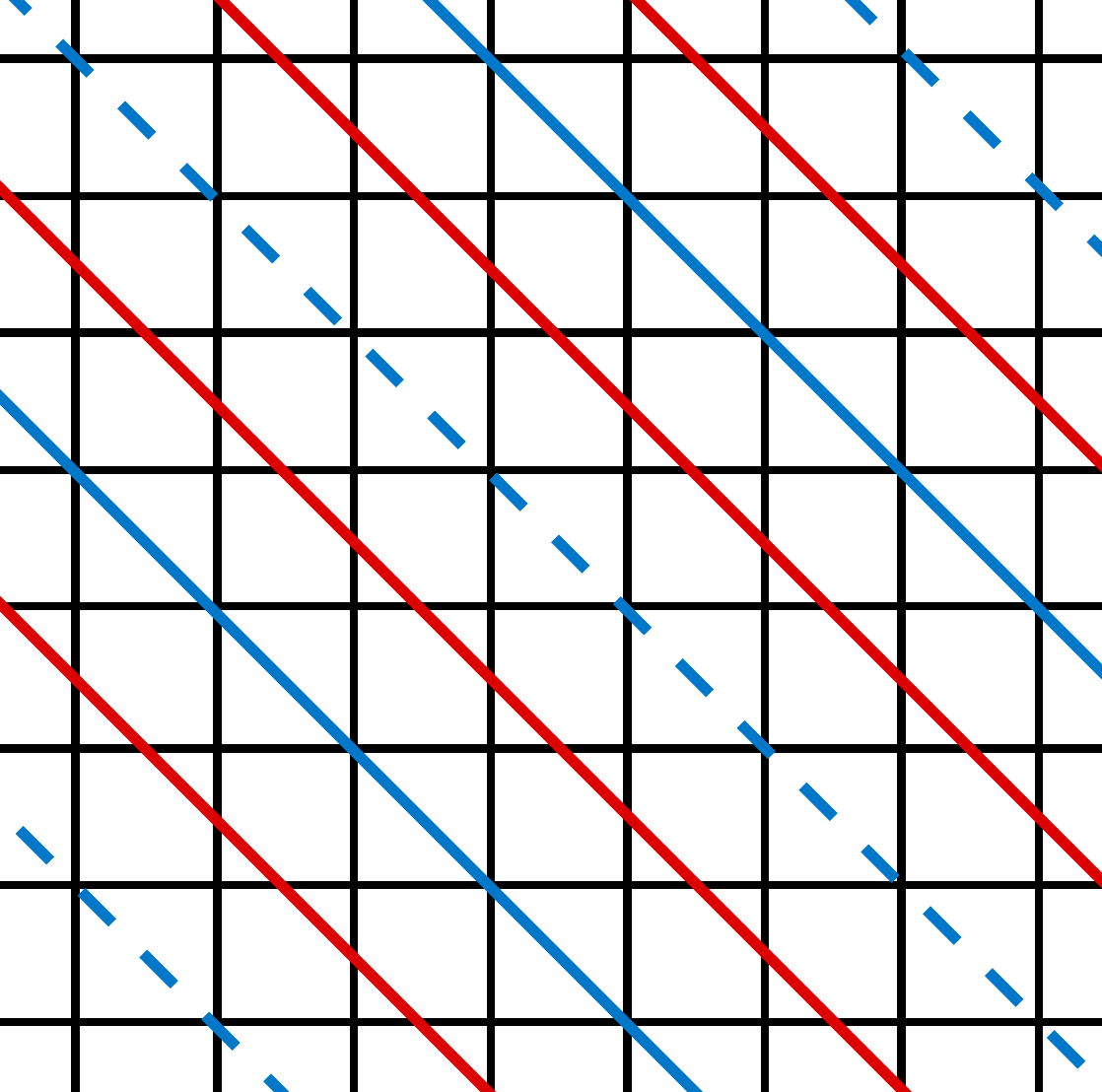}
\caption[Illustration of a uni-directional MDW generated by period-6 PDW]{Illustration of a uni-directional MDW generated by period-6 PDW. The line of maximum and minimum of the magnetization is shown as solid and dashed blue lines. The zero of magnetization is shown as red lines. Black lines shows the underlying lattice.}
\label{Fig: MDW}
\end{center}
\end{figure}

In this section we consider the consequences of symmetry breaking of the MDW, which is one of the composite orders associated with the PDW. We consider the case of commensurate PDW, and for concreteness we first discuss the case $P=2\pi/6$.  We have many different choices of phases corresponding to different relative positions between the lattice and the CDW/MDW.(see Appendix A for a detailed explanation of these phases.) Lattice translations change PDW phases only by multiples of $2\pi/6$. A generic choice breaks all lattice symmetry, but it may require the CDW/MDW to be pinned at a unnatural position. We focus on the case where the maximum and minimum of a uni-directional MDW at momentum $(P\hat{x},P\hat{y})$ is on site, as shown by the blue lines in Fig.~\ref{Fig: MDW}. 
We shall see that this choice preserves inversion about the origin, but breaks all mirrors perpendicular to the plane. The MDW has magnetization $\vec{M}\propto \cos(Px+Py)\hat{z}$ which breaks mirror symmetry along both (1,1) and (1,-1) since magnetization is odd under mirror. On the other band, we can consider the mirror plane passing through the lines of zero magnetization (shown in red in ~\ref{Fig: MDW}). The mirror symmetry is preserved for the magnetization which is odd in this case, but is broken by the lattice. Thus in this example all mirror planes normal to the c-axis are broken. The same conclusion holds for $P=2\pi/7$. The exception is $P=2\pi/8$ where the line of zero's pass through a lattice site and mirror symmetry is preserved. 

Incommensurate PDWs are slightly more complicated. For the case of YBCO, the PDW wavelengths changes with doping between 6 and 7 lattice spacing. Distorted by lattice, it is natural to relax the cosine waves into domains with period-6 PDW and domains with period-7 PDW. Our discussion of mirror symmetry breaking also applies to this relaxed incommensurate PDW.

So far, we have been focusing on simplified situations where every relative phase between two PDW order parameters are perfectly ordered.  However, as temperature decreases, different relative phases, hence different density waves can order in turn. Although fluctuating PDW gives the tendency of CDW and MDW in both directions, the energy functional may actually prefer a unidirectional MDW/CDW, with a shorter range MDW/CDW in the orthogonal direction, at least in a range of temperature. 
Ref.~\cite{sato2017thermodynamic} reported a nematic phase transition at the onset of the pseudogap. This is most clear in the case of the Hg compound which has a tetragonal structure and the nematicity is along the diagonal. This result may be explained if the MDW preferentially forms short range order at momentum $P\hat{x} + P\hat{y}$  at $\text{T}^*$ without the MDW at $P\hat{x}-P\hat{y}$, giving rise to a nematic transition. 

In Sec~\ref{subsec:mdwmoment}, we estimated the magnetic moment per plaquette (of the Cooper lattice) is at the order of $5\times 10^{-3}$ Bohr magneton. The moment through a half period of the MDW is larger by the corresponding area and we estimate the magnetic field generated by this moment to be $\sim 0.5$ Gauss. However, the magnetic field changes smoothly in the range of 6 or 7 lattice spacing. In NMR experiments, such a magnetic field profile gives a broadening of the resonance peak, instead of a shift of the peak, therefore hard to detect. But the MDW may be detectable by neutron scattering.

\section{Discussion}

In this chapter, we discuss the low-energy effective theory of the pseudogap, relevant for underdoped cuprates when $\text{T}^*>>\text{T}_\text{c}$, and for the high-field ground state. We disorder bidirectional pair density waves, but maintaining the descendant orbital magnetization and charge density waves to get a ground state of small electron pocket and a hidden bosonic Mott insulator. The fluctuating PDW provides a smooth background for diamagnetism and Nernst effect on top of fluctuating zero-momentum superconductivity, without producing excess DC conductivity. We present detailed comparison of the theoretical predictions and the experiments on ARPES and infrared conductivity. We found the peculiar spectroscopic features of the pseudogap is consistent with having a small-gap charge 2e boson at finite momentum, as in our proposal for the fluctuating PDW. From the measured infrared conductivity and the correlation length of PDW in the vortex halo, we estimate the boson gap to be about $20$meV. However, infrared conductivity and ARPES probes only the two particle continuum of two bosons or of a boson and an electron. A direct probe of a single charge 2e boson near $20$meV, momentum $2\pi/8\sim2\pi/6$ would provide direct evidence for our proposal. We also propose an orbital magnetization density wave in $(1,1)$ direction, with momentum $1/\sqrt{2}$ of the momentum of CDW. This MDW breaks time reversal, and it could explain the nematic transition at the onset of the pseudogap~\cite{sato2017thermodynamic}. We have not discussed how the pseudogap descends from the strange metal, but it would be very interesting to explore the relation between our model and possible theories of the strange metal.

\appendix

\chapter{Vertex Correction and Optical Response of Fulde-Ferrell State}

We present the derivation of Eq. (\ref{eq:corrected vertex}) and (\ref{eq: final results}) in this appendix. For simplicity, we define $\tilde{p}_{0}\equiv p_{0}-\epsilon'_{p}$. The Green's function given by Eq. (\ref{eq:mean field Green}) can then be written as:
\begin{equation}
G(p)=\frac{1}{\tilde{p}_{0}-\bar{\epsilon}_{p}\tau_{3}-\Delta_{p}\tau_{1}+isgn(p_{0})0^{+}}=\frac{\tilde{p}_{0}+\bar{\epsilon}_{p}\tau_{3}+\Delta_{p}\tau_{1}}{(\tilde{p}_{0}+isgn(p_{0})0^{+})^{2}-\delta_{p}^{2}}
\end{equation}
where we have neglected the diagonal self-energy correction since it is not important for our purpose. We are free to choose the `direction' of the pairing term in the $\tau_{1}-\tau_{2}$ plan since they are related by gauge symmetry. The temporal component of the self-consistent vertex $\Gamma_{t}$ in the limit $|\vec{\mathbf{q}}|\to 0$ ($q$ is the momentum of the external field) is determined directly by the Ward-Takahashi identity 
\begin{equation}
q_{\mu}\Gamma_{\mu}(p+q,p)=-e\tau_{3}G^{-1}(p)+eG^{-1}(p+q)\tau_{3}
\end{equation}
Where $q_{\mu}\Gamma_{\mu}$ is a shorthand for $\vec{\mathbf{q}}\cdot\vec{\mathbf{\Gamma}}-\omega\Gamma_{t}$. Note that there are additional $\tau_{3}$'s compared to the standard Ward-Takahashi identity in QED since the two components of the Nambu spinor carry opposite charges. If we assume the spatial components of $\Gamma$ does not diverge in the limit $|\vec{\mathbf{q}}|\to 0$, which can be verified latter, only the temporal component of $\Gamma$ contribute the left hand side, and we have
\begin{equation}
\Gamma_{t}([p_{0}+\omega,\vec{\mathbf{p}}],[p_{0}, \vec{\mathbf{p}}])=-(-e\tau_{3}G^{-1}(p)+eG^{-1}(p+q)\tau_{3})/\omega=-e(\tau_{3}+2i\Delta_{p}\tau_{2}/\omega)
\end{equation}
On the other hand, the spatial components of $\Gamma$ take some calculation, and they acquire a simple form only when the four-Fermion interaction has no momentum dependence near the Fermi surface. In this case $\lambda_{k}$ can be treated as a constant, and the self-consistent equation (Eq. (\ref{eq:mean field Green})) shows that $\Delta_{p}$ is also a constant near the Fermi surface. Plugging the mean field Green's function in Eq. (\ref{eq: vertex correction}), and shifting the momentum of the integration, we have
\begin{align}
&\Gamma_{\mu}([p_{0}+\omega,\vec{\mathbf{p}}],[p_{0}, \vec{\mathbf{p}}])=\gamma_{\mu}(\vec{\mathbf{p}})\nonumber\\
&+i\lambda\int\frac{d^{3}p}{(2\pi)^{3}}\frac{\tau_{3}(\tilde{p}_{0}+\omega+\bar{\epsilon}_{p}\tau_{3}+\Delta\tau_{1})\Gamma_{\mu}([p_{0}+\omega,\vec{\mathbf{p}}],[p_{0}, \vec{\mathbf{p}}])(\tilde{p}_{0}+\bar{\epsilon}_{p}\tau_{3}+\Delta\tau_{1})\tau_{3}}{((\tilde{p}_{0}+\omega+isgn(p_{0}+\omega)0^{+})^{2}-\delta_{p}^{2})((\tilde{p}_{0}+isgn(p_{0})0^{+})^{2}-\delta_{p}^{2})}
\label{extremely long}
\end{align}
It is clear from the equation above that the vertex correction has no p dependence, this is of course only true when we ignore the momentum dependence of the four-Fermion interaction. In this case, we can write the self-consistent vertex as
\begin{equation}
\Gamma_{\mu}([p_{0}+\omega,\vec{\mathbf{p}}],[p_{0}, \vec{\mathbf{p}}])=\gamma_{\mu}(\vec{\mathbf{p}})-e\Gamma_{\mu}^{0}\mathbb{1}-e\sum_{i=1}^{3}\Gamma_{\mu}^{i}\tau_{i}
\label{eq:Gammai}
\end{equation}
where $\Gamma^{0}$ and $\Gamma^{i}$ are functions of $\omega$, and $\gamma_{\mu}(\vec{\mathbf{p}})$ is given by Eq. (\ref{eq:paramagnetic current}) and (\ref{eq:bare vertex}). The next step is to plug Eq. (\ref{eq:Gammai}) into Eq. (\ref{extremely long}), compute the matrix multiplication in the numerator, carry out the integral of $p_{0}$ using the residue theorem and solve $\Gamma^{0}$ and $\Gamma^{i}$. Note that there are 4 poles of $p_{0}$ in the complex plane, whose imaginary parts depend on the spatial momentum $\vec{\mathbf{p}}$. If $\vec{\mathbf{p}}$ lies in the `unpaired region', the two eigenenergies $E_{p}^{\pm}$ are of the same sign, so the four poles locate at the same side of the real axis. Then we know the integral must be zero since we can complete the contour on the other side including none of the residues. This observation confirms our statement that only the `paired region' in the B.Z. contribute to the optical conductivity. After all these laborious calculation, we arrive at the self-consistent equation for $\vec{\mathbf{\Gamma}}^{0}$ and $\vec{\mathbf{\Gamma}}^{i}$ (the spatial components of $\Gamma^{0}$ and $\Gamma^{i}$). We showed that, by direct calculation, the integral in Eq. (\ref{extremely long}) has no identity component, thus $\vec{\mathbf{\Gamma}}^{0}=0$. On the other hand, $\vec{\mathbf{\Gamma}}^{i}$ satisfies
\begin{align}
\left(\begin{array}{c}
 \vec{\mathbf{\Gamma}}^{1}\\ \vec{\mathbf{\Gamma}}^{2}\\ \vec{\mathbf{\Gamma}}^{3}
\end{array}\right)= \lambda
\left(\begin{array}{ccc}
2I(\bar{\epsilon}_{p}^{2}) & -i\omega I(\bar{\epsilon}_{p}) & -2\Delta I(\bar{\epsilon}_{p})\\
i\omega I(\bar{\epsilon}_{p}) & 2I(\delta_{p}^{2}) & -i\omega\Delta I(1)\\
2\Delta I(\bar{\epsilon}_{p}) & -i\omega\Delta I(1) & -2\Delta^{2}I(1)
\end{array}\right)
\left(\begin{array}{c}
 \vec{\mathbf{\Gamma}}^{1}\\ \vec{\mathbf{\Gamma}}^{2}\\ \vec{\mathbf{\Gamma}}^{3}
\end{array}\right) & \nonumber\\
+ \lambda
\left(\begin{array}{c}
 -2\Delta I(\bar{\epsilon}_{p}\vec{\mathbf{v}}_{2})\\
 -i\omega\Delta I(\vec{\mathbf{v}}_{2})\\ 
 -2\Delta^{2}I(\vec{\mathbf{v}}_{2})
\end{array}\right) & \label{eq:matrix}\\
\text{where}\ I(f(\vec{\mathbf{p}}))\equiv \int_{\text{paired}}\frac{d^{2}\vec{\mathbf{p}}}{(2\pi)^{2}}\frac{f(p)}{\delta_{p}(\omega-2\delta_{p}+isgn(\omega)0^{+})(\omega+2\delta_{p}+isgn(\omega)0^{+})} &
\label{eq:If, path integral}
\end{align}
If we further assume the pairing gap $\Delta$ and the frequency $\omega$ is much smaller than the band width, only a thin shell near $\bar{\epsilon}_{p}=0$ contribute to the integral. In this limit $I(\bar{\epsilon}_{p})\sim 0,\ I(\bar{\epsilon}_{p}\vec{\mathbf{v}}_{2})\sim 0$, so we have $\vec{\mathbf{\Gamma}}^{1}\sim 0$, $\vec{\mathbf{\Gamma}}^{2}$ and $\vec{\mathbf{\Gamma}}^{3}$ satisfies
\begin{eqnarray}
\left(\begin{array}{c}
 \vec{\mathbf{\Gamma}}^{2}\\ \vec{\mathbf{\Gamma}}^{3}
\end{array}\right)= \lambda
\left(\begin{array}{cc}
2I(\delta_{p}^{2}) & -i\omega\Delta I(1)\\
-i\omega\Delta I(1) & -2\Delta^{2}I(1)
\end{array}\right)
\left(\begin{array}{c}
 \vec{\mathbf{\Gamma}}^{2}\\ \vec{\mathbf{\Gamma}}^{3}
\end{array}\right) - \lambda I(\vec{\mathbf{v}}_{2})
\left(\begin{array}{c}
 i\omega\Delta \\ 
 2\Delta^{2}
\end{array}\right)
\end{eqnarray}{}
In addition, the mean field gap equation (Eq. (\ref{eq:familiar gap equation})) gives us
\begin{equation}
4\lambda I(\delta_{p}^{2})-\lambda\omega^{2}I(1)=-\lambda I(\omega^{2}-4\delta_{p}^{2})=-2\lambda\int_{\text{paired}}\frac{d^{2}\vec{\mathbf{p}}}{(2\pi)^{2}}\frac{1}{2\sqrt{\bar{\epsilon}_{p}^{2} + \Delta^{2}}}=2
\end{equation}
Using this identity, we can easily find
\begin{eqnarray}
\left\{\begin{array}{l}
\vec{\mathbf{\Gamma}}^{2}=\frac{2i\Delta I(\vec{\mathbf{v}}_{2})}{\omega I(1)}\\
\vec{\mathbf{\Gamma}}^{3}=0
\end{array}\right.
\end{eqnarray}
So the corrected vertex is
\begin{equation}
\Gamma_{\mu}([p_{0}+\omega,\vec{\mathbf{p}}],[p_{0}, \vec{\mathbf{p}}])=-e[\tau_{3}+2i\Delta\tau_{2}/\omega,\ \vec{\mathbf{v}}_{1}(\vec{\mathbf{p}})\mathbb{1} + \vec{\mathbf{v}}_{2}(\vec{\mathbf{p}})\tau_{3} + 2i\Delta I(\vec{\mathbf{v}}_{2})\tau_{2}/\omega I(1)]
\end{equation}
We are now ready to calculate the paramagnetic response function $P_{\mu\nu}$. For simplicity, define
\begin{equation}
\langle f,h\rangle\equiv-i\int\frac{d^{3}p}{(2\pi)^{3}}Tr[f(p,p')G_{p'}h(p',p)G_{p}]
\end{equation}
Then we have
\begin{eqnarray}
P_{ij}&=&\langle \gamma_{i},\Gamma_{j}\rangle\\
&=& e^{2}\langle v_{1i}(\vec{\mathbf{p}})\mathbb{1} + v_{2i}(\vec{\mathbf{p}})\tau_{3} ,v_{1j}(\vec{\mathbf{p}})\mathbb{1} + v_{2j}(\vec{\mathbf{p}})\tau_{3} + 2i\Delta I(v_{2j})\tau_{2}/\omega I(1)\rangle\\
&=& e^{2}\langle v_{2i}(\vec{\mathbf{p}})\tau_{3},v_{2j}(\vec{\mathbf{p}})\tau_{3}\rangle + (2i\Delta I(v_{2j})/\omega I(1))e^{2}\langle v_{2i}(\vec{\mathbf{p}})\tau_{3}, \tau_{2}\rangle
\end{eqnarray}
where we have used the fact that the identity component of the vertex does not contribute to the integral, which can be verified explicitly. Integrating out $p_{0}$ we have
\begin{equation}
P_{ij}=4e^{2}\Delta^{2}\left[I(v_{2i}v_{2j})-I(v_{2i})I(v_{2j})/I(1)\right]
\end{equation}
This result leads to the result for optical conductivity in Eq. (\ref{eq: final results}). We would like to remind the readers again that Eq. (\ref{eq: final results}) holds only for $\omega>0$ if we define the integral $I(f(\vec{\mathbf{p}}))$ as in Eq. (\ref{eq:If, path integral}), this is due to the difference between path integral and retarded response. It holds for both positive and negative $\omega$ if we replace the infinitesimal imaginary part $isgn(\omega)0^{+}$ in the integral $I(f(\vec{\mathbf{p}}))$ by $i0^{+}$.

\chapter{More on Incommensurate Pair Density Wave Bands}

For uniform PDW state, we calculate the band structure by diagonalizing a BdG Hamiltonian $H(k)$ for each momentum $k$. At each $k$, we need to use a $81*2=162$ basis:$\Psi_k=(\psi_\uparrow(k),\psi_\downarrow^\dagger(-k))$.  $\psi_\sigma(k)$ is a collection of $9\times9=81$ electron annihilation operators: $c_{k'}$ with momenta $k'=k+m \mathbf{P_x}+n\mathbf{P_y}$ where $\mathbf{P_x}\approx(0.14\times{2\pi},0)$ and $\mathbf{P_y}\approx(0,0.14\times{2\pi})$, $m,n=-4,-3,-2,-1,0,1,2,3,4$. We set a large truncation for m and n to better capture the effect of subsidiary CDW generated by PDW. In this basis, we rewrite the mean field Hamiltonian in Eq.~\ref{Eq: long range PDW mean field} at momentum $k$ as

\begin{align}
H_k =& \sum_{m,n}\e_{k+m \mathbf{P_x}+n\mathbf{P_y}} c^{\dagger}_{k+m \mathbf{P_x}+n\mathbf{P_y},\uparrow}c_{k+m \mathbf{P_x}+n\mathbf{P_y},\uparrow}\nonumber\\
&-\sum_{m,n}\e_{-k-m \mathbf{P_x}-n\mathbf{P_y}} c_{-k-m \mathbf{P_x}-n\mathbf{P_y},\downarrow}c^{\dagger}_{-k-m \mathbf{P_x}-n\mathbf{P_y},\downarrow}\nonumber\\ 
&+\sum_{m,n}2\D(\cos(k_x+m P_x+nP_y - P_x/2) - \cos(k_y+m P_x+n P_y))\nonumber\\
&\ \ \ \ \ \ \ \ \ \cdot c_{k+m \mathbf{P_x}+n\mathbf{P_y},\uparrow}c_{-k -m \mathbf{P_x}-n\mathbf{P_y} + \mathbf{P_x},\downarrow}+ h.c. \nonumber\\
&+\sum_{m,n}2\D(\cos(k_x+m P_x+nP_y + P_x/2) - \cos(k_y+m P_x+nP_y))\nonumber\\ 
&\ \ \ \ \ \ \ \ \ \cdot c_{k+m \mathbf{P_x}+n\mathbf{P_y},\uparrow}c_{-k -m \mathbf{P_x}-n\mathbf{P_y} - \mathbf{P_x},\downarrow}+ h.c. \nonumber\\
&+\sum_{m,n}2\D(\cos(k_x+m P_x+nP_y) - \cos(k_y+m P_x+n P_y- P_y/2))\nonumber\\
&\ \ \ \ \ \ \ \ \ \cdot c_{k+m \mathbf{P_x}+n\mathbf{P_y},\uparrow}c_{-k -m \mathbf{P_x}-n\mathbf{P_y} + \mathbf{P_y},\downarrow}+ h.c. \nonumber\\
&+\sum_{m,n}2\D(\cos(k_x+m P_x+nP_y) - \cos(k_y+m P_x+nP_y+ P_y/2))\nonumber\\
&\ \ \ \ \ \ \ \ \ \cdot c_{k+m \mathbf{P_x}+n\mathbf{P_y},\uparrow}c_{-k -m \mathbf{P_x}-n\mathbf{P_y} - \mathbf{P_y},\downarrow}+ h.c.,
\end{align}

where $\D = 45$meV. For the bare band dispersion $\e_k$, we use a tight banding model on square lattice with nearest neighbor hopping $t=0.21$eV, second neighbor hopping $t_p=-0.047$eV, third neighbor hopping $t_{pp}=0.04$eV and fourth neighbor hopping $t_{ppp}=-0.01$eV.
\bea
\e_k = -2t(\cos(k_x)+\cos(k_y)) - 4t_p\cos(k_x)\cos(k_y) - 2t_{pp}(\cos(2k_x)+\cos(2k_y))\nonumber\\
 - 4t_{ppp}(\cos(2k_x)\cos(k_y) + \cos(k_x)\cos(2k_y)) - \e_0
\eea 
We fix the chemical potential $\e_0$ self-consistently to match the hole doping.

\chapter{Numerical Simulation of d-wave Vortex Halo}
We did exact diagonalization to simulate Local Density of State(LDoS) inside Vortex Halo. Our Hamiltonian for PDW-Driven Model is:
\begin{align}
	&H_P=H_0\nonumber\\
	&+\sum_{\mathbf x,\mathbf{\mu}}F_d(\mu)\left(|\Delta_D|e^{i\theta_d+i\theta}+\left(\sum_a|\Delta_{P_a}|e^{i\theta_a+i\theta_d}\sin(\frac{1}{2}\mathbf{Q_a}\cdot (\mathbf{x+\frac{\mu}{2}}))\right)\right)c^\dagger_\uparrow(\mathbf x)c^\dagger_\downarrow(\mathbf{x+\mu})\nonumber\\
	&+h.c.
	\label{eq:pdw_real}
\end{align}
where $\mathbf \mu=\hat{x}$ or $\hat{y}$ labels two different kinds of nearest neighbor bond. $F_d(\hat x)=1$  and $F_d(\hat y)=-1$. $a$ means $x$ or $y$. We used $|\Delta_{P_x}|=|\Delta_{P_y}|=30$meV  at vortex center in our calculation, away from vortex center the PDW profile is
\begin{equation}
	\Delta_P(r)=30e^{1-\sqrt{r^2+\xi^2}/\xi} meV
\end{equation}
with $\xi=15$

Our Hamiltonian for CDW-Driven Model is:
\begin{align}
	H_C=H_0 + & \sum_{\mathbf x,\mathbf{\mu}}F_d(\mu)|\Delta_D|e^{i\theta_d+i\theta}c^\dagger_\uparrow(\mathbf x)c^\dagger_\downarrow(\mathbf{x+\mu})\nonumber\\
	+ &\sum_{\mathbf x,\mathbf{\mu}}F_s(\mu)\left(\sum_a|\Delta_{C_a}|e^{i\theta_a}\sin(\frac{1}{2}\mathbf{Q_a}\cdot (\mathbf{x+\frac{\mu}{2}}))\right)\sum_\sigma c^\dagger_\sigma(\mathbf x)c_\sigma(\mathbf{x+\mu})+h.c.
\end{align}
where $F_s(\hat x)=F_s(\hat y)=1$ is a $s$ wave form factor. We used $|\Delta_{C_x}|=|\Delta_{C_y}|=30$meV at vortex center in our calculation. Away from vortex center CDW has a profile similar to PDW-Driven model:
\begin{equation}
	\Delta_C(r)=30e^{1-\sqrt{r^2+\xi^2}/\xi} meV
\end{equation}

For both PDW-Driven and CDW-Driven model, we use $|\Delta_D|=20$meV far away from vortex core and $\Delta_D(r,\theta)=20 \frac{r}{\sqrt{r^2+r_0^2}}$ meV near vortex core. We add one d-wave vortex to a $100 a \times 100 a$ square lattice with open boundary condition. $\mathbf{Q_x/2}=(\frac{2\pi}{8},0)$ and $\mathbf{Q_y/2}=(0,\frac{2\pi}{8})$.

 After Exact Diagonalization, we can easily get on-site LDoS at any energy:
\begin{equation}
	\rho(\mathbf x, \omega)=\sum_{E,\sigma} \delta(\omega-E)\psi^*_E(\mathbf x;\sigma)\psi_E(\mathbf x;\sigma)
\end{equation}
where $E$ labels all energy levels and $\psi_E(x;\sigma)$ is the wavefunction for $\mathbf x$ site and spin $\sigma$ at energy level $E$.

For STM experiment, LDoS at Oxygen site is actually more important. In our simple one band model, we can define bond LDoS:
\begin{equation}
	\rho_{\mu}(\mathbf x, \omega)=\sum_{E,\sigma} \delta(\omega-E)\left(\psi^*_E(\mathbf x;\sigma)\psi_E(\mathbf {x+\mu};\sigma)+\psi^*_E(\mathbf {x+\mu};\sigma)\psi_E(\mathbf x;\sigma)\right)
\end{equation}
where $\mu=\hat x$ or $\hat y$.

It's then easy to define $s$ wave Bond LDoS as 
\begin{equation}
	\rho_d(\mathbf x,\omega)=\rho_{\hat x}(\mathbf x,\omega)+\rho_{\hat y}(\mathbf x, \omega)
\end{equation}
and $d$ wave Bond LDoS as
\begin{equation}
	\rho_s(\mathbf x,\omega)=\rho_{\hat x}(\mathbf x,\omega)-\rho_{\hat y}(\mathbf x, \omega)
\end{equation}

For PDW-Driven model, we found $\rho_d$ dominates and therefore we only show $d$ wave Bond DoS in the main text.  For our CDW-Driven model, it's dominated by $s$ wave CDW as an input and we show $s$ wave CDW in the main text.

\chapter{Symmetry of the Fluctuating PDW State}\label{Appendix: Symmetry of the fluctuating PDW state}

Before we discuss the symmetry of fluctuating PDW states, it is helpful to have in mind a specific pairing form factor in real space. We choose a local d-wave form factor. Define

\bea
S[(m,n), (m',n')] = c_{m,n,\uparrow}c_{m',n',\downarrow} - c_{m,n,\downarrow}c_{m',n',\uparrow}\nonumber \\
b_{m,n} = S[(m,n),(m+1,n)] + S[(m,n),(m-1,n)]\nonumber\\
- S[(m,n),(m,n+1)] - S[(m,n),(m,n-1)]\ \ \ 
\eea
where $(m,n)$ labels a Cu site in $\text{CuO}_2$ plane. $S[(m,n),(m',n')]$ represents a singlet pairing between two sites; $b_{m,n}$ represents d-wave pairing on nearest-neighbor bounds. (The following analysis is not restricted to this specific form.) A simple Hamiltonian with 4 PDWs can be

\bea
H = \sum_{m,n}\sum_{\ \vec{p} = P\hat{x}, P\hat{y}, -P\hat{x}, -P\hat{y}}\D_{\vec{p}} \,e^{i\vec{p}\cdot (m,n)} b_{m,n} + h.c.\ \ 
\eea
In order to gain pairing energy from all anti-nodal fermions, and for the approximate $C_4$ symmetry of $\text{CuO}_2$ plane, we assume the 4 PDW amplitudes have approximately equal amplitude. At low temperature, we assume only the overall superconducting phase of the 4 PDW order parameters is fluctuating. Relative phases between every pair of PDW order parameters are all ordered.

Time reversal symmetry maps $(\D_{P\hat{x}}, \D_{P\hat{y}}, \D_{-P\hat{x}}, \D_{-P\hat{y}})$ to $(\D^*_{-P\hat{x}}, \D^*_{-P\hat{y}}, \D^*_{P\hat{x}}, \D^*_{P\hat{y}})$. Time reversal invariance requires that these two set of phases differ only by an overall $U(1)_\text{charge}$ transformation.

\bea
\text{Time reversal: }(\D_{P\hat{x}}, \D_{P\hat{y}}, \D_{-P\hat{x}}, \D_{-P\hat{y}}) =\nonumber\\
e^{i\phi}(\D^*_{-P\hat{x}}, \D^*_{-P\hat{y}}, \D^*_{P\hat{x}}, \D^*_{P\hat{y}})
\eea
Similarly, invariance under inversion (about (0,0)), and Mirror along (1,-1) direction (passing through (0,0)) requires
\bea\label{Eq: inversion}
\text{Inversion about (0,0): }(\D_{P\hat{x}}, \D_{P\hat{y}}, \D_{-P\hat{x}}, \D_{-P\hat{y}}) \nonumber\\
=e^{i\phi'}(\D_{-P\hat{x}}, \D_{-P\hat{y}}, \D_{P\hat{x}}, \D_{P\hat{y}})\ \\
\text{Mirror along (1,-1): }(\D_{P\hat{x}}, \D_{P\hat{y}}, \D_{-P\hat{x}}, \D_{-P\hat{y}}) \nonumber\\
=e^{i\phi''}(\D_{-P\hat{y}}, \D_{-P\hat{x}}, \D_{P\hat{y}}, \D_{P\hat{x}})\ \label{Eq: Mirror}
\eea
where we have chose the mirror passing through $(0,0)$.

Last, under translation, $(x,y)\rightarrow (x,y) + (a,b),\ (a,b)\in\mathbb{R}^2$, 
\bea
(\D_{P\hat{x}}, \D_{P\hat{y}}, \D_{-P\hat{x}}, \D_{-P\hat{y}}) \rightarrow\nonumber\\
(e^{iPa}\D_{P\hat{x}}, e^{iPb}\D_{P\hat{y}}, e^{-iPa}\D_{-P\hat{x}}, e^{-iPb}\D_{-P\hat{y}})
\eea

To the second order of PDW amplitudes, CDW and MDW at momentum $P\hat{x}+P\hat{y}$ are generated:

\bea
\rho_{P\hat{x} + P\hat{y}} = c(\D_{P\hat{x}}\D^*_{-P\hat{y}} + \D_{P\hat{y}}\D^*_{-P\hat{x}}),\ c\in\mathbb{R}\\
M_{P\hat{x} + P\hat{y}} = id(\D_{P\hat{x}}\D^*_{-P\hat{y}} - \D_{P\hat{y}}\D^*_{-P\hat{x}}),\ d\in\mathbb{R}
\eea
where $\rho$ is charge density, $M \equiv \hat{z}\cdot\nabla\times\vec{j}$ is the orbital magnetization in $\hat{z}$ direction. Time reversal symmetry and inversion symmetry of the theory requires $c$ and $d$ to be real and exclude other free parameters. To give an example of this symmetry argument, we analyze the coefficients of MDW. By momentum and charge conservation, and that the magnetization is real in real space, the most general form of MDW at the second order is
\bea
M_{P\hat{x} + P\hat{y}} = d_1\D_{P\hat{x}}\D^*_{-P\hat{y}} + d_2\D_{P\hat{y}}\D^*_{-P\hat{x}}\\
M_{-P\hat{x} - P\hat{y}} = d_2^*\D_{-P\hat{x}}\D^*_{P\hat{y}} + d_1^*\D_{-P\hat{y}}\D^*_{P\hat{x}}
\eea
Consider the time reversal partner of the system, with pairing amplitude
$(\tilde\D_{P\hat{x}}, \tilde\D_{P\hat{y}}, \tilde\D_{-P\hat{x}}, \tilde\D_{-P\hat{y}})$ $= e^{i\phi}(\D^*_{-P\hat{x}}, \D^*_{-P\hat{y}}, \D^*_{P\hat{x}}, \D^*_{P\hat{y}})$.

\bea
\tilde M_{P\hat{x} + P\hat{y}} = d_1\tilde \D_{P\hat{x}}\tilde \D^*_{-P\hat{y}} + d_2\tilde \D_{P\hat{y}}\tilde \D^*_{-P\hat{x}}\nonumber\\
=d_1\D^*_{-P\hat{x}}\D_{P\hat{y}} + d_2\D^*_{-P\hat{y}}\D_{P\hat{x}}\\
\tilde M_{-P\hat{x} - P\hat{y}} = d_2^*\tilde \D_{-P\hat{x}}\tilde \D^*_{P\hat{y}} + d_1^*\tilde \D_{-P\hat{y}}\tilde \D^*_{P\hat{x}}\nonumber\\
= d_2^*\D^*_{P\hat{x}}\D_{-P\hat{y}} + d_1^*\D^*_{P\hat{y}}\D_{-P\hat{x}}
\eea
Since $\tilde M(x) = -M(x)$, we know that $d_1 = -d_2$. Similar arguments for inversion requires $d_1 = d_2^*$. Thus $d_1=-d_2 =id, \ d\in\mathbb{R}$. Similarly, time reversal symmetry of the theory requires the density wave generated in the leading order at momentum $2P\hat{x}$ and $2P\hat{y}$ are pure CDW with no magnetization.

In the limit PDW wavelength is much larger than the lattice spacing, we can use two lattice translation and $U(1)_\text{charge}$ to continuously change 3 of the 4 phases of the PDW amplitudes. In this limit, the only nontrivial phase is

\bea
e^{i\theta} \equiv \frac{\D_{P\hat{y}} \D_{-P\hat{y}} }{\D_{P\hat{x}} \D_{-P\hat{x}} }
\eea
This phase determines whether we have CDW or MDW at momentum $P\hat{x}+P\hat{y}$, and it affects the band structure (Fig.~\ref{Fig: Bogoliubov bands}). Time reversal symmetry forbids MDW, and requires $\theta = 0$, hence a CDW at momentum $P\hat{x}+P\hat{y}$. However, such a CDW is not observed experimentally. We postulate the opposite scenario, $\theta = \pi$, with only MDW at momentum $P\hat{x}+P\hat{y}$, which breaks time reversal. In the long-wavelength limit, inversion symmetry and mirror symmetry are always preserved. We can always find an inversion center and a mirror by translation. In the main text we consider further the case of finite wavevector P.

\chapter{Finite-size Extrapolation of Boson and Fermion Gap}\label{Appendix: DMRG}

\begin{figure}[h]
\begin{center}
\includegraphics[width=0.6\linewidth]{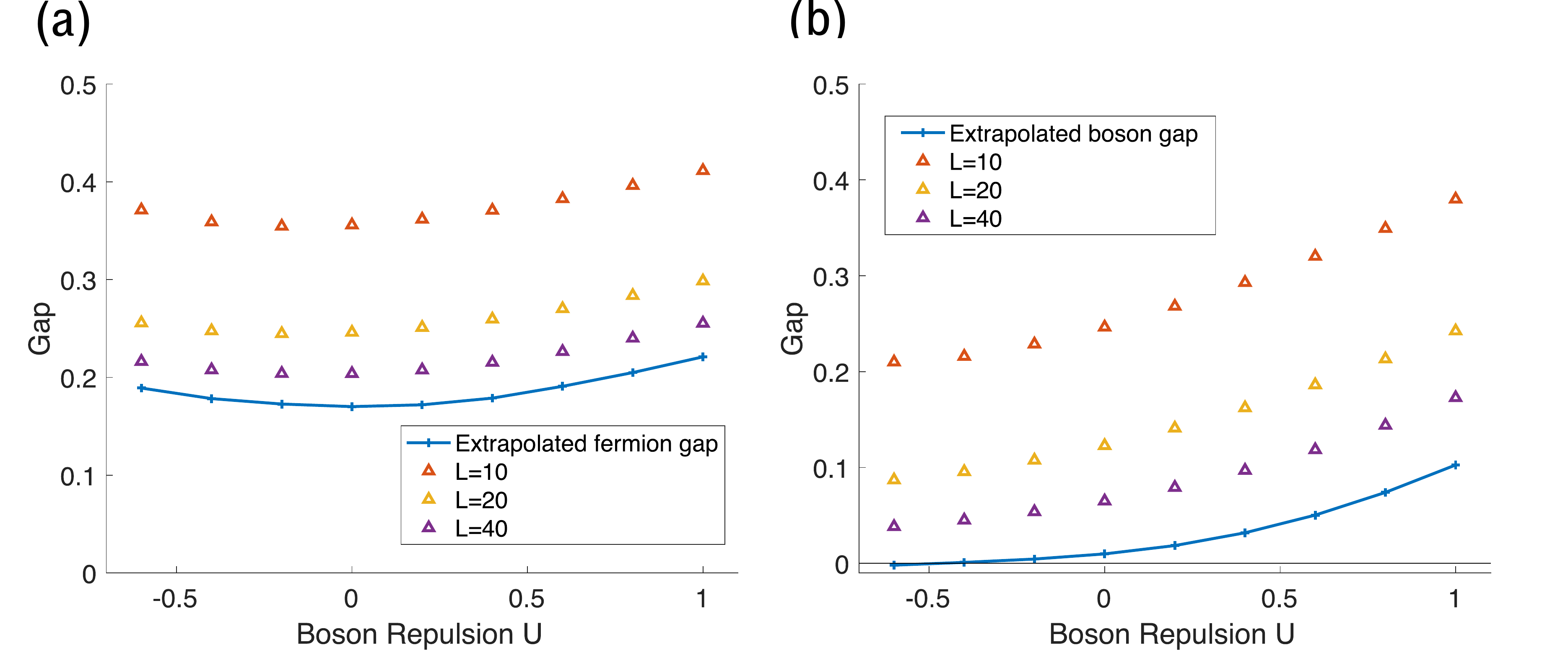}
\caption[Extrapolation of fermion and boson gap of the 1D model]{(a) Extrapolation of fermion gap, $t=1.0, p= 0.5$. (b) Extrapolation of boson gap, $t=1.0, p= 0.5$.}
\label{Fig: DMRG extrapolation p05}
\end{center}
\end{figure}

We compute boson gaps and fermion gaps of the 1D model in Sec.~\ref{subsection: 1D numerics} (as a function of the boson repulsion $U$) on system with length $L=10,20,40$, and then fit the gap to the form

\bea
E(L) = E_{\infty} + a/L + b/L^2
\eea
to get the thermodynamic gap $E_{\infty}$. Fig.~\ref{Fig: DMRG extrapolation p05} shows finite-size gaps together with extrapolated gaps for $p=0.5$.

\begin{singlespace}
\bibliography{main}

\newcommand{\noopsort}[1]{} \newcommand{\printfirst}[2]{#1}
  \newcommand{\singleletter}[1]{#1} \newcommand{\switchargs}[2]{#2#1}
\begin{thebibliography}{100}

\bibitem{Agt2PhysRevB.91.054502}
D.~F. Agterberg, Drew~S. Melchert, and M.~K. Kashyap.
\newblock Emergent loop current order from pair density wave superconductivity.
\newblock {\em Phys. Rev. B}, 91:054502, Feb 2015.

\bibitem{agterberg2020physics}
Daniel~F Agterberg, JC~S{\'e}amus Davis, Stephen~D Edkins, Eduardo Fradkin,
  Dale~J Van~Harlingen, Steven~A Kivelson, Patrick~A Lee, Leo Radzihovsky,
  John~M Tranquada, and Yuxuan Wang.
\newblock The physics of pair-density waves: Cuprate superconductors and
  beyond.
\newblock {\em Annual Review of Condensed Matter Physics}, 11:231--270, 2020.

\bibitem{agterberg2015checkerboard}
Daniel~F Agterberg and Julien Garaud.
\newblock Checkerboard order in vortex cores from pair-density-wave
  superconductivity.
\newblock {\em Physical Review B}, 91(10):104512, 2015.

\bibitem{agterberg2019review}
DF~Agterberg, John~C Davis, S~Edkins, E~Fradkin, D~van Harlingen, S~Kivelson,
  P~Lee, L~Radzihovsky, J~Tranquada, and Y~Wang.
\newblock Physics of pair-density-waves.
\newblock {\em ArXiv: 1904.09687. submitted to Annual Reviews of Condensed
  Matter Physics}, 2019.

\bibitem{agterberg2008dislocations}
DF~Agterberg and H~Tsunetsugu.
\newblock Dislocations and vortices in pair-density-wave superconductors.
\newblock {\em Nature Physics}, 4(8):639, 2008.

\bibitem{aji2009quantum}
Vivek Aji and CM~Varma.
\newblock Quantum criticality in dissipative quantum two-dimensional {XY} and
  {A}shkin-{T}eller models: Application to the cuprates.
\newblock {\em Physical Review B}, 79(18):184501, 2009.

\bibitem{PhysRevLett.96.147003}
A.~S. Alexandrov.
\newblock Normal-state diamagnetism of charged bosons in cuprate
  superconductors.
\newblock {\em Phys. Rev. Lett.}, 96:147003, Apr 2006.

\bibitem{PhysRev.115.2}
P.~W. Anderson.
\newblock New approach to the theory of superexchange interactions.
\newblock {\em Phys. Rev.}, 115:2--13, Jul 1959.

\bibitem{anderson1987resonating}
Philip~W Anderson.
\newblock The resonating valence bond state in $\text{{La}}_2\text{{CuO}}_4$
  and superconductivity.
\newblock {\em science}, 235(4793):1196--1198, 1987.

\bibitem{anderson2000brainwashed}
Philip~W Anderson.
\newblock Brainwashed by feynman?
\newblock {\em Physics Today}, 53(2):11--12, 2000.

\bibitem{PhysRevB.60.1654}
Leon Balents, Matthew P.~A. Fisher, and Chetan Nayak.
\newblock Dual order parameter for the nodal liquid.
\newblock {\em Phys. Rev. B}, 60:1654--1667, Jul 1999.

\bibitem{PhysRevB.77.174502}
Shirit Baruch and Dror Orgad.
\newblock Spectral signatures of modulated $d$-wave superconducting phases.
\newblock {\em Phys. Rev. B}, 77:174502, May 2008.

\bibitem{RevModPhys.77.721}
D.~N. Basov and T.~Timusk.
\newblock Electrodynamics of high-$\text{{T}}_\text{c}$ superconductors.
\newblock {\em Rev. Mod. Phys.}, 77:721--779, Aug 2005.

\bibitem{bednorz1986possible}
J~George Bednorz and K~Alex M{\"u}ller.
\newblock Possible hight c superconductivity in the ba- la- cu- o system.
\newblock {\em Zeitschrift f{\"u}r Physik B Condensed Matter}, 64(2):189--193,
  1986.

\bibitem{PhysRevLett.99.127003}
E.~Berg, E.~Fradkin, E.-A. Kim, S.~A. Kivelson, V.~Oganesyan, J.~M. Tranquada,
  and S.~C. Zhang.
\newblock Dynamical layer decoupling in a stripe-ordered
  high-$\text{{T}}_\text{c}$ superconductor.
\newblock {\em Phys. Rev. Lett.}, 99:127003, Sep 2007.

\bibitem{berg1NatPhys2009charge}
Erez Berg, Eduardo Fradkin, and Steven~A Kivelson.
\newblock Charge-4e superconductivity from pair-density-wave order in certain
  high-temperature superconductors.
\newblock {\em Nature Physics}, 5(11):830, 2009.

\bibitem{berg2009charge}
Erez Berg, Eduardo Fradkin, and Steven~A Kivelson.
\newblock Charge-4e superconductivity from pair-density-wave order in certain
  high-temperature superconductors.
\newblock {\em Nature Physics}, 5(11):830, 2009.

\bibitem{PhysRevB.79.064515}
Erez Berg, Eduardo Fradkin, and Steven~A. Kivelson.
\newblock Theory of the striped superconductor.
\newblock {\em Phys. Rev. B}, 79:064515, Feb 2009.

\bibitem{berg2009striped}
Erez Berg, Eduardo Fradkin, Steven~A Kivelson, and John~M Tranquada.
\newblock Striped superconductors: how spin, charge and superconducting orders
  intertwine in the cuprates.
\newblock {\em New Journal of Physics}, 11(11):115004, 2009.

\bibitem{berg22009NTPhysstriped}
Erez Berg, Eduardo Fradkin, Steven~A Kivelson, and John~M Tranquada.
\newblock Striped superconductors: how spin, charge and superconducting orders
  intertwine in the cuprates.
\newblock {\em New Journal of Physics}, 11(11):115004, 2009.

\bibitem{Yeh1}
AD~Beyer, MS~Grinolds, ML~Teague, S~Tajima, and N-C Yeh.
\newblock Scanning tunneling spectroscopic evidence for magnetic-field--induced
  microscopic orders in the high-tc superconductor
  $\text{{YBa}}_2\text{{Cu}}_3\text{{O}}_{7-\delta}$.
\newblock {\em EPL (Europhysics Letters)}, 87(3):37005, 2009.

\bibitem{bianchi2003possible}
A.~Bianchi, R.~Movshovich, C.~Capan, P.~G. Pagliuso, and J.~L. Sarrao.
\newblock Possible {F}ulde-{F}errell-{L}arkin-{O}vchinnikov superconducting
  state in
  ${{\mathrm{C}}\mathrm{e}{\mathrm{C}}\mathrm{o}{\mathrm{I}}\mathrm{n}}_{5}$.
\newblock {\em Phys. Rev. Lett.}, 91:187004, Oct 2003.

\bibitem{blackburn2013x}
E.~Blackburn, J.~Chang, M.~H\"ucker, A.~T. Holmes, N.~B. Christensen, Ruixing
  Liang, D.~A. Bonn, W.~N. Hardy, U.~R\"utt, O.~Gutowski, M.~v. Zimmermann,
  E.~M. Forgan, and S.~M. Hayden.
\newblock X-ray diffraction observations of a charge-density-wave order in
  superconducting ortho-ii
  ${{\mathrm{YBa}}}_{2}{{\mathrm{Cu}}}_{3}{{\mathrm{O}}}_{6.54}$ single
  crystals in zero magnetic field.
\newblock {\em Phys. Rev. Lett.}, 110:137004, Mar 2013.

\bibitem{BlancoPhysRevB.90.054513}
S.~Blanco-Canosa, A.~Frano, E.~Schierle, J.~Porras, T.~Loew, M.~Minola,
  M.~Bluschke, E.~Weschke, B.~Keimer, and M.~Le~Tacon.
\newblock Resonant x-ray scattering study of charge-density wave correlations
  in ${{\mathrm{YBa}}}_{2}{{\mathrm{Cu}}}_{3}{{\mathrm{O}}}_{6+x}$.
\newblock {\em Phys. Rev. B}, 90:054513, Aug 2014.

\bibitem{bouadim2011single}
Karim Bouadim, Yen~Lee Loh, Mohit Randeria, and Nandini Trivedi.
\newblock Single-and two-particle energy gaps across the disorder-driven
  superconductor--insulator transition.
\newblock {\em Nature Physics}, 7(11):884, 2011.

\bibitem{bourges2017comment}
P.~Bourges, Y.~Sidis, and L.~Mangin-Thro.
\newblock Comment on ``no evidence for orbital loop currents in charge-ordered
  ${{\mathrm{YBa}}}_{2}{{\mathrm{Cu}}}_{3}{{\mathrm{O}}}_{6+x}$ from polarized
  neutron diffraction''.
\newblock {\em Phys. Rev. B}, 98:016501, Jul 2018.

\bibitem{bourges2011novel}
Philippe Bourges and Yvan Sidis.
\newblock Novel magnetic order in the pseudogap state of
  high-$\text{{T}}_\text{c}$ copper oxides superconductors.
\newblock {\em Comptes Rendus Physique}, 12(5-6):461--479, 2011.

\bibitem{boyack2016gauge}
Rufus Boyack, Brandon~M Anderson, Chien-Te Wu, and K~Levin.
\newblock Gauge invariant theories of linear response for strongly correlated
  superconductors.
\newblock {\em arXiv preprint arXiv:1602.02156}, 2016.

\bibitem{PhysRevLett.101.215301}
Aurel Bulgac and Michael~McNeil Forbes.
\newblock Unitary fermi supersolid: The {L}arkin-{O}vchinnikov phase.
\newblock {\em Phys. Rev. Lett.}, 101:215301, Nov 2008.

\bibitem{RevModPhys.76.263}
Roberto Casalbuoni and Giuseppe Nardulli.
\newblock Inhomogeneous superconductivity in condensed matter and qcd.
\newblock {\em Rev. Mod. Phys.}, 76:263--320, Feb 2004.

\bibitem{changNatureComm72016magnetic}
J~Chang, E~Blackburn, O~Ivashko, AT~Holmes, Niels~Bech Christensen,
  M~H{\"u}cker, Ruixing Liang, DA~Bonn, WN~Hardy, U~R{\"u}tt, et~al.
\newblock Magnetic field controlled charge density wave coupling in underdoped
  $\text{{YBa}}_2\text{{Cu}}_3\text{{O}}_{6+x}$.
\newblock {\em Nature communications}, 7:11494, 2016.

\bibitem{chang2012decrease}
J~Chang, N~Doiron-Leyraud, O~Cyr-Choiniere, G~Grissonnanche, F~Lalibert{\'e},
  E~Hassinger, J-Ph Reid, R~Daou, S~Pyon, T~Takayama, et~al.
\newblock Decrease of upper critical field with underdoping in cuprate
  superconductors.
\newblock {\em Nature Physics}, 8(10):751, 2012.

\bibitem{PhysRevB.94.205117}
Shubhayu Chatterjee and Subir Sachdev.
\newblock Fractionalized fermi liquid with bosonic chargons as a candidate for
  the pseudogap metal.
\newblock {\em Phys. Rev. B}, 94:205117, Nov 2016.

\bibitem{chen2019incoherent}
Su-Di Chen, Makoto Hashimoto, Yu~He, Dongjoon Song, Ke-Jun Xu, Jun-Feng He,
  Thomas~P Devereaux, Hiroshi Eisaki, Dong-Hui Lu, Jan Zaanen, et~al.
\newblock Incoherent strange metal sharply bounded by a critical doping in
  {Bi}2212.
\newblock {\em Science}, 366(6469):1099--1102, 2019.

\bibitem{chen2012classification}
Yu~Chen, Jinwu Ye, and Guangshan Tian.
\newblock Classification of a supersolid: Trial wavefunctions, symmetry
  breakings and excitation spectra.
\newblock {\em Journal of Low Temperature Physics}, 169(3-4):149--168, 2012.

\bibitem{PhysRevLett.81.4716}
Andrey~V. Chubukov and Dirk~K. Morr.
\newblock Spectral function of superconducting cuprates near optimal doping.
\newblock {\em Phys. Rev. Lett.}, 81:4716--4719, Nov 1998.

\bibitem{PhysRevLett.113.046402}
Philippe Corboz, T.~M. Rice, and Matthias Troyer.
\newblock Competing states in the $t$-${J}$ model: Uniform $d$-wave state
  versus stripe state.
\newblock {\em Phys. Rev. Lett.}, 113:046402, Jul 2014.

\bibitem{dai2017opticalconductivity}
Zhehao Dai and Patrick~A. Lee.
\newblock Optical conductivity from pair density waves.
\newblock {\em Phys. Rev. B}, 95:014506, Jan 2017.

\bibitem{dai2019loop}
Zhehao Dai and Adam Nahum.
\newblock Quantum criticality of loops with topologically constrained dynamics.
\newblock {\em arXiv preprint arXiv:1910.01136}, 2019.

\bibitem{dai2019pseudogap}
Zhehao Dai, T.~Senthil, and Patrick~A. Lee.
\newblock Modeling the pseudogap metallic state in cuprates: Quantum disordered
  pair density wave.
\newblock {\em Phys. Rev. B}, 101:064502, Feb 2020.

\bibitem{dai2018STM}
Zhehao Dai, Ya-Hui Zhang, T.~Senthil, and Patrick~A. Lee.
\newblock Pair-density waves, charge-density waves, and vortices in
  high-$\text{{T}}_\text{c}$ cuprates.
\newblock {\em Phys. Rev. B}, 97:174511, May 2018.

\bibitem{damascelli2003angle}
Andrea Damascelli, Zahid Hussain, and Zhi-Xun Shen.
\newblock Angle-resolved photoemission studies of the cuprate superconductors.
\newblock {\em Rev. Mod. Phys.}, 75:473--541, Apr 2003.

\bibitem{PhysRevLett.47.1556}
C.~Dasgupta and B.~I. Halperin.
\newblock Phase transition in a lattice model of superconductivity.
\newblock {\em Phys. Rev. Lett.}, 47:1556--1560, Nov 1981.

\bibitem{PhysRevLett.87.227001}
H.~Ding, J.~R. Engelbrecht, Z.~Wang, J.~C. Campuzano, S.-C. Wang, H.-B. Yang,
  R.~Rogan, T.~Takahashi, K.~Kadowaki, and D.~G. Hinks.
\newblock Coherent quasiparticle weight and its connection to high-
  $\text{{T}}_\text{c}$ superconductivity from angle-resolved photoemission.
\newblock {\em Phys. Rev. Lett.}, 87:227001, Nov 2001.

\bibitem{doiron2007quantum}
Nicolas Doiron-Leyraud, Cyril Proust, David LeBoeuf, Julien Levallois,
  Jean-Baptiste Bonnemaison, Ruixing Liang, DA~Bonn, WN~Hardy, and Louis
  Taillefer.
\newblock Quantum oscillations and the fermi surface in an underdoped
  high-$\text{{T}}_ \text{c}$ superconductor.
\newblock {\em Nature}, 447(7144):565, 2007.

\bibitem{PhysRevLett.117.187001}
Andreas Eberlein, Walter Metzner, Subir Sachdev, and Hiroyuki Yamase.
\newblock Fermi surface reconstruction and drop in the hall number due to
  spiral antiferromagnetism in high-$\text{{T}}_\text{c}$ cuprates.
\newblock {\em Phys. Rev. Lett.}, 117:187001, Oct 2016.

\bibitem{edkins2019magnetic}
Stephen~D Edkins, Andrey Kostin, Kazuhiro Fujita, Andrew~P Mackenzie, Hiroshi
  Eisaki, S~Uchida, Subir Sachdev, Michael~J Lawler, E-A Kim, JC~S{\'e}amus
  Davis, et~al.
\newblock Magnetic field--induced pair density wave state in the cuprate vortex
  halo.
\newblock {\em Science}, 364(6444):976--980, 2019.

\bibitem{edkins2018magnetic}
Stephen~D Edkins, Andrey Kostin, Kazuhiro Fujita, Andrew~P Mackenzie, Hiroshi
  Eisaki, Shin-Ichi Uchida, Subir Sachdev, Michael~J Lawler, Eun-Ah Kim,
  JC~Davis, et~al.
\newblock Magnetic-field induced pair density wave state in the cuprate vortex
  halo.
\newblock {\em Science}, 364:976, 2019.

\bibitem{Dominic}
Dominic Else.
\newblock private communication.

\bibitem{PhysRevLett.58.2794}
V.~J. Emery.
\newblock Theory of high-${{\mathrm{T}}}_{\mathrm{c}}$ superconductivity in
  oxides.
\newblock {\em Phys. Rev. Lett.}, 58:2794--2797, Jun 1987.

\bibitem{emery1995importance}
V.~J. Emery and S.~A. Kivelson.
\newblock Importance of phase fluctuations in superconductors with small
  superfluid density.
\newblock {\em Nature}, 374(6521):434, 1995.

\bibitem{feng2000signature}
DL~Feng, DH~Lu, KM~Shen, C~Kim, H~Eisaki, A~Damascelli, R~Yoshizaki, J-i
  Shimoyama, K~Kishio, GD~Gu, et~al.
\newblock Signature of superfluid density in the single-particle excitation
  spectrum of $\text{{Bi}}_2\text{{Sr}}_2\text{{CaCu}}_2\text{{O}}_{8+
  \delta}$.
\newblock {\em Science}, 289(5477):277--281, 2000.

\bibitem{PhysRevB.39.2756}
Matthew P.~A. Fisher and D.~H. Lee.
\newblock Correspondence between two-dimensional bosons and a bulk
  superconductor in a magnetic field.
\newblock {\em Phys. Rev. B}, 39:2756--2759, Feb 1989.

\bibitem{fradkin2015colloquium}
Eduardo Fradkin, Steven~A. Kivelson, and John~M. Tranquada.
\newblock Colloquium: Theory of intertwined orders in high temperature
  superconductors.
\newblock {\em Rev. Mod. Phys.}, 87:457--482, May 2015.

\bibitem{PhysRevB.66.054535}
M.~Franz, Z.~Te\ifmmode \check{s}\else \v{s}\fi{}anovi\ifmmode~\acute{c}\else
  \'{c}\fi{}, and O.~Vafek.
\newblock ${\mathrm{qed}}_{3}$ theory of pairing pseudogap in cuprates: From
  $d$-wave superconductor to antiferromagnet via an algebraic fermi liquid.
\newblock {\em Phys. Rev. B}, 66:054535, Aug 2002.

\bibitem{fulde1964superconductivity}
Peter Fulde and Richard~A. Ferrell.
\newblock Superconductivity in a strong spin-exchange field.
\newblock {\em Phys. Rev.}, 135:A550--A563, Aug 1964.

\bibitem{ZX1science350949gerber2015three}
Simon Gerber, H~Jang, H~Nojiri, S~Matsuzawa, H~Yasumura, DA~Bonn, R~Liang,
  WN~Hardy, Z~Islam, A~Mehta, et~al.
\newblock Three-dimensional charge density wave order in
  $\text{{YBa}}_2\text{{Cu}}_3\text{{O}}_{6.67}$ at high magnetic fields.
\newblock {\em Science}, 350(6263):949--952, 2015.

\bibitem{PhysRevB.55.3173}
V.~B. Geshkenbein, L.~B. Ioffe, and A.~I. Larkin.
\newblock Superconductivity in a system with preformed pairs.
\newblock {\em Phys. Rev. B}, 55:3173--3180, Feb 1997.

\bibitem{ghiringhelli2012long}
G~Ghiringhelli, M~Le~Tacon, M~Minola, S~Blanco-Canosa, C~Mazzoli, NB~Brookes,
  GM~De~Luca, A~Frano, DG~Hawthorn, F~He, et~al.
\newblock Long-range incommensurate charge fluctuations in $\text{{(Y, Nd)
  Ba}}_2\text{{Cu}}_3\text{{O}}_{6+ x}$.
\newblock {\em Science}, page 1223532, 2012.

\bibitem{grissonnanche2014direct}
G~Grissonnanche, O~Cyr-Choini{\`e}re, F~Lalibert{\'e}, S~Ren{\'e} De~Cotret,
  A~Juneau-Fecteau, S~Dufour-Beaus{\'e}jour, M-E Delage, D~LeBoeuf, J~Chang,
  BJ~Ramshaw, et~al.
\newblock Direct measurement of the upper critical field in cuprate
  superconductors.
\newblock {\em Nature communications}, 5:3280, 2014.

\bibitem{PhysRevB.93.064513}
G.~Grissonnanche, F.~Lalibert\'e, S.~Dufour-Beaus\'ejour, M.~Matusiak,
  S.~Badoux, F.~F. Tafti, B.~Michon, A.~Riopel, O.~Cyr-Choini\`ere, J.~C.
  Baglo, B.~J. Ramshaw, R.~Liang, D.~A. Bonn, W.~N. Hardy, S.~Kr\"amer,
  D.~LeBoeuf, D.~Graf, N.~Doiron-Leyraud, and Louis Taillefer.
\newblock Wiedemann-franz law in the underdoped cuprate superconductor
  ${{\mathrm{YBa}}}_{2}{{\mathrm{Cu}}}_{3}{{\mathrm{O}}}_{y}$.
\newblock {\em Phys. Rev. B}, 93:064513, Feb 2016.

\bibitem{PhysRevLett.110.216405}
Emanuel Gull, Olivier Parcollet, and Andrew~J. Millis.
\newblock Superconductivity and the pseudogap in the two-dimensional hubbard
  model.
\newblock {\em Phys. Rev. Lett.}, 110:216405, May 2013.

\bibitem{hamidian2016atomic}
MH~Hamidian, Stephen~David Edkins, Chung~Koo Kim, James~C Davis, AP~Mackenzie,
  H~Eisaki, S~Uchida, MJ~Lawler, E-A Kim, Subir Sachdev, et~al.
\newblock Atomic-scale electronic structure of the cuprate d-symmetry form
  factor density wave state.
\newblock {\em Nature Physics}, 12(2):150, 2016.

\bibitem{PhysRevLett.106.226402}
N.~Harrison and S.~E. Sebastian.
\newblock Protected nodal electron pocket from multiple-$\mathbf{Q}$ ordering
  in underdoped high temperature superconductors.
\newblock {\em Phys. Rev. Lett.}, 106:226402, May 2011.

\bibitem{hashimoto2014energy}
Makoto Hashimoto, Inna~M Vishik, Rui-Hua He, Thomas~P Devereaux, and Zhi-Xun
  Shen.
\newblock Energy gaps in high-transition-temperature cuprate superconductors.
\newblock {\em Nature Physics}, 10(7):483, 2014.

\bibitem{he2011single}
Rui-Hua He, M~Hashimoto, H~Karapetyan, JD~Koralek, JP~Hinton, JP~Testaud,
  V~Nathan, Y~Yoshida, Hong Yao, K~Tanaka, et~al.
\newblock From a single-band metal to a high-temperature superconductor via two
  thermal phase transitions.
\newblock {\em Science}, 331(6024):1579--1583, 2011.

\bibitem{heSci3312011single}
Rui-Hua He, M~Hashimoto, H~Karapetyan, JD~Koralek, JP~Hinton, JP~Testaud,
  V~Nathan, Y~Yoshida, Hong Yao, K~Tanaka, et~al.
\newblock From a single-band metal to a high-temperature superconductor via two
  thermal phase transitions.
\newblock {\em Science}, 331(6024):1579--1583, 2011.

\bibitem{he2015establishing}
Yan He and Hao Guo.
\newblock Establishing the gauge invariant linear response of fermionic
  superfluids with pair fluctuations: A diagrammatic approach.
\newblock {\em arXiv preprint arXiv:1505.04080}, 2015.

\bibitem{himeda2002}
A~Himeda, T~Kato, and M~Ogata.
\newblock Stripe states with spatially oscillating d-wave superconductivity in
  the two-dimensional t - t' -{J} model.
\newblock {\em Physical review letters}, 88(11):117001, 2002.

\bibitem{husain2019crossover}
AA~Husain, Matteo Mitrano, Melinda~S Rak, SI~Rubeck, Bruno Uchoa, John
  Schneeloch, Ruidan Zhong, Genda~D Gu, and Peter Abbamonte.
\newblock Crossover of charge fluctuations across the strange metal phase
  diagram.
\newblock {\em arXiv preprint arXiv:1903.04038}, 2019.

\bibitem{ZX2PNAS11314647jang2016ideal}
H~Jang, W-S Lee, H~Nojiri, S~Matsuzawa, H~Yasumura, L~Nie, AV~Maharaj,
  S~Gerber, Y-J Liu, A~Mehta, et~al.
\newblock Ideal charge-density-wave order in the high-field state of
  superconducting ybco.
\newblock {\em Proceedings of the National Academy of Sciences},
  113(51):14645--14650, 2016.

\bibitem{PhysRevB.98.140505}
Hong-Chen Jiang, Zheng-Yu Weng, and Steven~A. Kivelson.
\newblock Superconductivity in the doped $\mathit{t}\ensuremath{-}\mathit{J}$
  model: Results for four-leg cylinders.
\newblock {\em Phys. Rev. B}, 98:140505, Oct 2018.

\bibitem{PhysRevB.69.212509}
A.~Kaminski, S.~Rosenkranz, H.~M. Fretwell, J.~Mesot, M.~Randeria, J.~C.
  Campuzano, M.~R. Norman, Z.~Z. Li, H.~Raffy, T.~Sato, T.~Takahashi, and
  K.~Kadowaki.
\newblock Identifying the background signal in angle-resolved photoemission
  spectra of high-temperature cuprate superconductors.
\newblock {\em Phys. Rev. B}, 69:212509, Jun 2004.

\bibitem{keimer2015quantum}
B~Keimer, SA~Kivelson, MR~Norman, S~Uchida, and J~Zaanen.
\newblock From quantum matter to high-temperature superconductivity in copper
  oxides.
\newblock {\em Nature}, 518(7538):179, 2015.

\bibitem{koren2016observation}
Gad Koren and Patrick~A Lee.
\newblock Observation of two distinct pairs fluctuation lifetimes and
  supercurrents in the pseudogap regime of cuprate junctions.
\newblock {\em Physical Review B}, 94(17):174515, 2016.

\bibitem{larkin1965inhomogeneous}
AI~Larkin and IUN Ovchinnikov.
\newblock Inhomogeneous state of superconductors(production of superconducting
  state in ferromagnet with fermi surfaces, examining green function).
\newblock {\em Soviet Physics-JETP}, 20:762--769, 1965.

\bibitem{larkin2008fluctuation}
AI~Larkin and AA~Varlamov.
\newblock Fluctuation phenomena in superconductors.
\newblock In {\em Superconductivity}, pages 369--458. Springer, 2008.

\bibitem{laughlin1983anomalous}
Robert~B Laughlin.
\newblock Anomalous quantum hall effect: an incompressible quantum fluid with
  fractionally charged excitations.
\newblock {\em Physical Review Letters}, 50(18):1395, 1983.

\bibitem{leboeuf2007electron}
David LeBoeuf, Nicolas Doiron-Leyraud, Julien Levallois, Ramzy Daou, J-B
  Bonnemaison, NE~Hussey, Luis Balicas, BJ~Ramshaw, Ruixing Liang, DA~Bonn,
  et~al.
\newblock Electron pockets in the fermi surface of hole-doped
  high-$\text{{T}}_\text{c}$ superconductors.
\newblock {\em Nature}, 450(7169):533--536, 2007.

\bibitem{PhysRevX.4.031017}
Patrick~A. Lee.
\newblock Amperean pairing and the pseudogap phase of cuprate superconductors.
\newblock {\em Phys. Rev. X}, 4:031017, Jul 2014.

\bibitem{PhysRevB.99.035132}
Patrick~A. Lee.
\newblock Proposal to measure the pair field correlator of a fluctuating pair
  density wave.
\newblock {\em Phys. Rev. B}, 99:035132, Jan 2019.

\bibitem{RevModPhys.78.17}
Patrick~A. Lee, Naoto Nagaosa, and Xiao-Gang Wen.
\newblock Doping a mott insulator: Physics of high-temperature
  superconductivity.
\newblock {\em Rev. Mod. Phys.}, 78:17--85, Jan 2006.

\bibitem{PhysRevLett.98.067006}
Sung-Sik Lee, Patrick~A. Lee, and T.~Senthil.
\newblock Amperean pairing instability in the u(1) spin liquid state with fermi
  surface and application to
  $\ensuremath{\kappa}\mathrm{\text{\ensuremath{-}}}(\mathrm{BEDT}\mathrm{\text{\ensuremath{-}}}\mathrm{TTF}{)}_{2}{{\mathrm{Cu}}}_{2}(\mathrm{CN}{)}_{3}$.
\newblock {\em Phys. Rev. Lett.}, 98:067006, Feb 2007.

\bibitem{PhysRevLett.99.067001}
Q.~Li, M.~H\"ucker, G.~D. Gu, A.~M. Tsvelik, and J.~M. Tranquada.
\newblock Two-dimensional superconducting fluctuations in stripe-ordered
  ${{\mathrm{La}}}_{1.875}{{\mathrm{Ba}}}_{0.125}{{\mathrm{CuO}}}_{4}$.
\newblock {\em Phys. Rev. Lett.}, 99:067001, Aug 2007.

\bibitem{liao2010spin}
Yean-an Liao, Ann Sophie~C Rittner, Tobias Paprotta, Wenhui Li, Guthrie~B
  Partridge, Randall~G Hulet, Stefan~K Baur, and Erich~J Mueller.
\newblock Spin-imbalance in a one-dimensional fermi gas.
\newblock {\em Nature}, 467(7315):567--569, 2010.

\bibitem{mahan2013many}
Gerald~D Mahan.
\newblock {\em Many-particle physics}.
\newblock Springer Science \& Business Media, 2013.

\bibitem{Matsuda2unpublished}
Matsuda and et.al.
\newblock unpublished.

\bibitem{mayaffre2014evidence}
H~Mayaffre, S~Kr{\"a}mer, M~Horvati{\'c}, C~Berthier, K~Miyagawa, K~Kanoda, and
  VF~Mitrovi{\'c}.
\newblock Evidence of andreev bound states as a hallmark of the fflo phase in
  $\ensuremath{\kappa}\mathrm{\text{\ensuremath{-}}}(\mathrm{BEDT}\mathrm{\text{\ensuremath{-}}}\mathrm{TTF}{)}_{2}{{\mathrm{Cu}}}_{2}(\mathrm{NCS}{)}_{2}$.
\newblock {\em Nature Physics}, 10(12):928--932, 2014.

\bibitem{PhysRevX.8.041010}
B.~Michon, A.~Ataei, P.~Bourgeois-Hope, C.~Collignon, S.~Y. Li, S.~Badoux,
  A.~Gourgout, F.~Lalibert\'e, J.-S. Zhou, Nicolas Doiron-Leyraud, and Louis
  Taillefer.
\newblock Wiedemann-franz law and abrupt change in conductivity across the
  pseudogap critical point of a cuprate superconductor.
\newblock {\em Phys. Rev. X}, 8:041010, Oct 2018.

\bibitem{mitrano2018anomalous}
M~Mitrano, AA~Husain, S~Vig, A~Kogar, MS~Rak, SI~Rubeck, J~Schmalian, B~Uchoa,
  J~Schneeloch, R~Zhong, et~al.
\newblock Anomalous density fluctuations in a strange metal.
\newblock {\em Proceedings of the National Academy of Sciences},
  115(21):5392--5396, 2018.

\bibitem{PhysRevB.90.014503}
S.~J. Moon, Y.~S. Lee, A.~A. Schafgans, A.~V. Chubukov, S.~Kasahara,
  T.~Shibauchi, T.~Terashima, Y.~Matsuda, M.~A. Tanatar, R.~Prozorov,
  A.~Thaler, P.~C. Canfield, S.~L. Bud'ko, A.~S. Sefat, D.~Mandrus, K.~Segawa,
  Y.~Ando, and D.~N. Basov.
\newblock Infrared pseudogap in cuprate and pnictide high-temperature
  superconductors.
\newblock {\em Phys. Rev. B}, 90:014503, Jul 2014.

\bibitem{mott1937discussion}
NF~Mott and R~Peierls.
\newblock Discussion of the paper by de {B}oer and {V}erwey.
\newblock {\em Proceedings of the Physical Society}, 49(4S):72, 1937.

\bibitem{nambu1960quasi}
Yoichiro Nambu.
\newblock Quasi-particles and gauge invariance in the theory of
  superconductivity.
\newblock {\em Physical Review}, 117(3):648, 1960.

\bibitem{nie2014quenched}
Laimei Nie, Gilles Tarjus, and Steven~Allan Kivelson.
\newblock Quenched disorder and vestigial nematicity in the pseudogap regime of
  the cuprates.
\newblock {\em Proceedings of the National Academy of Sciences},
  111(22):7980--7985, 2014.

\bibitem{Normanunpublished}
M.~Norman and J.C. Davis.
\newblock unpublished.

\bibitem{PhysRevB.57.R11089}
M.~R. Norman and H.~Ding.
\newblock Collective modes and the superconducting-state spectral function of
  ${{\mathrm{Bi}}}_{2}{{\mathrm{Sr}}}_{2}{{\mathrm{CaCu}}}_{2}{{\mathrm{O}}}_{8}$.
\newblock {\em Phys. Rev. B}, 57:R11089--R11092, May 1998.

\bibitem{Note1}
In momentum space, there are two amplitudes $A^x_a$ and $A^y_a$ at momentum
  $\protect \mathbf {Q_a/2}$ which correspond to density waves in $x$ bond and
  $y$ bond. Here $a$ denotes $x$ or $y$: $\protect \mathbf {Q_x/2}=(\protect
  \frac {2\pi }{8},0)$ and $\protect \mathbf {Q_y/2}=(\protect \frac {2\pi
  }{8},0)$. The definition currently used by the community is to define
  $A^x_a\pm A^y_a$ as the s/d wave component. However, under $C_4$ rotation,
  $A^x_x$ transforms to $A^y_y$. Therefore the current definiton of s/d wave
  form factor is not related to symmetry and generallly they should be mixed.
  An alternative definition of s vs d wave component is $A^x_x \pm A^y_y$,
  which is related to the C4 rotation around a particular reference point.
  However, this definition may not be very useful because if we shift the
  reference point by half of the period in one direction, what we would define
  as d wave would becaome s wave.

\bibitem{PhysRevB.73.094503}
Vadim Oganesyan, David~A. Huse, and S.~L. Sondhi.
\newblock Theory of diamagnetic response of the vortex liquid phase of
  two-dimensional superconductors.
\newblock {\em Phys. Rev. B}, 73:094503, Mar 2006.

\bibitem{PepinPhysRevB.90.195207}
C.~P\'epin, V.~S. de~Carvalho, T.~Kloss, and X.~Montiel.
\newblock Pseudogap, charge order, and pairing density wave at the hot spots in
  cuprate superconductors.
\newblock {\em Phys. Rev. B}, 90:195207, Nov 2014.

\bibitem{peskin1995introduction}
M.E. Peskin and D.V. Schroeder.
\newblock {\em An Introduction to Quantum Field Theory}.
\newblock Advanced book classics. Addison-Wesley Publishing Company, 1995.

\bibitem{PhysRevLett.99.117004}
Daniel Podolsky, Srinivas Raghu, and Ashvin Vishwanath.
\newblock Nernst effect and diamagnetism in phase fluctuating superconductors.
\newblock {\em Phys. Rev. Lett.}, 99:117004, Sep 2007.

\bibitem{proust2019remarkable}
Cyril Proust and Louis Taillefer.
\newblock The remarkable underlying ground states of cuprate superconductors.
\newblock {\em Annual Review of Condensed Matter Physics}, 10:409--429, 2019.

\bibitem{proust2016fermi}
Cyril Proust, Baptiste Vignolle, Julien Levallois, S~Adachi, and Nigel~E
  Hussey.
\newblock Fermi liquid behavior of the in-plane resistivity in the pseudogap
  state of $\text{{YBa}}_2\text{{Cu}}_4\text{{O}}_8$.
\newblock {\em Proceedings of the National Academy of Sciences},
  113(48):13654--13659, 2016.

\bibitem{puchkov1996pseudogap}
AV~Puchkov, DN~Basov, and T~Timusk.
\newblock The pseudogap state in high-superconductors: an infrared study.
\newblock {\em Journal of Physics: Condensed Matter}, 8(48):10049, 1996.

\bibitem{ramshaw2011angle}
BJ~Ramshaw, Baptiste Vignolle, James Day, Ruixing Liang, WN~Hardy, Cyril
  Proust, and DA~Bonn.
\newblock Angle dependence of quantum oscillations in
  $\text{{YBa}}_2\text{{Cu}}_3\text{{O}}_{6.59}$ shows free-spin behaviour of
  quasiparticles.
\newblock {\em Nature Physics}, 7(3):234, 2011.

\bibitem{robinson2019anomalies}
Neil~J Robinson, Peter~D Johnson, T~Maurice Rice, and Alexei~M Tsvelik.
\newblock Anomalies in the pseudogap phase of the cuprates: Competing ground
  states and the role of umklapp scattering.
\newblock {\em Reports on Progress in Physics}, 82(12):126501, 2019.

\bibitem{sachdev_2011}
Subir Sachdev.
\newblock {\em Quantum Phase Transitions}.
\newblock Cambridge University Press, 2 edition, 2011.

\bibitem{sato2017thermodynamic}
Y~Sato, S~Kasahara, H~Murayama, Y~Kasahara, E-G Moon, T~Nishizaki, T~Loew,
  J~Porras, B~Keimer, T~Shibauchi, et~al.
\newblock Thermodynamic evidence for a nematic phase transition at the onset of
  the pseudogap in $\text{{YBa}}_2\text{{Cu}}_3\text{{O}}_y$.
\newblock {\em Nature Physics}, 13(11):1074, 2017.

\bibitem{schrieffer1964theory}
JR~Schrieffer.
\newblock {\em Theory of superconductivity}.
\newblock WA Benjamin, Inc., New York, 1964.

\bibitem{sebastian2011quantum}
Suchitra~E Sebastian, Neil Harrison, and Gilbert~G Lonzarich.
\newblock Quantum oscillations in the high-$\text{{T}}_\text{c}$ cuprates.
\newblock {\em Philosophical Transactions of the Royal Society A: Mathematical,
  Physical and Engineering Sciences}, 369(1941):1687--1711, 2011.

\bibitem{sebastian2012towards}
Suchitra~E Sebastian, Neil Harrison, and Gilbert~G Lonzarich.
\newblock Towards resolution of the fermi surface in underdoped
  high-$\text{{T}}_\text{c}$ superconductors.
\newblock {\em Reports on Progress in Physics}, 75(10):102501, 2012.

\bibitem{PhysRevB.79.245116}
T.~Senthil and Patrick~A. Lee.
\newblock Synthesis of the phenomenology of the underdoped cuprates.
\newblock {\em Phys. Rev. B}, 79:245116, Jun 2009.

\bibitem{shekhter2013bounding}
Arkady Shekhter, BJ~Ramshaw, Ruixing Liang, WN~Hardy, DA~Bonn, Fedor~F
  Balakirev, Ross~D McDonald, Jon~B Betts, Scott~C Riggs, and Albert Migliori.
\newblock Bounding the pseudogap with a line of phase transitions in
  $\text{{YBa}}_2\text{{Cu}}_3\text{{O}}_{6+\delta}$.
\newblock {\em Nature}, 498(7452):75, 2013.

\bibitem{PhysRevB.96.134510}
W.~Tabis, B.~Yu, I.~Bialo, M.~Bluschke, T.~Kolodziej, A.~Kozlowski,
  E.~Blackburn, K.~Sen, E.~M. Forgan, M.~v. Zimmermann, Y.~Tang, E.~Weschke,
  B.~Vignolle, M.~Hepting, H.~Gretarsson, R.~Sutarto, F.~He, M.~Le~Tacon,
  N.~Bari\ifmmode \check{s}\else \v{s}\fi{}i\ifmmode~\acute{c}\else \'{c}\fi{},
  G.~Yu, and M.~Greven.
\newblock Synchrotron x-ray scattering study of charge-density-wave order in
  ${{\mathrm{HgBa}}}_{2}{{\mathrm{CuO}}}_{4+\ensuremath{\delta}}$.
\newblock {\em Phys. Rev. B}, 96:134510, Oct 2017.

\bibitem{tajimaPRL862001c}
S.~Tajima, T.~Noda, H.~Eisaki, and S.~Uchida.
\newblock $\mathit{c}$-axis optical response in the static stripe ordered phase
  of the cuprates.
\newblock {\em Phys. Rev. Lett.}, 86:500--503, Jan 2001.

\bibitem{tallon2001doping}
Jeffery~L Tallon and JW~Loram.
\newblock The doping dependence of {T}*--what is the real
  high-$\text{{T}}_\text{c}$ phase diagram?
\newblock {\em Physica C: Superconductivity}, 349(1-2):53--68, 2001.

\bibitem{timusk1999pseudogap}
Tom Timusk and Bryan Statt.
\newblock The pseudogap in high-temperature superconductors: an experimental
  survey.
\newblock {\em Reports on Progress in Physics}, 62(1):61, 1999.

\bibitem{tranquada1995jm}
JM~Tranquada, BJ~Sternlieb, JD~Axe, Y~Nakamura, and S~Uchida.
\newblock Evidence for stripe correlations of spins and holes in copper oxide
  superconductors.
\newblock {\em nature}, 375(6532):561--563, 1995.

\bibitem{TsuiStormer}
D.~C. Tsui, H.~L. Stormer, and A.~C. Gossard.
\newblock Two-dimensional magnetotransport in the extreme quantum limit.
\newblock {\em Phys. Rev. Lett.}, 48:1559--1562, May 1982.

\bibitem{PhysRevB.95.201112}
A.~M. Tsvelik.
\newblock Ladder physics in the spin fermion model.
\newblock {\em Phys. Rev. B}, 95:201112, May 2017.

\bibitem{tu2019}
Wei-Lin Tu and Ting-Kuo Lee.
\newblock Evolution of pairing orders between pseudogap and superconducting
  phases of cuprate superconductors.
\newblock {\em Scientific Reports}, 9:1719, 2019.

\bibitem{PhysRevB.43.7942}
S.~Uchida, T.~Ido, H.~Takagi, T.~Arima, Y.~Tokura, and S.~Tajima.
\newblock Optical spectra of
  ${{\mathrm{La}}}_{2\mathrm{\ensuremath{-}}\mathit{x}}$${{\mathrm{Sr}}}_{\mathit{x}}$${{\mathrm{CuO}}}_{4}$:
  Effect of carrier doping on the electronic structure of the
  ${{\mathrm{CuO}}}_{2}$ plane.
\newblock {\em Phys. Rev. B}, 43:7942--7954, Apr 1991.

\bibitem{PhysRevLett.89.287001}
Iddo Ussishkin, S.~L. Sondhi, and David~A. Huse.
\newblock Gaussian superconducting fluctuations, thermal transport, and the
  nernst effect.
\newblock {\em Phys. Rev. Lett.}, 89:287001, Dec 2002.

\bibitem{varma2006theory}
CM~Varma.
\newblock Theory of the pseudogap state of the cuprates.
\newblock {\em Physical Review B}, 73(15):155113, 2006.

\bibitem{varma1987charge}
CM~Varma, S~Schmitt-Rink, and Elihu Abrahams.
\newblock Charge transfer excitations and superconductivity in “ionic”
  metals.
\newblock {\em Solid state communications}, 62(10):681--685, 1987.

\bibitem{wang2018}
Y.~{Wang}, S.~D. {Edkins}, M.~H. {Hamidian}, J.~C. {S{\'e}amus Davis},
  E.~{Fradkin}, and S.~A. {Kivelson}.
\newblock {Pair Density Waves in Superconducting Vortex Halos}.
\newblock {\em ArXiv e-prints}, February 2018.

\bibitem{PhysRevLett.95.247002}
Yayu Wang, Lu~Li, M.~J. Naughton, G.~D. Gu, S.~Uchida, and N.~P. Ong.
\newblock Field-enhanced diamagnetism in the pseudogap state of the cuprate
  ${{\mathrm{Bi}}}_{2}{{\mathrm{Sr}}}_{2}{{\mathrm{Ca}}}{{\mathrm{Cu}}}_{2}{{\mathrm{O}}}_{8+\ensuremath{\delta}}$
  superconductor in an intense magnetic field.
\newblock {\em Phys. Rev. Lett.}, 95:247002, Dec 2005.

\bibitem{PhysRevB.73.024510}
Yayu Wang, Lu~Li, and N.~P. Ong.
\newblock Nernst effect in high-$\text{{T}}_\text{c}$ superconductors.
\newblock {\em Phys. Rev. B}, 73:024510, Jan 2006.

\bibitem{PhysRevLett.88.257003}
Yayu Wang, N.~P. Ong, Z.~A. Xu, T.~Kakeshita, S.~Uchida, D.~A. Bonn, R.~Liang,
  and W.~N. Hardy.
\newblock High field phase diagram of cuprates derived from the nernst effect.
\newblock {\em Phys. Rev. Lett.}, 88:257003, Jun 2002.

\bibitem{WangPhysRevLett.114.197001}
Yuxuan Wang, Daniel~F. Agterberg, and Andrey Chubukov.
\newblock Coexistence of charge-density-wave and pair-density-wave orders in
  underdoped cuprates.
\newblock {\em Phys. Rev. Lett.}, 114:197001, May 2015.

\bibitem{JulienNature477191wu2011magnetic}
Tao Wu, Hadrien Mayaffre, Steffen Kr{\"a}mer, Mladen Horvati{\'c}, Claude
  Berthier, WN~Hardy, Ruixing Liang, DA~Bonn, and Marc-Henri Julien.
\newblock Magnetic-field-induced charge-stripe order in the high-temperature
  superconductor $\text{{YBa}}_2\text{{Cu}}_3\text{{O}}_y$.
\newblock {\em Nature}, 477(7363):191, 2011.

\bibitem{wu2015incipient}
Tao Wu, Hadrien Mayaffre, Steffen Kr{\"a}mer, Mladen Horvati{\'c}, Claude
  Berthier, WN~Hardy, Ruixing Liang, DA~Bonn, and Marc-Henri Julien.
\newblock Incipient charge order observed by nmr in the normal state of
  $\text{{YBa}}_2\text{{Cu}}_3\text{{O}}_y$.
\newblock {\em Nature communications}, 6:6438, 2015.

\bibitem{wu2013emergence}
Tao Wu, Hadrien Mayaffre, Steffen Kr{\"a}mer, Mladen Horvati{\'c}, Claude
  Berthier, Philip~L Kuhns, Arneil~P Reyes, Ruixing Liang, WN~Hardy, DA~Bonn,
  et~al.
\newblock Emergence of charge order from the vortex state of a high-temperature
  superconductor.
\newblock {\em Nature communications}, 4:2113, 2013.

\bibitem{yamada1998PRB57doping}
K~Yamada, CH~Lee, K~Kurahashi, J~Wada, S~Wakimoto, S~Ueki, H~Kimura, Y~Endoh,
  S~Hosoya, G~Shirane, et~al.
\newblock Doping dependence of the spatially modulated dynamical spin
  correlations and the superconducting-transition temperature in
  $\text{{La}}_{2- x}\text{{Sr}}_{x}\text{{CuO}}_4$.
\newblock {\em Physical Review B}, 57(10):6165, 1998.

\bibitem{PhysRevB.73.174501}
Kai-Yu Yang, T.~M. Rice, and Fu-Chun Zhang.
\newblock Phenomenological theory of the pseudogap state.
\newblock {\em Phys. Rev. B}, 73:174501, May 2006.

\bibitem{Yeh2}
N-C Yeh and AD~Beyer.
\newblock Unconventional low-energy excitations of cuprate superconductors.
\newblock {\em International Journal of Modern Physics B}, 23(22):4543--4577,
  2009.

\bibitem{yuPNAS126672016magnetic}
Fan Yu, Max Hirschberger, Toshinao Loew, Gang Li, Benjamin~J Lawson, Tomoya
  Asaba, JB~Kemper, Tian Liang, Juan Porras, Gregory~S Boebinger, et~al.
\newblock Magnetic phase diagram of underdoped
  $\text{{YBa}}_2\text{{Cu}}_3\text{{O}}_y$ inferred from torque magnetization
  and thermal conductivity.
\newblock {\em Proceedings of the National Academy of Sciences},
  113(45):12667--12672, 2016.

\bibitem{ZelliQtmOscPhysRevB.86.104507}
M.~Zelli, Catherine Kallin, and A.~John Berlinsky.
\newblock Quantum oscillations in a $\ensuremath{\pi}$-striped superconductor.
\newblock {\em Phys. Rev. B}, 86:104507, Sep 2012.

\bibitem{PhysRevB.37.3759}
F.~C. Zhang and T.~M. Rice.
\newblock Effective hamiltonian for the superconducting cu oxides.
\newblock {\em Phys. Rev. B}, 37:3759--3761, Mar 1988.

\bibitem{HsieNaturePhysicshzhao2017global}
L~Zhao, CA~Belvin, R~Liang, DA~Bonn, WN~Hardy, NP~Armitage, and D~Hsieh.
\newblock A global inversion-symmetry-broken phase inside the pseudogap region
  of $\text{{YBa}}_2\text{{Cu}}_3\text{{O}}_y$.
\newblock {\em Nature Physics}, 13(3):250, 2017.

\bibitem{zheng2017stripe}
Bo-Xiao Zheng, Chia-Min Chung, Philippe Corboz, Georg Ehlers, Ming-Pu Qin,
  Reinhard~M Noack, Hao Shi, Steven~R White, Shiwei Zhang, and Garnet Kin-Lic
  Chan.
\newblock Stripe order in the underdoped region of the two-dimensional hubbard
  model.
\newblock {\em Science}, 358(6367):1155--1160, 2017.

\bibitem{Julien2arXivzhou2017spin}
Rui Zhou, Michihiro Hirata, Tao Wu, Igor Vinograd, Hadrien Mayaffre, Steffen
  Kr{\"a}mer, Arneil~P Reyes, Philip~L Kuhns, Ruixing Liang, WN~Hardy, et~al.
\newblock Spin susceptibility of charge-ordered
  $\text{{YBa}}_2\text{{Cu}}_3\text{{O}}_y$ across the upper critical field.
\newblock {\em Proceedings of the National Academy of Sciences},
  114(50):13148--13153, 2017.

\end{thebibliography}
\bibliographystyle{plain}
\end{singlespace}

\end{document}